\documentclass[a4paper,11pt]{article}
\pdfoutput=1

\usepackage{jheppub,fancyhdr,graphicx,epstopdf,color,amsmath,cases,slashed,hyperref}
\usepackage{makecell}
\usepackage{caption, subcaption}  
\usepackage{dcolumn}   
\usepackage{amsmath, amssymb}        
\usepackage{color,latexsym, array,multirow,  verbatim, enumerate,cancel}
\usepackage{slashed}
\usepackage{bigcenter}
\usepackage[utf8x]{inputenc}
\usepackage{footnote}
\makesavenoteenv{tabular}
\makesavenoteenv{table}

\def\issue(#1,#2,#3){{\bf #1}, #2 (#3)}

\catcode`\@=11
\def\lsim{\mathrel{\mathpalette\@versim<}}
\def\gsim{\mathrel{\mathpalette\@versim>}}
\def\@versim#1#2{\vcenter{\offinterlineskip
\ialign{$\m@th#1\hfil##\hfil$\crcr#2\crcr\sim\crcr } }}
\catcode`\@=12

\catcode`@=12

\newcommand{\met}{$\slashed{E}_T$}

\newcommand{\newc}{\newcommand}
\newc{\wt}{\widetilde}
\newc{\ra}{\rightarrow}

\def\beq {\begin{equation}}
\def\eeq {\end{equation}}
\def\bi {\begin{itemize}}
\def\ei {\end{itemize}}
\def\bea {\begin{eqnarray}}
\def\eea {\end{eqnarray}}

\def \met{\slashed{E}_T}

\newcommand{\br}{\begin{eqnarray}}
\newcommand{\er}{\end{eqnarray}}
\newcommand{\be}{\begin{equation}}
\newcommand{\ee}{\end{equation}}

\newcommand{\ch}{\widetilde \chi^\pm}

\def \chonep {{\wt\chi_1^+}}

\def \ch2p {{\wt\chi_2^+}}
\def \chonem {{\wt\chi_1^-}}
\def \ch2m {{\wt\chi_2^-}}

\def \chonepm{{\wt\chi_1}^{\pm}}

\def \chonemp{{\wt\chi_1}^{\mp}}

\newc{\dmchi}{\Delta m_{\wt\chi}}

\def \chtwopm{{\wt\chi_2}^{\pm}}
\def \chtwomp{{\wt\chi_2}^{\mp}}

\def \chkpm{{\wt\chi_k}^{\pm}}
\def \chkmp{{\wt\chi_k}^{\mp}}

\def \lspi{\wt\chi_i^0}

\def \lspj{\wt\chi_j^0}

\def \lspone{\wt\chi_1^0}

\def \lsptwo{\wt\chi_2^0}

\def \lspthree{\wt\chi_3^0}

\def \lspfour{\wt\chi_4^0}

\def\issue(#1,#2,#3){{\bf #1}, #2 (#3)}

\hypersetup{
   colorlinks=true,       
   linkcolor=blue,        
   citecolor=blue,         
   filecolor=magenta,      
   }

\title{Searching for heavy Higgs in supersymmetric final states at the LHC}
  
\author[a]{Amit Adhikary,}
\author[a]{Biplob Bhattacherjee,}
\author[a]{Rohini M. Godbole,}
\author[a]{Najimuddin Khan,}
\author[b]{Suchita Kulkarni}
\affiliation[a]{Centre for High Energy Physics, Indian Institute of Science, Bengaluru - 560012, India}
\affiliation[b]{Institut f{\"u}r Hochenergiephysik, {\"O}sterreichische Akademie der Wissenschaften,\\ Nikolsdorfer Gasse 18, 1050 Wien, Austria}
\emailAdd{amitadhikary@iisc.ac.in}
\emailAdd{biplob@iisc.ac.in}
\emailAdd{rohini@iisc.ac.in} 
\emailAdd{najimuddink@iisc.ac.in}
\emailAdd{suchita.kulkarni@oeaw.ac.at}
\abstract
{
In this work, we analyse and demonstrate possible strategies to explore extended Higgs sector of the Minimal Supersymmetric Standard Model (MSSM). In particular we concentrate on heavy Higgs decays to electroweakinos. We analyse the Higgs to electroweakino  decays in the allowed MSSM parameter space after taking into account 13 TeV LHC searches for supersymmetric particles and phenomenological constraints such as flavour physics, Higgs measurements and dark matter constraints. We explore some novel aspects of these Higgs decays. The final states resulting from Higgs to electroweakino decays will have backgrounds arising from the Standard Model as well as direct electroweakino production at the LHC. We demonstrate explicit kinematical differences between Higgs to electroweakino decays and associated backgrounds. Furthermore, we demonstrate for a few specific example points, optimised analysis search strategies at the high luminosity LHC (HL-LHC) run. Finally, we comment on possible search strategies for heavy Higgs decays to exotic final states, where the lightest chargino is long lived and leads to a disappearing track at the LHC.} 

\begin{document}

\maketitle

\section{Introduction}
\label{intro}

In the search for extensions of the Standard Model, the pursuit of extended Higgs sector remains an important avenue to determine whether the Standard Model (SM) Higgs is the only elementary scalar or is a part of family. Such heavy Higgs is being searched for at the LHC via its decays to the SM final states. The Minimal Supersymmetric Standard Model (MSSM) is an example where such an extended Higgs sector necessarily arises and contains two Higgs doublets~\cite{Dreesbook, Baerbook, Martin:1997ns, Djouadi:2005gj}. The MSSM Higgs sector at tree level can be described completely by only two parameters, \textit{viz.} the ratio of the vacuum expectation value (vev) of the two Higgs doublets, $\tan\beta$ and the pseudoscalar Higgs mass, $M_A$. There are five Higgses in this model as compared to only the one in the SM. These are two neutral scalar Higgses (h,H), one neutral pseudoscalar Higgs (A) and two charged Higgses ($H^\pm$). The neutral heavy Higgs boson can be produced at the LHC via the gluon-gluon fusion (ggH) or the bottom quark annihilation (bbH). Production of heavy Higgs boson in association with a vector boson and via the vector boson fusion channels are suppressed due to alignment limit implied by observed properties of the SM Higgs. The relative strength of the ggH and bbH production modes is determined by $\tan\beta$. A similar consideration also determines the dominant decay mode of the heavy Higgs. Note that $WW/ZZ$ final states are suppressed due to alignment limit. At large $\tan\beta$, heavy Higgs coupling to down-type quark and leptons become important. Therefore, in this case, the $b\bar{b}$ fusion production rate ($b\bar{b} \to H/A$) dominates over the gluon fusion mode and the branching ratio for $H/A \to \tau^+ \tau^-$ becomes large.  Thus the channel, $b\bar{b} \to H/A \to \tau^+ \tau^-$ probes low $M_A$ and high $\tan\beta$ parameter space~\cite{Aaboud:2017sjh}. These probes are further complemented with $gg \to H/A \to hh$~\cite{CMS-PAS-HIG-17-008, CMS-PAS-HIG-17-009, Sirunyan:2017djm} and $gg \to H/A \to t\bar{t}$~\cite{Aaboud:2017hnm, Aaboud:2018mjh} searches in the low $\tan\beta$ parameter space. 

Despite negative results from these searches, a large part of heavy Higgs parameter space remains allowed, particularly in the intermediate $\tan\beta$ regime. This intermediate $\tan\beta$ regime is specially difficult to probe via the Standard Model (SM) final states as in this case the heavy Higgs branching ratio to SM final states is overtaken by that to supersymmetric final states, if kinematically allowed. Among supersymmetric particles, the most interesting sector is the electreoweakino sector of the MSSM. Due to comparatively weaker LHC limits, the heavy Higgs to decays into these final states are still possible. The MSSM electroweakino sector consists of four neutralinos and two charginos. The neutralinos ($\lspone, \lsptwo, \lspthree, \lspfour$) are mixtures of the gauginos \textit{i.e.} bino ($\tilde{B^0}$), wino ($\tilde{W^0}$) and higgsinos \textit{i.e.} $\tilde{H_u^0}, \tilde{H_d^0}$. Similarly the charged components of the gaugino \textit{i.e.} $\tilde{W^\pm}$ and higgsino \textit{i.e.} $\tilde{H_u^+}, \tilde{H_d^-}$ mix to form 2 chargino mass eigenstates ($\chonepm, \chtwopm$). We refer to all of them collectively as electroweakinos. The heavy Higgs can therefore decay into any combination of these four neutralinos or two charginos depending on phase space and couplings. This gives rise to multiple heavy Higgs decay modes. We collectively label such decays of heavy Higgs to supersymmetric (susy) final state as Higgs to susy decays throughout this work.

Complementary to the searches in the decays of heavy Higgs (see for example ref.~\cite{Adhikary:2018ise}), multiple LHC searches for direct electroweakino production exist. These target electroweakino production via SM mediators i.e. $Z, W, \gamma$ and decays into SM final states in association with missing energy ($\met$). Such production modes however require off-shell SM mediators as the collider searches have constrained the electroweakino masses above 100 GeV over a large region of parameter space. Production of electroweakinos via heavy Higgs decays on the other hand can target on-shell heavy Higgs as a mediator, yielding distinct kinematics in the final states. This presents another opportunity to search for extended Higgs sector beyond the SM final states discussed before. To exploit the kinematic features and suggest further search strategies for heavy Higgs sector is the main aim of our work. 

There have been several studies on the phenomenological aspects of heavy Higgs decays to electroweakinos~\cite{Bisset:2007mk, Bisset:2007mi, Arhrib:2011rp, Gunion:1987ki, GUNION1988445, Djouadi:1996mj, Belanger:2000tg, Bisset:2000ud, Choi:2002zp, Charlot_2006, Li:2013nma,Belanger:2015vwa, Ananthanarayan:2015fwa, Djouadi:2015jea, Barman:2016kgt, Gori:2018pmk, Baum:2019uzg, Profumo:2017ntc, Bahl:2018zmf, Bahl:2019ago}. Many of these analyses demonstrated the importance of mono-X (X = $j, W^{\pm}, h, Z, \gamma$) final states while exploring heavy Higgs to susy decays, and demonstrated the LHC potential to do so. In particular, a recent study~\cite{Gori:2018pmk}, demonstrates the reach of HL-LHC for heavy Higgs decays to susy particles in dilepton plus missing energy final state within the MSSM. This study uses the so called clustered transverse mass $m_{CT}$ variable for discrimination between the Higgs to susy signal and the SM backgrounds. It considers only ggH production mode and finally, it does not include direct production of susy backgrounds which also leads to mono-X final states. Going beyond these previous studies, in this work we add the backgrounds arising from direct susy production, consider ggH and bbH Higgs production modes separately and demonstrate the HL-LHC potential to probe Higgs to susy decays in multiple final states.

We employ a strategy similar to the one considered by a recent CMS search~\cite{Sirunyan:2017qfc}. The principle difference between our studies and that of the search is that we target the ggH and bbH production modes separately. We furthermore study the effects of additional b-jets in signal distributions. We work in the framework of specific susy models, which are not considered in the CMS analysis. It will none-the-less be interesting to recast the CMS search to understand the reach for models considered in this work. This is beyond the scope of the work and we leave it for future studies. 

With respect to signal over background optimisation, a particularly interesting situation arises while analysing Higgs decays to chargino. In general within the MSSM parameter space, the lightest chargino can be long lived~\cite{Giudice:1998xp, Randall:1998uk}, particularly if it is wino-like. Should the heavy Higgs branching ratio to chargino be large, it can lead to heavy stable charged particles or disappearing tracks at the LHC, which have very little background from other sources in detector. In the final part of our work, we elaborate such possible decay modes and suggest a few strategies for searches. 

The plan of the paper is as follows: we investigate a few benchmark scenarios and discuss Higgs to susy cross sections in section~\ref{sec:BS}. In section~\ref{sec:setup}, we discuss the numerical setup of our 19 dimensional MSSM parameter space scan and demonstrate the cross sections for mono-X final states, we also explore salient kinematical differences between signal and background distributions. From here onwards we specifically look at the impact of resonance mediated susy production on the event kinematics, we explicitly demonstrate the impact of presence of a resonance. In section~\ref{sec:collider} we propose for a few benchmark points, optimised set of cuts leading to several different significances. We furthermore present a benchmark study of LLP in section~\ref{sec:llp}. Finally in section~\ref{sec:conclusion}, we conclude. 

\section{Benchmark scenarios and their features}
\label{sec:BS}
The relative hierarchy of the higgsino-gaugino mass parameters affect the gaugino-higgsino content of the electroweakino mass eigenstates.  The heavy higgs decays to electroweakinos, if kinematically allowed, are enhanced if both the gaugino and higgsino content are sizable, as this maxmimises the couplings of the heavy Higgs  with the electroweakinos. The same hierarchy also affects the direct production of electroweakinos due to its effects on their couplings to the SM particles as well as their masses. We comment on the relative importance of the two modes, one where electroweakinos are produced via heavy Higgs decays and the other direct electroweakino production. To illustrate possible Higgs to susy decay modes and resulting final state at the LHC, in this section we consider three different gaugino - higgsino mass hierarchies. These hierarchies are responsible for generating either bino - higgsino or  wino - higgssino-like light electroweakinos. We also compare and contrast this with direct production of electroweakinos via SM mediators.

To illustrate this case by case, we select a few benchmark scenario in the MSSM parameter space and discuss some salient features of the Higgs and electroweakino sector. No experimental constraints are applied at this point for benchmark choices. We have however taken care to keep the lightest Higgs boson mass within $122-128$ GeV~\cite{Allanach:2004rh}. An investigation of allowed parameter space by including all experimental constraints will be discussed in section~\ref{sec:setup}. Here, we use {\tt Suspect2}~\cite{Djouadi:2002ze} to generate the mass spectrum of susy particles. The MSSM parameters are chosen as follows,

\begin{equation} 
\begin{gathered}
M_A~=~1~{\rm TeV},~4~<tan\beta~<~20,~M_3~=~5~{\rm TeV},~A_t~=~-5~{\rm TeV},\\
A_{e,\mu,\tau,u,d,c,s,b}~=~0,~M_{\tilde{e}_L,\tilde{\mu}_L,\tilde{\tau}_L,\tilde{e}_R,\tilde{\mu}_R,\tilde{\tau}_R}~=~5~{\rm TeV},\\
M_{\tilde{Q}_{1_L},\tilde{Q}_{2_L},\tilde{Q}_{3_L},}~=~5~{\rm TeV},~M_{\tilde{u}_R,\tilde{d}_R,\tilde{c}_R,\tilde{s}_R,\tilde{t}_R,\tilde{b}_R,}~=~5~{\rm TeV}.
\end{gathered}
\end{equation} 
Keeping these parameters fixed, we change the gaugino and higgsino mass parameters which results in different possible scenarios, as we discuss below. We would like to mention here that the large values of $A_t$ parameter may give rise to charge and colour breaking minima (see for example ref.~\cite{Chowdhury:2013dka}). Of course this is taken care of in the spectrum generator {\tt Suspect2}.

\subsection*{Case-1} In this scenario, we fix the wino mass parameter $M_2$ at 1500 GeV and vary the higgsino and bino masses. This results in a mixed bino - higgsino like scenario. 
\subsubsection*{Case-1a :}
\label{case1a}
\begin{figure}
\centering
\includegraphics[scale=0.37]{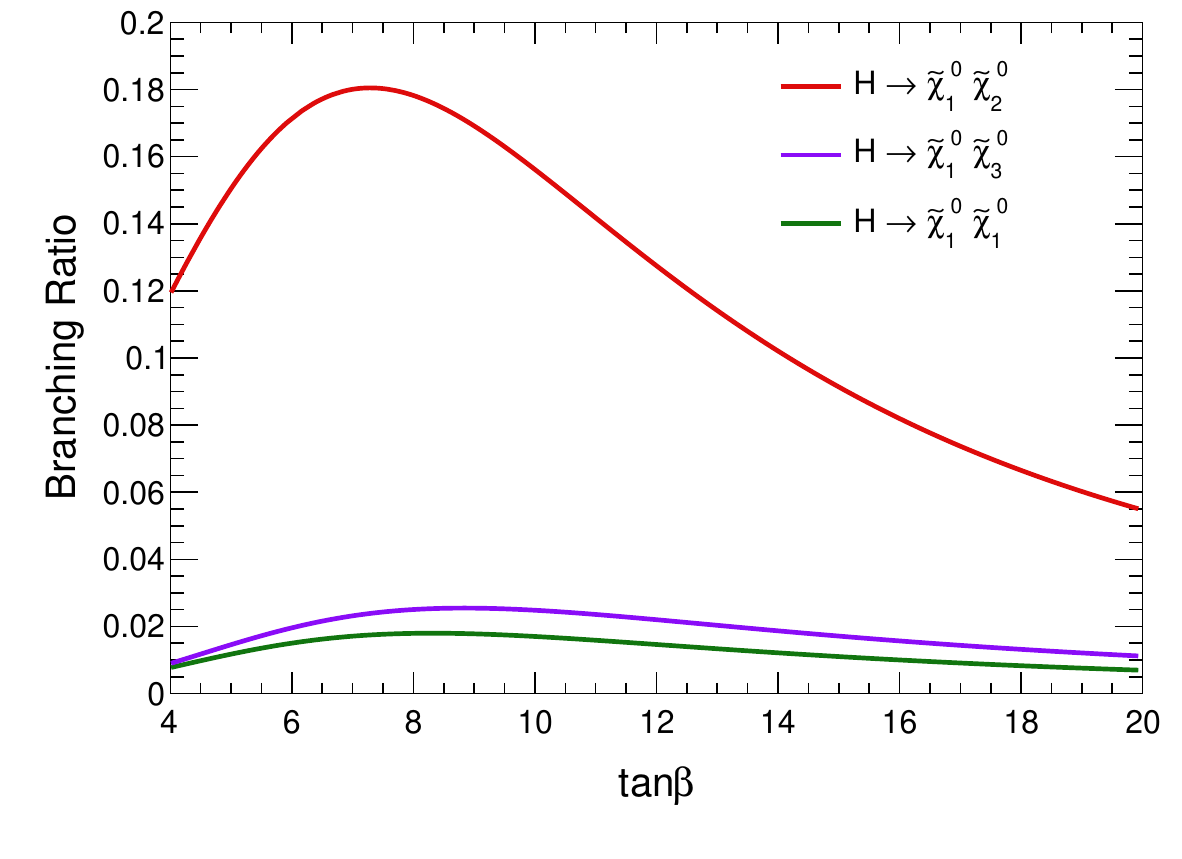}
\includegraphics[scale=0.37]{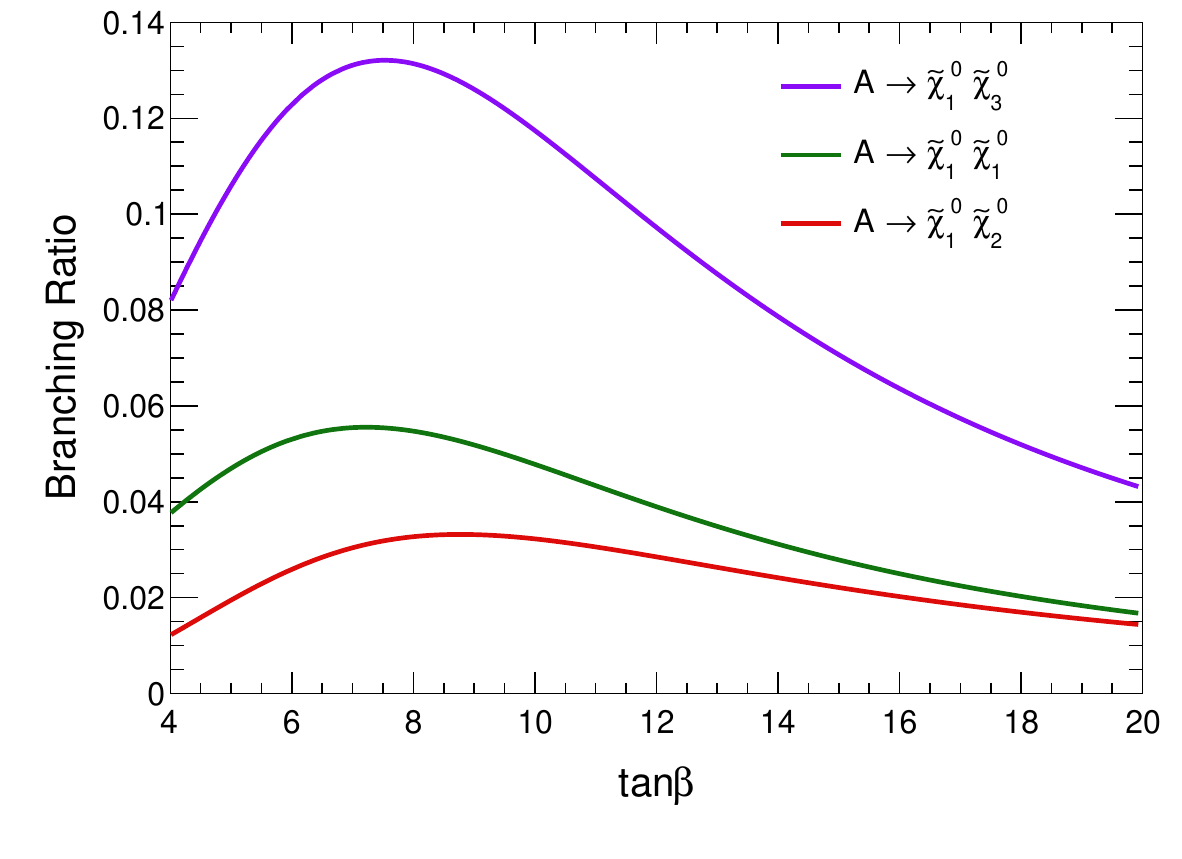}\\
\includegraphics[scale=0.37]{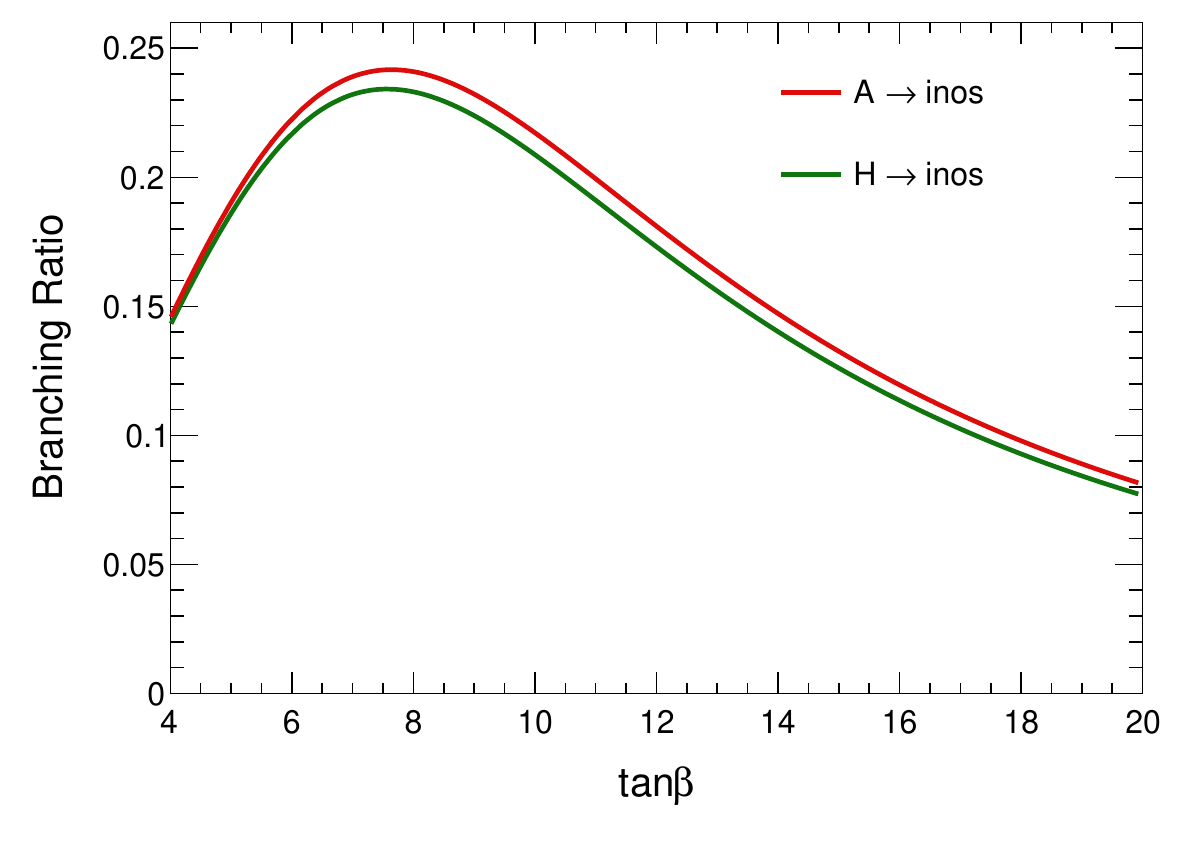}
\includegraphics[scale=0.37]{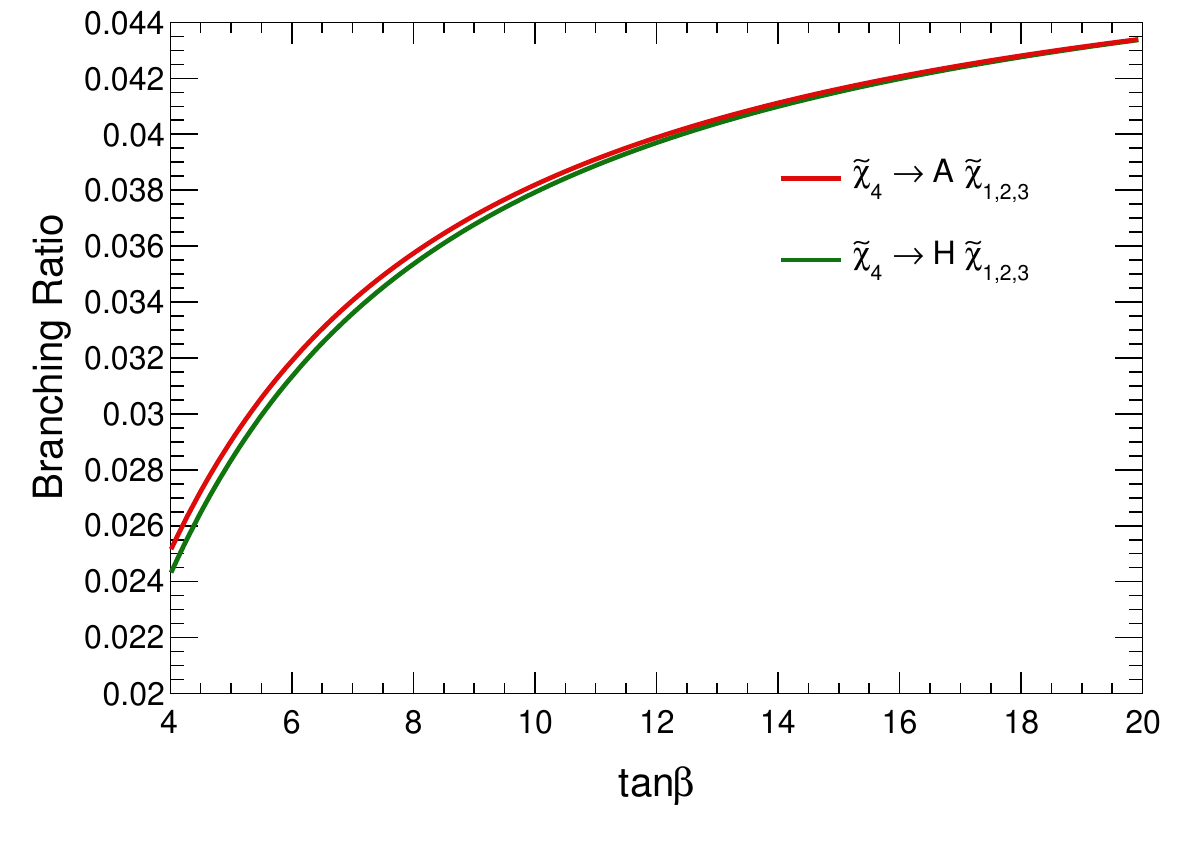}
\caption{\it Dominant branching ratios for heavy CP-even and CP-odd Higgs decays to individual electroweakino modes (top panel), the total branching ratio for two heavy Higgs (bottom left), branching ratio of heaviest neutralino decays to heavy Higgs (bottom right) for the case-1a. The wino, bino, higgsino mass parameters $M_2, M_1, \mu$ are fixed at 1500, 450 and 370 GeV and $\tan\beta$ is varied.}
\label{BS:1a}
\end{figure}

The higgsino and bino mass parameters are chosen as $\mu~=~450$ GeV and $M_1~=~370$ GeV. Therefore, the LSP ($\lspone$) becomes bino-higgsino mixture (bino fraction 89.72~\%, higgsino fraction 10.26~\%), the 2nd lightest neutralino ($\lsptwo$) is higgsino-like (higgsino fraction 99.83~\%) and the $\lspthree$ is mixed state of bino and higgsino (bino fraction 10.16~\% and  higgsino fraction 89.56~\%). In Fig \ref{BS:1a} we plot the resulting branching ratios as a function of $\tan\beta$. For phase space reasons, the heavy Higgs decays to susy particle always involves one $\lspone$ while the other can be $\lsptwo/\lspthree$. The $H\to\lspone\lsptwo$ and $A\to\lspone\lspthree$ branching fractions can be as large as $\sim 19\%$ and $\sim 13\%$ respectively, depending on the value of $\tan\beta$. Owing to the large bino-fraction, the branching ratio for the $\lspone$ pair production can be at most $\sim 2\%$ ($5\%$) from H (A) decay.  

Direct production of such neutral electroweakinos is however suppressed as neutralino coupling to Z requires a purely higgsino like nature, the only available channel here being neutralino production via SM Higgs. 

These electroweakinos further decay to $\lspone$ in association with a SM final state. The $\lsptwo$ and $\lspthree$ decay via a $Z$ boson with a $100\%$ branching ratio due to available phase space and coupling structure. This gives rise to mono-$Z+\met$ signature at the LHC. 

Along with the decays of heavy Higgs to electroweakino final states, it is also interesting to note that the heavy Higgs itself can be produced in cascade decays of heavier electroweakinos. In the benchmark scenario considered, the heaviest neutralino, $\lspfour$ being heavier than the heavy Higgses, (H/A) can decay via $\lspfour\to(H/A)~+~(\lspone/\widetilde{\chi}_{ 2,3}^{0})$ final state. This branching ratio can reach up to $\sim 4\%$ in this scenario. This process is important because if this branching is significant then this can contribute to the production of heavy Higgs.  At the same time, current limits on heavy Higgs mass requires the progenator electroweakinos to be heavy thus limiting the reach of LHC for such processes. These processes might nonetheless be interesting at future 100 TeV colliders. A detailed investigation is however beyond the scope of this work and we leave it for future considerations.

\subsubsection*{Case-1b :}
\label{case1b}

\begin{figure}
\centering
\includegraphics[scale=0.37]{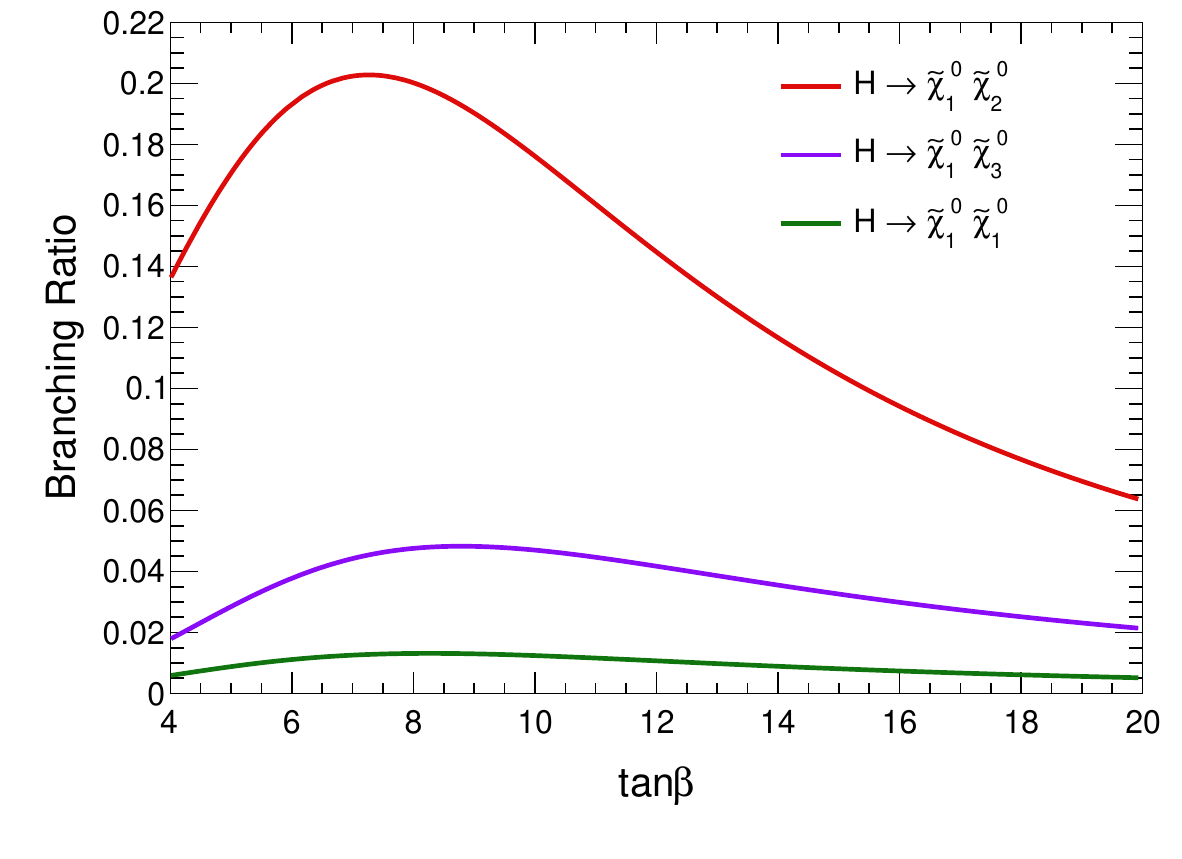}
\includegraphics[scale=0.37]{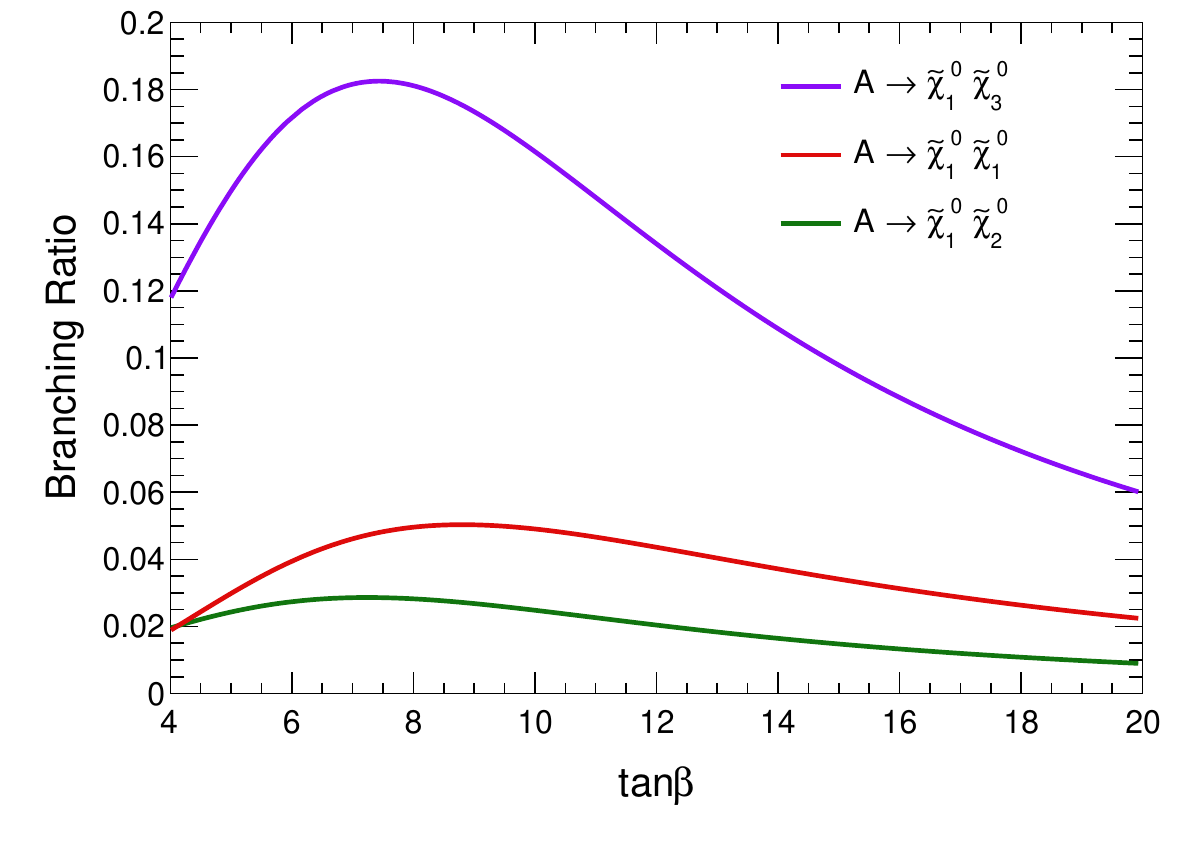}\\
\includegraphics[scale=0.37]{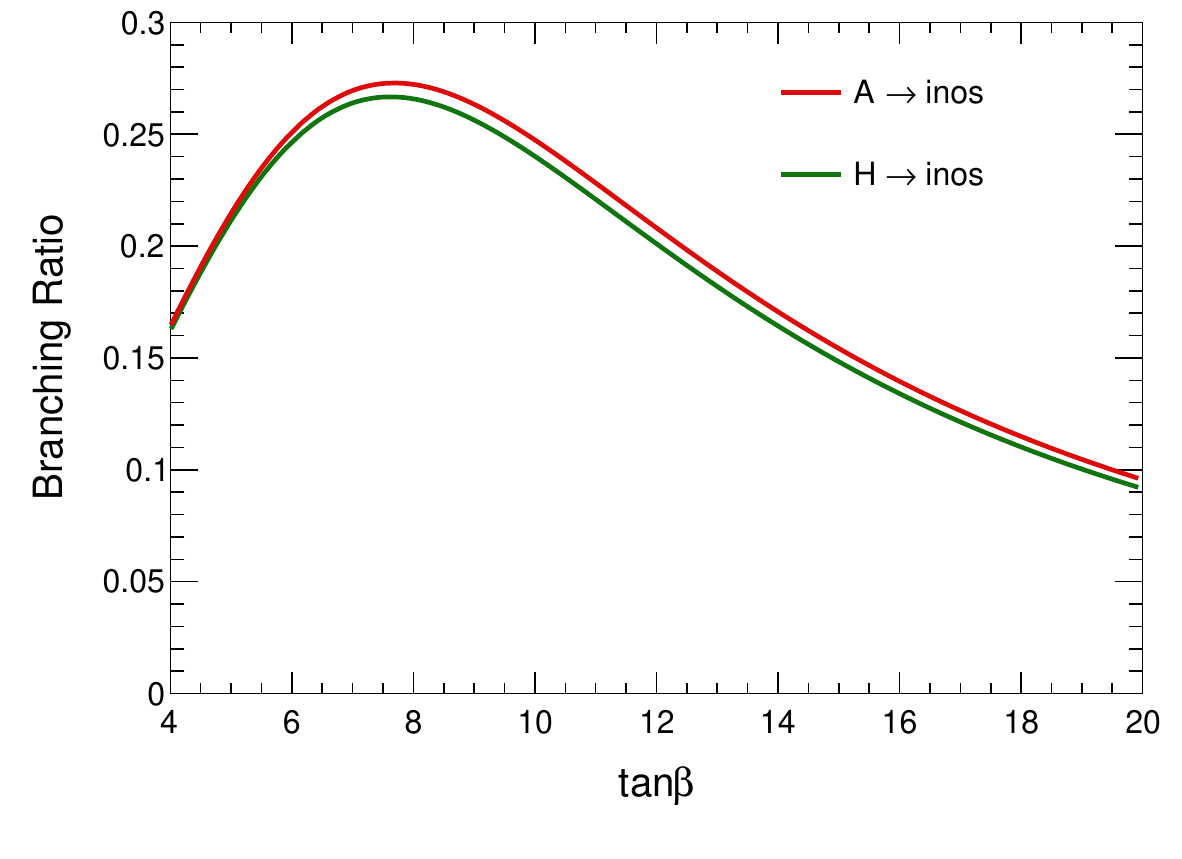}
\includegraphics[scale=0.37]{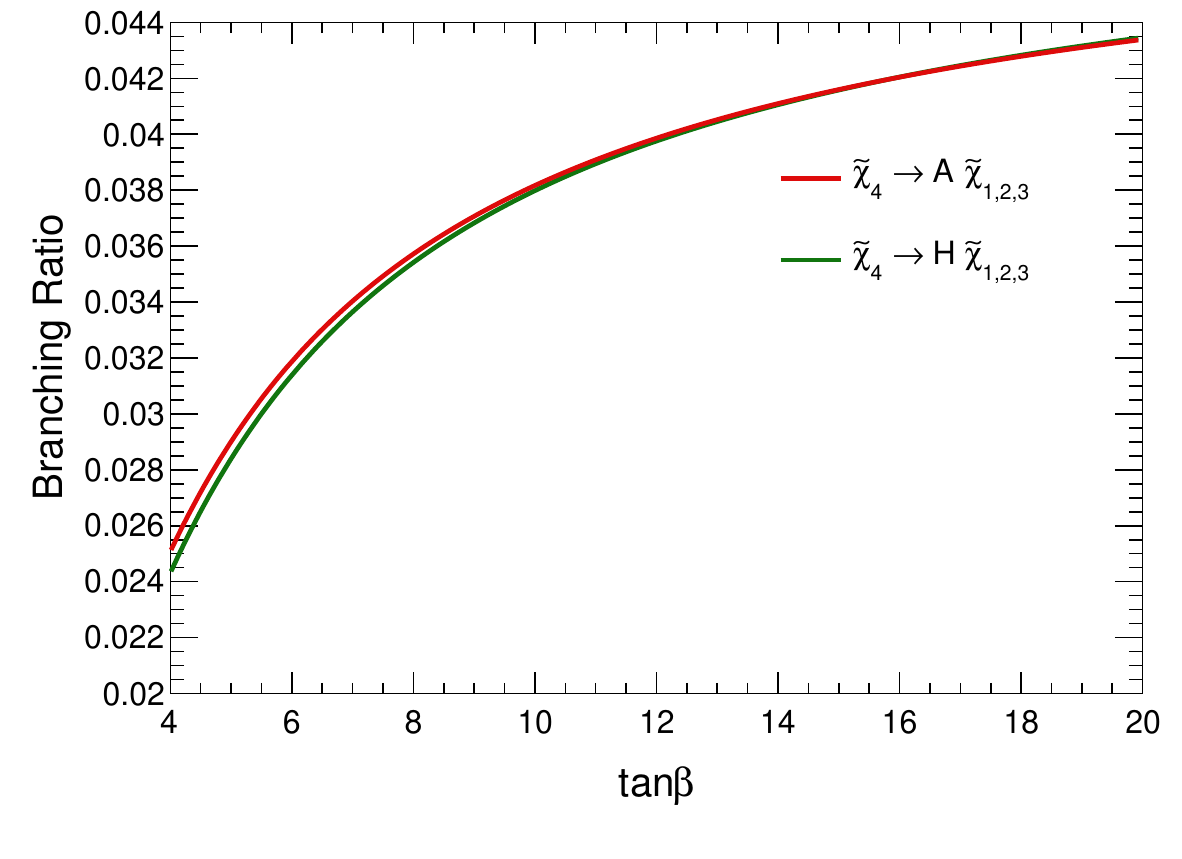}
\caption{\it Dominant branching ratios for heavy CP-even and CP-odd Higgs decays to individual electroweakino modes (top panel), the total branching ratio for two heavy Higgs (bottom left), branching ratio of heaviest neutralino decays to heavy Higgs (bottom right) for the case-1b. The wino, bino, higgsino mass parameters $M_2, M_1, \mu$ are fixed at 1500, 300 and 450 GeV and $\tan\beta$ is varied.}
\label{BS:1b}
\end{figure}

In this scenario, the higgsino mass parameter is fixed at $\mu~=~450$ GeV and the bino mass parameter is chosen to be $M_1~=~300$ GeV. Hence like in the previous case, gaugino and higgsino composition are similar except $\lspone$ becomes bino-like (bino fraction 95.80~\%). The main difference in this scenario as compared to the previous one is the difference between $M_1$ and $\mu$ mass parameter, which is more than the SM Higgs mass. We display the heavy Higgs to electroweakino branching ratios as a function of $\tan\beta$ in Fig.~\ref{BS:1b}. The heavier neutralinos \textit{viz.} $\lsptwo$ and $\lspthree$ can also decay to the LSP via SM Higgs boson. Since $\lsptwo$ is higgsino-like, it dominantly decays to $\lsptwo \to \lspone+Z$ (BR $\sim 98.17~\%$) which will give rise to the mono-$Z+\met$ final state, however since $\lspthree$ is admixture of bino and higgsino state it dominantly decays to $\lspthree \to \lspone+h$ (BR $\sim 93.98~\%$). This can lead to a mono-h $+\met$ signature at the collider. 

\subsubsection*{Case-1c :}
\label{case1c}

\begin{figure}
\centering
\includegraphics[scale=0.37]{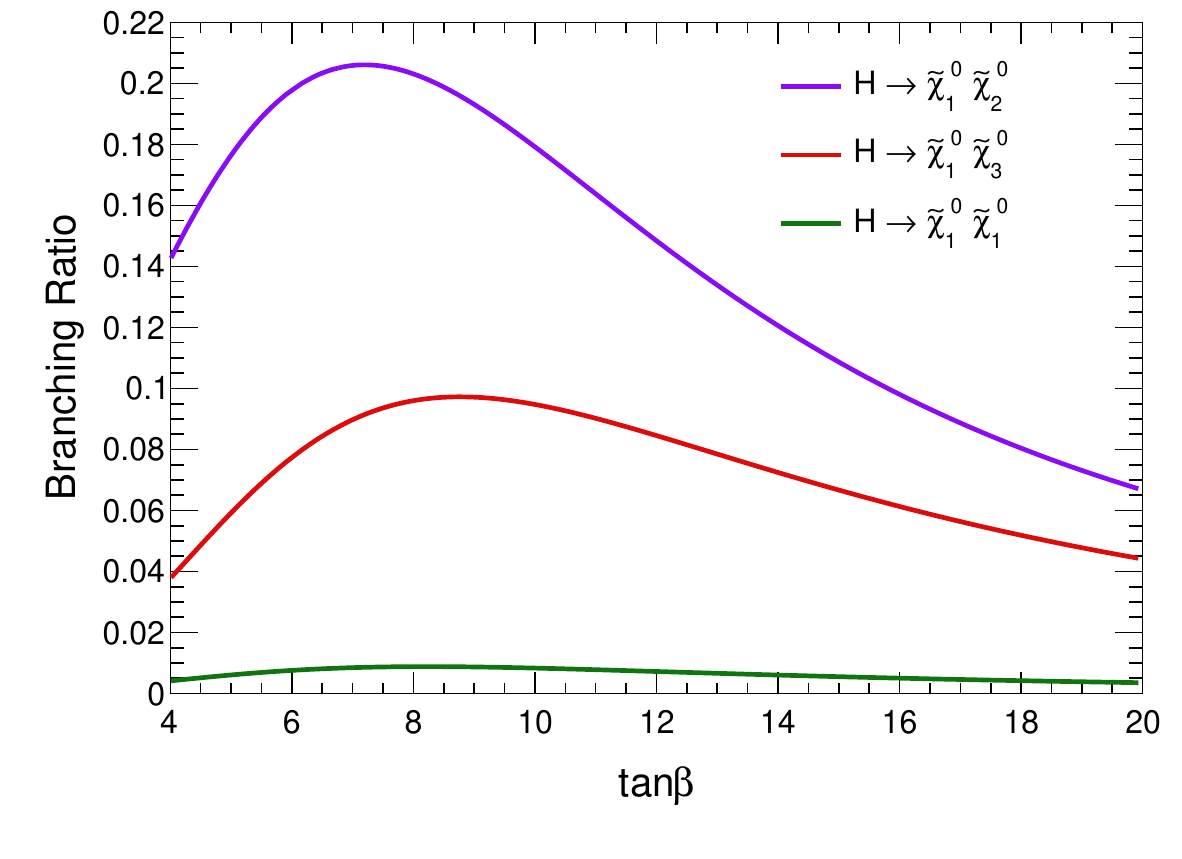}
\includegraphics[scale=0.37]{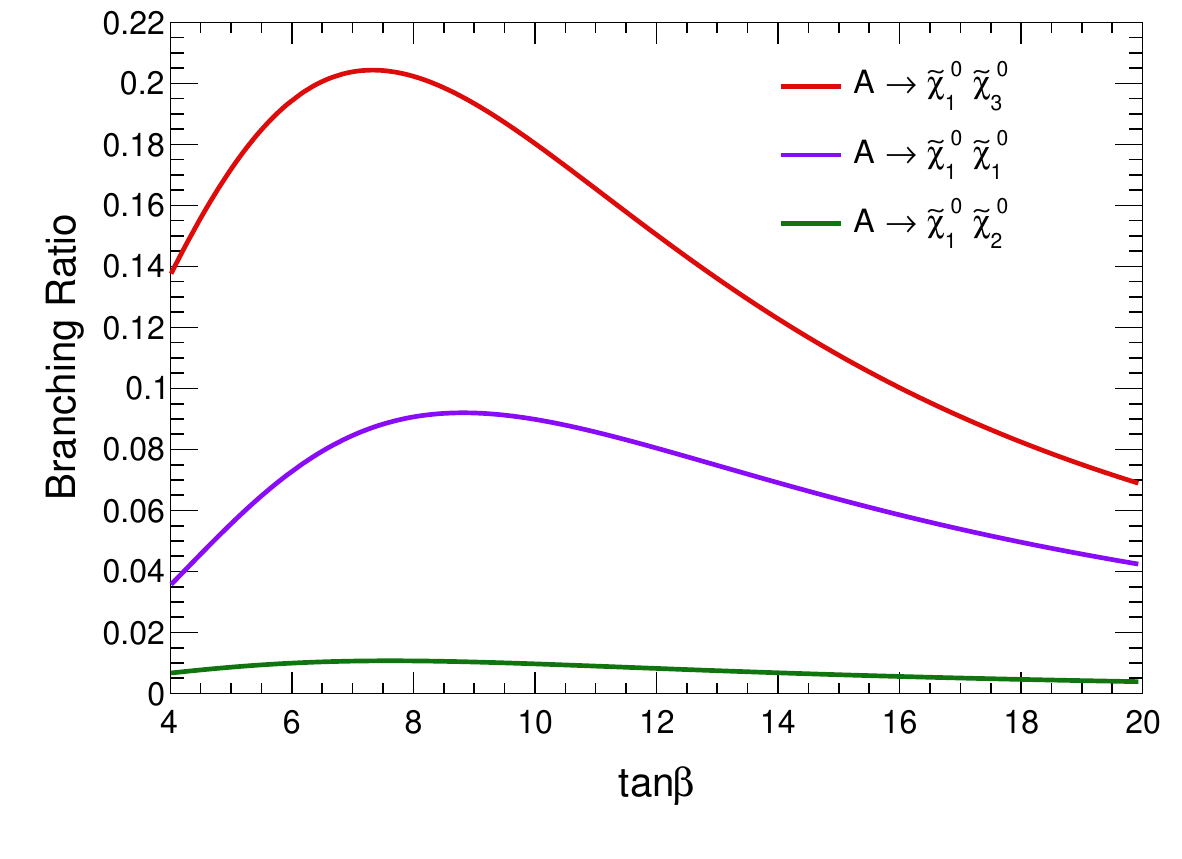}\\
\includegraphics[scale=0.37]{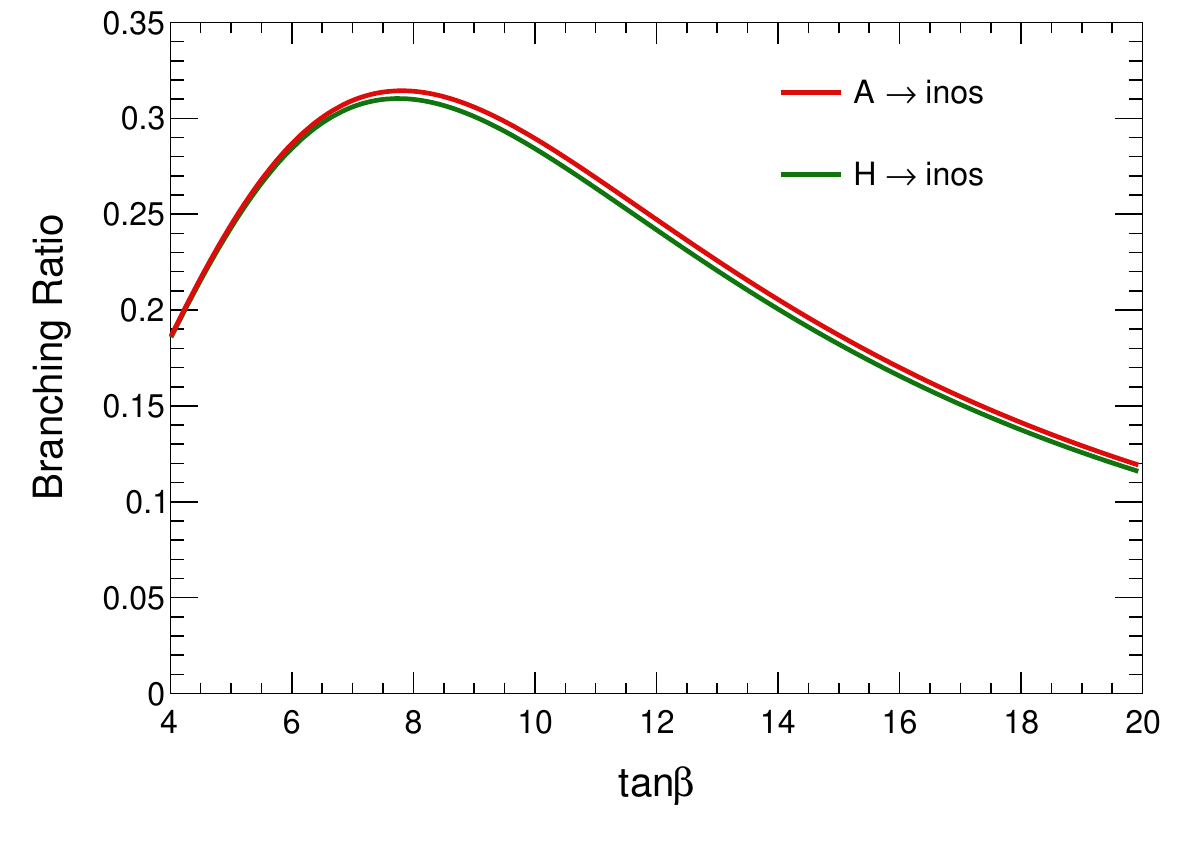}
\includegraphics[scale=0.37]{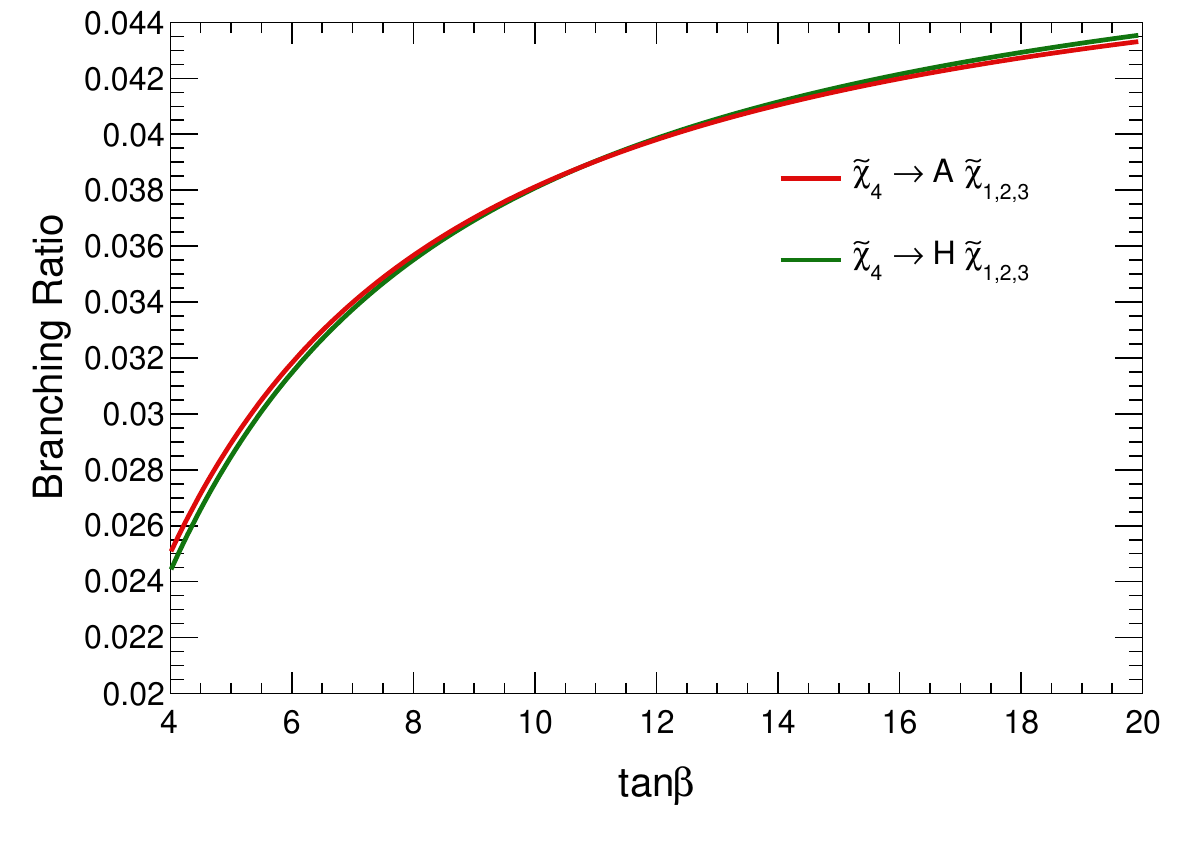}
\caption{\it Dominant branching ratios for heavy CP-even and CP-odd Higgs decays to individual electroweakino modes (top panel), the total branching ratio for two heavy Higgs (bottom left), branching ratio of heaviest neutralino decays to heavy Higgs (bottom right) for the case-1c. The wino, bino, higgsino mass parameters $M_2, M_1, \mu$ are fixed at 1500, 100 and 450 GeV and $\tan\beta$ is varied.}
\label{BS:1c}
\end{figure}

In the final variation of case-1, we change the bino mass parameter to even lower value, $M_1~=~100$ GeV. Here we get a bino-like LSP (bino fraction 98.73~\%) and two higgsino state, \textit{viz.} $\lsptwo,\lspthree$ (higgsino fraction 98.53~\%,~99.80~\%  respectively). As the neutralinos in this case are pure states as compared to the cases discussed previously, direct production of pure higgsino state is now possible. The following decay processes can give rise to mono-X final state topologies, \textit{viz.}

$$pp \to \lsptwo \lsptwo,~\lsptwo\to\lspone+h$$
$$pp \to \lsptwo \lspthree,~\lsptwo\to\lspone+h,~\lspthree\to\lspone+Z$$
$$pp \to \lspthree \lspthree,~\lspthree\to\lspone+Z.$$
Here, the $\lsptwo,\lspthree$ can decay to $\lspone$ with rate, $BR(\lsptwo\to\lspone h)\sim 85.93~\%$ and $BR(\lspthree\to\lspone Z)\sim 88.13~\%$. The branching ratios of heavy Higgs to electroweakinos are shown in Fig.~\ref{BS:1c}. The generic features for this scenario remain the same as in cast-1b, however in this hierarchy, the $\lsptwo$ decays to h while the $\lspthree$ decays to Z final state. 

\subsubsection*{Case-2 :}
\label{case2}
\begin{figure}
\centering
\includegraphics[scale=0.37]{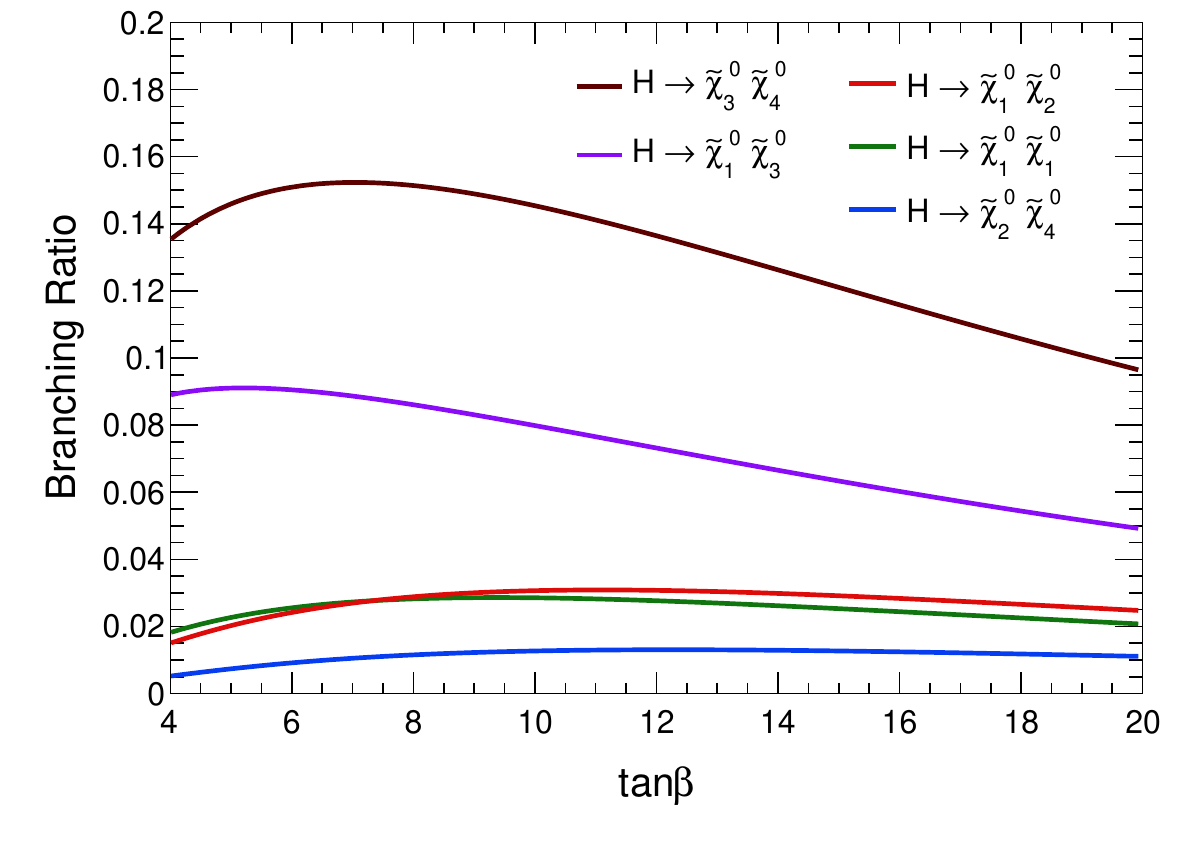}
\includegraphics[scale=0.37]{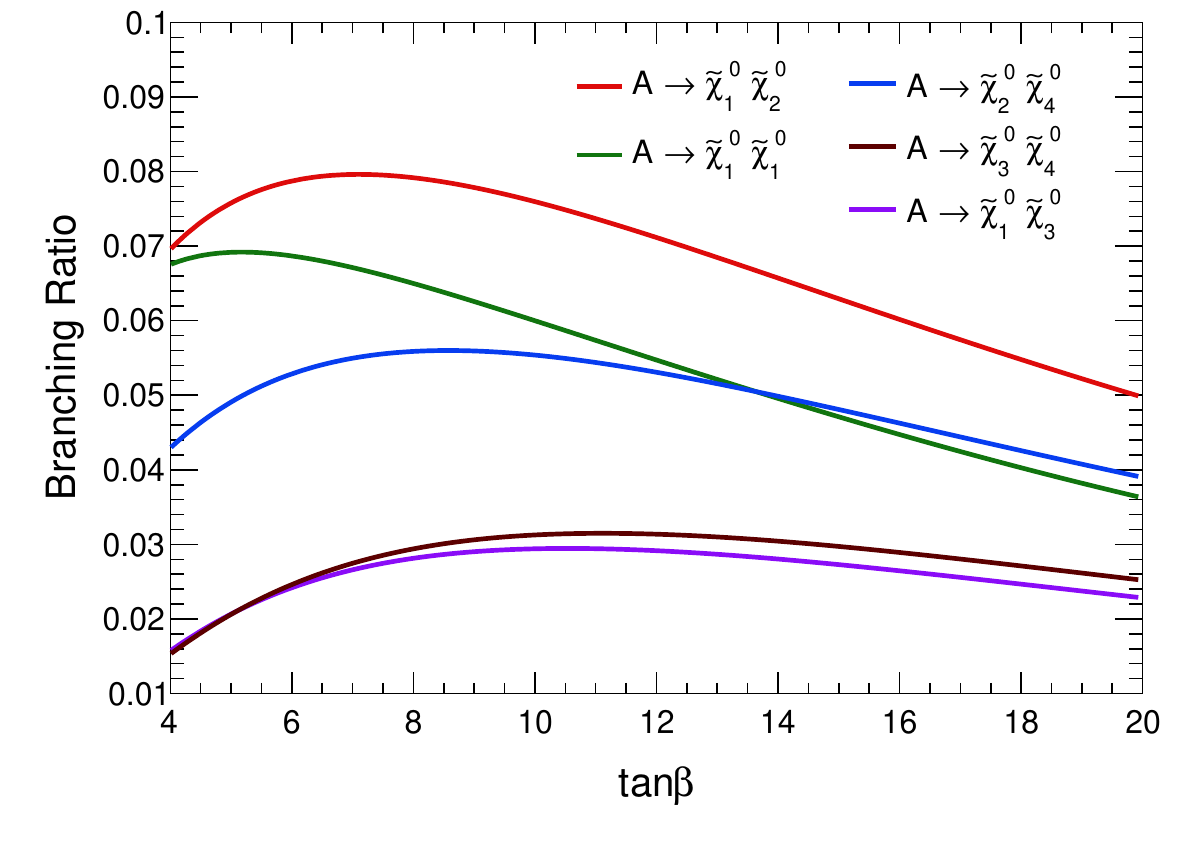}\\
\includegraphics[scale=0.37]{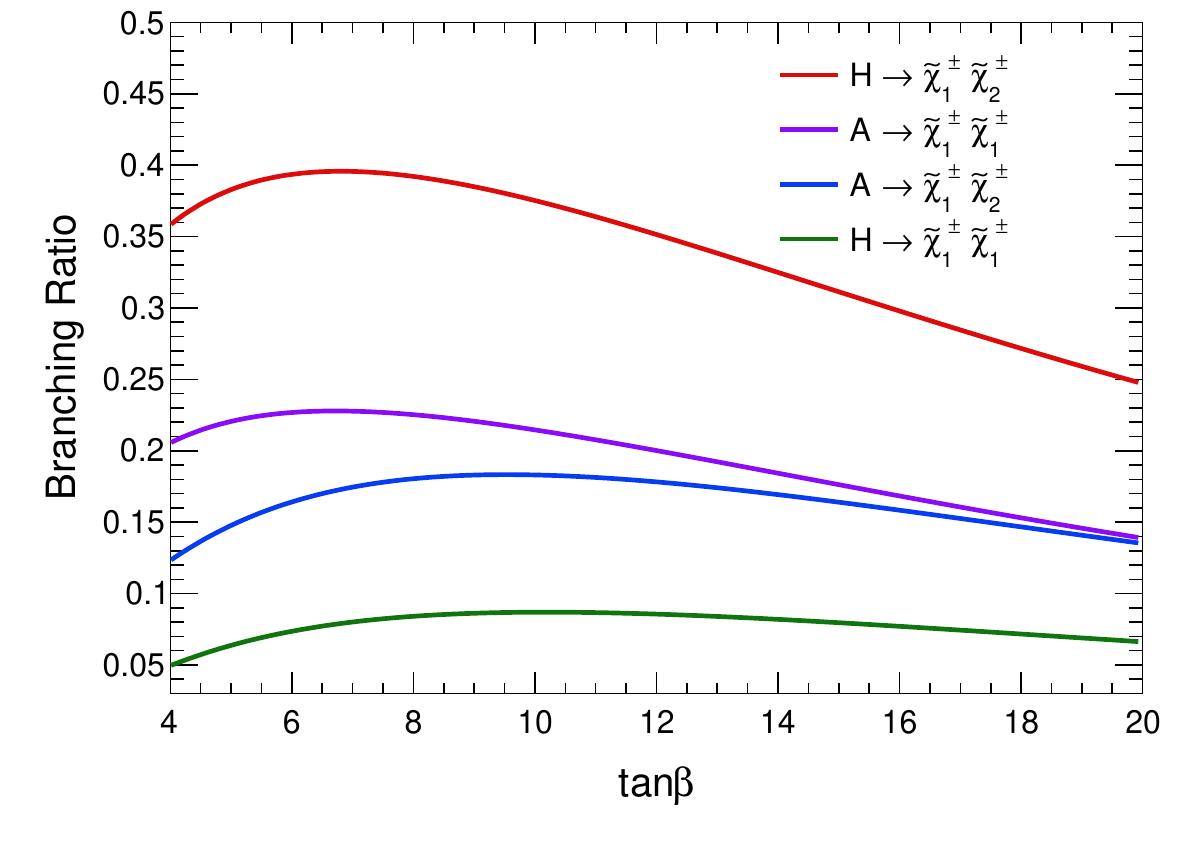}
\includegraphics[scale=0.37]{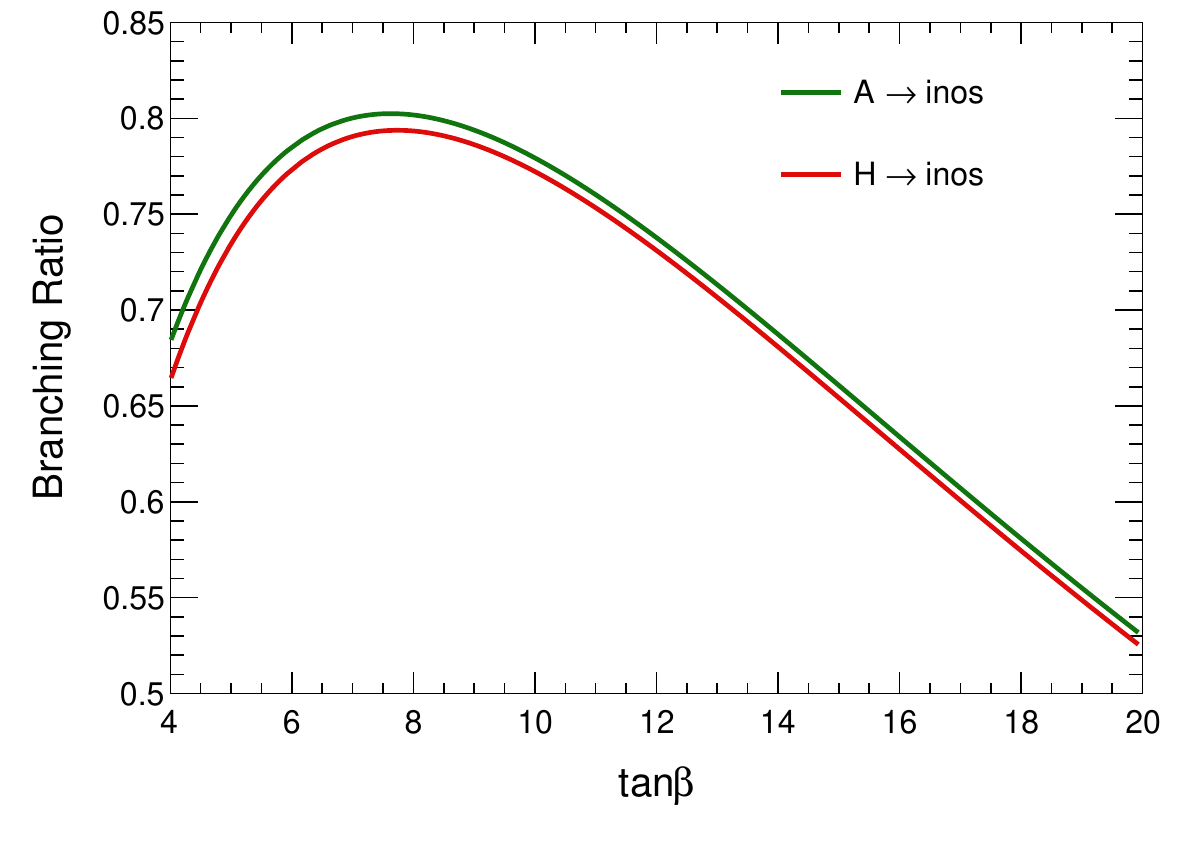}\\
\caption{\it Dominant branching ratios for heavy CP-even and CP-odd Higgs decays to individual electroweakino modes (top panel), the total branching ratio for two heavy Higgs (bottom right), branching ratio of heavy Higgs decaying to charginos (bottom left) for the case-2. The wino, bino, higgsino mass parameters $M_2, M_1, \mu$ are fixed at $400$, $300$ and $350$ GeV and $\tan\beta$ is varied.}
\label{BS:2}
\end{figure}

As opposed to previous benchmark where we considered a mixed bino-higgsino benchmark, we now discuss the phenomenological properties of a mixed higgsino - gaugino scenario. This is acquired by considering near degenerate $M_1, M_2,, \mu$. We select the mass parameters as $M_1=300$ GeV, $\mu=350$ GeV and $M_2=400$ GeV, hence resulting in electroweakinos which are mixed state of both the gaugino and higgsino components. As we fix heavy Higgs (H/A) mass at $1$ TeV, it can decay to any neutralino/chargino pair. The dominant branching ratios are shown in Fig~\ref{BS:2}. The heavy Higgs dominantly decays to electroweakinos with maximum branching ratio of $80\%$  depending on the $\tan\beta$ which is an advantage for searching for heavy Higgs via resonant electroweakino production. It should also be noted that because of the degenerate soft mass parameters, direct production of pure higgsino or gaugino state is highly suppressed. This scenario therefore shows the following unique final state properties. The $\lsptwo$ and $\lspthree$ can decay to the LSP via 2-body decay as well as 3-body decay. The 3-body decays with appreciable branching ratio are the following. 
$$\widetilde{\chi}_{ 2}^{0}~(\widetilde{\chi}_{ 3}^{0}) \to \lspone + \ell^{+} + \ell^{-}~(BR\sim 6.15~(9.84)~\%)$$
$$\widetilde{\chi}_{ 2}^{0}~(\widetilde{\chi}_{ 3}^{0}) \to \lspone + q + \bar{q}~(BR\sim 40.47~(64.52)~\%)$$
$$\widetilde{\chi}_{ 2}^{0}~(\widetilde{\chi}_{ 3}^{0}) \to \lspone + \nu + \bar{\nu}~(BR\sim 12.36~(19.71)~\%)$$
where $\ell=e,\mu,\tau$; $q=u,d,s,c,b$ and $\nu=\nu_e,\nu_\mu,\nu_\tau$.
The two body decay (via loop) includes,
$$\widetilde{\chi}_{ 2}^{0}~(\widetilde{\chi}_{ 3}^{0}) \to \lspone + \gamma~(BR\sim 3.09~(1.86\times10^{-2})~\%)$$
These can give rise to different possible final state signatures at the collider, \textit{viz.} (2/3/4)-lepton $+\met$, 2-lepton + jets $+\met$, 2-lepton $+~\gamma+\met$, 2$\gamma+\met$ etc. The lightest chargino \textit{i.e.} $\chonepm$ also decays via 3-body decay, \textit{viz.}
$$\chonepm \to \lspone +~q~+~\bar{q}^\prime~(BR\sim 66.80~\%)$$
$$\chonepm \to \lspone + \ell + \nu~(BR\sim 33.28~\%)$$
where $q=u,c$ and $q^\prime=d,s$. So, we can get multi-lepton $+~\met$, multi-jet $+~\met$ and lepton + jets $+\met$ final state from chargino pair production. This scenario therefore demonstrates a rich structure of final states which can potentially be probed at the LHC.

\subsubsection*{Case-3 :}
\label{case3}

Instead of decoupled wino-like state as done in case-1, if we rather consider wino-like LSP ($\lspone$), the lightest chargino can have longer lifetime. Heavy Higgs decays to such long lived states give rise to charged tracks at the collider and present another interesting set of collider signatures. To obtain such a long lived chargino, we fix $$M_1=1000~{\rm GeV},~M_2=300~{\rm GeV},~\mu=500~{\rm GeV}.$$ It should be noted that the chargino must contain some fraction of higgsino as the heavy Higgs only decays to an admixture of gaugino higgsino states. This necessitates considering low $\mu$ as well as $M_2$. We plot the resulting branching ratios in Fig.~\ref{BS:3b}. The branching ratio of heavy CP-even Higgs to pair of light charginos is about 6\% while the branching ratio of CP-odd Higgs to chargino pair can be as large as 14\% for this choice of parameters.

\begin{figure}
\centering
\includegraphics[scale=0.37]{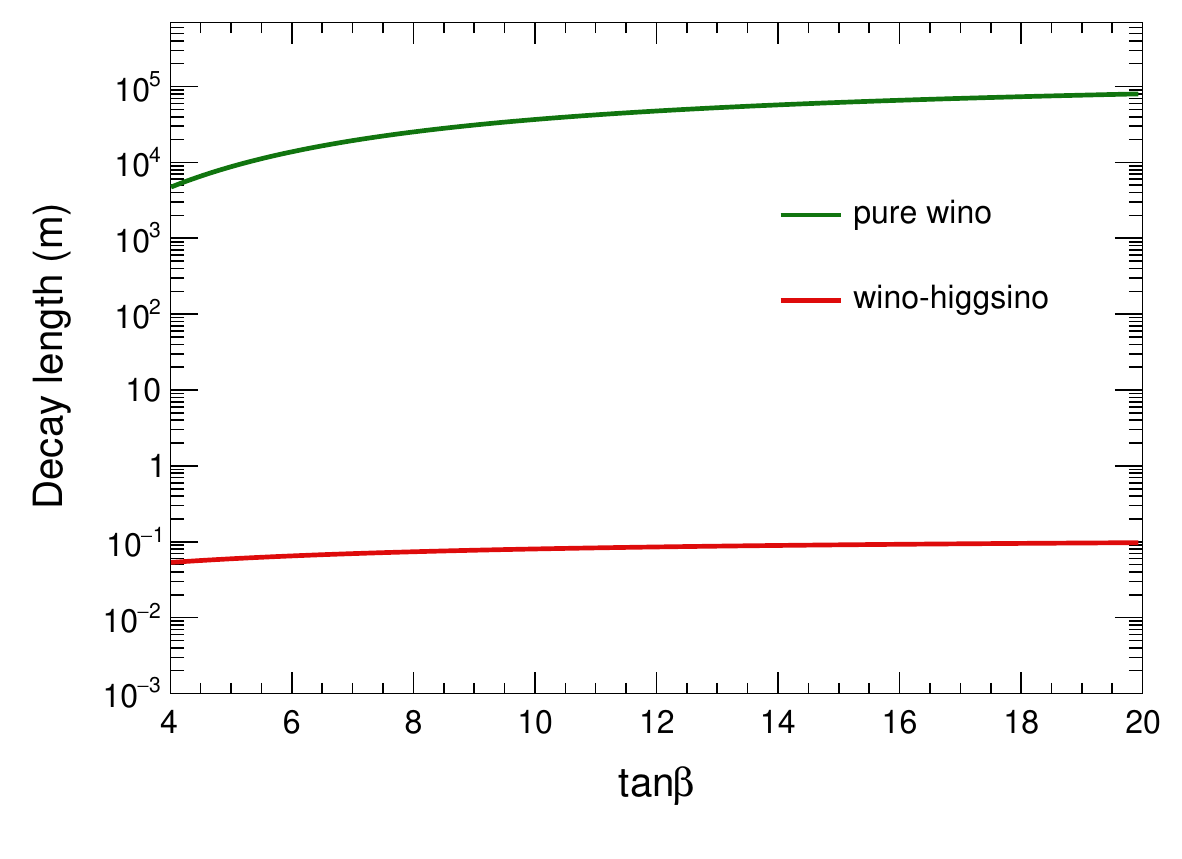}
\includegraphics[scale=0.37]{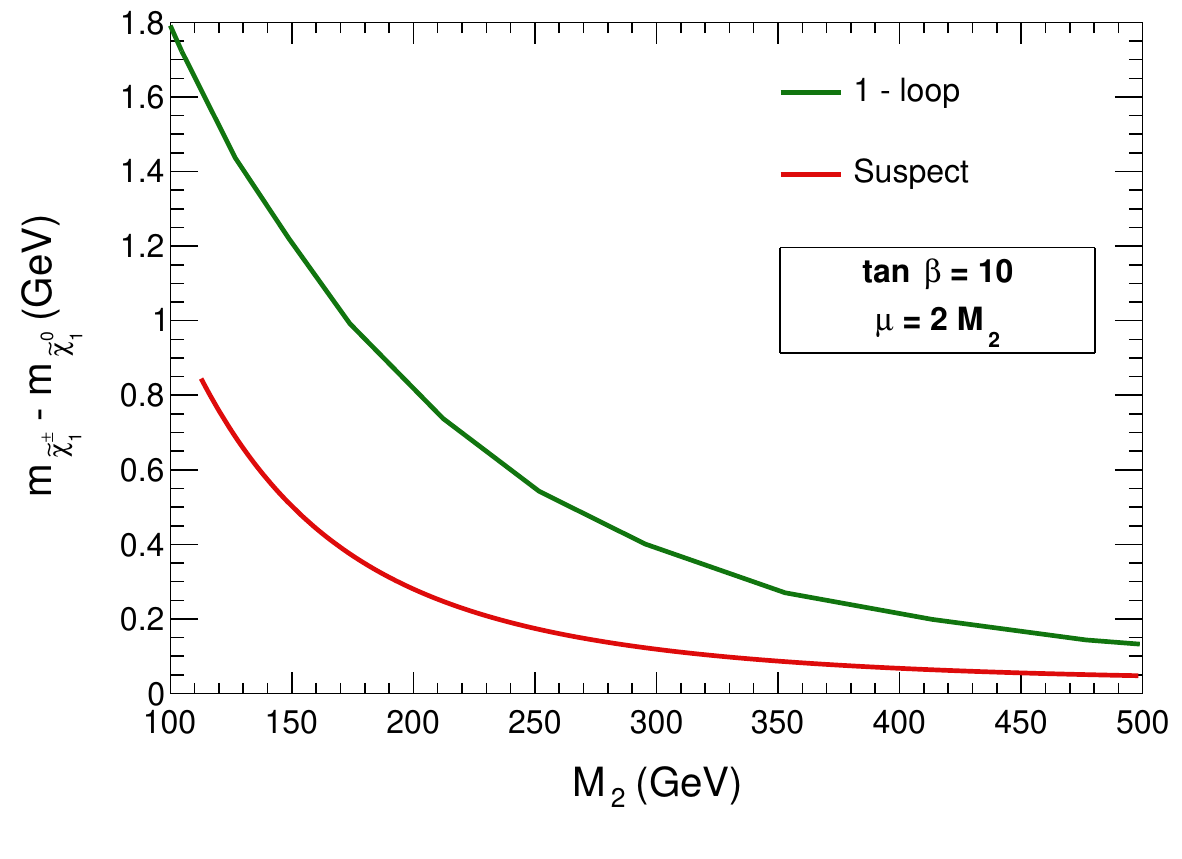}
\caption{\it Decay length of a charged wino in pure wino state (green) and mixed state of wino-higgsino (red) in Suspect2 (left), and comparison of mass splitting ($\Delta m_{\chonepm-\lspone}$) between Suspect2 and 1-loop result~\cite{GHERGHETTA199927} (right).}
\label{BS:3a}
\end{figure}

\begin{figure}
\centering
\includegraphics[scale=0.42]{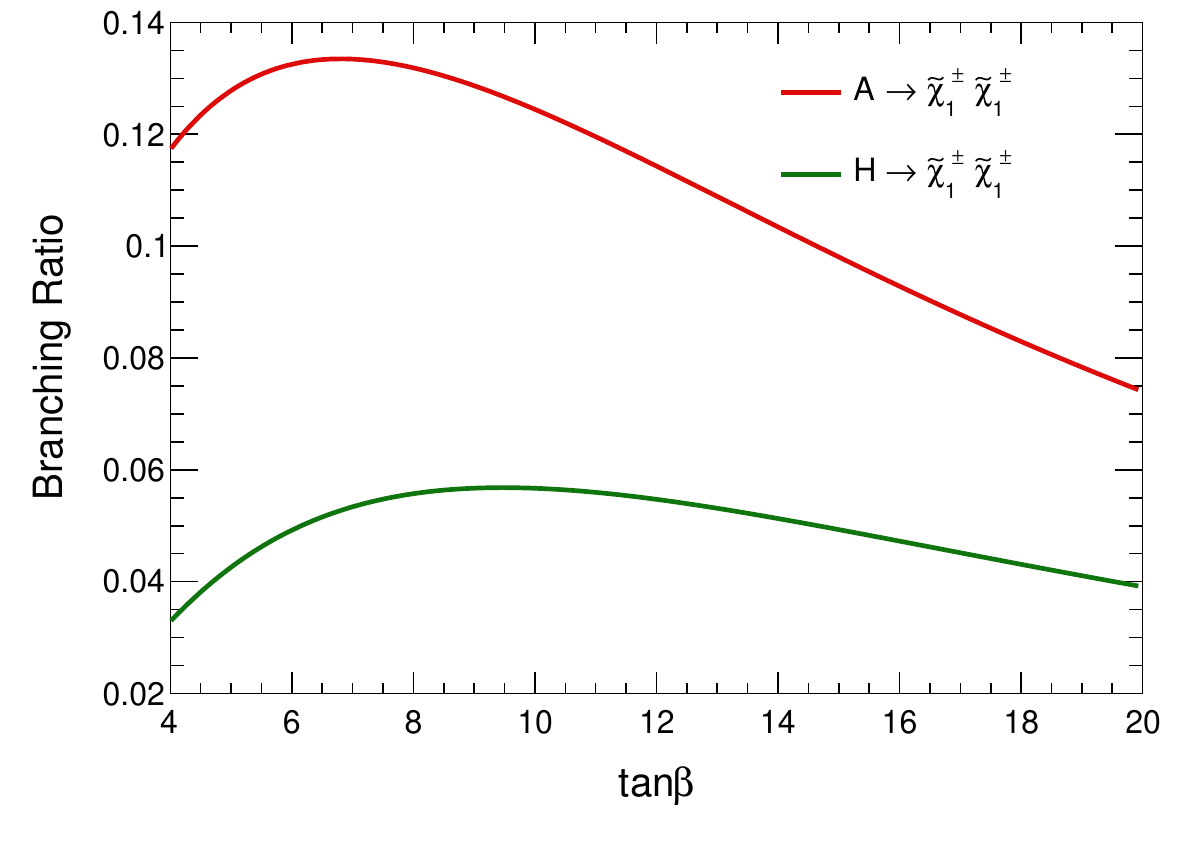}
\caption{\it Branching ratio for the decay of heavy Higgs to lightest chargino pair as a function of $\tan\beta$ with fixed, bino, wino and higgsino parameters at $M_1=1000~{\rm GeV},~M_2=300~{\rm GeV},~\mu=500~{\rm GeV}$.}
\label{BS:3b}
\end{figure}

Computing the correct chargino lifetime for such analysis however non-trivial. From theoretical calculation~\cite{GHERGHETTA199927, Ibe:2012sx, Gladyshev:2008ag}, the wino-like and higgsino-like chargino can have a decay length $\sim$ a few cm and $\sim$ mm$-\mu$m respectively. In case of the wino-like chargino, the mass difference between the charged and neutral wino state at the tree level is suppressed by a factor of $\sim \frac{m_W^4}{\Lambda^3}$ where $\Lambda \sim \mu, M_1$, ($\mu, M_1 \gg m_W$). Here the mass splitting arises from loop-corrections to make the chargino long-lived. On the other hand, the mass splitting between the higgsino-like chargino and neutralino arises at the tree level and the one-loop corrections are generally small. Predictions for lifetime from spectrum generator depend on whether such loop corrections are accounted for. Since the loop corrections to the chargino and neutralino mass matrix are absent in {\tt Suspect2}, the mass difference between wino-like $\chonepm$ and $\lspone$ is negligibly small and results in a large the chargino decay length, $\sim$ km range (green line in left plot of Fig.~\ref{BS:3a}). In the right plot of Fig.~\ref{BS:3a}, we compare this mass splitting between the lightest chargino and neutralino, generated by {\tt Suspect2} with the actual one-loop result where the 1-loop data has been taken from the Fig.3 of \cite{GHERGHETTA199927} for the case of $\mu=2M_2$. The figure shows that loop correction can make a significant difference to the chargino decay length. However in this work, we have used Suspect2 which does not include these loop corrections.

\section{pMSSM random scan}
\label{sec:setup}
It has been previously shown that the heavy neutral Higgs boson in the MSSM has a significant branching ratio to susy final states~\cite{Belanger:2015vwa, Kulkarni:2017xtf, Arbey:2013jla, Barman:2016kgt}. It is however important to analyse the phenomenologically viable parameter space and understand the signal cross sections of heavy Higgs to susy final states. In order to achieve this, a large scan within 19 dimensional pMSSM using random scan was performed. Table~\ref{tab:scan-ranges} details the ranges of the scan. The points resulting from the random scan were compatible with dark matter direct detection constraints~\cite{Aprile:2018dbl}, the flavour physics constraints~\cite{Amhis:2016xyh, Aaij:2017vad}, LEP constraints~\cite{Abbiendi:2003sc} and Higgs signal strengths as well as searches for heavy Higgs at the colliders. These constraints were checked using  {\tt micromegas}~\cite{Belanger:2018mqt}, {\tt HiggsSignals}, {\tt HiggsBounds}~\cite{Bechtle:2013wla, Bechtle:2013gu, Bechtle:2011sb, Bechtle:2008jh, Bechtle:2013xfa}. It is important to note that no constraints on the relic density of the dark matter were applied. In doing so, we allow for the possibility of a non-thermal history in the early Universe. The resulting points were then passed through {\tt SUSY-AI}~\cite{Caron:2016hib} and {\tt SModelS-1.2.2}~\cite{Kraml:2013mwa,Ambrogi:2017neo,Ambrogi:2018ujg,Heisig:2018kfq,Dutta:2018ioj} in order to test against the LHC 8 and 13 TeV constraints.  {\tt SModelS} contains up to 36 fb$^{-1}$ results from the susy searches at ATLAS~\cite{Aaboud:2016nwl, Aaboud:2016lwz, Aaboud:2016zdn, Aad:2016tuk, Aaboud:2018kya, Aaboud:2018zjf, Aaboud:2018ujj, Aaboud:2018sua, ATLAS-CONF-2013-007, ATLAS-CONF-2013-061, ATLAS-CONF-2013-089, Aad:2014wea, Aad:2013wta, Aad:2013ija, Aad:2014mha, Aad:2014pda, Aad:2014vma, Aad:2014nua, Aad:2014kra, Aad:2014bva, Aad:2014lra, Aad:2014qaa, Aad:2014nra, Aad:2015jqa, Aad:2015gna} and CMS~\cite{CMS-PAS-EXO-16-036, CMS-PAS-SUS-16-022, CMS-PAS-SUS-16-052, CMS-PAS-SUS-17-004, Sirunyan:2275490, Sirunyan:2261105, Sirunyan:2285878, Sirunyan:2260986, Sirunyan:2264381, Sirunyan:2264382, Sirunyan:2284431, Sirunyan:2290511, Sirunyan:2286124, Sirunyan:2272346, Sirunyan:2282000, Sirunyan:2293644, Sirunyan:2275103, Sirunyan:2274031, Sirunyan:2291344, Sirunyan:2269047, Sirunyan:2291416, Chatrchyan:1545325, Khachatryan:1987723, CMS-PAS-SUS-13-015, CMS-PAS-SUS-13-016, CMS-PAS-SUS-13-018, CMS-PAS-SUS-13-023, Chatrchyan:1546693, Chatrchyan:1527115, Chatrchyan:1696925, Khachatryan:1984165, Khachatryan:1704963, CMS-PAS-SUS-13-007, Chatrchyan:1567175, Chatrchyan:1662652, Chatrchyan:1631468, Khachatryan:1989788, Khachatryan:1976453, Khachatryan:2117955}, therefore, the most recent updates of susy searches with higher luminosity have not been accounted for. For {\tt SModelS}, the production cross sections were computed with {\tt Pythia8}~\cite{Sjostrand:2006za,Sjostrand:2014zea}, the branching ratios with {\tt SUSY-HIT}~\cite{Djouadi:2006bz}, NLO corrections to the production cross sections were evaluated using {\tt NLL-FAST}~\cite{Beenakker:1996ch,Beenakker:1997ut,Kulesza:2008jb,Kulesza:2009kq,Beenakker:2009ha,Beenakker:2010nq,Beenakker:2011fu}. Finally, it should be noted that there is a basic difference in the way {\tt SUSY-AI} and {\tt SModelS} evaluates LHC constraints. {\tt SUSY-AI} uses machine learning techniques to infer the viability of a MSSM parameter point based on the existing public results from ATLAS pMSSM analysis~\cite{Aad:2015baa}. {\tt SModelS} on the other hand uses simplified model technology to decompose the input spectra into the corresponding simplified model topologies. It compares the theory cross sections resulting from decomposition procedure with the corresponding experimental results. While {\tt SUSY-AI} is more robust than {\tt SModelS} in this aspect, {\tt SModelS} contains a more comprehensive and updated database of results compared to {\tt SUSY-AI}. Drawing outright comparison between the two codes is therefore non-trivial.

\begin{table}[htb!]
\centering
\begin{tabular}{|c|c||c|c| }
\hline
Parameter & range & Parameter & range\\
 \hline
$M_1$ &    [1, 1000]                                      & $M_A$ &    [100, 2000]\\
$M_2$ &    [100, 1000]                                    & $\mu$ &    [0, 1000]\\
$M_3$ &    [700, 5000]                                    & $\tan\beta$ &   [1, 60]\\
$m_{{\tilde e_R},{\tilde \mu_R}}$ &    3000               & $m_{{\tilde e_L},{\tilde \mu_L}}$ &    3000 \\
$m_{\tilde \tau_R}$ &    [80, 2000]                       & $m_{\tilde \tau_L}$ &    [80,2000]\\
$m_{\tilde q3_L}$ &    [500, 10000]                       & $m_{{\tilde q_{1L}},{\tilde q_{2L}}}$ &   3000 \\
$m_{\tilde t_R}$ &    [500,10000]                         & $m_{\tilde b_R}$ &    [500,10000] \\
$A_b$ &   [-2000, 2000]                                   & $A_t$ &    [-10000, 10000]\\
$A_{\tau}$ &   [-2000, 2000]                              & $A_{u,d,e}$ & 0  \\
$m_{{\tilde u_R},{\tilde d_R},{\tilde c_R},{\tilde s_R}}$  &    3000          &&\\
\hline
\end{tabular}
\caption{\it The ranges of MSSM parameters searched by random scan. The mass scales are in GeV units.}
\label{tab:scan-ranges}
\end{table}
 
For direct detection, we utilised the latest results form the XENON1T collaboration~\cite{Aprile:2018dbl}, while the theory parameter space was appropriately rescaled by $\zeta$ where $\zeta$ is defined by $\zeta = \Omega h^2_{central}/\Omega h^2_{theory}$, where $\Omega h^2_{central} = 0.1189$~\cite{Aghanim:2018eyx}. Here, we use a parametric form of the direct detection cross section similar to~\cite{Belanger:2015vwa}. We would like to mention here that we also do mcmc scan in the MSSM parameter space which give similar results.
 
\begin{table}[htb!]
\centering
\begin{tabular}{|c|c|}
\hline
Constraint name & Range \\ \hline
$B \rightarrow X_s \gamma$ &   $[2.583, 4.057]\times 10^{-4}$ \\
$B_s \rightarrow \mu \mu$ &   $[1.2912, 4.8974]\times 10^{-9}$  \\ \hline
\end{tabular}
\caption{\it Flavour physics constraints used in our random scan. We consider $10\%$ uncertainty around central value as theoretical error which is added in quadrature with the experimental error to get the total error. This range is obtained by including two times this total error ($2\sigma$).}
\label{tab:FC}
\end{table}

\subsection{Scan results}
\label{sec:scan-result}

The results of our scan in particular show that $\tan\beta\gtrsim 20$ is ruled out for a heavy Higgs mass  $m_A < 1$ TeV. On the other hand, all values of $5 < \tan\beta < 60$ are allowed for $m_A \gtrsim 1.65$ TeV.  We will not further discuss the features of the parameter space, however this exercise demonstrates that particularly in the electroweak sector a large parameter space remains unconstrained by the current experimental searches (see Appendix~\ref{sec:appendixA} for details).

\begin{figure}[htb!]
\centering
\includegraphics[scale=0.40]{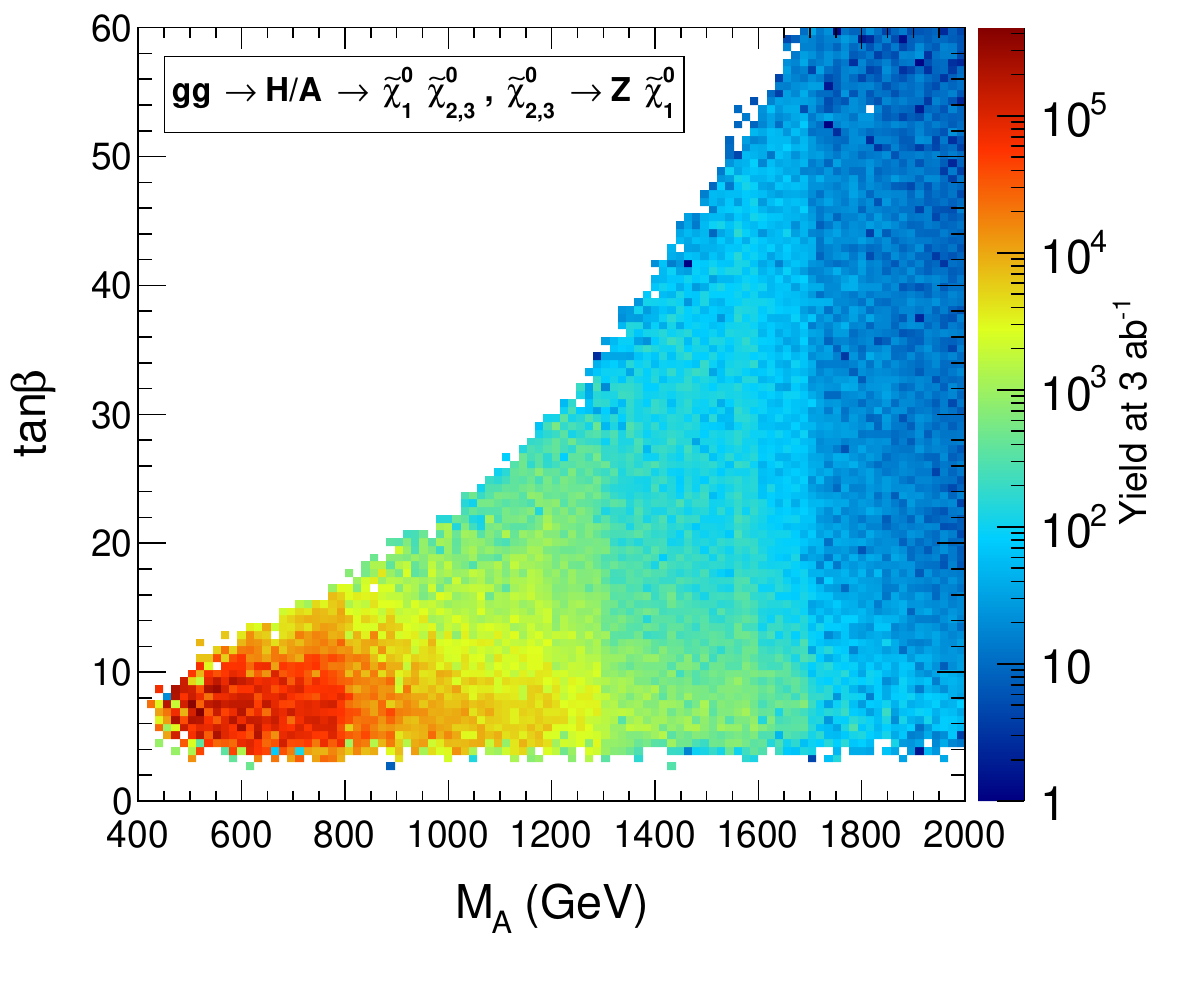}\includegraphics[scale=0.40]{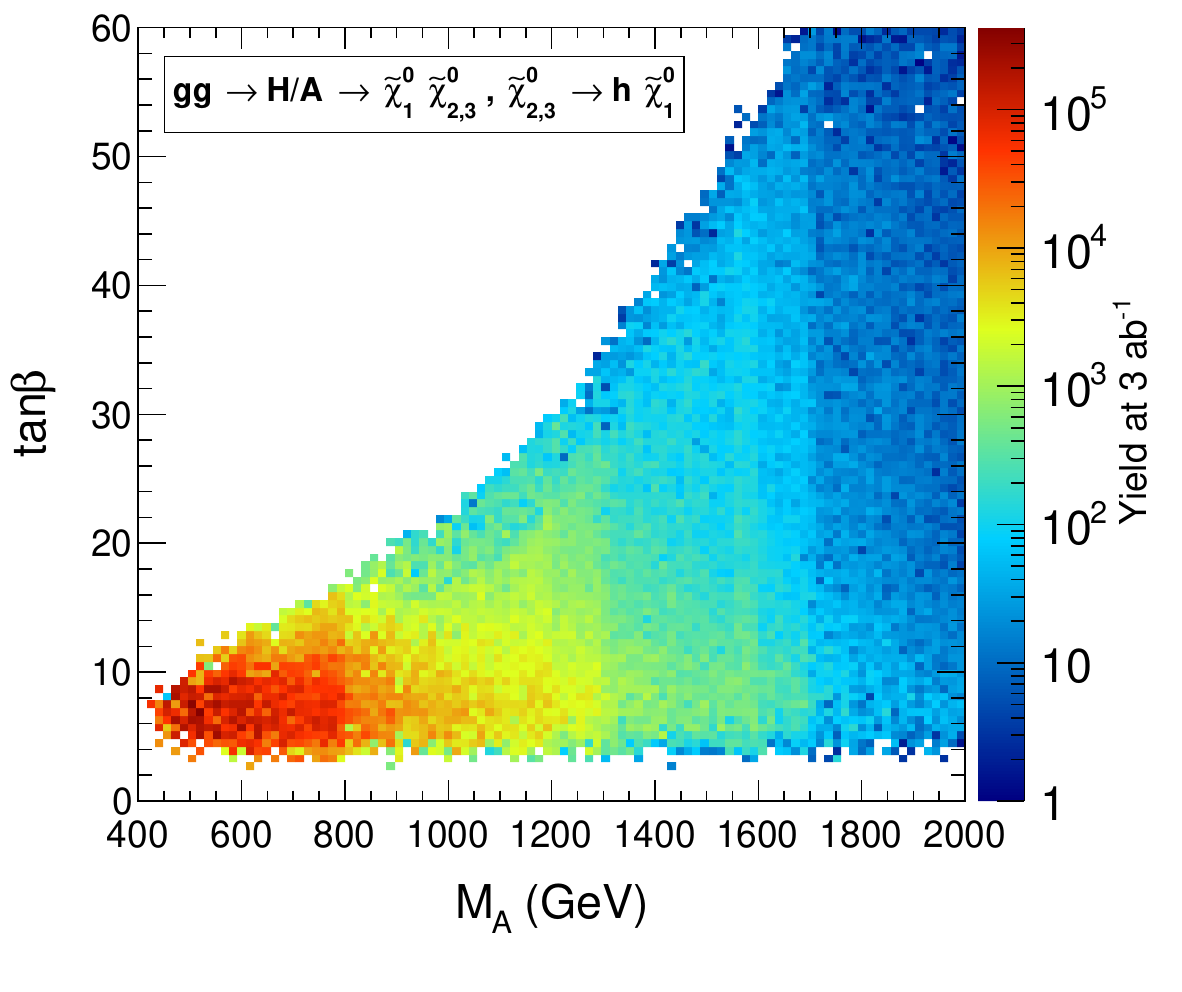}\\
\includegraphics[scale=0.40]{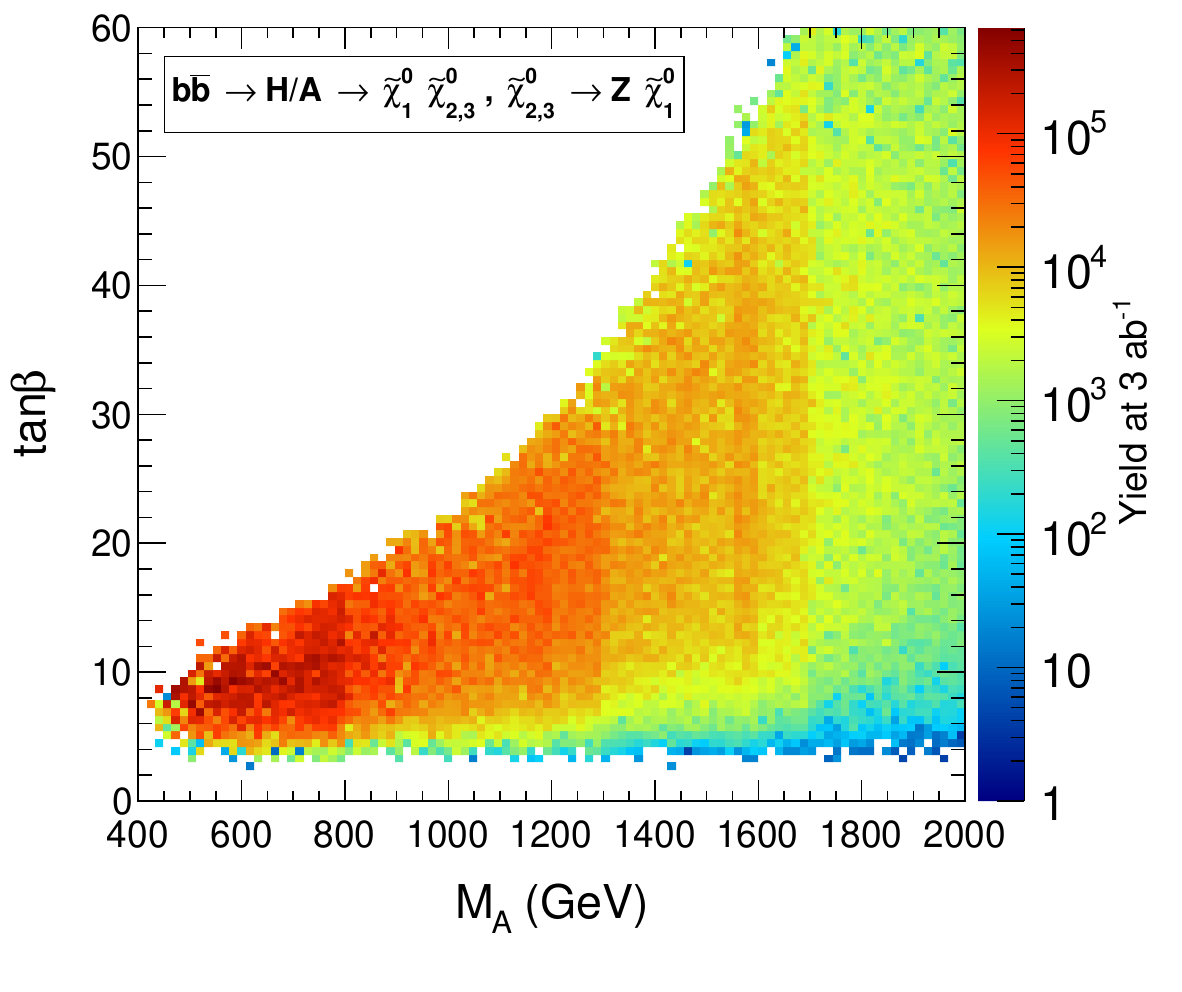}\includegraphics[scale=0.40]{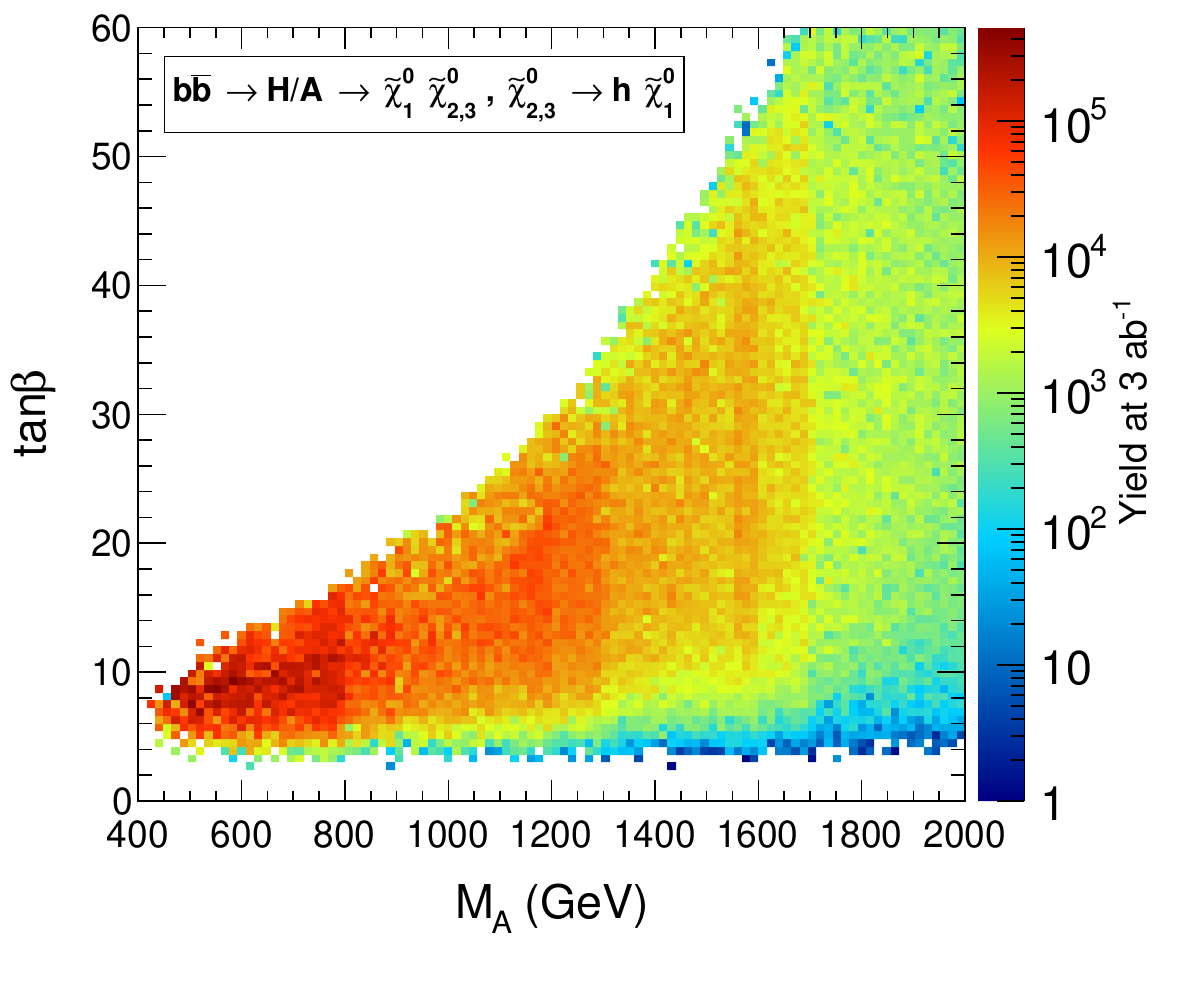}
\caption{\it Event yield at 3 ab$^{-1}$ of integrated luminosity for all points passing phenomenological constraints for gluon fusion process leading to mono-Z/h final state (top left, top right panel) while the bbH production mode leading to mono-Z/h final state (bottom left, right panel). The plots demonstrate that bbH production in general leads to high event yields for moderate $\tan\beta$ due to enhanced Yukawa couplings. }
\label{fig:scancs}
\end{figure}

With this allowed parameter space as a base for our further studies, we concentrate on mono-X signatures at HL-LHC. As has been demonstrated before~\cite{Barman:2016kgt,Kulkarni:2017xtf}, the heavy Higgs decays to susy particles mostly lead to mono-X final states after accounting for phenomenological constraints, hence we expect these signatures to be the most promising ones in the search for heavy Higgs to susy decays. More concretely, we calculate the yield at HL-LHC configuration for the following processes, $$gg\to H/A\to \lspone~ \widetilde{\chi}_{ 2,3}^{0},~\widetilde{\chi}_{ 2,3}^{0}\to \lspone~(h/Z),$$ $$b\bar{b}\to H/A\to \lspone~ \widetilde{\chi}_{ 2,3}^{0},~\widetilde{\chi}_{ 2,3}^{0}\to \lspone~(h/Z).$$ In Fig.~\ref{fig:scancs}, we show event yield at 3 ab$^{-1}$ for the mono-Z and mono-h final states arising from aforementioned channels as a function of $m_A$ and $\tan\beta$. We compute the heavy Higgs production cross section using {\tt SusHi}~\cite{Harlander:2016hcx}. Furthermore, we divide event yields in two categories corresponding to ggH and bbH Higgs production processes. The production cross section is in general higher for the $gg$ initiated Higgs prodction in the low $m_A$ and low $\tan\beta$ region, therefore leads to higher event yield. Since heavy Higgs coupling to the down type quarks is proportional to $\tan\beta$, the event yield in case of $b\bar{b}$ initiated production increases for larger $\tan\beta$ values. In general the event yield reaches up to $10^5$ events in ggH mode for Higgs masses less than 1 TeV, for bbH mode even larger masses can yield substantial $(\sim 10^4)$ number of events. This motivates development of separate dedicated search strategies for the ggH and bbH modes, where for the bbH mode the presence of additional b-jets could be exploited to gain sensitivity as compared to the ggH mode.

\begin{figure}[htb!]
\centering
\hspace*{-0.3in}
\includegraphics[width=.45\linewidth]{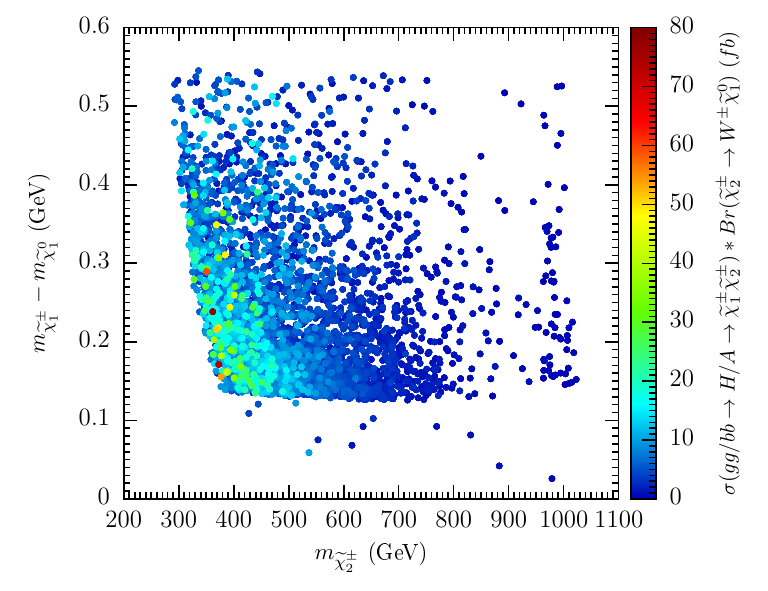} \includegraphics[width=.45\linewidth]{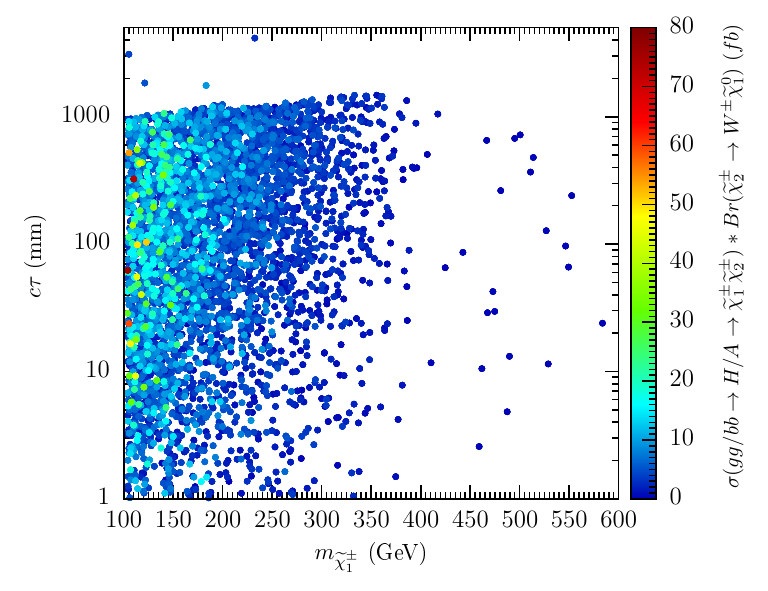}
\hspace*{-0.3in}
\includegraphics[width=.45\linewidth]{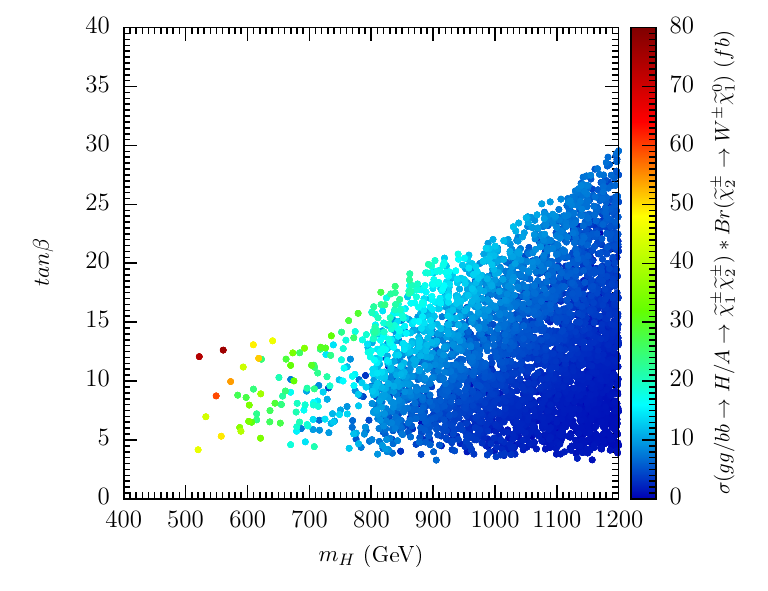}
\caption{\it The event yield for $\chonepm$ production from heavy Higgs in $m_{\chtwopm}$ - $(m_{\chonepm} - m_{\lspone})$, decay length - $m_{\chonepm}$ and $m_H$ - $\tan\beta$ plane for all points passing phenomenological constraints. The chargino - neutralino mass difference does not include loop corrections discussed in the text. Event yields as a function of heavier chargino and chargino - neutralino mass difference (top left), as a function of lightest chargino mass and lifetime (top right) and as a function of heavy Higgs mass and $\tan\beta$ are displayed. }
\label{fig:llp_yield}
\end{figure}

The above discussed processes correspond to prompt decays of neutralinos. In correspondence to the case-3 in section~\ref{sec:BS}, we also compute the production cross section for heavy Higgs decays to long-lived $\chonepm$ at the HL-LHC. For this, we consider the process, $$gg/b\bar{b}\to (H/A)\to \chonepm~\chtwomp,~\chtwomp\to W^{\mp}~\lspone.$$ We show the chargino production cross section times $Br(\chtwomp\to W^{\mp}~\lspone)$ in the plane of $m_{\chtwopm}$ vs $(m_{\chonepm} - m_{\lspone})$ (top left), $m_{\chonepm}$ vs decay length (top right) and $m_H$ vs $\tan\beta$ (bottom panel) in Fig.~\ref{fig:llp_yield}. Of particular importance here is the impact of two loop corrections on the chargino - neutralino mass splitting and associated change in the chargino decay length. We have used {\tt Suspect2} to compute the MSSM mass spectrum, this version does not include loop corrections to chargino masses. For small chargino - neutralino mass difference these corrections are particularly important. Therefore, while the qualitative features of this final state are robust, the quantitative estimates in particular for the chargino lifetimes are subject to change and have not been accounted for within this work~\footnote{We have also checked our scan results using the most recent version of {\tt Suspect3-beta}~\cite{suspect3} which accounts for full one-loop and dominant two-loop radiative corrections to the masses of electroweakinos. We get similar results as with {\tt Suspect2}. However, the version {\tt Suspect3-beta} is under development. Therefore, we do not use this in our analysis.}. Fig.~\ref{fig:llp_yield} demonstrates the chargino - neutralino mass difference (top left), the chargino decay length (top right), and associated cross section where the heavier chargino decays to a W boson (bottom panel).  It can be seen that in general the cross sections for these processes are large and given the rather low background for searches involving disappearing track and heavy stable charged particles, such final states present an interesting avenue for heavy Higgs searches. 

\begin{figure}[htb!]
\centering
\hspace*{-0.3in}
\includegraphics[width=0.50\linewidth]{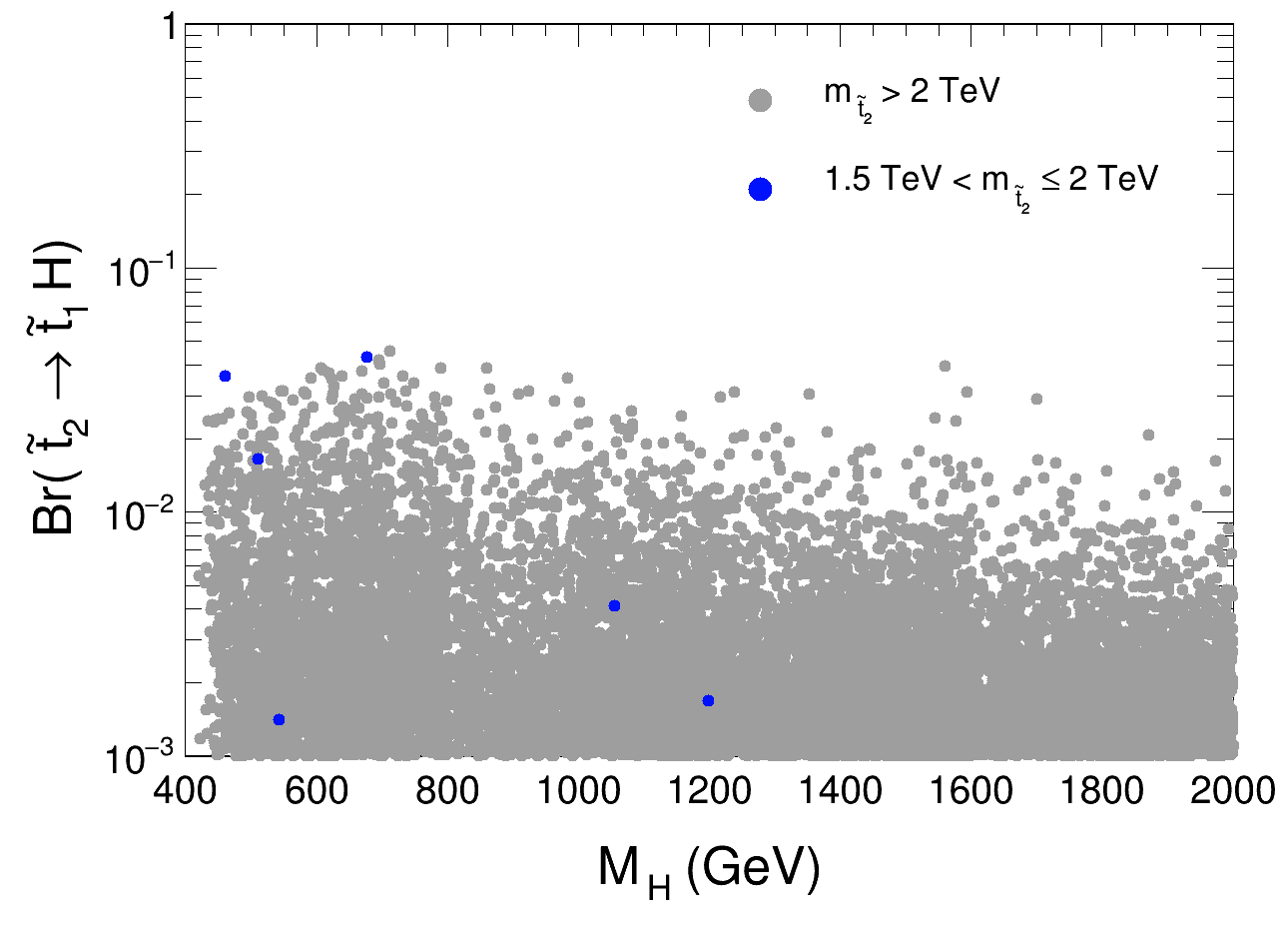}\includegraphics[width=0.50\linewidth]{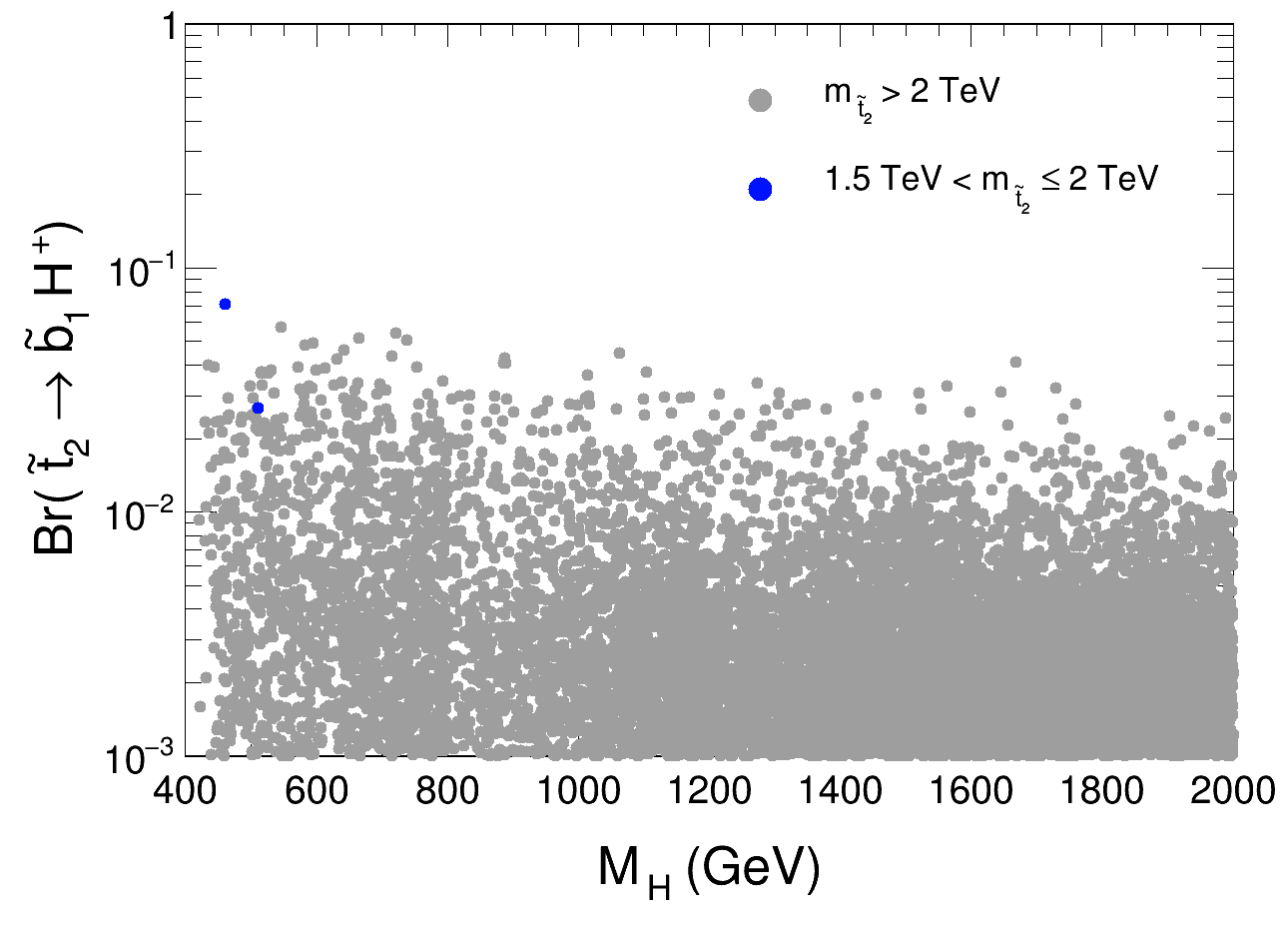}\\
\hspace*{-0.3in}
\includegraphics[width=0.50\linewidth]{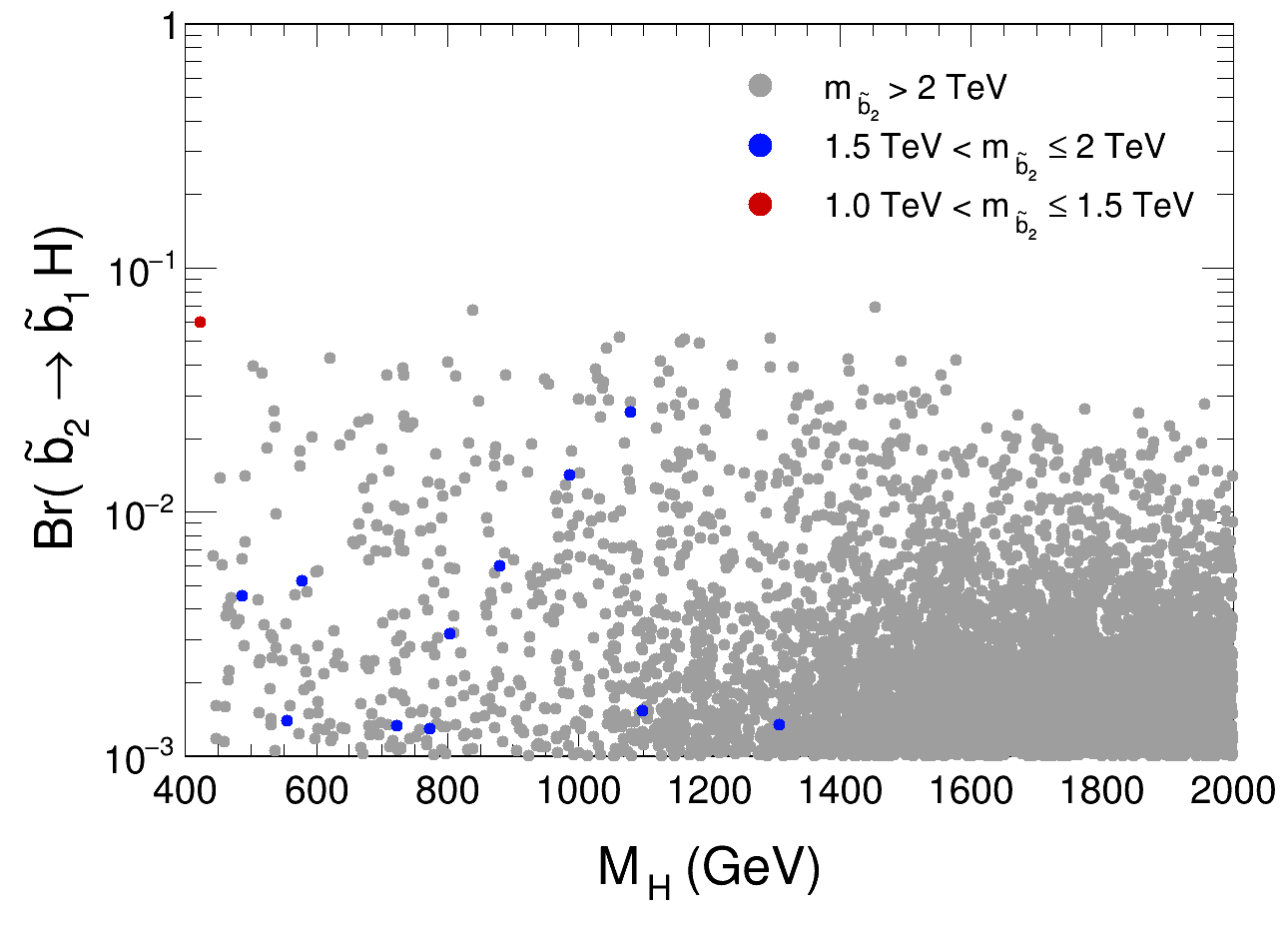}\includegraphics[width=0.50\linewidth]{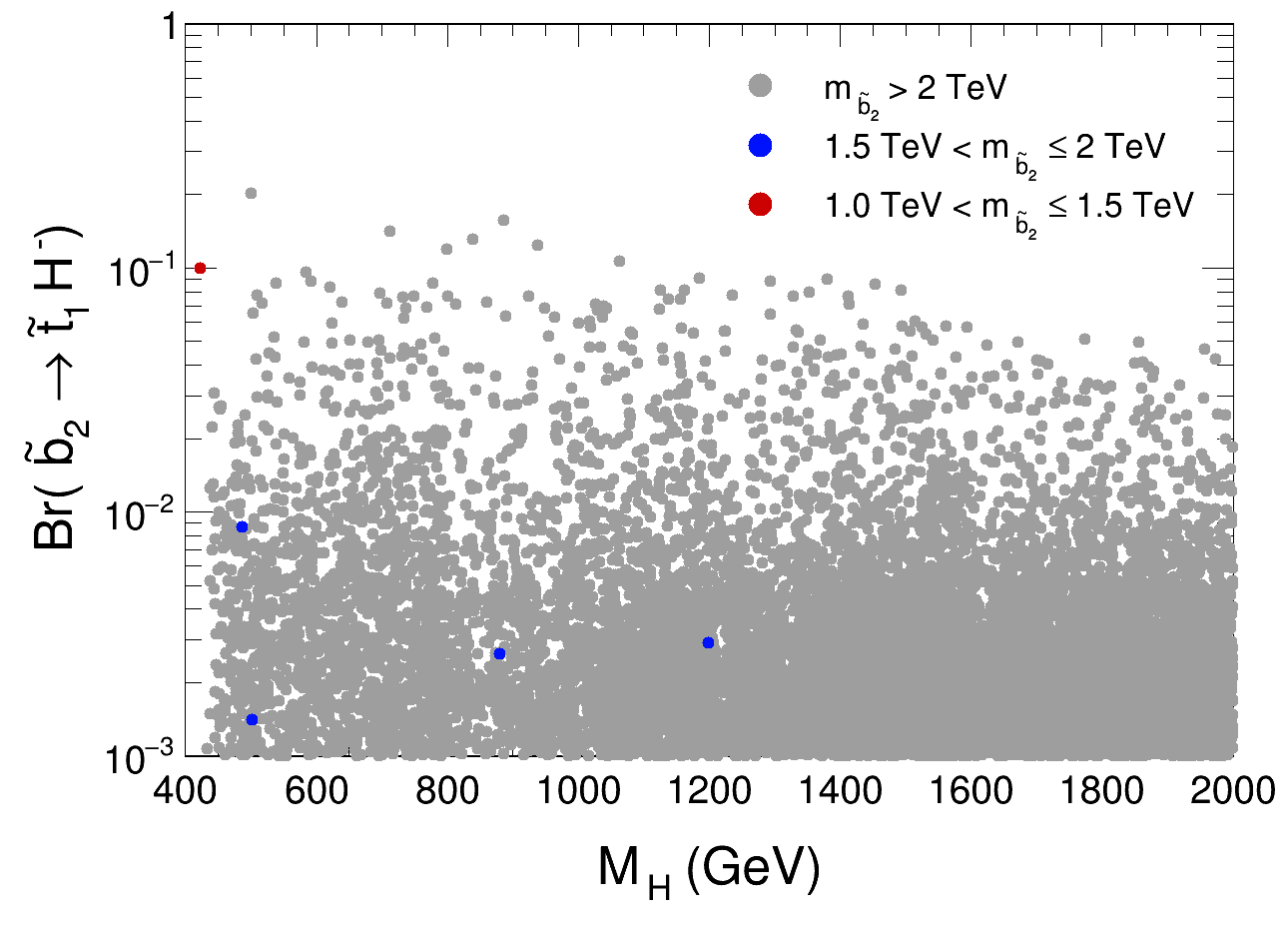}
\caption{\it Branching ratios of heavy Higgs production from heavy squarks for all points passing phenomenological constraints. Points in grey are all points for which we find non-zero branching ratio to heavy Higgs final state, we highlight in blue points for which heavy stop/sbottom mass is less than 2 TeV. }
\label{Brheavysq}
\end{figure}

Apart from decays of heavy Higgs to susy final states, it is also possible to produce heavy Higgs from decays of susy particles. We alluded to this possibility in section~\ref{sec:BS} while discussing several benchmark scenarios for heavy Higgs to electroweakino final states. It is also possible that heavy Higgs emerges from cascade decays of susy particles apart from electroweakinos. We explicitly demonstrate the branching ratios for $\widetilde{t_2} \to \widetilde{t_1}~H$, $\widetilde{t_2} \to \widetilde{b_1}~H^+$, $\widetilde{b_2} \to \widetilde{b_1}~H$ and $\widetilde{b_2} \to \widetilde{t_1}~H^-$. We calculate these branching ratios with {\tt Susyhit} for the parameter space regions which satisfy all of the previously discussed experimental constraints. In Fig.~\ref{Brheavysq}, we show the branching ratios for above mentioned decay modes~\footnote{For this plot we do not sum up CP-even and CP-odd Higgs in final state, in case that is done, the branching ratios will be almost twice as large for these processes.}. For convenience we highlight points where heavy stop/sbottom masses are less than 2 TeV, while plotting all points which have a non-zero branching ratio to heavy Higgs (grey). The decay rate can reach up to $8-10\%$, however this generally requires a stop/sbottom heavier than 2 TeV, thus limiting the reach of LHC for such processes. A detailed investigation of this analysis is beyond the scope of this study however we stress the need of characterising this parameter space further and understand potential for heavy Higgs final states at the LHC. These present also an additional opportunity for exploring heavy Higgs and susy sectors at the 100 TeV collider.

\subsection{Mono-Z event kinematics}
\label{sec:monozkin}
\begin{figure}[htb!]
  \includegraphics[width=0.50\linewidth]{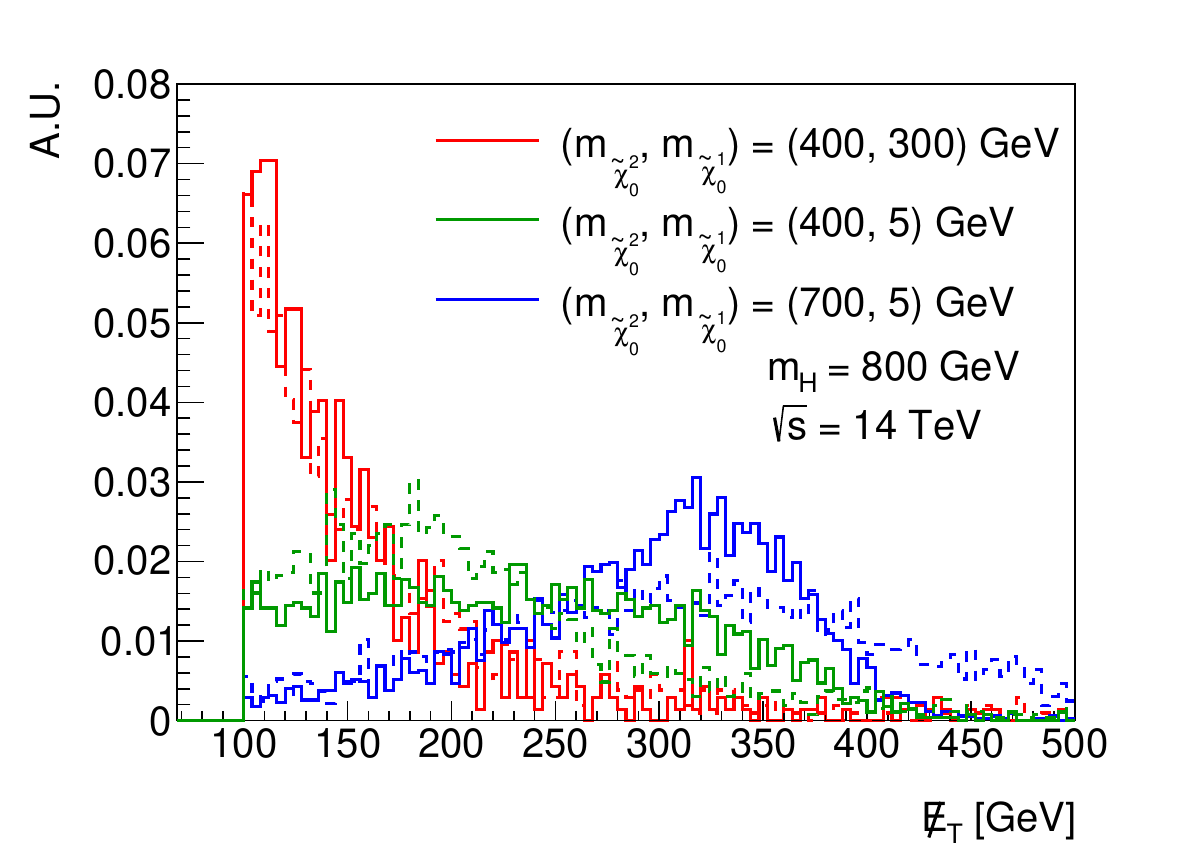}
    \includegraphics[width=0.50\linewidth]{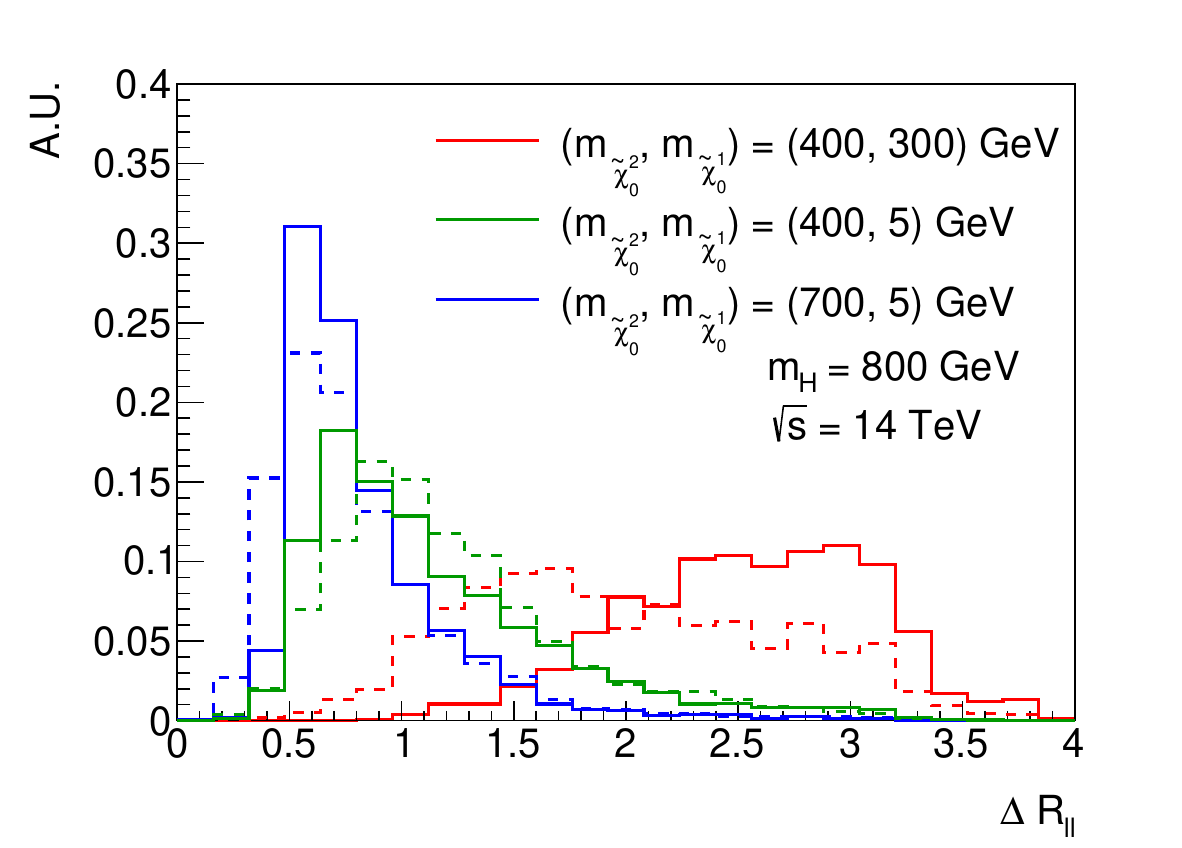}
\caption{Normalised distributions for missing energy of the events (left) and $\Delta R$ between two leptons (right) with preselections as defined in the text for three different benchmark points. The dashed lines represent production via SM mediated processes and the solid lines represent production via heavy Higgs channel. }
  \label{fig:kin_met_deltaR}
\end{figure}
Having motivated an in-depth analysis of mono-X final states in the previous subsection, we now turn our attention to understanding the salient kinematical features for these final states emerging via the decay of heavy Higgs. We will rely on these kinematic features for analysis optimization in the following sections. We expect the kinematics to be different for electroweakino resulting from decays of heavy Higgs against the ones produced via SM mediators. The reason being the heavy Higgs is produced on-shell and hence should leave an imprint of resonant production on the final state, the SM mediators responsible for direct electroweakino production on the other hand are always off-shell. In order to understand potential differences between the Standard Model (SM) mediated and the Higgs mediated production of the electorweakinos, in this section, we present several kinematic features for a few benchmark scenarios. We choose a fixed heavy Higgs mass at 800 GeV, and vary the masses of $\lsptwo$ and  $\lspone$. We simulate the Higgs boson production, subsequent decays and hadronise the events using {\tt Pythia8}. We consider only gluon fusion channel. The heavy Higgs is decayed to $\lsptwo+\lspone$ using 100\% branching ratio. We further simulate decays of $\lsptwo$ to Z+$\lspone$ also using 100\% branching ratios, furthermore, Z is decayed only to electron and muon final states. The events are passed through {\tt Delphes}~\cite{deFavereau:2013fsa} using default ATLAS card and the resulting final state in dilepton + $\met$ is analysed. For all the distributions below, we choose events with exactly two leptons, less than equal to 2 jets in the final state with $p_T(j_1) > 30$ GeV, and  $\met > 100$ GeV.
\begin{figure}[htb!]
  \includegraphics[width=0.50\linewidth]{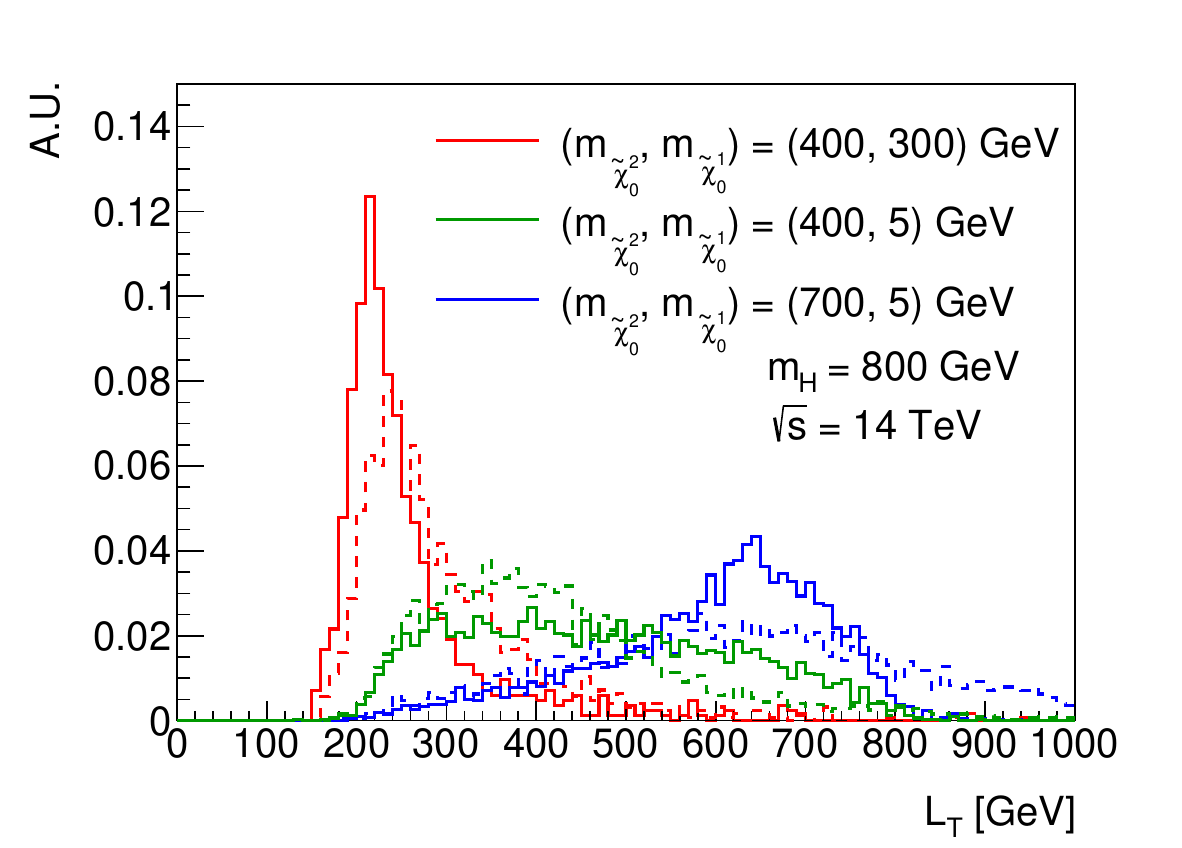}
  \includegraphics[width=0.50\linewidth]{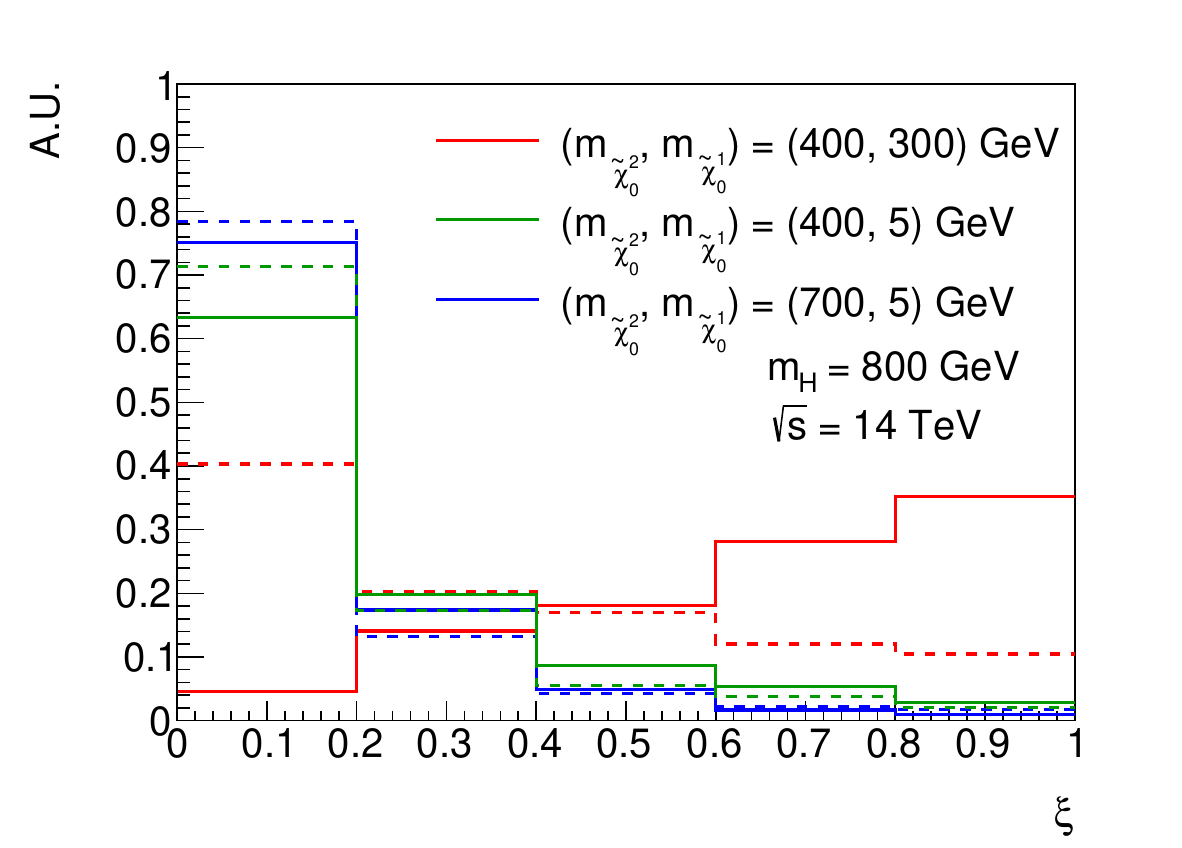}
\caption{Normalised distributions for $L_T$ of the events with preselections as defined in the text (left) and $\xi$ (right) for three different benchmark points. The dashed lines represent production via SM mediated processes and the solid lines represent production via heavy Higgs channel. }
  \label{fig:kin_epsilon_lt}
\end{figure}

In figure~\ref{fig:kin_met_deltaR}, we plot the missing energy (left) and the $\Delta R_{\ell\ell}$~\footnote{The distance between two particles \textit{a} and \textit{b} in the $\eta-\phi$ plane, $\Delta R_{ab}$ is defined as $\Delta R_{ab} = \sqrt{\Delta\eta_{ab}^2 + \Delta \phi_{ab}^2}$ where $\Delta\eta_{ab}$ is the distance in the pseudorapidity plane and $\Delta \phi_{ab}$ is the azimuthal angle separation between the particles \textit{a} and \textit{b}.} between two leptons (right) at reconstructed level. The dashed lines represent direct production of  $\lsptwo, \lspone$ while the solid lines represent the production of same particles except via the heavy Higgs boson whose mass is fixed at 800 GeV. Three different benchmark points are chosen such that the heavy Higgs always decays on shell to $\lsptwo, \lspone$. In addition care has been taken to choose different masses to represent a range of different boosts received by final state leptons. The combination ($m_{\lsptwo}, m_{\lspone}$) = (400, 300) GeV corresponds to the maximal allowed  ($\lsptwo, \lspone$)  masses such that the decays of Higgs boson to electroweakinos and electroweakino to Z are on shell~\footnote{It is possible that the electroweakinos can decay via off shell Z boson, however we do not consider this possibility and associated kinematics in this work.}. The missing energy distribution for direct electroweakino production or production via heavy Higgs is very similar. This is because the Z in the final state is almost produced at rest, this is reflected in the $\Delta R_{\ell\ell}$ distribution, which peaks for maximum values. When the mass difference between $\lsptwo, \lspone$ is increased drastically, for ($\lsptwo, \lspone$) = (400, 5), we find a very different situation. Here, the missing energy is generally harder, the corresponding $\Delta R_{\ell\ell}$ distribution shows a more collimated pair of leptons compared to the susy counterpart. Finally, we change the situation completely and consider the largest $\lsptwo$ mass allowed for on shell Higgs, which leads to combination ($\lsptwo, \lspone$) = (700, 5). In this case, the MET generated by the SM mediated process is somewhat harder than the corresponding Higgs mediated process. We thus see an interesting complementarity between kinematic distributions generated by SM mediated and Higgs mediated processes. Also, we present a discussion of the difference between parton level and detector level kinematic feature of $\met$ in Appendix~\ref{sec:appendixB}.


Finally, to conclude this discussion, we demonstrate in figure~\ref{fig:kin_epsilon_lt} two more kinematic variables which are derived using the basic measurable quantities in the dilepton plus missing energy final state events. The two quantities are $L_T$ which is the scalar sum of lepton $p_T$ and missing energy of the event and the $\xi$ defined by $\xi = |p_{T,\ell\ell}-\slashed{E}_T|/p_{T,\ell\ell}$~\cite{Barman:2016kgt, Sirunyan:2017qfc} where $p_{T,\ell\ell}$ is the transverse momentum of the dilepton system. While $L_T$ should give us an indication of the presence of any resonance, the variable $\xi$ is an indicator of the momentum imbalance in the system. For the $L_T$ distribution, ($\lsptwo, \lspone$) = (400, 300) the $L_T$ is soft, it peaks around 200 -- 300 GeV, with a long tail, the distributions for the Higgs and SM mediated processes are similar. For the other two points which correspond to large mass difference between $\lsptwo$ and $\lspone$, the $L_T$ distribution exhibits a clear end point for Higgs mediated process over the SM mediated process. Finally, the $\xi$ distribution shows an interesting dependence on the mass difference, for the benchmarks with high mass difference, the distribution peaks for low values, while for small mass difference it peaks for high values. For ($m_{\lsptwo}, m_{\lspone}$) = (400, 300) GeV, the leptons are produced with low $p_T$ and therefore the final state $\met$ is generated from the jets. Correspondingly, the $\xi$ distribution peaks near 1 as opposed to other two benchmark points.

\subsection{LLCP kinematics}
\label{sec:llpdecaykine}
\begin{figure}[htb!]
\centering
\includegraphics[width  = 0.49\linewidth]{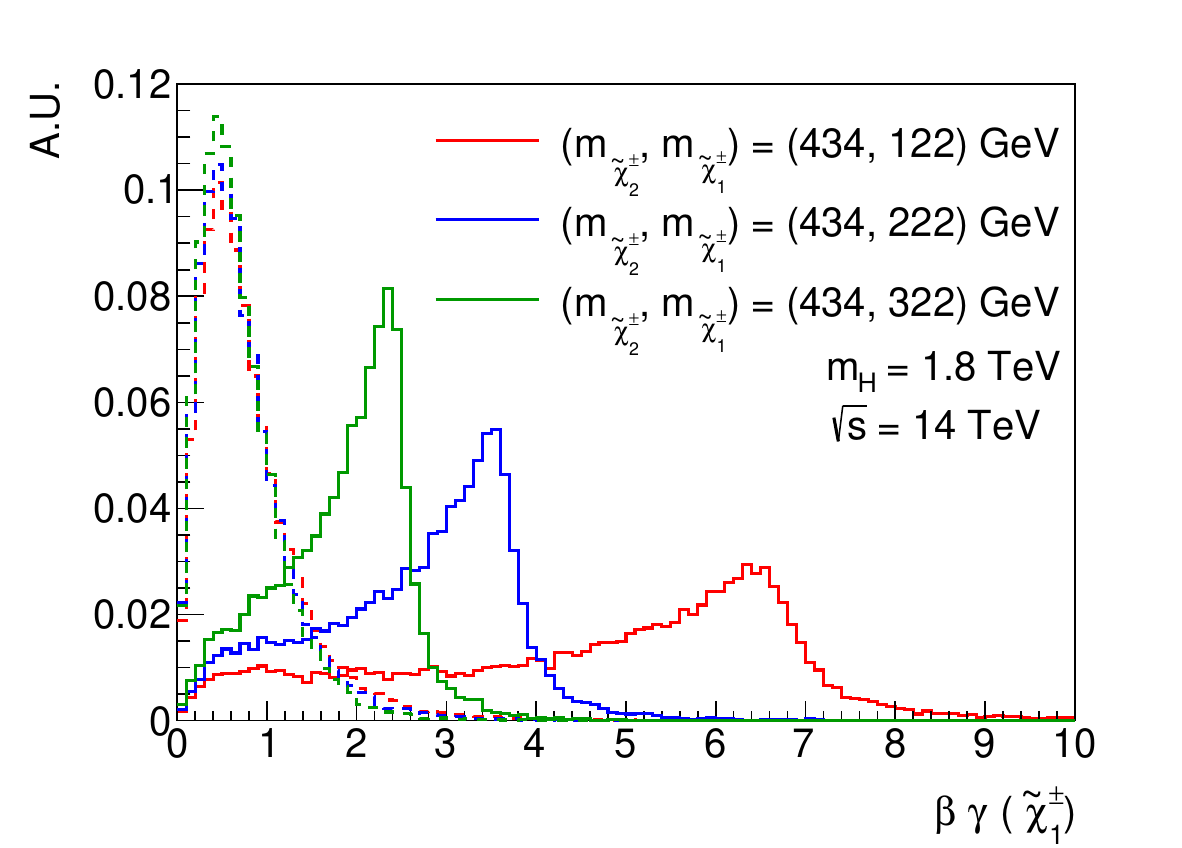}
\includegraphics[width  = 0.49\linewidth]{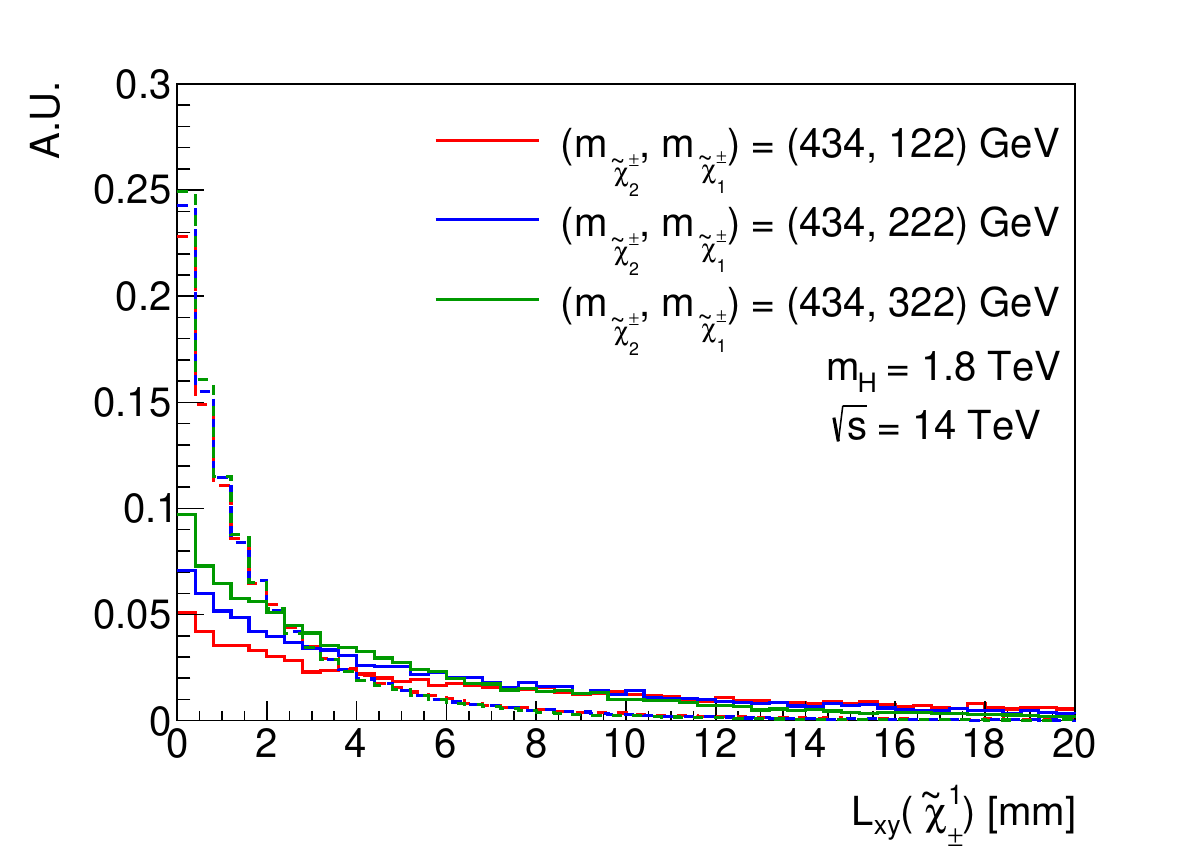}
\caption{The boost received by lightest chargino (left) and the resulting transverse displacement distribution (right) when chargino is produced in Drell-Yan processes (dashed lines) vs. chargino produced via decays of heavy Higgs (solid lines). The mass of heavy Higgs and heavier chargino is fixed at 1.8TeV and 434 GeV respectively, the mass of lighter chargino is varied, mass difference between $\chonepm, \lspone$ is fixed at 0.5 GeV. }
\label{llcp:kine}
\end{figure}

As discussed in section~\ref{sec:scan-result}, heavy Higgs decays to long lived charged particles (LLCP) can result in a substantial signal at the HL-LHC. In order to illustrate the salient kinematical differences between LLCP produced via decays of the heavy Higgs and via Drell-Yan process at the LHC, in this section we take a few benchmark points and analyse their kinematics. We use a similar setup as for the mono-Z final state in constructing the simplified model. Concretely, we generate the process $p p \to H \to \chonepm \chtwopm,~\chtwopm \to W^{\pm} \lspone,~\chonepm \to \lspone W^{*}$. The W boson has been decayed inclusively. For the Drell-Yan production of charginos, we simulate pair production of lightest chargino~\footnote{We neglect other production channels as $p p \to \chonepm \chonepm$ has the highest production cross section.}. We generate signal and hadronize with {\tt Pythia8}, we perform no detector simulation and present the kinematics at generator level. The heavy Higgs mass has been fixed to 1.8 TeV, and the mass of heavy chargino ($\chtwopm$) is fixed to 434 GeV. We then vary the mass of lighter chargino ($\chonepm$) (122, 222, 322 GeV) and fix the chargino lifetime to 3mm. For a chargino to be long lived, the mass difference between chargino and LSP must be small, we therefore fix the mass difference between $\chonepm, \lspone$ to be 0.5 GeV.

Figure~\ref{llcp:kine} (left) shows resulting chargino boost distributions without any cuts. It is clear that the boost received by the lightest chargino depends on the mass hierarchy between heavy Higgs, and chargino~\footnote{In principle it also depends on the mass of the heavy chargino, however it has been fixed for kinematic studies.}. The lightest chargino with a mass of $\sim 100$ GeV produced via decay of heavy Higgs (solid red line) is maximally boosted given the large mass difference between heavy Higgs and the chargino. This boost gets smaller and smaller as the mass of the chargino increases to $\sim 200$ GeV (solid blue) and $\sim 300$ GeV (solid green). For the chargino pair production via the Drell-Yan process (dashed lines), the boost is much smaller compared to the heavy Higgs decays case and the three masses show no significant differences. Such varied boost distribution results in different transverse decay lengths of chargino as depicted in Figure~\ref{llcp:kine} (right). It demonstrates that the chargino arising from heavy Higgs decays traverse longer distances through the detector.


\section{Collider analysis}
\label{sec:collider}
\begin{figure}[htb!]
\centering
\includegraphics[trim=0 260 0 70,clip,width=\textwidth]{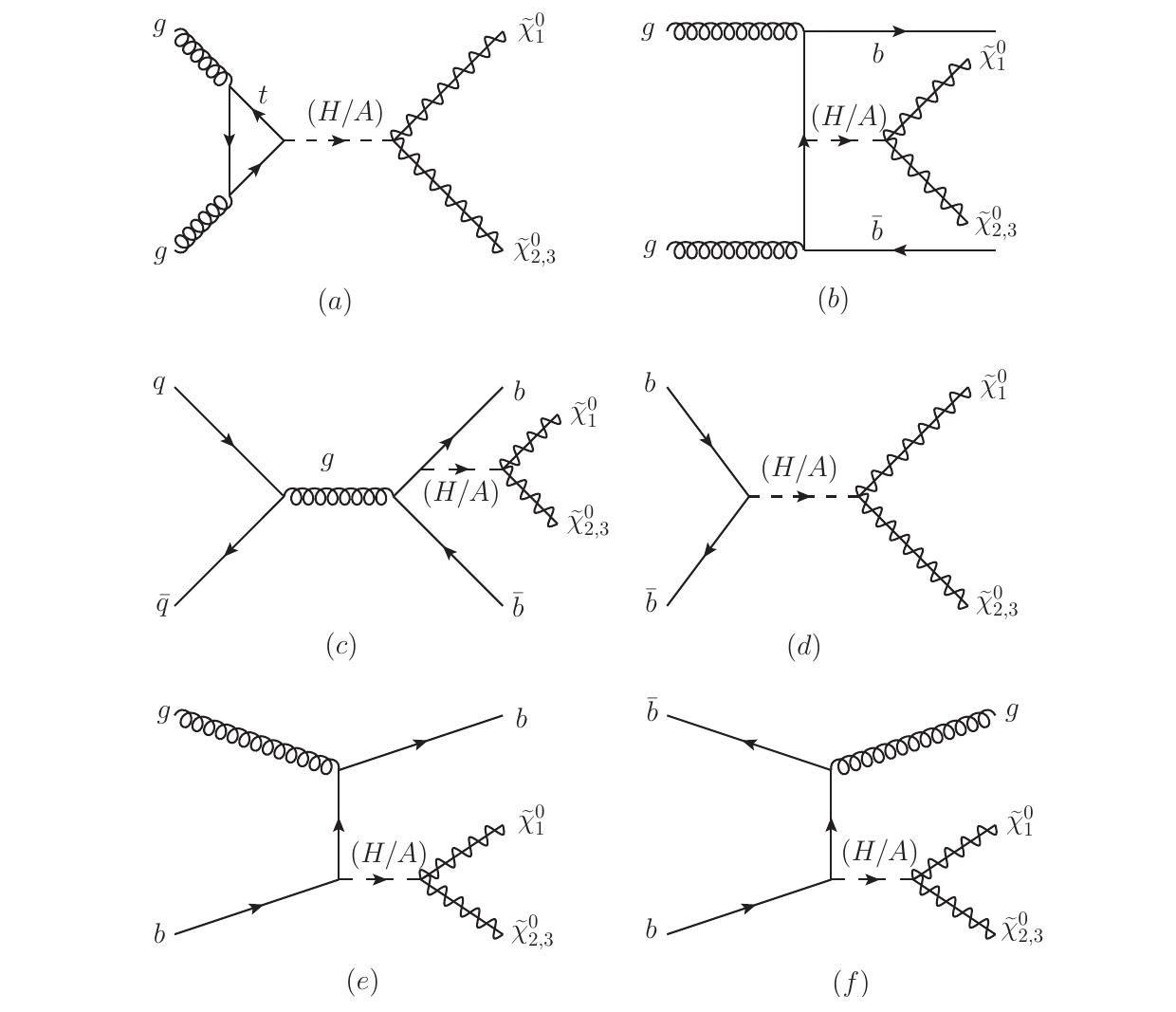}

\caption{\it The Feynman diagram of signal processes from (a) $gg$ fusion process; 4F (b) $gg\to b\bar{b}H$ and (c) $q\bar{q} \to b\bar{b}H$ production; 5F (d) $b\bar{b}\to H$ (LO), (e) $gb\to bH$ and (f) $b\bar{b}\to gH$ production. The  $\widetilde{\chi}_{ 2,3}^{0}$ then decays to $\widetilde{\chi}_{ 2,3}^{0} \to \lspone+(Z/h)$.}
\label{FD:sig}
\end{figure}

\begin{figure}[htb!]
\centering
\includegraphics[trim=0 400 0 70,clip,width=\textwidth]{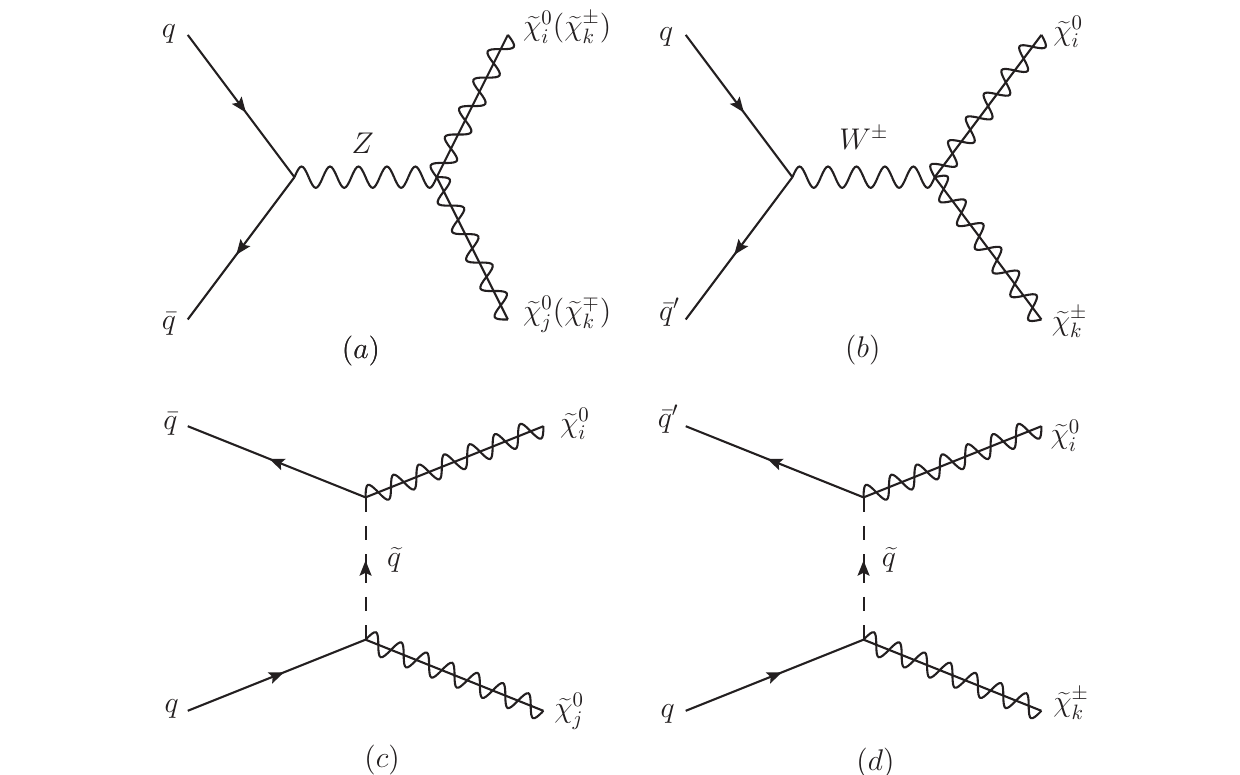}
\caption{\it The Feynman diagram of susy background production \textit{e.g} $pp\to \lspi \lspj$ ($\chkpm\chkmp$) and $pp\to  \lspi \chkpm$ with $i,j=1,2,3,4$ and $k=1,2$, in the s-channel ((a) and (b) respectively) and in the t-channel ((c) and (d) respectively). Here, the s-channel is the dominant production mode because the squarks in the t-channel are heavier for our chosen benchmark points.}
\label{FD:susybkg}
\end{figure}

As shown in the previous section, the mono-X signatures arising from heavy Higgs decays can be of an interest at the LHC. In this section, we will mainly focus on $pp\to H/A \to \lspone+(\widetilde{\chi}_{ 2,3}^{0}),~(\widetilde{\chi}_{ 2,3}^{0}) \to \lspone+(Z/h)$ which leads to mono-$Z$ and mono-$h$ final state. Furthermore, the Z and h bosons can decay to several different SM final states. Among them, we choose 3 possible decay modes for our analysis, mainly in terms of cleanliness and/or larger branching ratio \textit{viz.} (a) $Z\to \ell\ell$, (b) $h\to b\bar{b}$ and (c) $h\to \gamma\gamma$ (Fig.~\ref{FD:sig}~\footnote{JaxoDraw~\cite{Binosi:2008ig} has been used to generate all the Feynman diagrams in this paper.}), which leads to $\ell\ell+\met$, $b\bar{b}+\met$ and $\gamma\gamma+\met$ final states respectively. In order to uncover such a signal at colliders, it needs to be discriminated against not just the SM backgrounds but electroweakino production via SM mediators which leads to the same finals states, potentially with different kinematics. To this end, we consider two kinds of background processes \textit{viz.} (i) the usual SM background production \textit{e.g.} $pp\rightarrow VV,~t\bar{t},~Vh,~t\bar{t}h,~t\bar{t}V$ etc. where $V$ denotes $W^\pm$ and $Z$ boson, and, (ii) susy production via SM mediators, which leads to the same final state as the signal processes. The second kind of background \textit{i.e.} susy production, mainly comes from the direct production of the electroweakino pairs via the $Z$ and $W$ bosons in the s-channel or via squarks in the t-channel \textit{e.g.} $pp\to \lspi \lspj, \lspi \chkpm$ and $\chkpm\chkmp$, where $i,j=1,2,3,4$ and $k=1,2$ as depicted in Fig.~\ref{FD:susybkg}. We would like to mention here that these susy backgrounds have so far been not accounted for in the existing mono-$Z$ and mono-$h$ phenomenology arising from decays of heavy Higgs~\cite{Barman:2016kgt, Gori:2018pmk}. 

We generate SM backgrounds with generator level cuts as specified in the Appendix~\ref{sec:appendixC}, while the susy backgrounds are generated without any generator level cuts. 

In order to demonstrate possible optimisation of analysis, we choose the following two benchmark points (BPs)~\footnote{Since the LSP is lighter here, we check against CRESST-II\cite{Angloher:2015ewa} limits before choosing these benchmark points.} in the allowed MSSM parameter space, given in Table~\ref{tab:susy_bpt} where we also mention the branching ratios and cross sections relevant for our mono-$Z$ and mono-$h$ analysis. The common soft parameters for both of the BPs are:
\begin{equation} 
\begin{gathered}
M_1 = 5~{\rm GeV},~M_2 = 1.1~{\rm TeV},~\mu = 243.2~{\rm GeV},~M_3 = 2~{\rm TeV},\\
M_{\tilde{Q}_{1_L},\tilde{Q}_{2_L}} = M_{\tilde{u}_R,\tilde{d}_R,\tilde{c}_R,\tilde{s}_R} = M_{\tilde{e}_L,\tilde{\mu}_L,\tilde{e}_R,\tilde{\mu}_R} = 3~{\rm TeV},~M_{\tilde{Q}_{3_L},} = 4.9~{\rm TeV},\\
A_t =-3.7~{\rm TeV},~A_b =-1.1~{\rm TeV},~A_\tau =-1.5~{\rm TeV},~A_{e,\mu,u,d,c,s} = 0,\\
M_{\tilde{\tau}_L} = 961.5~{\rm GeV},~M_{\tilde{\tau}_R} = 1.1~{\rm TeV},~M_{\tilde{t}_R} = 5.9~{\rm TeV},~M_{\tilde{b}_R,} = 2~{\rm TeV}
\end{gathered}
\label{eq:prompt_common}
\end{equation} 

\begin{table}[htb!]
\begin{bigcenter}
\scalebox{0.8}{
\begin{tabular}{|c|c|c|c|c|c|} 
\hline
Benchmark                                                                    & Parameters                               & Mass (GeV)                 & Branching                                  & Processes  &  Cross-section (fb) \\ 
Points                                                                       &                                          &                            & Ratio (\%)
&  &\\ \hline\hline


       &                       & $m_{\lspone}=3.23$     & BR$(H (A) \to \lspone\lsptwo)=11.00~(16.11)$      & $gg \to H (A)$                  & $14.76~(29.84)$ \\ 
                                                                             
       & $M_A = 650~{\rm GeV}$ & $m_{\lsptwo}=251.17$   & BR$(H (A) \to \lspone \lspthree)=15.25~(9.46)$   & $b\bar{b} \to H (A)$            & $111.37~(111.86)$\\

       & $\tan\beta=10.8$      & $m_{\lspthree}=255.55$ & BR$(\lsptwo \to \lspone Z)=50.74$     & $pp \to b\bar{b}H (A)$          & $43.00~(43.20)$\\
       
      &     eq.(\ref{eq:prompt_common})    & $m_{\chonepm}=248.20$  & BR$(\lsptwo \to \lspone h)=49.26$     &  $4F~b\bar{b}H$ (NLO)    & $79.30$ \\

BP1    &        &  $m_{H}=650.31$ & BR$(\lspthree \to \lspone Z)=70.84$      & $pp \to \lspone\lsptwo$     & $18.30$ \\

       &                       &          &   BR$(\lspthree \to \lspone h)=29.16$ & $pp \to \lsptwo\lsptwo$     & $4.71\times 10^{-5}$ \\

       &                       &                        &  BR$(\chonepm \to \lspone W^\pm)= 100$  & $pp \to \chonepm\lsptwo$    & $217.90$      \\ 
       
       &                       &                        &  & $pp \to \lspone\lspthree$   & $19.70$      \\
       
       &                       &                        &                                       & $pp \to \lspthree\lspthree$ & $9.82\times 10^{-3}$ \\ 
       
       &                       &                        &                                       & $pp \to \chonepm\lspthree$  & $210.20$       \\  

       &                       &                        &                                       & $pp \to \chonepm\lspone$  & $27.08$       \\ 
       
       &                       &                        &                                       & $pp \to \lsptwo\lspthree$  & $107.00$       \\ 
       
       &                       &                        &                                       & $pp \to \chonep\chonem$  & $126.00$       \\ \hline \hline


       &                       & $m_{\lspone}=3.39$     & BR$(H (A) \to \lspone\lsptwo)=10.84~(15.31)$      & $gg \to H (A)$                  & $5.00~(12.03)$ \\ 
                                                                             
       & $M_A = 750~{\rm GeV}$ & $m_{\lsptwo}=251.18$   & BR$(H (A) \to \lspone \lspthree)=14.26~(9.43)$   & $b\bar{b} \to H (A)$            & $70.21~(70.41)$\\

       & $\tan\beta=12.1$      & $m_{\lspthree}=255.69$ & BR$(\lsptwo \to \lspone Z)=51.80$     & $pp \to b\bar{b}H (A)$          & $22.00~(22.06)$\\
       
BP2    &      eq.(\ref{eq:prompt_common})   & $m_{\chonepm}=248.34$  & BR$(\lsptwo \to \lspone h)=48.20$     & $4F~b\bar{b}H$ (NLO)    & $47.45$ \\

      &                         & $m_{H}=750.22$  & BR$(\lspthree \to \lspone Z)=70.00$     & $pp \to \lspone\lsptwo$     & $17.50$ \\

       &                       &                         &  BR$(\lspthree \to \lspone h)=30.00$   & $pp \to \lsptwo\lsptwo$     & $4.72\times 10^{-5}$ \\

       &                       &                        & BR$(\chonepm \to \lspone W^\pm)= 100$   & $pp \to \chonepm\lsptwo$    & $217.90$      \\ 
       
       &                       &                        & & $pp \to \lspone\lspthree$   & $19.40$      \\
       
       &                       &                        &                                       & $pp \to \lspthree\lspthree$ & $9.85\times 10^{-3}$ \\ 
       
       &                       &                        &                                       & $pp \to \chonepm\lspthree$  & $209.10$       \\ 
            
       &                       &                        &                                       & $pp \to \chonepm\lspone$  & $26.95$       \\ 
       
       &                       &                        &                                       & $pp \to \lsptwo\lspthree$  & $107.00$       \\ 
       
       &                       &                        &                                       & $pp \to \chonep\chonem$  & $126.00$       \\ \hline \hline

\end{tabular}
}
\caption{\it Two benchmark points to study the mono-$Z$ $+\met$ and mono-$h$ $+\met$ final state.}
\label{tab:susy_bpt}
\end{bigcenter}
\end{table}

These BPs are allowed by all the experimental constraints except the relic density constraint. The BPs are chosen such that the heavy Higgs bosons have a significant branching ratio into susy final states. Given the hierarchy of the bino, wino and higgsino mass parameters ( $M_1<\mu<M_2$), the LSP is primarily bino-like whereas the $\lsptwo$ and $\lspthree$ contains mostly the neutral higgsino components. Also, the $\chonepm$ contain the charged higgsino fields, while the heavy electroweakinos \textit{i.e.} $\lspfour$ and $\chtwopm$, becomes wino-like due to large value of $M_2$ parameter.

In the following subsections, we perform an optimised analysis for our chosen benchmark points for mono-$Z$ and mono-$h$ search channels. Besides, we divide our analysis in two parts, b-veto category where there is no extra b-tagged jet along with the final state particles and b-tag category where we demand the presence of an additional b-jet. We further analyse the b-tag analysis in two different ways. The first is via Santander matching of the 4F and 5F $b\bar{b}H$ production and the second is by generating the NLO $4F~b\bar{b}H$ production. We describe these two procedures in the following.

In case of the 4F and 5F matching, we generate the heavy Higgs signal events \textit{i.e.} $pp\to H\to \lspone \lsptwo$ and $pp\to H\to \lspone \lspthree$ with $\widetilde{\chi}_{ 2,3}^{0}$ decaying via $\widetilde{\chi}_{ 2,3}^{0} \to \lspone~+~Z,~Z\to \ell\ell$ and $\widetilde{\chi}_{ 2,3}^{0} \to \lspone~+~h,~(h\to b\bar{b}~{\rm and}~h\to \gamma\gamma)$, in three different production modes separately, namely, the gluon-gluon fusion (diagram (a) in Fig~\ref{FD:sig}), bbH fusion process in 4F scheme and the 5F scheme in the remaining figures. It is important to carefully generate and match the cross sections in the 4 flavour (4F) and five flavour (5F) schemes. The reason is as follows. The cross section in the 4F scheme at leading order (LO), receives contribution from two QCD processes, \textit{viz.} gluon fusion production, $gg\rightarrow b\bar{b}H$ and quark anti-quark annihilation, $q\bar{q}\rightarrow b\bar{b}H$ (diagram (b) and (c) respectively in Fig~\ref{FD:sig}). The heavy Higgs production in both of the cases are accompanied by two $b$-quarks. However, in case of collinear splitting of a gluon into two bottom quarks, the logarithmic terms in the 4F inclusive cross section becomes very large which has the form of $\sim \textrm{ln}(\frac{\mu_F}{m_b})$, where $\mu_F$ is the factorisation scale. Once these large logarithmic terms are absorbed in the parton distribution function (PDF) of the bottom quark, the theory remains perturbative. This is done by the re-summation of these terms at all orders in the perturbation theory which makes the basis of 5F scheme.  The heavy Higgs is produced in the 5F scheme at LO mainly via the QCD process, $b\bar{b}\rightarrow H$ (diagram (d) in Fig.~\ref{FD:sig}) with no extra parton in the final state. The processes where heavy Higgs is produced with a quark or gluon, become important when we demand an additional b-jets in the final state along with the heavy Higgs decay products, \textit{e.g}, $gb\to bH$ and $b\bar{b}\to gH$ (diagram (e) and (f) in  Fig.~\ref{FD:sig}). Also, the process $gg\rightarrow b\bar{b}H$ is the LO process in the 4F scheme which shows up in 5F scheme at the NNLO order. The cross section in these two scheme does not match when calculated upto a fixed order because the perturbative expansion is different in the two schemes. The cross section in the 4F scheme is known up to NLO accuracy in QCD~\cite{Dittmaier:2003ej, Dawson:2005vi, Dawson:2004wq, Dawson:2003kb}. On the other hand, the 5F scheme calculation is available up to NNLO in QCD~\cite{Harlander:2003ai}. The processes $gb\to bH$ and $b\bar{b}\to gH$ has been derived up to NLO order in QCD~\cite{Dawson:2004sh} and the electroweak (EW)~\cite{Dawson:2010yz}. Matching the two schemes thus removes the potentially overlapping part of the cross section and accounts for genuine bottom PDF inside the proton.

The inclusive cross-section in the two schemes agrees very well with an appropriate choice of factorisation and renormalisation scale~\cite{Maltoni:2003pn, Boos:2003yi, Plehn:2002vy}. The cross-section in these two schemes are multiplied by their respective weight factor and added together to get the total inclusive cross-section of the $b\bar{b}H$ process. This is known as Santander matching~\cite{Harlander:2011}. The weight factors for such procedure depend logarithmically on the heavy Higgs mass ($m_H$) and the bottom quark mass ($m_b$)~\footnote{The pole mass of the bottom quark, $m_b=4.78$ GeV is used which enters in the logarithmic terms during re-summation.}. The matched cross-section is given by: $$\sigma^{\text{matched}}=\frac{\sigma^{\text{4FS}}+w\sigma^{\text{5FS}}}{1+w},$$ where the $\sigma^{\text{4FS}}$ and $\sigma^{\text{5FS}}$ are the cross sections in the 4F and 5F scheme respectively, and the weight factor $w$, is defined as: $$w=\text{ln}\frac{m_H}{m_b}-2.$$

We use {\tt MadGraph-2.6.5}~\cite{Alwall:2014hca} to generate the signal and SM, susy background events at tree level (LO).  For showering and hadronisation of the signal and background events, we use {\tt Pythia-8}~\cite{Sjostrand:2014zea} with {\tt CTEQ6l1} PDF. We furthermore process the events through {\tt Delphes-3.4.1}~\cite{deFavereau:2013fsa} to take into account the detector effects. We use the default ATLAS card with updated b-tagging efficiency and mis-tagging efficiency of a light or c-jet as a b-jet as a function of the jet transverse momentum, $p_T$~\cite{Sirunyan:2017ezt}. Jets are reconstructed using {\tt FastJet-3.2.1}~\cite{Cacciari:2011ma}  with anti-kt algorithm with $\Delta R = 0.4$. The NNLO cross-section for the signal production from gluon fusion and $b\bar{b}H$ process in 5F scheme are calculated using {\tt SusHi-1.6.1}~\cite{Harlander:2016hcx}, which calculates Higgs cross sections in gluon fusion and bottom-quark annihilation at hadron colliders in the SM and various BSM models. It is important to note that  {\tt SusHi} does not generate events, but only predicts differential or integrated cross section upon user request. We thus use these cross sections for normalisation purposes only. Furthermore, as described above, we take care of matching between 4F and 5F scheme using Santander matching. For the 4F scheme $b\bar{b}H$ process we use the cross section at LO from Madgraph~\footnote{The cross section in the 4F scheme of $b\bar{b}H$ process is calculated in MSSM via Madgraph by using the SLHA file corresponding to the benchmark point as parameter card.}. We use {\tt Prospino2}~\cite{Beenakker:1996ed} to calculate the cross section of the susy backgrounds at NLO. In Table~\ref{tab:susy_bpt}, we show the production cross section and the branching ratio of the relevant susy final states. In principle, there are other susy processes \textit{viz.} $pp\to \chonepm \lspfour, ~\chonepm \chonemp,~\chtwopm \chtwomp$, which can also contribute to the background, however owing to the negligible production cross section and branching ratios for these processes, we exclude them from our analysis.

\begin{figure}[htb!]
\centering
\includegraphics[scale=0.6]{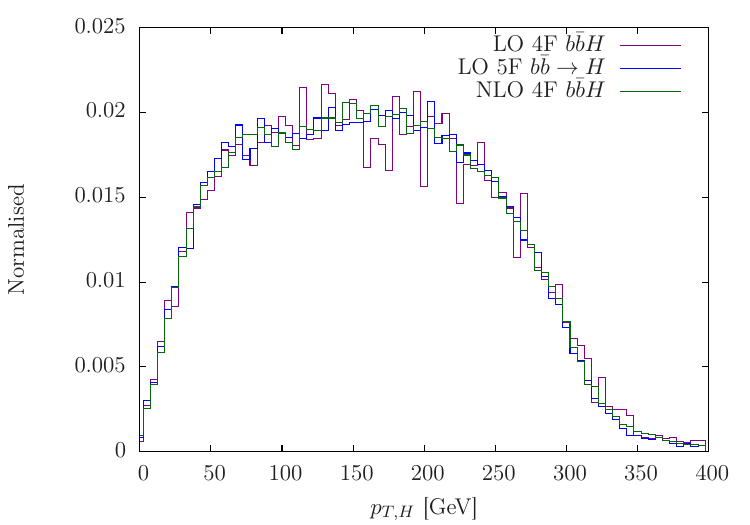}\includegraphics[scale=0.6]{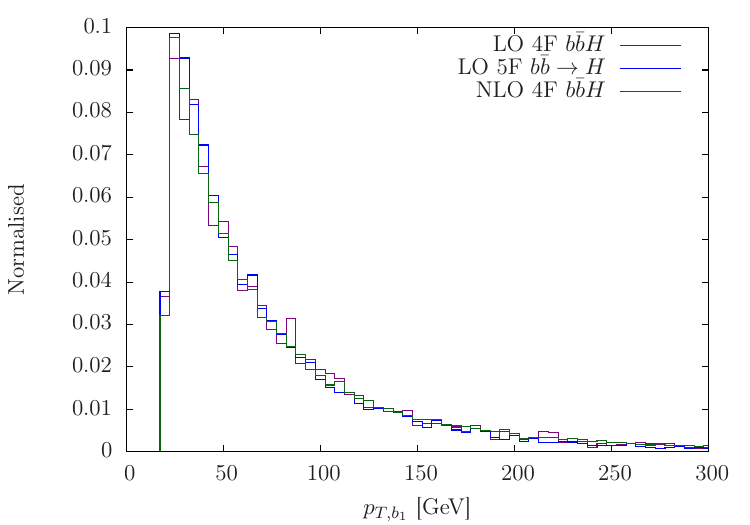}
\caption{The normalised distributions of $p_{T,H}$ and $p_{T,b_1}$ for $m_H=650$ GeV. The plots represent a comparison between LO 4F, 5F $b\bar{b}H$ and NLO 4F $b\bar{b}H$ process.}
\label{fig:nlo4Fcompare}
\end{figure}

In case of b-tag analysis, since we generate the $5F$ $b\bar{b}\to H$ process at tree level (LO), the extra b-jet comes from the parton shower which may not be in the hard regime. To simulate the kinematic effects of such hard b-tagged jet in the final state, we also perform an analogous analysis by generating the $4F$ $b\bar{b}H$ process at next-to-leading order (NLO) without merging the 4F and 5F $b\bar{b}H$ production. We closely follow \cite{Wiesemann:2014ioa} to generate the NLO $4F$ $b\bar{b}H$ process for our benchmark points. In Fig.~\ref{fig:nlo4Fcompare}, we compare the kinematic distributions of Higgs boson and the hardest b-jet $p_{T}$ in the LO 4F and 5F $b\bar{b}H$ process, and also the NLO $4F~b\bar{b}H$ production.   These distributions do not change for 4F and 5F $b\bar{b}H$ process, and the NLO $4F~b\bar{b}H$ production, in the kinematic region which corresponds to our signal region~\footnote{To validate our signal generation, we have also cross-checked the kinematic distributions of $p_{T,h}$ and $p_{T,b_1}$ for the Standard Model (SM) $125$ GeV Higgs boson. The differences in kinematic distributions for b-jet $p_T$ obtained via 4F LO and 4F NLO samples seem to decrease with the increasing mass of the Higgs boson.}.   

\subsection{The mono-Z final state}
\label{sec:monoz}
In this section, we analyse the di-lepton $+\met$ channel coming from the mono-Z final state \textit{viz.} $pp\to H\to \lspone~+~\widetilde{\chi}_{ 2,3}^{0} \to \lspone~+~(\lspone~+~Z,~Z\to \ell\ell)$ in the context of the future HL-LHC run with center of mass energy, $\sqrt{s} = 14$ TeV and an integrated luminosity of $3~{\rm ab}^{-1}$. Here, $\ell$ refers to electron, muon and tau.   
\subsubsection{The $\ell\ell+\met$ channel}
\label{sec:llmet}
We have discussed at the beginning of the collider analysis that our proposed signal needs to be discriminated against the SM as well as susy background processes. We therefore generate the following six susy backgrounds depending upon the production cross-section as given in Table~\ref{tab:susy_bpt}, \textit{viz.} $pp\to \lspone \widetilde{\chi}_{ 2,3}^{0},~\chonepm \widetilde{\chi}_{ 2,3}^{0},~\chonepm\chonemp~\text{and}~\lsptwo\lspthree$. The $pp\to \lspone \widetilde{\chi}_{ 2,3}^{0}$ and $pp\to \chonepm \widetilde{\chi}_{ 2,3}^{0}$ backgrounds are generated with $\widetilde{\chi}_{ 2,3}^{0}$ decaying via Z boson where leptons contain $e,\mu$ and $\tau$~\footnote{We include $\tau$ while simulating both signal and background processes.}.  For the $pp\to \chonepm\chonemp$, the $\chonepm$ decays to W boson and LSP.  The W boson further decays to lepton ($e,\mu$ and $\tau$) and a neutrino. In case of $pp\to \lsptwo\lspthree$, there are two Z bosons coming from the neutralinos, where one of them decays into leptons and the other decays into neutrinos which gives rise to di-lepton $+\met$ final state. 

The dominant SM backgrounds contributing to this channel are $ZZ$ and $WZ$. We generate them up to additional 3 jets for the $ZZ$ background and 2 jets for the $WZ$ background, matched via MLM scheme~\cite{Mangano:2006rw}. This extra jet contains gluon, light quarks, c-quark and bottom quark. The next dominant backgrounds are $VVV$ where $V=W,Z$ boson and $t\bar{t}Z$. These backgrounds are generated with no extra jet in the final state. In addition, we simulate the $t\bar{t}$ background where the $W$ bosons coming from top-quark are decayed leptonically. There are other sub-dominant background processes like Drell-Yan production, $WW$, $t\bar{t}h$ and $t\bar{t}W$. In Drell-Yan production, the leptons are produced from a Z boson or an off-shell photon, \textit{i.e.} $pp\to Z/\gamma^*\to \ell\ell$ where $\ell$ contains electron, muon and tau lepton. (For details see Appendix~\ref{sec:appendixC}). We divide our analysis into b-veto and b-tag category in the following. 

\subsubsection*{A. b-veto category}
\label{sec:bvetoll}

We select events containing exactly two isolated~\footnote{We define an isolated electron (muon) as a lepton candidate where the fraction of energy deposited within a cone of $\Delta R < 0.5$ is less than 12\% (25\%) of the lepton $p_T$.}, same flavour and opposite sign leptons (electron or muon) with transverse momentum, $p_{T,\ell} > 20$ GeV and pseudorapidity, $|\eta_\ell|<2.47~(2.5)$ for electron (muon). We require the invariant mass of dilepton system, $ 76 < m_{\ell\ell} < 106$~\footnote{This will reduce the contamination from the backgrounds where leptons come from different sources rather than Z boson in signal event.} and the di-lepton system should be within the pseudorapidity range of $|\eta|<2.5$. We further veto events containing b-jets ($N_b=$ Number of b-tagged jets in the final state) with $p_{T} > 20$ GeV, $|\eta|<2.5$. To suppress backgrounds with high jet multiplicity further, we restrict the maximum number of light jets ($N_j$) with $p_{T,j} > 20$ GeV, $|\eta_j|<4.5$ in an event to be one. These correspond to basic trigger cuts. Next, we define more sophisticated variables over which we optimise our signal and background events.

In case of the signal events, the $\widetilde{\chi}_{ 2,3}^{0}$ comes from the decay of heavy Higgs boson and hence has nontrivial transverse momentum, $p_{T}$. For the benchmarks considered here, the mass difference between $\widetilde{\chi}_{ 2,3}^{0}$ and the decayed particle, $\lspone$ is larger than $m_{Z}$. Therefore, the $Z$ boson from $\widetilde{\chi}_{ 2,3}^{0}$ decay can be boosted, giving rise to collimated leptons with small $\Delta R_{\ell\ell}$. A similar feature appears for the susy backgrounds where $\widetilde{\chi}_{ 2,3}^{0}$ decays to $Z$ boson and $\lspone$, \textit{viz.} $pp\to \lspone \widetilde{\chi}_{ 2,3}^{0},~\chonepm \widetilde{\chi}_{ 2,3}^{0}~\text{and}~\lsptwo\lspthree$. Contrary to this observation, for the SM backgrounds, the leptons in the final state are not boosted, \textit{viz.} in diboson background the Z bosons are produced at rest. We also apply $\Delta \phi_{\ell\ell,\met}>2.1$ for both the benchmark point, which is the azimuthal angle separation between the di-lepton system and the missing transverse energy as an additional discriminating variable. This variable peaks at around $\pi$ for the signal events where the source of missing energy and the di-lepton system are going back-to-back, while it is distributed over the whole region for background events. Additionally, we use a large missing transverse energy cut of $\met > 180~(210)~{\rm GeV}$ for the 1st (2nd) benchmark point.

The final discriminating variable arises from genuine imbalance in the missing energy and visible system distribution created in different signal and background samples. We construct the following kinematic variable, \textit{viz.} $\xi$. We define $\xi$ as, $$\xi = \frac{|p_{T,\ell\ell}-\met|}{p_{T,\ell\ell}},$$ which is a measure of momentum imbalance in the system. For the signal and susy backgrounds, the di-lepton system is against $\met$, which leads to $\xi\sim 0$. For the SM backgrounds, the $\met$ is very small as compared to the transverse momentum of the two lepton system, giving rise to $\xi\sim 1$. 

\begin{figure}[htb!]
\centering
\includegraphics[scale=0.37]{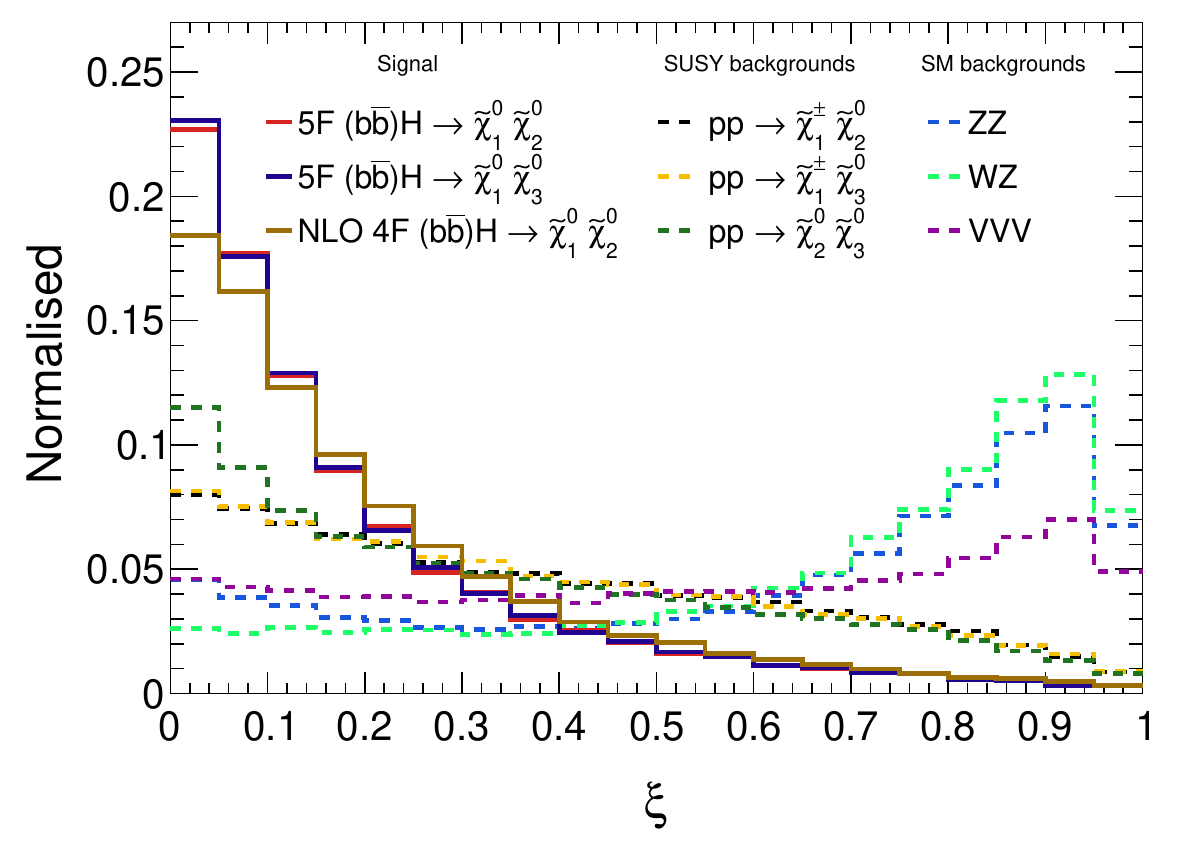}\includegraphics[scale=0.37]{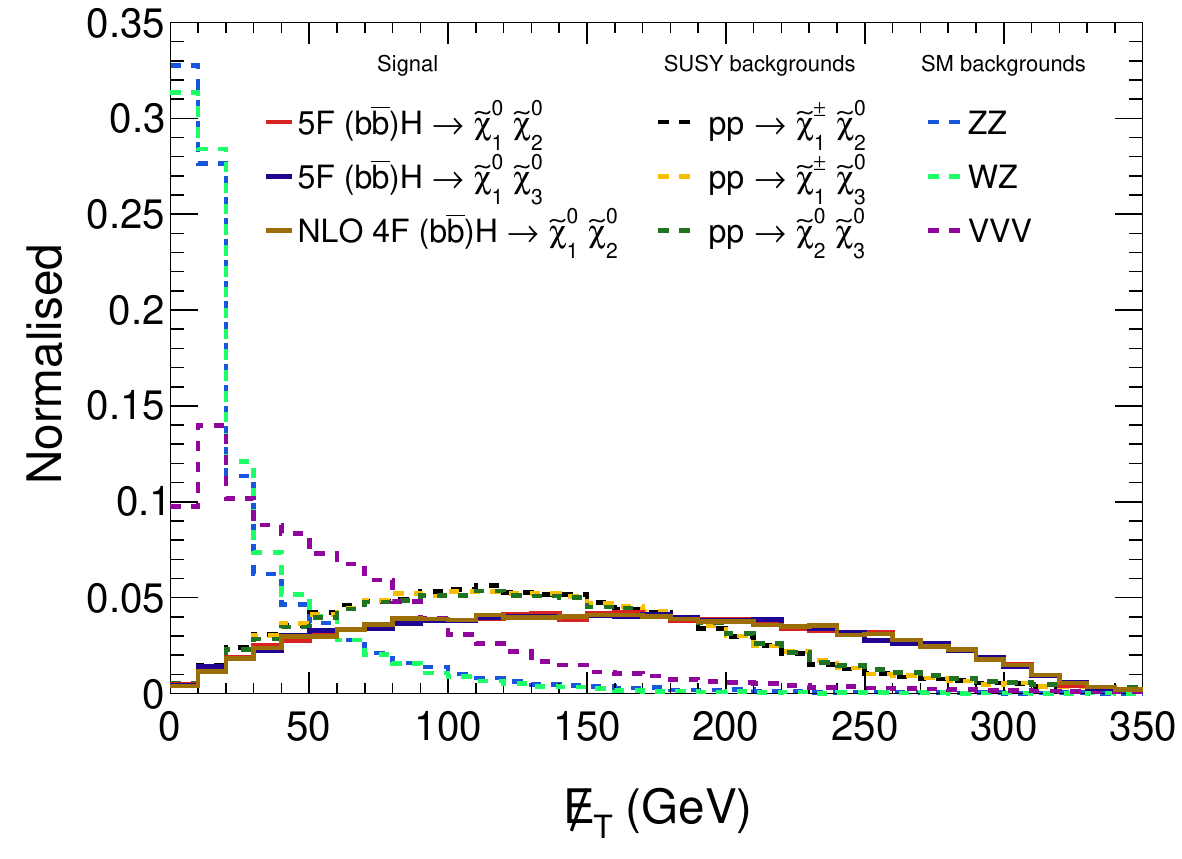}\\
\includegraphics[scale=0.37]{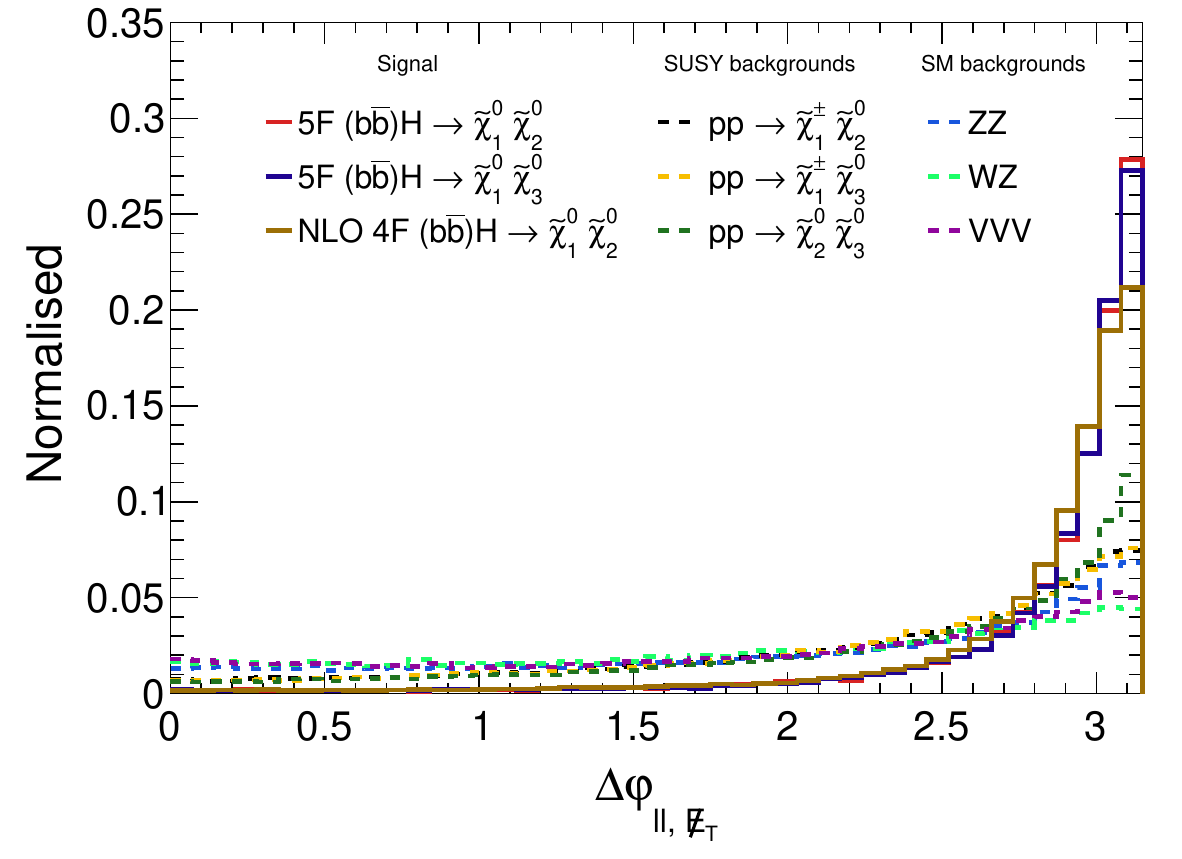}\includegraphics[scale=0.37]{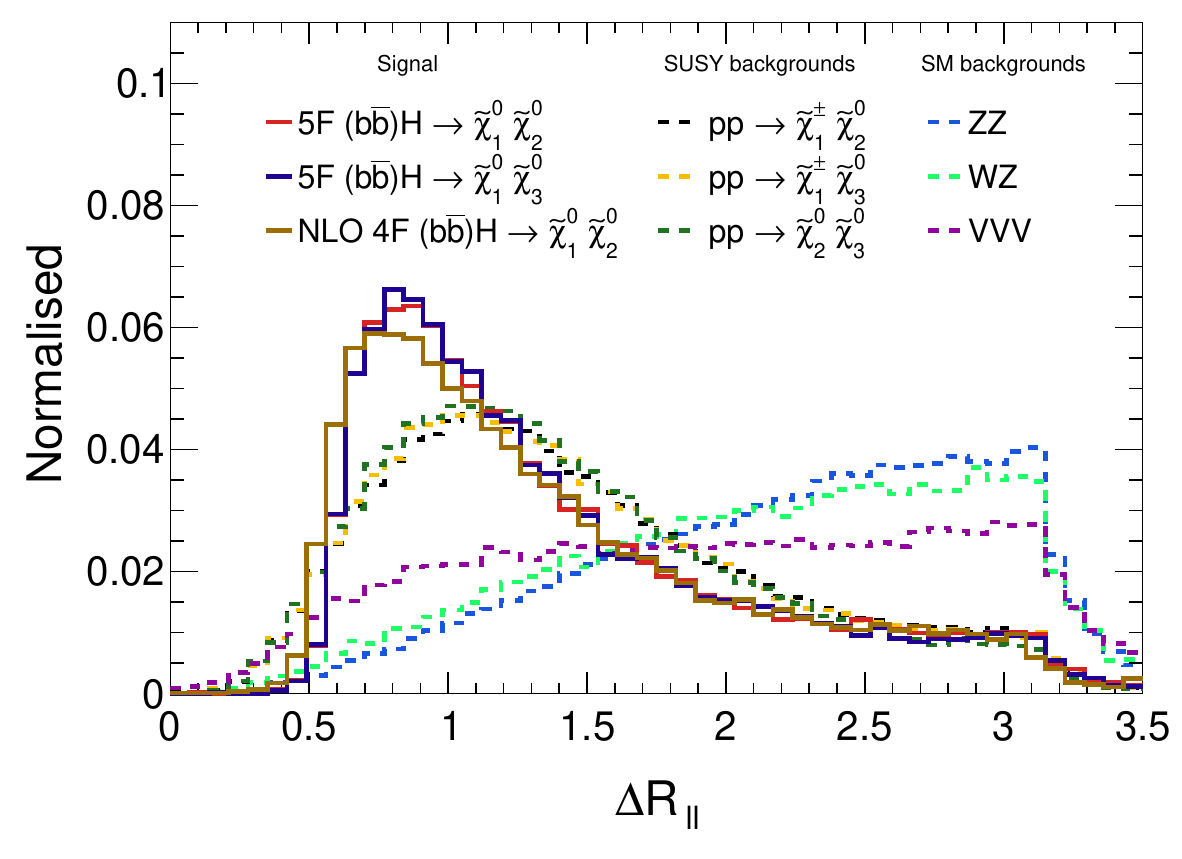}
\caption{The normalised distributions of $\xi$, $\met$, (top panel) $\Delta\phi_{\ell \ell,\met}$ and $\Delta R_{\ell \ell}$ (bottom panel) for the b-veto category of $\ell\ell+\met$ final state, after the basic trigger cuts corresponding to BP-1 scenario. We show the signal distributions for 5F $b\bar{b}\to H$ in $\lspone \lsptwo$ and $\lspone \lspthree$ final state, and NLO 4F $b\bar{b}H$ production in $\lspone \lsptwo$ final state, in solid line. Also, we show the susy as well as SM backgrounds in dashed line.} 
\label{fig:llplot}
\end{figure}

The discriminating power of $\met$ and $\Delta R_{\ell\ell}$ distributions for the resonant signal production and susy backgrounds was already seen in section \ref{sec:monozkin}. Normalised distributions for all four variables discussed above ($\met, \xi, \Delta R_{\ell\ell}, \Delta \phi_{\ell\ell,\met}$) after basic trigger cuts are shown in figure~\ref{fig:llplot}. They show the differential distributions of signal and background processes with respect to the corresponding discriminating variable. It can be clearly seen that there are three distinct classes of distributions, one corresponding to SM processes (dashed blue, green, red lines), second corresponding to susy backgrounds (dashed dark green, yellow, black lines) and finally the signal distributions (solid red, purple lines). These three classes of processes have different features in corresponding variables and it shows that an optimised analysis will be capable of discriminating among the three.

\begin{table}
\begin{center}
\begin{tabular}{|c|c|}\hline 
\multicolumn{2}{|c|}{Selection cuts} \\ \hline\hline
BP 1                                & BP 2 \\ \hline\hline
\multicolumn{2}{|c|}{$2\ell$, $N_b = 0$} \\
\multicolumn{2}{|c|}{$76.0 < m_{\ell\ell} < 106.0$} \\
\multicolumn{2}{|c|}{$|\eta_{\ell\ell}| < 2.5$} \\
\multicolumn{2}{|c|}{$N_j \leq 1$} \\ \hline
$\Delta R_{\ell \ell} < 1.3$        & $\Delta R_{\ell \ell} < 1.5$ \\
$\Delta\phi_{\ell \ell,\met} > 2.1$ & $\Delta\phi_{\ell \ell,\met} > 2.1$ \\
$\met > 180~{\rm GeV}$              & $\met > 210~{\rm GeV}$ \\
$\xi < 0.4$                         & $\xi < 0.3$ \\ \hline
\end{tabular}
\caption{\it The optimised selection cuts for the cut-based analysis in the b-veto category of $\ell\ell+\met$ channel.}
\label{tab:llcuts}
\end{center}
\end{table}

Using the observations described before we do a cut-based analysis optimising for $\Delta R_{\ell \ell}$, $\Delta\phi_{\ell \ell,\met}$, $\met$ and $\xi$ variables in favour of the signal events. We show the optimised cuts chosen for the two benchmark points in Table~\ref{tab:llcuts} along with the applied fixed cuts. We would like to mention here that the cuts obtained after optimisation of signal significance for each kind of signal sample, \textit{viz.} gluon-gluon fusion, $4F$, $5F$ scheme and NLO $4F~b\bar{b}H$, are similar. In order to speed up the process of optimisation, we use only samples produced using $5F$ scheme and use the same optimisation for gluon-gluon fusion, $4F$ scheme and NLO $4F~b\bar{b}H$ production. As the kinematics of the final state is largely independent of the scheme used to produce the Higgs bosons, this approach is justified. In the following, we present our analysis in two parts, first, the matched 4F, 5F $b\bar{b}H$ scheme and second, NLO $4F~b\bar{b}H$ signal process. The final significance in case of matched 4F, 5F $b\bar{b}H$ signal is computed by adding the correctly calculated matched $4F + 5F$ scheme cross section with gluon-gluon fusion cross section. For NLO $4F~b\bar{b}H$ signal, we add the NLO $4F~b\bar{b}H$ cross section with the gluon fusion cross section to get the final signal significance. We show the cut flow table with $5F$ $b\bar{b}\to H$ signal production along with the dominant backgrounds in Table~\ref{tab:llcutflow} after applying all the cuts in sequence. This demonstrates relative importance of each cut in reducing the backgrounds. In Table~\ref{tab:llyield}, we display the susy and SM background yields at the HL-LHC after the cut-based analysis. Here, we have checked that the sub-dominant backgrounds have negligible contribution and therefore neglect them in our final analysis. It is to be noted that the susy backgrounds contribute $\sim 10\%$ to the total background events. The number of signal events for the gluon fusion production, matched $b\bar{b}H$ process and the NLO $4F~b\bar{b}H$ production at $\sqrt{s}=14$ TeV with 3 ab$^{-1}$ of integrated luminosity after the cut-based analysis are shown in Table~\ref{tab:llsigni1} and Table~\ref{tab:llsigni2}. Finally, we calculate the statistical significance as $S/\sqrt{B}$ where S is the signal yield which is defined as $\sigma(pp\to H\to \lspone\lsptwo)\times BR(\lsptwo \to \lspone Z)\times BR(Z\to \ell\ell)\times \mathcal{L}\times \epsilon$ for the process $pp\to H\to \lspone\lsptwo,~\lsptwo \to \lspone Z$ where $\epsilon$ is the signal efficiency and $\mathcal{L}$ is the integrated luminosity. Similarly, $B$ represents the total background yield after the cut-based analysis. We quote the final signal significance in Table~\ref{tab:llsigni1} which is $6.57$ for the first benchmark point and $4.66$ for the second benchmark point in case of 4F, 5F matched $b\bar{b}H$ production. However, these significances drop upon adding the systematic uncertainty~\footnote{The signal significance formula changes with adding $x\%$ systematic uncertainty: $S/\sqrt{B+(0.01\times x \times B)^2}$.}, \textit{viz.} adding a $5\%$ systematic, the $S/\sqrt{B}$ changes to $1.24$ and $1.12$ respectively. Therefore, these results are consistent with any extrapolation of the current LHC results. We also tabulate the signal yield from NLO $4F~b\bar{b}H$ process as well as the final signal significance in Table~\ref{tab:llsigni2}. The signal significance is lower as compared to the matched 4F, 5F $b\bar{b}H$ analysis which results from the low production cross-section of the NLO $4F~b\bar{b}H$ process. We would like to mention that the final signal efficiencies after the cut-based analysis are similar from the combined 4F, 5F $b\bar{b}H$ production and the NLO $4F~b\bar{b}H$ signal process. In case of benchmark point 1, the final signal efficiency is $0.039$ from the NLO $4F~b\bar{b}H$ and $0.038$ for the matched 4F, 5F $b\bar{b}H$ production. The signal efficiencies are $0.036$ and $0.04$ from the NLO $4F~b\bar{b}H$ process and matched 4F, 5F $b\bar{b}H$ production for benchmark point 2 respectively.

\begin{table}[htb!]
\begin{bigcenter}
\scalebox{0.8}{%
\begin{tabular}{|c||c||c|c|c|c|c||c|c|c|c|}
\hline
 &  \multicolumn{10}{c|}{Event yield with $3 \; \textrm{ab}^{-1}$ of integrated luminosity} \\ \cline{2-11}
 Cut flow   & Signal  (BP 1) & \multicolumn{5}{c||}{susy Backgrounds} & \multicolumn{4}{c|}{SM Backgrounds} \\
\cline{3-11} 
 & $b\bar{b} \to H~(5F)$ & $\chonepm \lsptwo$ & $\chonepm \lspthree$ &  $\lsptwo \lspthree$ & $\lspone \lsptwo$ & $\lspone \lspthree$ & $ZZ$ & $WZ$ & $VVV$ & $t\bar{t}Z$ \\\hline\hline
 
$2\ell$                          & $2496$  & $7685$  & $10433$  & $4796$  & $743$ & $1124$ & $1103323$  & $2067690$  & $16251$ & $12885$ \\

\hline
$m_{\ell\ell}$                   & $2391$  & $7036$  & $9520$   & $4586$  & $711$ & $1076$ & $1084794$  & $1952657$  & $9626$  & $9046$  \\

\hline
$|\eta_{\ell\ell}|$              & $2336$  & $6829$  & $9244$   & $4464$  & $683$ & $1035$ & $914415$   & $1687769$  & $8865$  & $8728$  \\

\hline
$N_j$                            & $1872$  & $2154$  & $2946$   & $1555$  & $570$ & $863$  & $401278$   & $552658$   & $1528$  & $172$ \\

\hline
$\Delta R_{\ell \ell}$           & $1064$  & $951$   & $1332$   & $721$   & $176$ & $280$  & $29356$    & $28790$    & $286$   & $48$ \\

\hline
$\Delta\phi_{\ell \ell,\met}$    & $1056$  & $826$   & $1151$   & $667$   & $172$ & $274$  & $23326$    & $17223$    & $234$   & $34$ \\

\hline
$\met$                           & $738$   & $390$   & $552$    & $378$   & $78$  & $125$  & $7620$     & $2389$     & $98$    & $9.5$ \\
 
\hline
$\xi$                            & $720$   & $362$   & $515$    & $362$   & $75$  & $121$  & $7181$     & $2137$     & $88$    & $8.8$ \\ \hline

\end{tabular}}
\end{bigcenter}
\caption{ \it The cut-flow table for the benchmark point $1$ in the b-veto category of $\ell\ell + \met$ mode with 5F signal production and dominant backgrounds.}

\label{tab:llcutflow}
\end{table}

\begin{table}[htb!]
\begin{bigcenter}
\scalebox{0.8}{
\begin{tabular}{|c||c|c|c|c|c|c||c|c|c|c|c||c|}
\hline
BPs & \multicolumn{12}{c|}{Background yield at $3 \; \textrm{ab}^{-1}$ after all cuts} \\ \cline{2-13}
    & \multicolumn{6}{c||}{susy Backgrounds} &\multicolumn{5}{c||}{SM Backgrounds} & Total \\ \cline{2-12}
    
    & $\chonepm \lsptwo$ & $\chonepm \lspthree$ & $\lsptwo \lspthree$ & $\lspone \lsptwo$ & $\lspone \lspthree$ & $\chonepm \chonepm$ & $ZZ$ & $WZ$ & $VVV$ & $t\bar{t}Z$ & $t\bar{t}$ & Background \\\cline{1-12}
Order          & \multicolumn{6}{c||}{NLO~\cite{Beenakker:1996ed}} & LO & LO & LO & NLO~\cite{Lazopoulos:2008de} & NNLO~\cite{ttbarNNLO} & \\ \hline\hline 

BP 1 & $362$ & $515$ & $362$ & $75$ & $121$ & $23$ & $7181$ & $2137$ & $88$ & $8.8$ & $\sim 0$ & $10873$\\
\hline

BP 2 & $227$ & $305$ & $254$ & $48$ & $73$  & $6.8$  & $4440$ & $1131$ & $58$ & $5.1$ & $\sim 0$ & $6548$\\
\hline
\end{tabular}}
\end{bigcenter}
\caption{ \it The background yield at $14$ TeV with $3 \; \textrm{ab}^{-1}$ of integrated luminosity after the cut-based analysis for the two benchmark points.}
\label{tab:llyield}
\end{table}

\begin{table}[htb!]
\begin{bigcenter}\scalebox{0.68}{
\begin{tabular}{|c||c|c|c|c|c|c|c|c|c|c|c|c|c||c|c|c|}
\hline
 \multicolumn{10}{|c|}{Using LO 4F and 5F $b\bar{b}H$ process} \\\hline
 \hline
BPs & \multicolumn{6}{c||}{Signal rates at $3 \; \textrm{ab}^{-1}$ after all cuts} & \multicolumn{3}{c|}{Significance calculation}\\
\cline{2-10}

    & \multicolumn{3}{c|}{$pp \to H/A \to \lspone \lsptwo$} & \multicolumn{3}{c||}{$pp \to H/A \to \lspone \lspthree$} & Total signal, & Total background, & \multicolumn{1}{c|}{Significance, \tiny{$\dfrac{S}{\sqrt{B}}$}} \\\cline{2-7}

    & $pp \to b\bar{b}H$ & $b\bar{b} \to  H$ & $gg \to H$ & $pp \to b\bar{b}H$ & $b\bar{b} \to  H$ & \multicolumn{1}{c||}{$gg \to H$} & \multirow{2}{*}{$S=\dfrac{N_{4F}+wN_{5F}}{1+w}$} & B & \multicolumn{1}{c|}{\multirow{2}{*}{\makecell{without (with $5\%$)\\ systematics}}}\\
    
    & ($4F$) & ($5F$) & ($ggF$) & ($4F$) & ($5F$) & \multicolumn{1}{c||}{($ggF$)} & & (From Table~\ref{tab:llyield}) & \multicolumn{1}{c|}{} \\\cline{1-7}
    
 Order & LO & \multicolumn{2}{c|}{NNLO~\cite{Harlander:2016hcx}} & LO & \multicolumn{2}{c||}{NNLO~\cite{Harlander:2016hcx}} & \multicolumn{1}{c|}{$+~N_{ggF}$}  &  & \multicolumn{1}{c|}{} \\\hline\hline

 BP 1 & $55$ & $312$ & $55$ & $73$ & $408$ & \multicolumn{1}{c||}{$61$} & $685$ & $10873$  & \multicolumn{1}{c|}{$6.57$ ($1.24$)} \\\hline

 BP 2 & $22$ & $194$ & $20$ & $30$ & $236$ & \multicolumn{1}{c||}{$20$} & $377$ & $6548$  & \multicolumn{1}{c|}{$4.66$ ($1.12$)} \\\hline


    
    


 
\end{tabular}}
\end{bigcenter}
\caption{\it The signal yield along with signal significance for the b-veto category of $\ell\ell + \met$ final state. $N_{ggF}$, $N_{4F}$, $N_{5F}$ are the total event yield from the gluon fusion, $4F$, $5F$ production processes respectively. The $N_{4F}$ and $N_{5F}$ are later added according to the Santander matching as described in the text.}
\label{tab:llsigni1}
\end{table}

\begin{table}[htb!]
\begin{bigcenter}\scalebox{0.72}{
\begin{tabular}{|c||c|c|c|c|c|c|c|c|c|c|c|c|c||c|c|c|}
\hline
 \multicolumn{10}{|c|}{Using NLO 4F $b\bar{b}H$ process} \\\hline
  \hline
 BPs & \multicolumn{6}{c||}{Signal rates at $3 \; \textrm{ab}^{-1}$ after all cuts} & \multicolumn{3}{c|}{Significance calculation}\\
\cline{2-10}

    & \multicolumn{3}{c|}{$pp \to H/A \to \lspone \lsptwo$} & \multicolumn{3}{c||}{$pp \to H/A \to \lspone \lspthree$} & Total signal, & Total background, & \multicolumn{1}{c|}{Significance, \tiny{$\dfrac{S}{\sqrt{B}}$}} \\\cline{2-7}

   & \multicolumn{2}{c|}{$pp \to b\bar{b}H$ } & $gg \to H$ & \multicolumn{2}{c|}{$pp \to b\bar{b}H$ } & \multicolumn{1}{c||}{$gg \to H$} & $S=$ & B &  \multicolumn{1}{c|}{\multirow{2}{*}{\makecell{without (with $5\%$)\\ systematics}}}\\
    
   & \multicolumn{2}{c|}{($4F$) } & ($ggF$) & \multicolumn{2}{c|}{($4F$) } & \multicolumn{1}{c||}{($ggF$)} &  \multicolumn{1}{c|}{$N_{4F}^{NLO}+~N_{ggF}$} & (From Table~\ref{tab:llyield}) &  \\\cline{1-7}
    
 Order & \multicolumn{2}{c|}{NLO~\cite{Bonvini:2016fgf}} & NNLO~\cite{Harlander:2016hcx} & \multicolumn{2}{c|}{NLO~\cite{Bonvini:2016fgf}} & \multicolumn{1}{c||}{NNLO~\cite{Harlander:2016hcx}}  &  &   & \multicolumn{1}{c|}{} \\\hline\hline

BP 1 & \multicolumn{2}{c|}{$130$ } & $55$ &  \multicolumn{2}{c|}{$164$} & \multicolumn{1}{c||}{$61$} & $410$ & $10873$  & \multicolumn{1}{c|}{$3.93$ ($0.74$)} \\\hline

BP 2 & \multicolumn{2}{c|}{$72$ } & $20$ &  \multicolumn{2}{c|}{$91$} & \multicolumn{1}{c||}{$20$} & $203$ & $6548$  & \multicolumn{1}{c|}{$2.51$ ($0.6$)} \\\hline
 
\end{tabular}}
\end{bigcenter}
\caption{\it The signal yield along with signal significance for the b-veto category of $\ell\ell + \met$ final state for NLO $4F~b\bar{b}H$ process. $N_{ggF}$, $N_{4F}^{NLO}$ are the total event yield from the gluon fusion NLO $4F~b\bar{b}H$ production processes respectively.}
\label{tab:llsigni2}
\end{table}
\subsubsection*{B. b-tag category}
\label{sec:btagll}
The event selection in this case contains at least one b-tagged jet along with the two isolated leptons in the final state. We perform the cut-based analysis in a similar way as discussed in the previous section and show the selection cuts in Table~\ref{tab:bllcuts}. The signal and background yields after the cut-based analysis is tabulated in Table~\ref{tab:bllyield} and Table~\ref{tab:bllsigni}. The signal significance improves here over the b-veto analysis due to reduced background composition resulting from the extra b-jet requirement. The signal significance for the benchmark point 1 is $10.48\sigma$ from matched 4F, 5F $b\bar{b}H$ signal and $7.17\sigma$ from the NLO $4F~b\bar{b}H$ production, without any systematic uncertainty. This large $10.48\sigma$ and $7.17\sigma$ significance reduces to $6.80\sigma$ and $4.65\sigma$ after consideration of $5\%$ systematic uncertainty respectively. There are two important points to be noted here. First, this reduction in significance demonstrates the importance of including systematic uncertainty. Second, demanding additional b-jet activity can improve the signal to background discrimination. Also, the final signal efficiencies in case of BP1 (BP2) after applying all the cuts are $0.022~(0.027)$ from NLO $4F~b\bar{b}H$ process and $0.016~(0.022)$ from the cross-section weighted sum of the matched $4F$ and $5F$ $b\bar{b}H$ production.

\begin{table}[htb!]
\begin{center}
\begin{tabular}{|c|c|}\hline 
\multicolumn{2}{|c|}{Selection cuts} \\ \hline\hline
BP 1                                & BP 2 \\ \hline\hline
\multicolumn{2}{|c|}{$2\ell$, $N_b \geq 1$} \\
\multicolumn{2}{|c|}{$76.0 < m_{\ell\ell} < 106.0$} \\
\multicolumn{2}{|c|}{$|\eta_{\ell\ell}| < 2.5$} \\
\multicolumn{2}{|c|}{$N_j \leq 1$} \\ \hline
$\Delta R_{\ell \ell} < 1.3$        & $\Delta R_{\ell \ell} < 1.3$ \\
$\Delta\phi_{\ell \ell,\met} > 2.1$ & $\Delta\phi_{\ell \ell,\met} > 2.3$ \\
$\met > 160~{\rm GeV}$              & $\met > 170~{\rm GeV}$ \\
$\xi < 0.4$                         & $\xi < 0.8$ \\ \hline
\end{tabular}
\caption{\it The optimised selection cuts for the cut-based analysis in the b-tag category of $\ell\ell+\met$ channel.}
\label{tab:bllcuts}
\end{center}
\end{table}

\begin{table}[htb!]
\begin{bigcenter}
\scalebox{0.9}{
\begin{tabular}{|c||c|c|c|c|c|c||c|c|c|c|c|c||c|}
\hline
BPs & \multicolumn{12}{c|}{Background yield at $3 \; \textrm{ab}^{-1}$ after all cuts} \\ \cline{2-13}
    & \multicolumn{6}{c||}{susy Backgrounds} &\multicolumn{5}{c||}{SM Backgrounds} & \multicolumn{1}{c|}{Total} \\ \cline{2-12}
    
    & $\chonepm \lsptwo$ & $\chonepm \lspthree$ & $\lsptwo \lspthree$ & $\lspone \lsptwo$ & $\lspone \lspthree$ & $\chonepm \chonepm$ & $ZZ$ & $WZ$ & $VVV$ & $t\bar{t}Z$ & \multicolumn{1}{c||}{$t\bar{t}$} & \multicolumn{1}{c|}{Background} \\ \hline\hline 

BP 1 & $29.15$ & $43.09$ & $75.17$ & $1.31$ & $1.92$ & $0.2$ & $137.05$ & $151.18$ & $12.40$ & $84.70$ & \multicolumn{1}{c||}{$14.62$} & \multicolumn{1}{c|}{$550.79$}\\
\hline

BP 2 & $37.28$ & $43.23$ & $80.16$ & $1.46$ & $1.89$  & $\sim 0$  & $137.05$ & $100.79$ & $14.80$ & $98.57$ & \multicolumn{1}{c||}{$10.96$} & \multicolumn{1}{c|}{$526.19$}\\
\hline
\end{tabular}}
\end{bigcenter}
\caption{ \it The background yield at $14$ TeV with $3 \; \textrm{ab}^{-1}$ of integrated luminosity after the cut-based analysis for the two benchmark points in b-tag category.}
\label{tab:bllyield}
\end{table}

\begin{table}[htb!]
\begin{bigcenter}\scalebox{0.76}{
\begin{tabular}{|c||c|c|c|c|c|c|c|c|c|c|c|c|c||c|c|c|}
\hline
BPs & \multicolumn{6}{c||}{Signal rates at $3 \; \textrm{ab}^{-1}$ after all cuts} & \multicolumn{3}{c|}{Significance calculation}\\
\cline{2-10}

    & \multicolumn{3}{c|}{$pp \to H/A \to \lspone \lsptwo$} & \multicolumn{3}{c||}{$pp \to H/A \to \lspone \lspthree$} & Total signal, & Total background, & \multicolumn{1}{c|}{Significance, \tiny{$\dfrac{S}{\sqrt{B}}$}} \\\cline{2-7}

    & \multirow{2}{*}{$4F$} & \multirow{2}{*}{$5F$} & \multirow{2}{*}{$ggF$} & \multirow{2}{*}{$4F$} & \multirow{2}{*}{$5F$} & \multicolumn{1}{c||}{\multirow{2}{*}{$ggF$}} & S & B & \multicolumn{1}{c|}{\multirow{2}{*}{\makecell{without (with $5\%$)\\ systematics}}}\\
    
    &  &  &  &  &  & \multicolumn{1}{c||}{} &  & (From Table~\ref{tab:bllyield}) & \multicolumn{1}{c|}{} \\\hline\hline

 BP 1 & $37.51$ & $128.15$ & $1.75$ & $44.23$ & $168.93$ & \multicolumn{1}{c||}{$2.27$} & $246.06$ & $550.79$  & \multicolumn{1}{c|}{$10.48$ ($6.80$)} \\\hline

 BP 2 & $23.22$ & $100.89$ & $0.91$ & $25.31$ & $125.28$ & \multicolumn{1}{c||}{$0.86$} & $184.14$ & $526.19$  & \multicolumn{1}{c|}{$8.03$ ($5.28$)} \\\hline
 
  \multicolumn{10}{|c|}{Using NLO 4F $b\bar{b}H$ process} \\\hline

 BPs   & \multicolumn{2}{c|}{$4F$ NLO} & $ggF$ & \multicolumn{2}{c|}{$4F$ NLO} & \multicolumn{1}{c||}{$ggF$} & Total signal,  & Total background, & \multicolumn{1}{c|}{{\tiny$\dfrac{S}{\sqrt{B}}$} without (with}\\
    
    & \multicolumn{2}{c|}{} & & \multicolumn{2}{c|}{} & \multicolumn{1}{c||}{} &  \multicolumn{1}{c|}{$S$} & B & \multicolumn{1}{c|}{ $5\%$) systematics}\\\hline

 BP 1 & \multicolumn{2}{c|}{$72.43$ } & $1.75$ &  \multicolumn{2}{c|}{$91.79$} & \multicolumn{1}{c||}{$2.27$} & $168.24$ & $550.79$  & \multicolumn{1}{c|}{$7.17$ ($4.65$)} \\\hline

 BP 2 & \multicolumn{2}{c|}{$54.2$ } & $0.91$ &  \multicolumn{2}{c|}{$67.13$} & \multicolumn{1}{c||}{$0.86$} & $123.1$ & $526.19$  & \multicolumn{1}{c|}{$5.37$ ($3.53$)} \\\hline
 
\end{tabular}}
\end{bigcenter}
\caption{\it The signal yield for two benchmark points along with signal significance for the b-tag category of $\ell\ell + \met$ final state.}
\label{tab:bllsigni}
\end{table}
\subsection{The mono-h final state}
\label{sec:monoh}
As we have seen in Table~\ref{tab:susy_bpt}, the electroweakinos can decay to the SM Higgs with a substantial rate, and can potentially be probed via the mono-h final state. The signal processes are $pp\to H\to \lspone~+~\widetilde{\chi}_{ 2,3}^{0}\to \lspone~+~(\lspone~+~h)$. Here, we consider two possible decay modes of the SM Higgs for our analysis, \textit{viz.} $h\to b\bar{b}$ and $h\to \gamma\gamma$, which gives rise to $b\bar{b}+\met$ and $\gamma\gamma+\met$ final state respectively. The $b\bar{b}+\met$ channel has substantial rate (BR($h\to b\bar{b}$)$\sim 0.58$) but this channel is contaminated by huge QCD backgrounds. While the $\gamma\gamma+\met$ channel suffers from small production rate because of the very small decay rate of $h\to \gamma\gamma$ (BR($h\to \gamma\gamma$)$\sim 2.27\times 10^{-3}$) however has the advantage of being clean in terms of the background contamination. In the next two subsections, we do a cut-based analysis for these two channels.
\subsubsection{The $b\bar{b}+\met$ channel}
\label{sec:bbmet}
As with the previous optimisation procedure, there are two kinds of backgrounds to this channel, \textit{viz.} the backgrounds arising from the SM processes and the susy backgrounds. In case of susy backgrounds, we generate samples associated with large production cross-section (Table~\ref{tab:susy_bpt}), \textit{viz.} $pp\to \lspone~\widetilde{\chi}_{ 2,3}^{0},~\chonepm~\widetilde{\chi}_{ 2,3}^{0}~\text{and}~\lsptwo~\lspthree$. Each of the $\widetilde{\chi}_{ 2,3}^{0}$ can decay into a Z or h and $\chonepm$ decays into a W boson. Based on this, there could be 5 possible final state configurations, \textit{viz.} $hh$, $Zh$, $ZZ$, $Wh$ and $WZ$. We combine all of these decay configuration while generating this background. 

In addition, we generate the dominant irreducible SM backgrounds, \textit{viz.} $Zb\bar{b}$ and $t\bar{t}$. For the $Zb\bar{b}$, we decay the $Z$ boson to neutrinos to get a similar final state as the signal event. We separately generate the $t\bar{t}$ background in fully leptonic mode, in semi-leptonic mode, and in hadronic mode. We also generate the other subdominant SM backgrounds, \textit{i.e.} $Zh$, $Wh$, $t\bar{t}h$, $t\bar{t}Z$ and $t\bar{t}W$. The $Zh$ background is generated upon merging with two jets in the final state by employing the MLM merging scheme. We generate the $Wh$ background by merging with one extra parton in the final state where the SM Higgs is decayed to pair of bottom quarks and the $W$ boson decays leptonically. We also generate the $t\bar{t}+X$ backgrounds where $X=h,Z,W$ with no extra jets in the final state. (See Appendix~\ref{sec:appendixC} for details.)
 
\subsubsection*{A. b-veto category}
\label{sec:bvetobb}

\begin{figure}[htb!]
\centering
\includegraphics[scale=0.37]{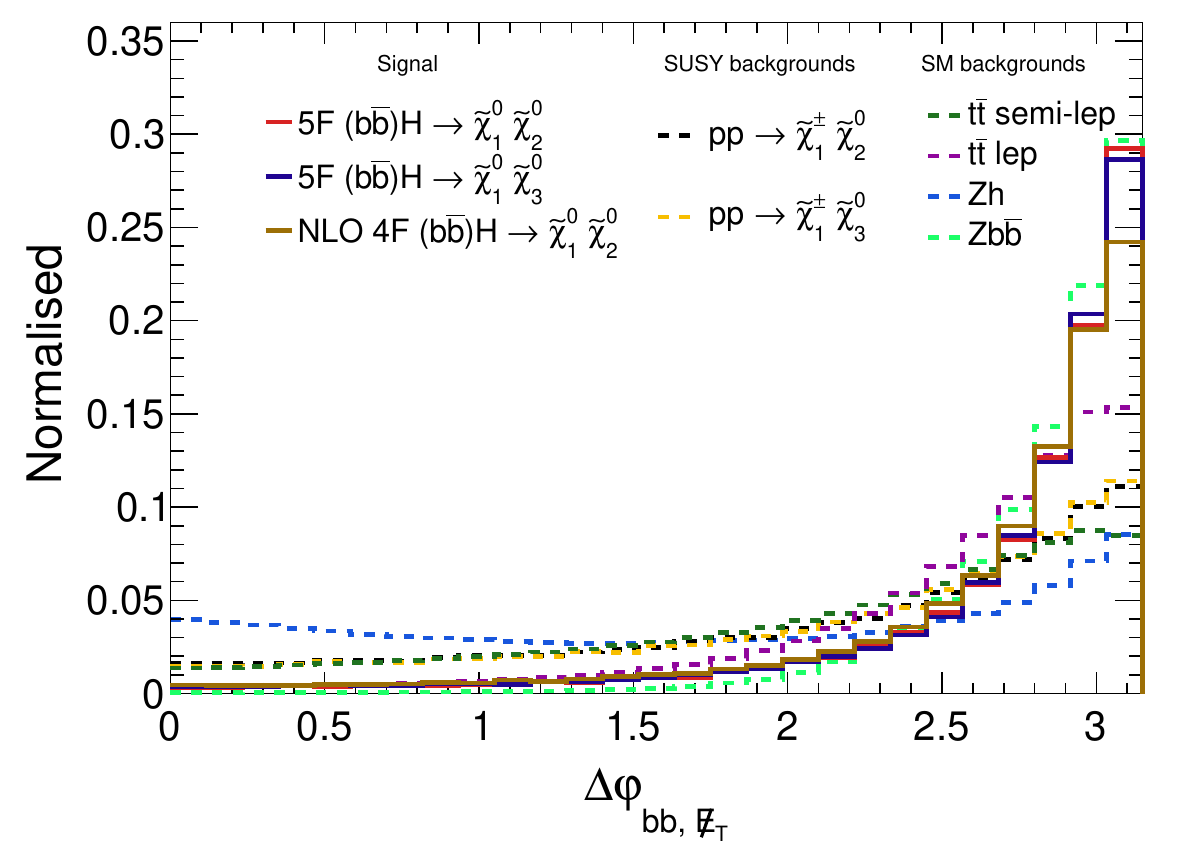}\includegraphics[scale=0.37]{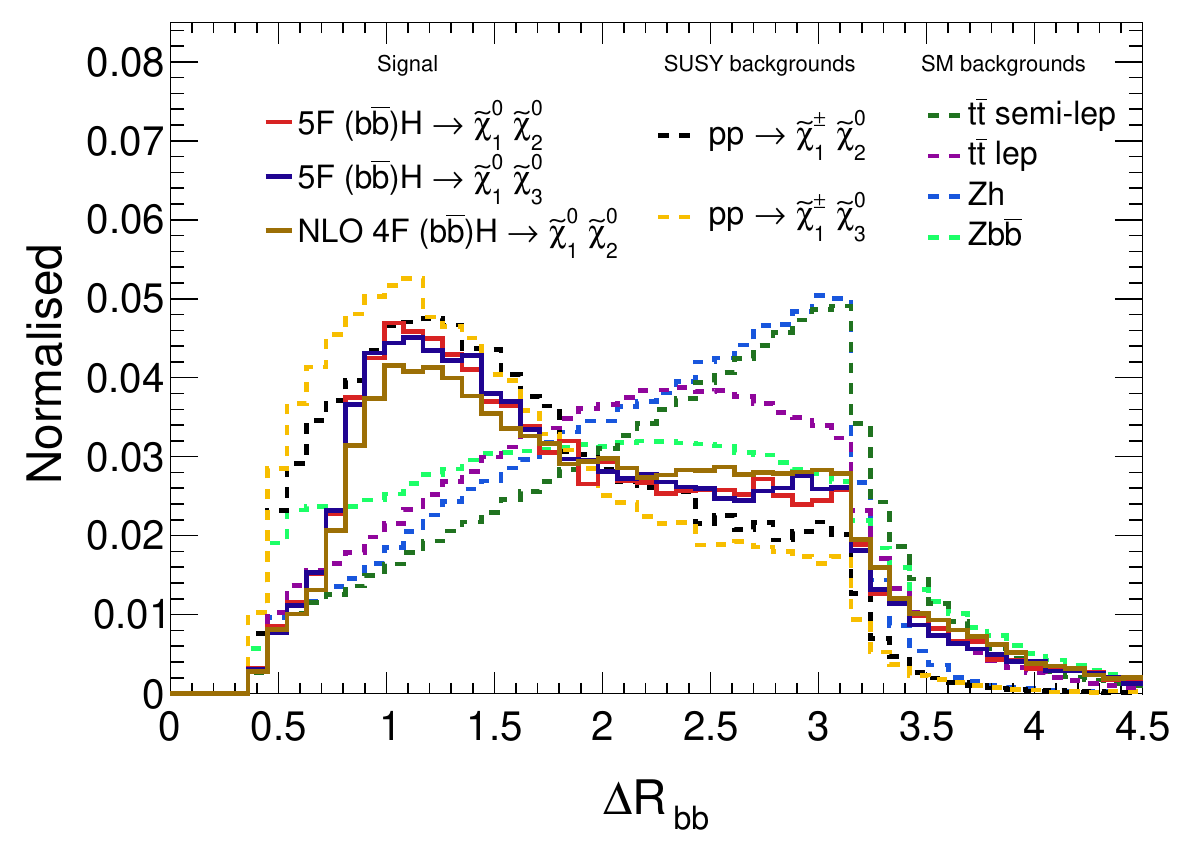}\\
\includegraphics[scale=0.37]{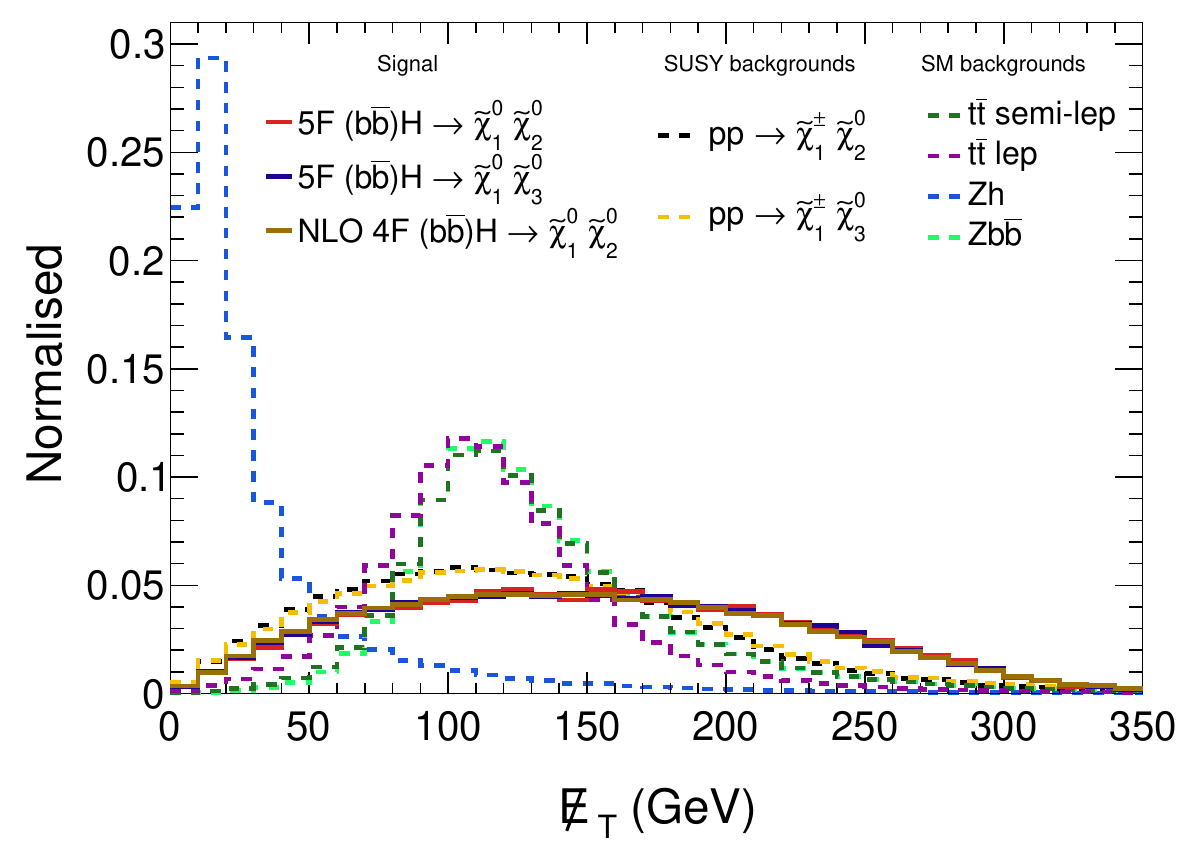}
\caption{\it The normalised distributions of $\Delta \phi_{bb,\met}$, $\Delta R_{bb}$ (top panel) and $\met$ (bottom panel) for the $b\bar{b}+\met$ final state after the basic trigger cuts for the BP 1 scenario.}
\label{fig:bbplot}
\end{figure}

The event selection for this analysis is governed by demanding that the event must have exactly two b-jets with $p_T > 20$ GeV and $|\eta| < 2.5$. We veto leptons ($N_\ell=$ Number of leptons in the final state) with $p_T > 20$ GeV and $|\eta| < 2.47~(2.5)$ (for $e$ ($\mu$)) in the final state to reduce the contamination from $t\bar{t}$, $Wh$ and $pp\to \chonepm~\widetilde{\chi}_{ 2,3}^{0}$ backgrounds where the final state contains leptons. The light jets are required to satisfy the transverse momentum of $p_T > 20$ GeV and pseudorapidity, $|\eta| <  4.5$. Finally, we construct the kinematic variables to perform the cut-based analysis. The invariant mass of the bottom pair will peak around the SM Higgs mass and we require $90 < m_{bb} < 130$ GeV~\cite{Adhikary:2017jtu}. We further require at most one light jet in the final state. These cuts define the set of basic trigger cuts in this analysis. The separation in the $\eta-\phi$ plane between the two b-jets measured as $\Delta R_{bb}$ is small for the signal event since the b-jets coming from SM Higgs are boosted, while this is not the case for the SM backgrounds. The missing transverse energy, $\met$ is large for the signal event as compared to the backgrounds as discussed previously (see section~\ref{sec:monoz}). Also, we construct the azimuthal angle separation, between the missing transverse momentum, $\met$ and the two b-jet system, \textit{viz.} $\Delta \phi_{b\bar{b},\met}$. We show the normalised distributions of these kinematic variables after trigger cuts for the signal and the dominant backgrounds in Fig.~\ref{fig:bbplot}. It can be seen that $\met$ and $\Delta R_{bb}$ are very strong discriminating variables between the signal and background processes.

\begin{table}
\begin{center}
\begin{tabular}{|c|c|}\hline 
\multicolumn{2}{|c|}{Selection cuts} \\ \hline\hline
BP 1                      & BP 2 \\ \hline\hline
\multicolumn{2}{|c|}{$2$ b-jet, $N_\ell = 0$} \\
\multicolumn{2}{|c|}{$90.0 < m_{bb} < 130.0$} \\
\multicolumn{2}{|c|}{$N_j \leq 1$} \\\hline
$0.4 < \Delta R_{bb} < 1.4$    & $0.4 < \Delta R_{bb} < 1.3$\\ 
$\Delta \phi_{bb,\met} > 2.8$  & $\Delta \phi_{bb,\met} > 2.6$\\ 
$\met > 180~{\rm GeV}$         & $\met > 210~{\rm GeV}$ \\ \hline
\end{tabular}
\caption{\it The selection cuts optimised in the $b\bar{b}+\met$ channel for the cut-based analysis.}
\label{tab:bbcuts}
\end{center}
\end{table}

With the discriminating variables ($\Delta R_{bb}$, $\Delta \phi_{bb,\met}$ and $\met$) explained above, we suggest a cut-based analysis optimising the signal over background. The final choice of cuts for these variables are listed in Table~\ref{tab:bbcuts}. As explained in the previous section, we use on the $5F$ scheme sample for performing the optimisation. Corresponding signal and dominant background yields after all the cuts in succession are displayed in Table~\ref{tab:bbcutflow}. The number of SM and susy background events corresponding to each benchmark points at the HL-LHC configuration are given in Table~\ref{tab:bbyield}. The final signal significance along with the signal yield is presented in Table~\ref{tab:bbsigni}. Here, the signal significance is slightly higher than the previous $\ell\ell+\met$ analysis (section~\ref{sec:bvetoll}). Since the $S/B$ ratio is very poor in this channel which reduces the significance drastically upon adding a systematic uncertainty, viz. for the 1st benchmark point $S/\sqrt{B}$ changes from $7.25$ to $0.86$ and from $4.7$ to $0.56$ by adding a $5\%$ systematic, in case of matched 4F, 5F $b\bar{b}H$ and NLO $4F~b\bar{b}H$ production respectively.

\begin{table}[htb!]
\begin{bigcenter}
\scalebox{0.8}{%
\begin{tabular}{|c||c||c|c|c|c|c||c|c|c|c|}
\hline
 &  \multicolumn{10}{c|}{Event yield with $3 \; \textrm{ab}^{-1}$ of integrated luminosity} \\ \cline{2-11}
 Cut flow   & Signal & \multicolumn{5}{c||}{susy Backgrounds} & \multicolumn{4}{c|}{SM Backgrounds} \\
\cline{3-11} 
 & $b\bar{b} \to H~(5F)$ & $\chonepm \lsptwo$ & $\chonepm \lspthree$ &  $\lsptwo \lspthree$ & $\lspone \lsptwo$ & $\lspone \lspthree$ & $t\bar{t}$ semi-lep & $t\bar{t}$ lep & $Zb\bar{b}$ & $Zh$ \\\hline\hline
 
$2$ b-jet                & $13462$ & $52229$  & $38069$  & $24940$  & $4249$ & $3440$ & $20986447$  & $6550973$  & $1819565$   & $401788$ \\

\hline
$m_{bb}$                 & $6157$  & $24030$  & $14348$  & $9650$   & $2077$ & $1335$ & $4703884$   & $1395299$  & $330348$    & $207187$ \\

\hline
$N_j$                    & $4468$  & $7433$   & $4317$   & $3808$   & $1687$ & $1081$ & $438444$    & $665118$   & $233317$    & $86475$ \\

\hline
$\Delta R_{bb}$          & $1861$  & $2427$   & $1542$   & $1184$   & $350$  & $264$  & $40719$     & $41208$    & $37335$     & $7195$ \\

\hline
$\Delta\phi_{bb,\met}$   & $1728$  & $1518$   & $959$    & $895$    & $312$  & $231$  & $27123$     & $24511$    & $32073$     & $4839$ \\

\hline
$\met$                   & $1305$  & $902$    & $575$    & $545$    & $181$  & $130$  & $7229$      & $1873$     & $13758$     & $2543$ \\
\hline
\end{tabular}}
\end{bigcenter}
\caption{ \it The cut-flow table for the benchmark point $1$ in the $b\bar{b} + \met$ mode with 5F signal production and dominant backgrounds.}

\label{tab:bbcutflow}
\end{table}

\begin{table}[htb!]
\begin{bigcenter}
\scalebox{0.65}{
\begin{tabular}{|c||c|c|c|c|c||c|c|c|c|c|c|c|c|c||c|}
\hline
BPs & \multicolumn{15}{c|}{Background yield at $3 \; \textrm{ab}^{-1}$ after all cuts} \\ \cline{2-16}
    & \multicolumn{5}{c||}{susy Backgrounds} &\multicolumn{9}{c||}{SM Backgrounds} & Total \\ \cline{2-15}
    
    & $\chonepm \lsptwo$ & $\chonepm \lspthree$ & $\lsptwo \lspthree$ & $\lspone \lsptwo$ & $\lspone \lspthree$ & $t\bar{t}$ had & $t\bar{t}$ semi-lep & $t\bar{t}$ lep & $Zb\bar{b}$ & $Zh$ & $Wh$ & $t\bar{t}h$ & $t\bar{t}Z$ & $t\bar{t}W$ & Background\\\cline{1-15}
    
\multirow{2}{*}{Order} & \multicolumn{5}{c||}{\multirow{2}{*}{NLO~\cite{Beenakker:1996ed}}} & \multicolumn{3}{c|}{\multirow{2}{*}{NNLO~\cite{ttbarNNLO}}} & \multirow{2}{*}{LO} & \multicolumn{2}{c|}{NNLO (QCD)+} & \multirow{2}{*}{NLO~\cite{bkg_twiki_cs}} & \multirow{2}{*}{NLO~\cite{Lazopoulos:2008de}} & \multirow{2}{*}{NLO~\cite{Campbell:2012dh}}&\\  

 & \multicolumn{5}{c||}{} & \multicolumn{3}{c|}{} &  & \multicolumn{2}{c|}{NLO (EW)~\cite{bkg_twiki_cs}} & &  &&\\ \hline\hline 

BP 1 & $902$ & $575$ & $545$ & $181$ & $130$ & $\sim 0$ & $7229$ & $1873$ & $13758$ & $2543$ & $19$ & $32$ & $98$ & $28$ & $27913$\\
\hline

BP 2 & $552$ & $408$ & $376$ & $111$ & $88$ & $\sim 0$ & $3439$ & $525$ & $7768$ & $1683$ & $12$ & $24$ & $65$ & $18$ & $15069$\\
\hline
\end{tabular}}
\end{bigcenter}
\caption{ \it The background yield at $14$ TeV with $3 \; \textrm{ab}^{-1}$ of integrated luminosity after the cut-based analysis for the two benchmark points.}
\label{tab:bbyield}
\end{table}

\begin{table}[htb!]
\begin{bigcenter}\scalebox{0.8}{
\begin{tabular}{|c||c|c|c|c|c|c|c|c|c|c|c|c|c||c|c|c|}
\hline
BPs & \multicolumn{6}{c||}{Signal rates at $3 \; \textrm{ab}^{-1}$ after all cuts} & \multicolumn{3}{c|}{Significance calculation}\\
\cline{2-10}

    & \multicolumn{3}{c|}{$pp \to H/A \to \lspone \lsptwo$} & \multicolumn{3}{c||}{$pp \to H/A \to \lspone \lspthree$} & Total signal, & Total background, & \multicolumn{1}{c|}{Significance, \tiny{$\dfrac{S}{\sqrt{B}}$}} \\\cline{2-7}

    & \multirow{2}{*}{$4F$} & \multirow{2}{*}{$5F$} & \multirow{2}{*}{$ggF$} & \multirow{2}{*}{$4F$} & \multirow{2}{*}{$5F$} & \multicolumn{1}{c||}{\multirow{2}{*}{$ggF$}} & S & B & \multicolumn{1}{c|}{\multirow{2}{*}{\makecell{without (with $5\%$)\\ systematics}}}\\
    
    &  &  &  &  &  & \multicolumn{1}{c||}{} & & (From Table~\ref{tab:bbyield}) & \multicolumn{1}{c|}{} \\\hline\hline
    
 BP 1 & $137$ & $859$ & $127$ & $73$ & $447$ & \multicolumn{1}{c||}{$59$} & $1211$ & $27913$  & \multicolumn{1}{c|}{$7.25$ ($0.86$)} \\\hline

 BP 2 & $74$  & $534$ & $52$  & $41$ & $294$ & \multicolumn{1}{c||}{$25$}  & $729$  & $15069$  & \multicolumn{1}{c|}{$5.94$ ($0.95$)} \\\hline
 
  \multicolumn{10}{|c|}{Using NLO 4F $b\bar{b}H$ process} \\\hline

 BPs   & \multicolumn{2}{c|}{$4F$ NLO} & $ggF$ & \multicolumn{2}{c|}{$4F$ NLO} & \multicolumn{1}{c||}{$ggF$} & Total signal, $S=$ & Total background, & \multicolumn{1}{c|}{{\tiny$\dfrac{S}{\sqrt{B}}$} without (with}\\
    
    & \multicolumn{2}{c|}{} & & \multicolumn{2}{c|}{} & \multicolumn{1}{c||}{} &  \multicolumn{1}{c|}{$N_{4F}^{NLO}+~N_{ggF}$} & B & \multicolumn{1}{c|}{ $5\%$) systematics}\\\hline

 BP 1 & \multicolumn{2}{c|}{$386$ } & $127$ &  \multicolumn{2}{c|}{$212$} & \multicolumn{1}{c||}{$59$} & $784$ & $27913$  & \multicolumn{1}{c|}{$4.7$ ($0.56$)} \\\hline

 BP 2 & \multicolumn{2}{c|}{$233$ } & $52$ &  \multicolumn{2}{c|}{$127$} & \multicolumn{1}{c||}{$25$} & $437$ & $15069$  & \multicolumn{1}{c|}{$3.56$ ($0.57$)} \\\hline
 
\end{tabular}}
\end{bigcenter}
\caption{\it The signal yield for two benchmark points along with signal significance for the $b\bar{b} + \met$ final state.}
\label{tab:bbsigni}
\end{table}

\subsubsection*{B. b-tag category}
\label{sec:btagbb}

\begin{figure}[htb!]
\centering
\includegraphics[scale=0.37]{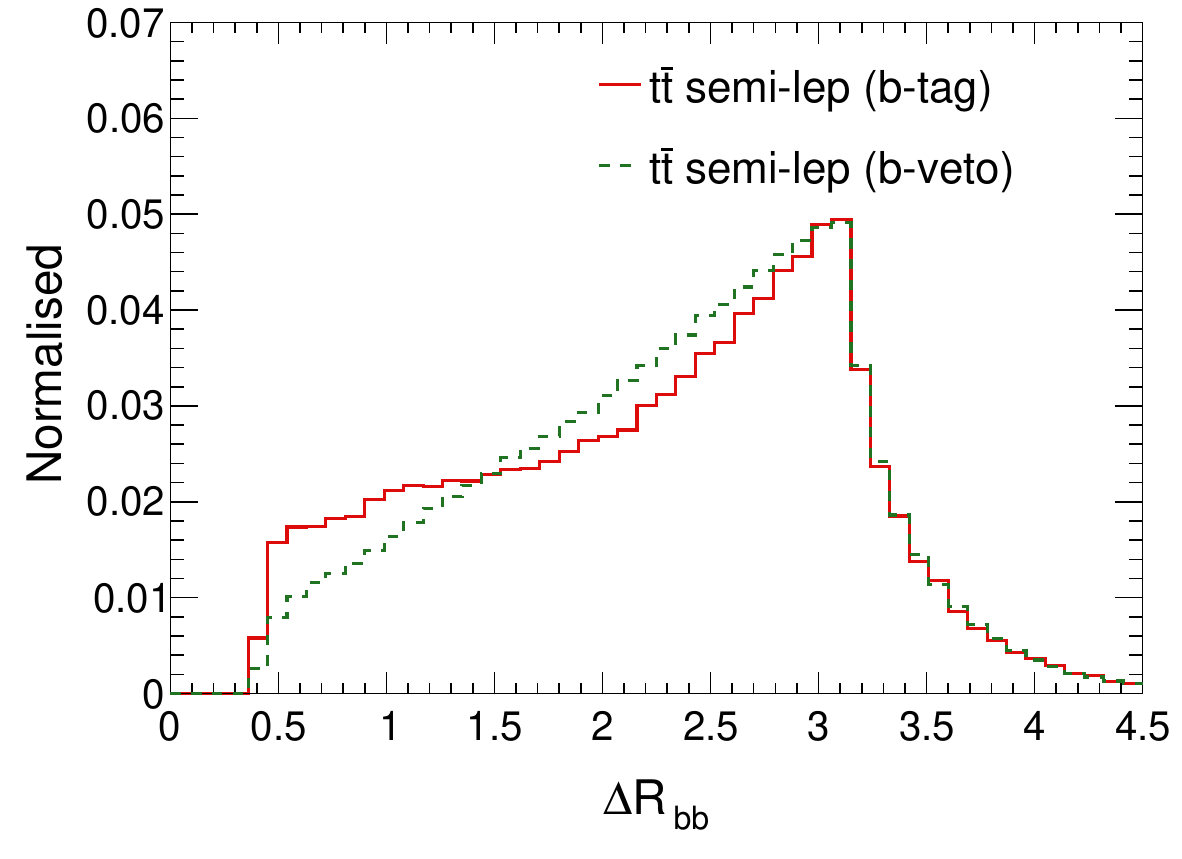}\includegraphics[scale=0.37]{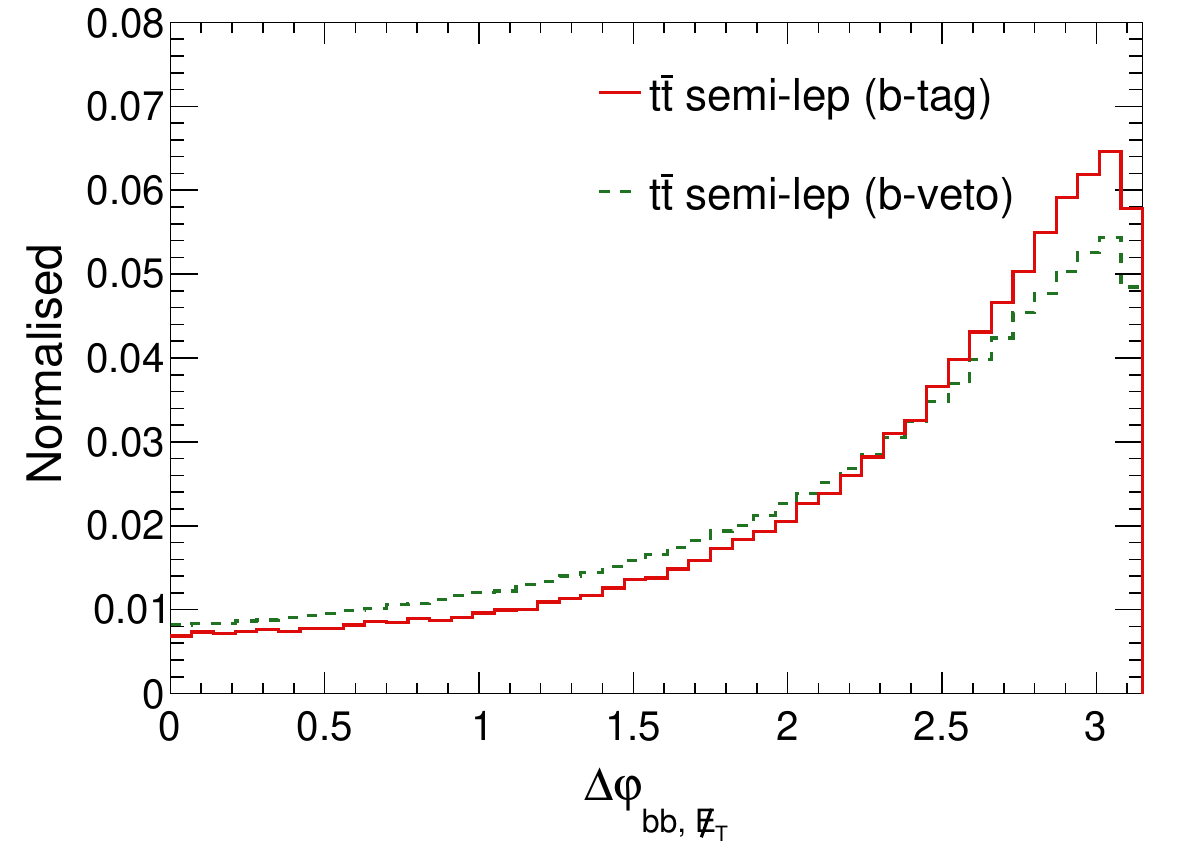}
\caption{\it The comparison between the normalised distributions of $\Delta R_{bb}$ (left) and $\Delta \phi_{bb,\met}$ (right panel) for $t\bar{t}$ semi-leptonic background in the b-veto and b-tag category of $b\bar{b}+\met$ final state.}
\label{fig:bbbcompare}
\end{figure}

\begin{figure}[htb!]
\centering
\includegraphics[scale=0.37]{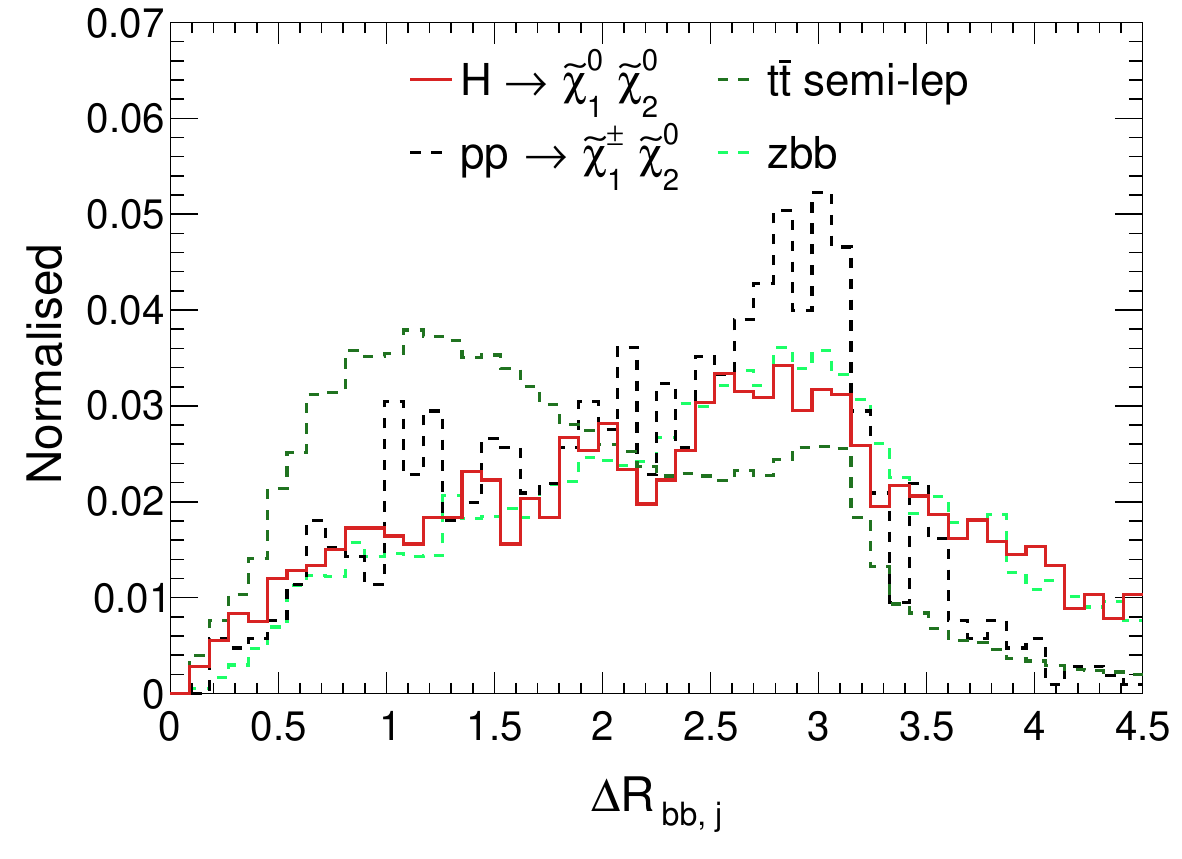}
\caption{\it The normalised distributions of $\Delta R_{bb,j}$ for the $t\bar{t}$ semi-leptonic background along with other dominant backgrounds in the b-tag category of $b\bar{b}+\met$ final state for benchmark point 1. Overlaid in solid is the distribution for the $5F$ scheme signal process.}
\label{fig:bbbdrbbj}
\end{figure}

In this channel, we demand at least three b-tagged jets with $p_T > 20$ GeV and $|\eta| < 2.5$ in the final state. Similar to the previous b-veto analysis, we reconstruct the kinematic variables for our cut-based analysis. We would like to mention here that the $t\bar{t}$ semi-leptonic background has an increased overlapping distribution in $\Delta R_{bb}, \Delta \phi_{bb,\met}$ variables with signal ($0.4 < \Delta R_{bb} < 1.4, \Delta \phi_{bb,\met} > 2.4$), making this background dominant, unlike in the b-veto case. We show a comparison of these distributions with the $t\bar{t}$ semi-leptonic background in the b-veto case in Fig.~\ref{fig:bbbcompare}. It is particularly interesting to understand the origin of this increased overlap. As this category demands 3 b-tagged jets, the extra b-jet in the $t \bar{t}$ background in semi-leptonic mode arises when a c-quark fakes as a b-jet in hadronic decays of W bosons originating from tops. This fake b-jet can have high transverse momentum and gets selected when reconstructing the two b-jet system. However, we note that there must be a light jet around this di-b-jet system, which originates from the $W$ boson decay for background processes. To catch this feature in our optimisation analysis, we first demand that events must have at most one light jet in the final state. If the events contain a light jet then the di-b-jet system is reconstructed with b-jets closest in the $\eta-\phi$ plane, \textit{i.e.} we select jets for which $\Delta R_{bb}$ is minimum. Using this $b\bar{b}$ system, we compute the distance between this b-jet system and the light jet, $\Delta R_{bb,j}$, as shown in Fig.~\ref{fig:bbbdrbbj}. This distribution is shifted towards low values of $\Delta R_{bb,j}$ for the $t\bar{t}$ semi-leptonic than the signal events. Therefore, we include this variable in our cut-based analysis and put a lower bound after optimisation. To construct the other variables, \textit{viz.} $m_{bb},~\Delta R_{bb},~\text{and}~\Delta \phi_{bb,\met}$, we take the two hardest b-jets in the event as before.  We show the optimised selection cuts in Table~\ref{tab:bbbcuts}. We quote the background yields and final signal significance along with signal yields after the cut-based analysis in Table~\ref{tab:bbbyield} and Table~\ref{tab:bbbsigni}. The tables demonstrate that the final significance is about $4$ and $3$ for both the benchmark points in case of matched 4F, 5F $b\bar{b}H$ and NLO $4F~b\bar{b}H$ production respectively. With respect to b-veto category, we obtain lesser significance for b-tag primarily because of a different background composition.

\begin{table}
\begin{center}
\begin{tabular}{|c|c|}\hline 
\multicolumn{2}{|c|}{Selection cuts} \\ \hline\hline
BP 1                      & BP 2 \\ \hline\hline
\multicolumn{2}{|c|}{$N_b \geq 3$, $N_\ell = 0$} \\
\multicolumn{2}{|c|}{$N_j \leq 1$} \\
\multicolumn{2}{|c|}{$90.0 < m_{bb} < 130.0$} \\\hline
$0.4 < \Delta R_{bb} < 1.5$    & $0.4 < \Delta R_{bb} < 1.4$\\ 
$\Delta \phi_{bb,\met} > 2.3$  & $\Delta \phi_{bb,\met} > 2.2$\\ 
$\met > 180~{\rm GeV}$         & $\met > 210~{\rm GeV}$ \\ 
$\Delta R_{bb,j} > 2.2$        & $\Delta R_{bb,j} > 1.7$\\ 
\hline
\end{tabular}
\caption{\it The selection cuts optimised for the b-tag $b\bar{b}+\met$ channel for the cut-based analysis.}
\label{tab:bbbcuts}
\end{center}
\end{table}

\begin{table}[htb!]
\begin{bigcenter}
\scalebox{0.7}{
\begin{tabular}{|c||c|c|c|c|c||c|c|c|c|c|c|c|c|c||c|}
\hline
BPs & \multicolumn{15}{c|}{Background yield at $3 \; \textrm{ab}^{-1}$ after all cuts} \\ \cline{2-16}
    & \multicolumn{5}{c||}{susy Backgrounds} &\multicolumn{9}{c||}{SM Backgrounds} & Total \\ \cline{2-15}
    
    & $\chonepm \lsptwo$ & $\chonepm \lspthree$ & $\lsptwo \lspthree$ & $\lspone \lsptwo$ & $\lspone \lspthree$ & $t\bar{t}$ had & $t\bar{t}$ semi-lep & $t\bar{t}$ lep & $Zb\bar{b}$ & $Zh$ & $Wh$ & $t\bar{t}h$ & $t\bar{t}Z$ & $t\bar{t}W$ & Background\\\hline\hline 

BP 1 & $45.16$ & $32.31$ & $143.41$ & $2.79$ & $1.64$ & $\sim 0$ & $2947.89$ & $470.61$ & $741.51$ & $40.13$ & $0.41$ & $164.13$ & $44.72$ & $6.24$ & $4640.95$\\
\hline

BP 2 & $28.16$ & $18.48$ & $83.34$ & $1.88$ & $1.88$ & $\sim 0$ & $1784.78$ & $235.30$ & $527.80$ & $34.40$ & $0.26$ & $58.68$ & $43.44$ & $4.16$ & $2822.56$\\
\hline
\end{tabular}}
\end{bigcenter}
\caption{ \it The background yield at $14$ TeV with $3 \; \textrm{ab}^{-1}$ of integrated luminosity after the cut-based analysis for the two benchmark points in the b-tag category.}
\label{tab:bbbyield}
\end{table}

\begin{table}[htb!]
\begin{bigcenter}\scalebox{0.8}{
\begin{tabular}{|c||c|c|c|c|c|c|c|c|c|c|c|c|c||c|c|c|}
\hline
BPs & \multicolumn{6}{c||}{Signal rates at $3 \; \textrm{ab}^{-1}$ after all cuts} & \multicolumn{3}{c|}{Significance calculation}\\
\cline{2-10}

    & \multicolumn{3}{c|}{$pp \to H/A \to \lspone \lsptwo$} & \multicolumn{3}{c||}{$pp \to H/A \to \lspone \lspthree$} & Total signal, & Total background, & \multicolumn{1}{c|}{Significance,} \\\cline{2-7}

    & \multirow{2}{*}{$4F$} & \multirow{2}{*}{$5F$} & \multirow{2}{*}{$ggF$} & \multirow{2}{*}{$4F$} & \multirow{2}{*}{$5F$} & \multicolumn{1}{c||}{\multirow{2}{*}{$ggF$}} & S & B & \multicolumn{1}{c|}{\multirow{2}{*}{$\dfrac{S}{\sqrt{B}}$}}\\
    
    &  &  &  &  &  & \multicolumn{1}{c||}{} &  & (From Table~\ref{tab:bbbyield}) & \multicolumn{1}{c|}{} \\\hline\hline

 BP 1 & $55.64$ & $206.50$ & $4.00$ & $33.05$ & $115.00$ & \multicolumn{1}{c||}{$1.96$} & $267.96$ & $4640.95$  & \multicolumn{1}{c|}{$3.93~(1.11)$} \\\hline

 BP 2 & $34.41$ & $158.05$ & $2.03$ & $18.67$ & $82.06$ & \multicolumn{1}{c||}{$0.71$} & $196.74$ & $2822.56$  & \multicolumn{1}{c|}{$3.70~(1.30)$} \\\hline
 
  \multicolumn{10}{|c|}{Using NLO 4F $b\bar{b}H$ process} \\\hline

 BPs   & \multicolumn{2}{c|}{$4F$ NLO} & $ggF$ & \multicolumn{2}{c|}{$4F$ NLO} & \multicolumn{1}{c||}{$ggF$} & Total signal, & Total background, & \multicolumn{1}{c|}{{\tiny$\dfrac{S}{\sqrt{B}}$} without (with}\\
    
    & \multicolumn{2}{c|}{} & & \multicolumn{2}{c|}{} & \multicolumn{1}{c||}{} &  \multicolumn{1}{c|}{ $S$} & B & \multicolumn{1}{c|}{ $5\%$) systematics}\\\hline

 BP 1 & \multicolumn{2}{c|}{$133.01$ } & $4$ &  \multicolumn{2}{c|}{$66.94$} & \multicolumn{1}{c||}{$1.96$} & $205.91$ & $4640.95$  & \multicolumn{1}{c|}{$3.02$ ($0.85$)} \\\hline

 BP 2 & \multicolumn{2}{c|}{$89.27$ } & $2.03$ &  \multicolumn{2}{c|}{$46.0$} & \multicolumn{1}{c||}{$0.71$} & $138.01$ & $2822.56$  & \multicolumn{1}{c|}{$2.6$ ($0.92$)} \\\hline
 
\end{tabular}}
\end{bigcenter}
\caption{\it The signal yield along with signal significance for the b-tag $b\bar{b} + \met$ final state.}
\label{tab:bbbsigni}
\end{table}
\subsubsection{The $\gamma\gamma+\met$ channel}
\label{sec:gagamet}

Finally, we turn our focus on the $\gamma\gamma+\met$ final state which is clean in terms of the background contamination with the disadvantage of having very low event yield as compared to the other search channels discussed earlier. We generate the following susy backgrounds, \textit{viz.} $pp\to \lspone~\widetilde{\chi}_{ 2,3}^{0},~\chonepm~\widetilde{\chi}_{ 2,3}^{0}~\text{and}~\lsptwo~\lspthree$. For all of these backgrounds, the $\widetilde{\chi}_{ 2,3}^{0}$ is decayed to SM Higgs boson and it further decays to a photon pair. The dominant SM backgrounds are $Zh$, $Wh$ and $Z\gamma\gamma$. We generate the $Zh$ and $Wh$ background upon merging with one additional jet in the final state where the SM Higgs decays to $\gamma\gamma$. For the $Z\gamma\gamma$, we decay the $Z$ boson into neutrinos and merge with one extra jet in the final state. Also, we generate $t\bar{t}h$ with $h\to \gamma\gamma$ which is a subdominant background to this final state. For details see Appendix~\ref{sec:appendixC}.

\subsubsection*{A. b-veto category}
\label{sec:bvetogaga}

\begin{figure}[htb!]
\centering
\includegraphics[scale=0.37]{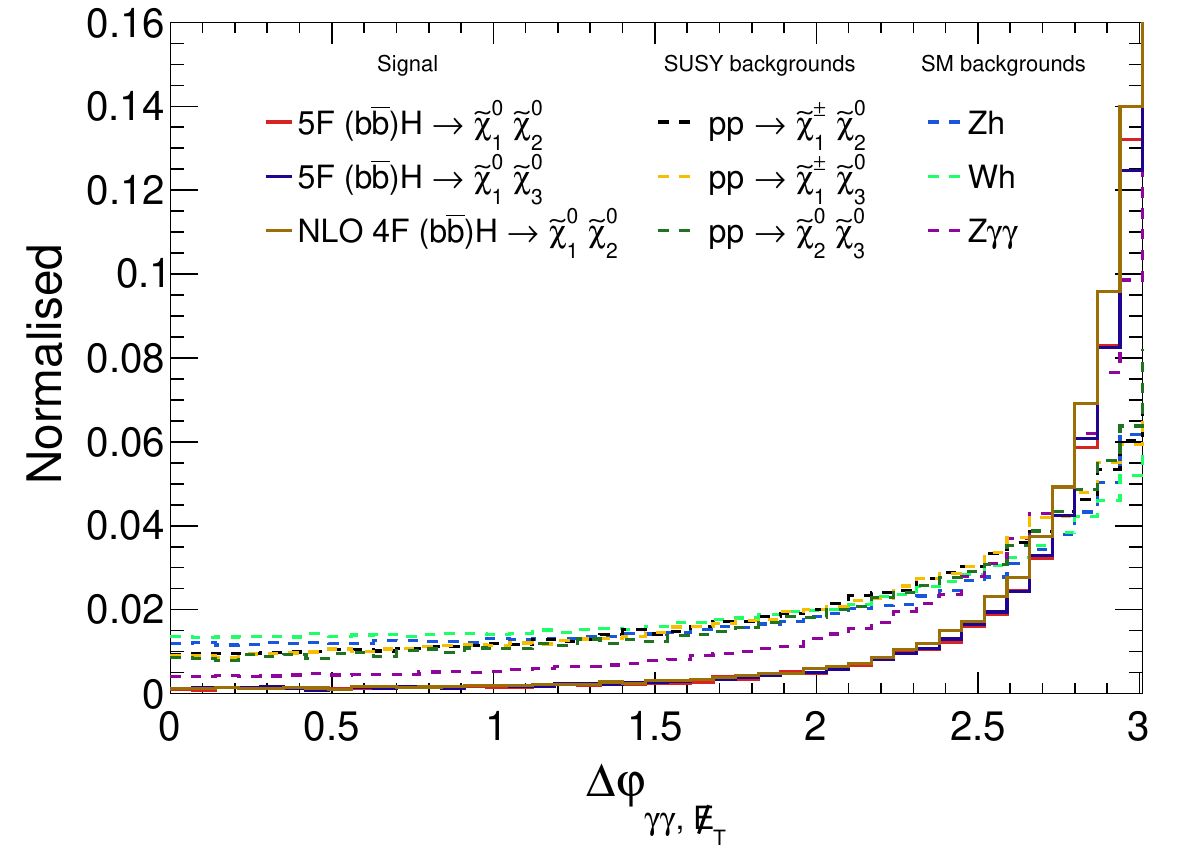}\includegraphics[scale=0.37]{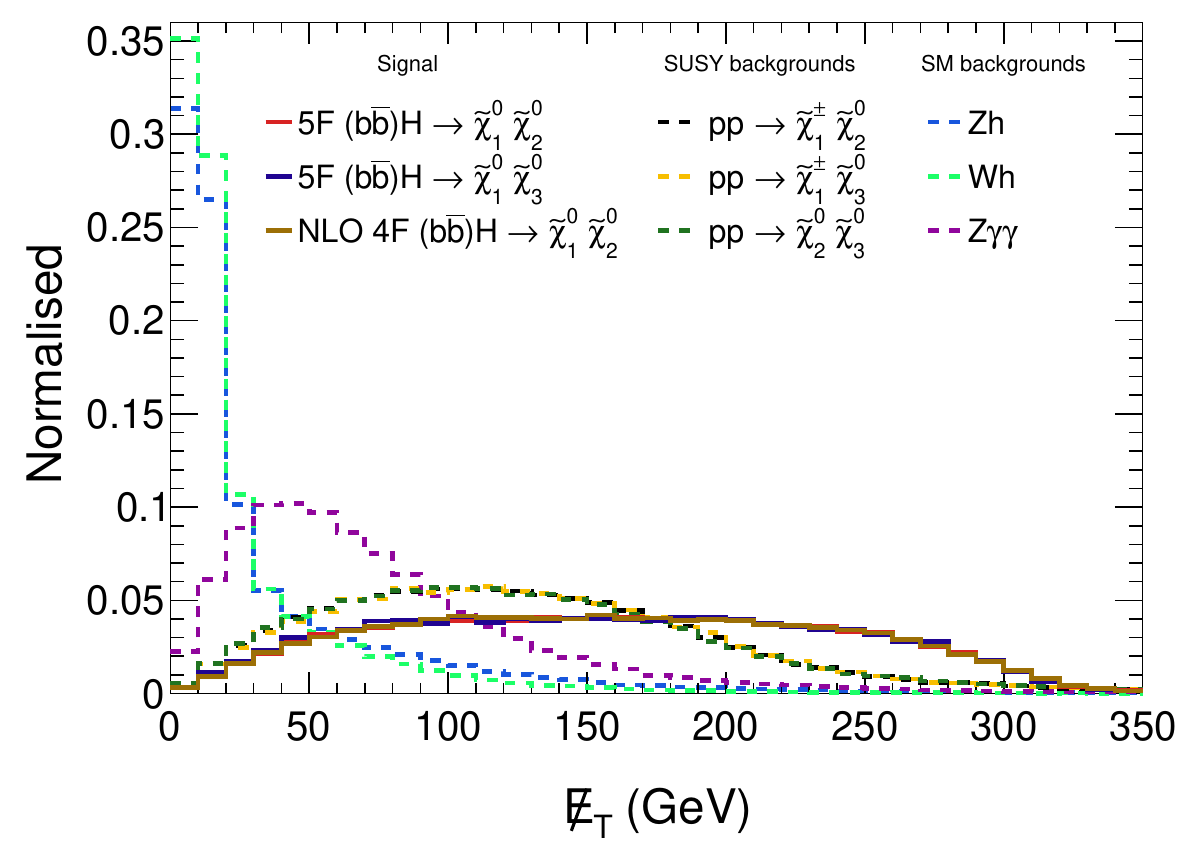}\\
\caption{\it The normalised distributions of $\Delta \phi_{\gamma\gamma,\met}$ (left) and $\met$ (right) for the $\gamma\gamma+\met$ final state after the basic trigger cuts.}
\label{fig:aaplot}
\end{figure}

In this category, the selected event must contain exactly two photons with $p_T > 30$ GeV and $|\eta|<2.5$ along with no b-jets satisfying $p_T > 20$ GeV and $|\eta|<2.5$ in the final state. We veto events which contain leptons with $p_T > 20$ GeV and $|\eta|<2.47~(2.5)$ (for electron (muon)) in the final state to reduce the $pp\to \chonepm~\widetilde{\chi}_{ 2,3}^{0}$, $Zh$, $Wh$ and $t\bar{t}h$ backgrounds. Since photon has very clean signature with excellent mass resolution at the LHC, we restrict the di-photon invariant mass within ($122,128$) GeV. The number of maximum light jets with $p_T > 20$ GeV and $|\eta|<4.5$ are restricted to be one to reduce the backgrounds with multiple jets in the final state. Similar to the previous sections, we optimise the signal over backgrounds with missing energy, $\met$ and, the azimuthal angle separation between the di-photon system and $\met$, $\Delta \phi_{\gamma\gamma,\met}$. The normalised distribution of these variables for signal and dominant background events are shown in Fig.~\ref{fig:aaplot}. 

The results of the optimised cuts for both of the benchmark points are displayed in Table~\ref{tab:aacuts}. The cut flow table for the benchmark point 1 is shown in Table~\ref{tab:aacutflow}. We present the number of SM and susy background events at $\sqrt{s}=14$ TeV with 3 ab$^{-1}$ of integrated luminosity corresponding to each benchmark point in Table~\ref{tab:aayield}. Also, we display the final signal significance in Table~\ref{tab:aasigni}. The signal significance is very poor in this channel. However, the signal to background ratio, $S/B$ is large here which reduces the effect of adding systematic uncertainty on the final signal significance, \textit{viz.} a $5\%$ systematic changes the significance from $1.74$ to $1.48$ and $1.11$ to $0.95$ for the first benchmark point in case of matched 4F, 5F $b\bar{b}H$ and NLO $4F~b\bar{b}H$ production respectively.

\begin{table}
\begin{center}
\begin{tabular}{|c|c|}\hline 
\multicolumn{2}{|c|}{Selection cuts} \\ \hline\hline
BP 1                      & BP 2 \\ \hline\hline
\multicolumn{2}{|c|}{$2\gamma$, $N_{\ell,b} = 0$} \\ 
\multicolumn{2}{|c|}{$122.0 < m_{\gamma\gamma} < 128.0$} \\
\multicolumn{2}{|c|}{$N_j \leq 1$} \\ \hline
$\met > 150~{\rm GeV}$    & $\met > 190~{\rm GeV}$ \\ 
$\Delta \phi_{\gamma\gamma,\met} > 2.7$    & $\Delta \phi_{\gamma\gamma,\met} > 2.5$ \\ \hline
\end{tabular}
\caption{\it The selection cuts optimised in the $\gamma\gamma+\met$ channel for the cut-based analysis.}
\label{tab:aacuts}
\end{center}
\end{table}

\begin{table}[htb!]
\begin{bigcenter}
\scalebox{0.8}{%
\begin{tabular}{|c||c||c|c|c|c|c||c|c|c|c|}
\hline
 &  \multicolumn{10}{c|}{Event yield with $3 \; \textrm{ab}^{-1}$ of integrated luminosity} \\ \cline{2-11}
 Cut flow   & Signal & \multicolumn{5}{c||}{susy Backgrounds} & \multicolumn{4}{c|}{SM Backgrounds} \\
\cline{3-11} 
 & $b\bar{b} \to H~(5F)$ & $\chonepm \lsptwo$ & $\chonepm \lspthree$ &  $\lsptwo \lspthree$ & $\lspone \lsptwo$ & $\lspone \lspthree$ & $Zh$ & $Wh$ & $Z\gamma\gamma$ & $t\bar{t}h$ \\\hline\hline
 
$2\gamma$                           & $55.50$  & $236.06$  & $138.57$  & $158.83$  & $26.57$ & $13.25$ & $1782.53$  & $2761.95$  & $1663.45$ & $175.10$ \\

\hline
$m_{\gamma\gamma}$                  & $54.44$  & $227.10$  & $133.13$  & $153.93$  & $26.01$ & $12.97$ & $1751.64$  & $2702.96$  & $329.98$  & $167.78$  \\

\hline
$N_j$                               & $43.67$  & $72.26$   & $42.49$   & $54.49$   & $16.05$ & $8.12$  & $791.14$   & $1122.31$   & $259.10$  & $2.53$ \\

\hline
$\met$                              & $23.15$  & $23.59$   & $14.08$   & $18.63$   & $6.62$  & $3.28$  & $54.68$    & $31.17$     & $19.05$   & $0.39$ \\

\hline
$\Delta\phi_{\gamma\gamma,\met}$    & $22.18$  & $17.17$   & $10.32$   & $14.80$   & $5.70$  & $2.83$  & $51.08$    & $29.25$    & $15.90$   & $0.28$ \\\hline

\end{tabular}}
\end{bigcenter}
\caption{ \it The cut-flow table for the benchmark point $1$ in the $\gamma\gamma + \met$ mode with 5F signal production and backgrounds.}
\label{tab:aacutflow}
\end{table}

\begin{table}[htb!]
\begin{bigcenter}
\scalebox{0.8}{
\begin{tabular}{|c||c|c|c|c|c||c|c|c|c||c|}
\hline
BPs & \multicolumn{10}{c|}{Background yield at $3 \; \textrm{ab}^{-1}$ after all cuts} \\ \cline{2-11}
    & \multicolumn{5}{c||}{susy Backgrounds} &\multicolumn{4}{c||}{SM Backgrounds} & Total \\ \cline{2-10}
    
    & $\chonepm \lsptwo$ & $\chonepm \lspthree$ & $\lsptwo \lspthree$ & $\lspone \lsptwo$ & $\lspone \lspthree$ & $Zh$ & $Wh$ & $Z\gamma\gamma$ & $t\bar{t}h$ & Background, \\\cline{1-10}
    
\multirow{2}{*}{Order} & \multicolumn{5}{c||}{\multirow{2}{*}{NLO~\cite{Beenakker:1996ed}}} & \multicolumn{2}{c|}{NNLO (QCD)+} & \multirow{2}{*}{LO} & \multirow{2}{*}{NLO~\cite{bkg_twiki_cs}} & B  \\

 & \multicolumn{5}{c||}{} & \multicolumn{2}{c|}{NLO (EW)~\cite{bkg_twiki_cs}} &   &    &   \\ \hline\hline

BP 1 & $17.17$ & $10.32$ & $14.80$ & $5.70$ & $2.83$ & $51.08$ & $29.25$ & $15.90$ & $0.28$ & $147.33$ \\
\hline

BP 2 & $10.90$ & $6.23$ & $9.35$ & $5.22$ & $2.57$  & $30.02$  & $14.82$ & $7.83$ & $0.15$ & $87.09$ \\
\hline
\end{tabular}}
\end{bigcenter}
\caption{ \it The background yield at $14$ TeV with $3 \; \textrm{ab}^{-1}$ of integrated luminosity after the cut-based analysis for the two benchmark points.}
\label{tab:aayield}
\end{table}

\begin{table}[htb!]
\begin{bigcenter}\scalebox{0.8}{
\begin{tabular}{|c||c|c|c|c|c|c|c|c|c|c|c|c|c||c|c|c|}
\hline
BPs & \multicolumn{6}{c||}{Signal rates at $3 \; \textrm{ab}^{-1}$ after all cuts} & \multicolumn{3}{c|}{Significance calculation}\\
\cline{2-10}

    & \multicolumn{3}{c|}{$pp \to H/A \to \lspone \lsptwo$} & \multicolumn{3}{c||}{$pp \to H/A \to \lspone \lspthree$} & Total signal, & Total background, & \multicolumn{1}{c|}{Significance,} \\\cline{2-7}

    & \multirow{2}{*}{$4F$} & \multirow{2}{*}{$5F$} & \multirow{2}{*}{$ggF$} & \multirow{2}{*}{$4F$} & \multirow{2}{*}{$5F$} & \multicolumn{1}{c||}{\multirow{2}{*}{$ggF$}} & S & B & \multicolumn{1}{c|}{\multirow{1}{*}{{\tiny$\dfrac{S}{\sqrt{B}}$} without (with}}\\
    
    &  &  &  &  &  & \multicolumn{1}{c||}{} &  & (From Table~\ref{tab:aayield}) & \multicolumn{1}{c|}{$5\%$) systematics} \\\hline\hline

 BP 1 & $2.51$ & $14.53$ & $2.37$ & $1.27$ & $7.65$ & \multicolumn{1}{c||}{$1.22$} & $21.07$ & $147.33$  & \multicolumn{1}{c|}{$1.74~(1.48)$} \\\hline

 BP 2 & $1.00$ & $7.75$ & $0.81$ & $0.57$ & $4.35$ & \multicolumn{1}{c||}{$0.72$} & $11.03$ & $87.09$  & \multicolumn{1}{c|}{$1.18~(1.07)$} \\\hline
 
  \multicolumn{10}{|c|}{Using NLO 4F $b\bar{b}H$ process} \\\hline

 BPs   & \multicolumn{2}{c|}{$4F$ NLO} & $ggF$ & \multicolumn{2}{c|}{$4F$ NLO} & \multicolumn{1}{c||}{$ggF$} & Total signal,  & Total background, & \multicolumn{1}{c|}{{\tiny$\dfrac{S}{\sqrt{B}}$} without (with}\\
    
    & \multicolumn{2}{c|}{} & & \multicolumn{2}{c|}{} & \multicolumn{1}{c||}{} &  \multicolumn{1}{c|}{$S$} & B & \multicolumn{1}{c|}{ $5\%$) systematics}\\\hline

 BP 1 & \multicolumn{2}{c|}{$6.41$ } & $2.37$ &  \multicolumn{2}{c|}{$3.46$} & \multicolumn{1}{c||}{$1.22$} & $13.46$ & $147.33$  & \multicolumn{1}{c|}{$1.11$ ($0.95$)} \\\hline

 BP 2 & \multicolumn{2}{c|}{$3.34$ } & $0.81$ &  \multicolumn{2}{c|}{$1.77$} & \multicolumn{1}{c||}{$0.72$} & $6.64$ & $87.09$  & \multicolumn{1}{c|}{$0.71$ ($0.64$)} \\\hline
 
\end{tabular}}
\end{bigcenter}
\caption{\it The signal yield along with signal significance for the $\gamma\gamma + \met$ final state.}
\label{tab:aasigni}
\end{table}

\subsubsection*{B. b-tag category}
\label{sec:btaggaga}

Here, we select events with exactly two photon and atleast one extra b-jet in the final state meeting the basic cuts mentioned in the previous section~\ref{sec:bvetogaga}. The event yield further reduces by demanding the extra b-jet. Next, we do a cut-based analysis with the kinematic variables, $\met$ and $\Delta \phi_{\gamma\gamma,\met}$ for the matched $4F$ and $5F$ $b\bar{b}H$ process, and also with the signal process, NLO $4F~b\bar{b}H$. The selection cuts are presented in Table~\ref{tab:baacuts}. We then present in Table~\ref{tab:baasigni} the final signal significance calculated using the backgrounds given in Table~\ref{tab:baayield}. 

\begin{table}[htb!]
\begin{center}
\begin{tabular}{|c|c|}\hline 
\multicolumn{2}{|c|}{Selection cuts} \\ \hline\hline
BP 1                      & BP 2 \\ \hline\hline
\multicolumn{2}{|c|}{$2\gamma$, $N_{\ell} = 0$, $N_{b} \geq 1$} \\
\multicolumn{2}{|c|}{$122.0 < m_{\gamma\gamma} < 128.0$} \\
\multicolumn{2}{|c|}{$N_j \leq 1$} \\ \hline
$\met > 200~{\rm GeV}$    & $\met > 230~{\rm GeV}$ \\ 
$\Delta \phi_{\gamma\gamma,\met} > 2.0$    & $\Delta \phi_{\gamma\gamma,\met} > 2.1$ \\ \hline
\end{tabular}
\caption{\it The selection cuts optimised in the b-tag $\gamma\gamma+\met$ channel for the cut-based analysis.}
\label{tab:baacuts}
\end{center}
\end{table}

\begin{table}[htb!]
\begin{bigcenter}
\scalebox{0.8}{
\begin{tabular}{|c||c|c|c|c|c||c|c|c|c||c|}
\hline
BPs & \multicolumn{10}{c|}{Background yield at $3 \; \textrm{ab}^{-1}$ after all cuts} \\ \cline{2-11}
    & \multicolumn{5}{c||}{susy Backgrounds} &\multicolumn{4}{c||}{SM Backgrounds} & Total \\ \cline{2-10}
    
    & $\chonepm \lsptwo$ & $\chonepm \lspthree$ & $\lsptwo \lspthree$ & $\lspone \lsptwo$ & $\lspone \lspthree$ & $Zh$ & $Wh$ & $Z\gamma\gamma$ & $t\bar{t}h$ & Background, B \\\hline\hline 

BP 1 & $0.63$ & $0.37$ & $2.29$ & $0.13$ & $0.06$ & $0.33$ & $0.33$ & $0.24$ & $1.36$ & $5.74$ \\
\hline

BP 2 & $0.32$ & $0.19$ & $1.28$ & $0.15$ & $0.08$  & $0.15$  & $0.25$ & $0.16$ & $0.72$ & $3.30$ \\
\hline
\end{tabular}}
\end{bigcenter}
\caption{ \it The background yield at $14$ TeV with $3 \; \textrm{ab}^{-1}$ of integrated luminosity after the cut-based analysis for the two benchmark points in b-tag category.}
\label{tab:baayield}
\end{table}

\begin{table}[htb!]
\begin{bigcenter}\scalebox{0.8}{
\begin{tabular}{|c||c|c|c|c|c|c|c|c|c|c|c|c|c||c|c|c|}
\hline
BPs & \multicolumn{6}{c||}{Signal rates at $3 \; \textrm{ab}^{-1}$ after all cuts} & \multicolumn{3}{c|}{Significance calculation}\\
\cline{2-10}

    & \multicolumn{3}{c|}{$pp \to H/A \to \lspone \lsptwo$} & \multicolumn{3}{c||}{$pp \to H/A \to \lspone \lspthree$} & Total signal, & Total background, & \multicolumn{1}{c|}{Significance,} \\\cline{2-7}

    & \multirow{2}{*}{$4F$} & \multirow{2}{*}{$5F$} & \multirow{2}{*}{$ggF$} & \multirow{2}{*}{$4F$} & \multirow{2}{*}{$5F$} & \multicolumn{1}{c||}{\multirow{2}{*}{$ggF$}} & S & B & \multicolumn{1}{c|}{\multirow{1}{*}{{\tiny$\dfrac{S}{\sqrt{B}}$} without (with } }\\
    
    &  &  &  &  &  & \multicolumn{1}{c||}{} &  & (From Table~\ref{tab:baayield}) & \multicolumn{1}{c|}{$5\%$) systematics} \\\hline\hline
    
 BP 1 & $1.11$ & $3.86$ & $0.068$ & $0.58$ & $2.10$ & \multicolumn{1}{c||}{$0.034$} & $4.97$ & $5.74$  & \multicolumn{1}{c|}{$2.07~(2.07)$} \\\hline

 BP 2 & $0.54$ & $2.47$ & $0.023$ & $0.31$ & $1.30$ & \multicolumn{1}{c||}{$0.02$} & $3.09$ & $3.30$  & \multicolumn{1}{c|}{$1.70~(1.70)$} \\\hline
 
  \multicolumn{10}{|c|}{Using NLO 4F $b\bar{b}H$ process} \\\hline

 BPs   & \multicolumn{2}{c|}{$4F$ NLO} & $ggF$ & \multicolumn{2}{c|}{$4F$ NLO} & \multicolumn{1}{c||}{$ggF$} & Total signal, & Total background, & \multicolumn{1}{c|}{{\tiny$\dfrac{S}{\sqrt{B}}$} without (with}\\
    
    & \multicolumn{2}{c|}{} & & \multicolumn{2}{c|}{} & \multicolumn{1}{c||}{} &  \multicolumn{1}{c|}{$S$} & B & \multicolumn{1}{c|}{ $5\%$) systematics}\\\hline

 BP 1 & \multicolumn{2}{c|}{$2.52$ } & $0.068$ &  \multicolumn{2}{c|}{$1.33$} & \multicolumn{1}{c||}{$0.034$} & $3.95$ & $5.74$  & \multicolumn{1}{c|}{$1.65$ ($1.64$)} \\\hline

 BP 2 & \multicolumn{2}{c|}{$1.51$ } & $0.023$ &  \multicolumn{2}{c|}{$0.81$} & \multicolumn{1}{c||}{$0.02$} & $2.36$ & $3.30$  & \multicolumn{1}{c|}{$1.3$ ($1.29$)} \\\hline
 
\end{tabular}}
\end{bigcenter}
\caption{\it The signal yield along with signal significance for the b-tag $\gamma\gamma + \met$ final state.}
\label{tab:baasigni}
\end{table}

\section*{Summary prompt final states}
\label{sec:summcollider}

We end our discussion of Higgs to susy decays in prompt final state with a summary. We evaluate the prospect of discovering heavy Higgs in electroweakino decays at the HL-LHC. The signature for such decays are mono-$Z+\met$ and mono-$h+\met$ final state. The leptonic decay of $Z$ boson in mono-$Z+\met$ final state gives rise to $\ell\ell+\met$ channel. In case of mono-$h+\met$ final state, we consider the decay of the SM Higgs into $bb$ and $\gamma\gamma$ final state which gives rise to $b\bar{b}+\met$ and $\gamma\gamma+\met$ channel respectively. Further, we divide our analysis into b-veto and b-tag category to account for extra b-jet requirement in case of bottom pair fusion production mode of heavy Higgs. The $\ell\ell+\met$ channel can give rise to promising signature at the collider with higher signal significance. The result improves up to $\sim 70\%$ upon considering b-tag category. For the case of $b\bar{b}+\met$ channel the signal to background ratio is very poor, mainly because of large $Zb\bar{b}$ and $t\bar{t}$ background. To probe heavy Higgs in this channel one has to look for better ways to reduce these backgrounds. The $\gamma\gamma+\met$ channel suffers from smaller event rate in spite of being a clean final state. This channel might be very important search channel at higher energy colliders like HE-LHC or 100 TeV collider. We would like to stress that our work for the first time, demonstrates the importance and the impact of susy backgrounds for Higgs to susy decays. The production cross section of susy backgrounds can be comparable or larger than the signal processes considered here. This depends on exact details of parameter space, most importantly, the composition of electroweakinos. The relative importance of signal and susy background process for few benchmark scenarios can be seen in Table~\ref{tab:susy_bpt}. Apart from an overall increase in total background cross sections due to susy processes, we observe an increased overlap in the kinematic distributions e.g. $\Delta R_{bb}$ and $\met$ as shown Fig.~\ref{fig:bbplot}. These features of susy backgrounds altogether can lead to an appreciable amount of contribution in the total background. Hence one must appropriately take into account these susy backgrounds into the analysis while searching for heavy Higgs decays in mono-X signatures at the collider. 

\section{Long lived charged particle (LLCP)}
\label{sec:llp}
Along with the prompt final states discussed in last sections, it is also possible that heavy Higgs decays into long lived charginos (with path lengths of few centimetres in the detector). Such long lived chargino can then decay with a soft final state in the detector, leaving the so called disappearing track. In this section, we turn our focus on these scenarios. Recall that the kinematic feature of these type of final states were discussed in section~\ref{sec:llpdecaykine}. As we have discussed earlier, for a winolike LSP the mass gap between $\lspone$ and $\chonepm$ is very small and the $\chonepm$ becomes long lived charged particle (LLCP). We discuss the features of these LLCP and prospects of observing them at collider in the following sections. It should be noted that the heavy neutral Higgs will decay to pair of charginos. In case heavy Higgs decays to pair of light chargino, the existing disappearing track searches are applicable. However as shown previously in section~\ref{sec:scan-result}, heavy Higgs can have a significant branching ratio to $\chonepm \chtwopm$. The $\chtwopm$ decays promptly with visible final states (e.g. leptons, photons or jets). While the existing disappearing track searches~\cite{Aaboud:2017mpt, Sirunyan:2018ldc} are sensitive to presence of additional jets in the final state, they veto energetic leptons. In addition to LLCP production via neutral Higgs, we also review charged Higgs signature which can give rise to disappearing track. This search channel also suffers from same problem of generating additional hits inside tracker which are usually vetoed out in the existing search strategies. 

Before we proceed, a comment about choice of benchmarks in this section is in order. In section~\ref{sec:BS}, we discussed the impact of loop corrections on the predicted chargino decay length. In this work, we have not taken into account such loop corrections. However, the results being discussed depend on the mass hierarchy between the heavy Higgs and the chargino. Including the loop corrections will largely impact the chargino decay length while the mass hierarchy will largely be unaltered. Therefore, we consider that the choice of our benchmark points is justified irrespective of the missing loop corrections to chargino decay lengths.

\subsection{Decay fraction of LLCP at various tracker ranges in the detector}
\label{sec:llpdecayfrac}
\begin{figure}[htb!]
\centering
\includegraphics[trim=0 480 0 70,clip,width=10 cm]{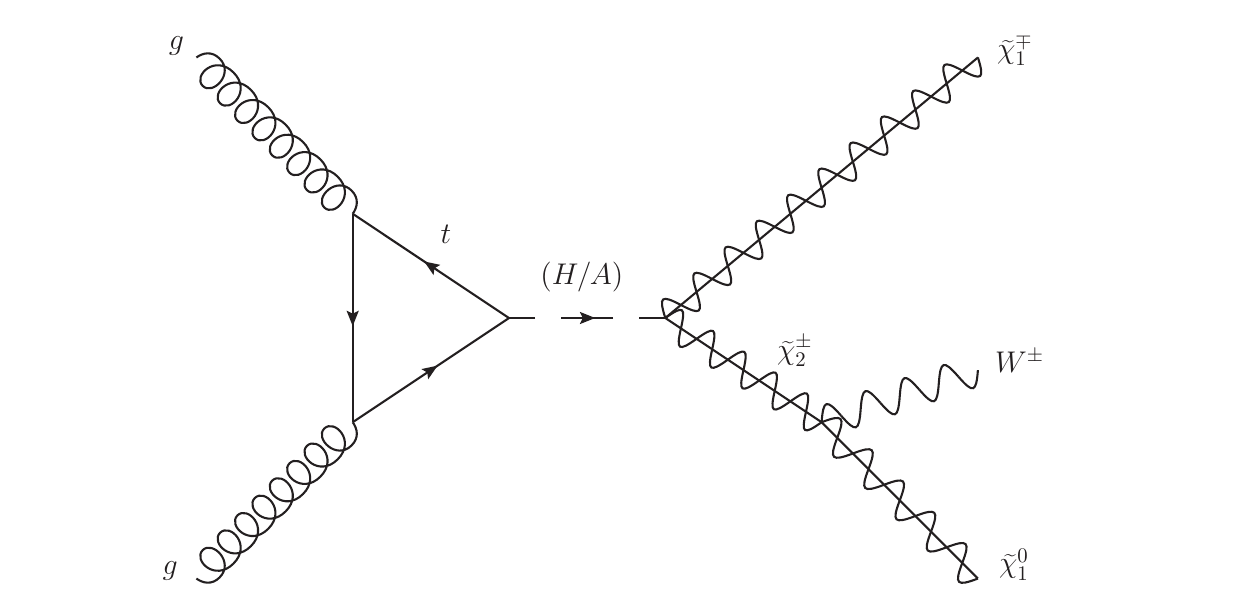}
\caption{The Feynman diagram for the production of $\chonemp$ from $pp\to H\to \chonemp\chtwopm$, $\chtwopm\to W^{\pm}\lspone$ process.}
\label{FD:llcp}
\end{figure}

\begin{figure}[htb!]
\centering
\includegraphics[scale=0.37]{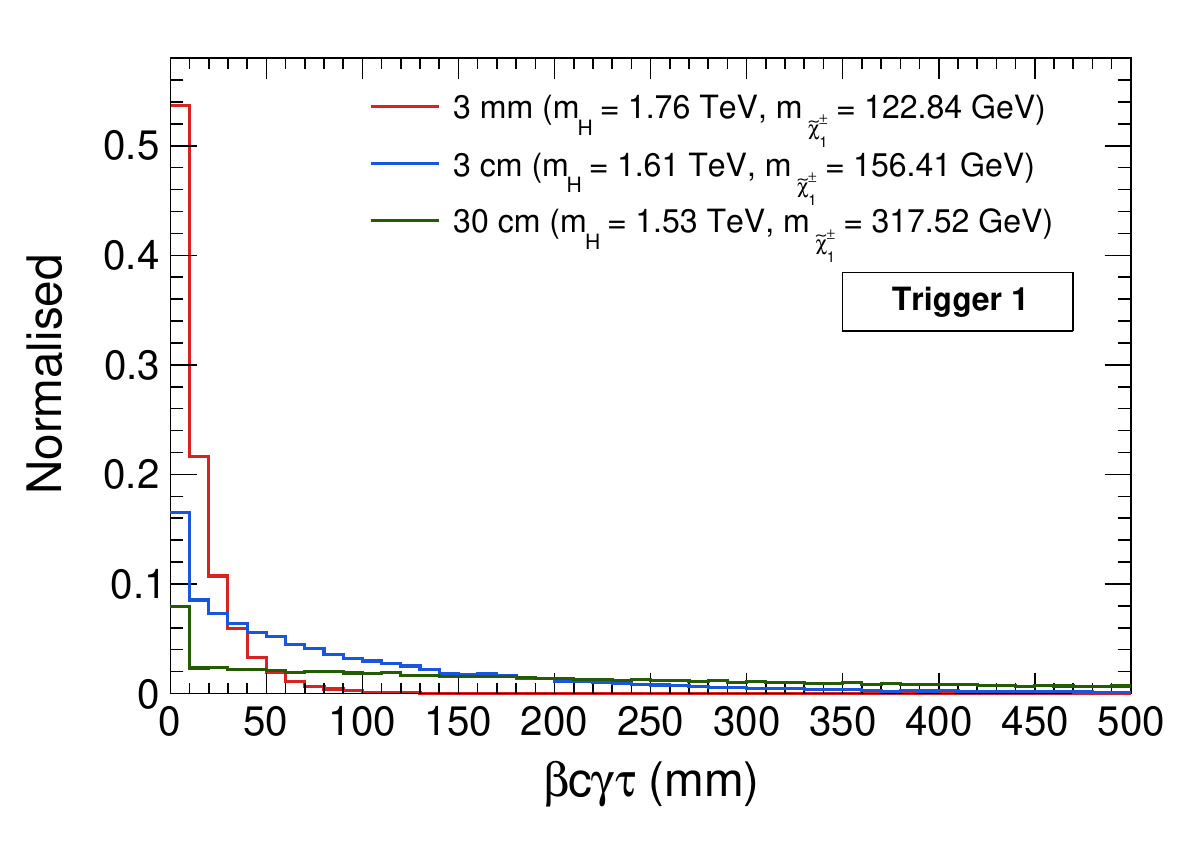}\includegraphics[scale=0.37]{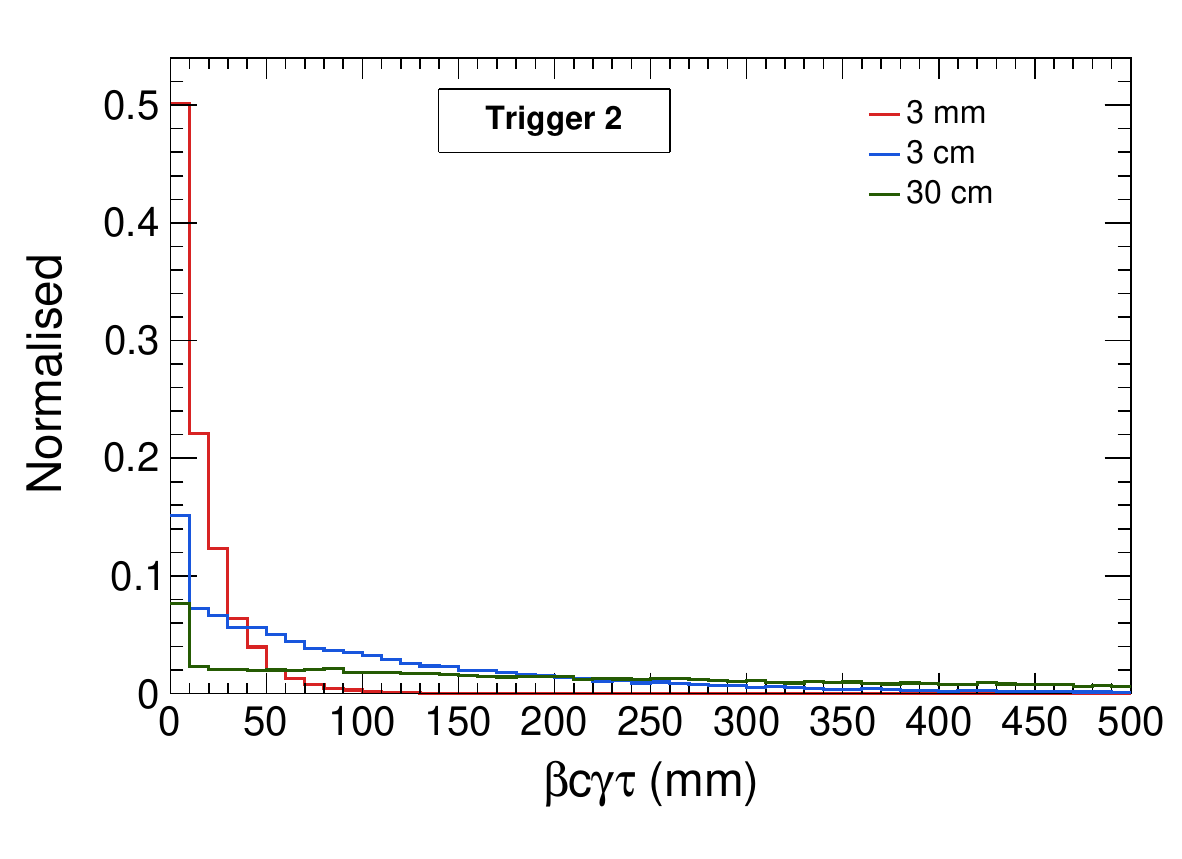}\\
\caption{\it The normalised distribution of mean decay length for the three LLCP scenario after requiring Trigger 1 (left) and Trigger 2 (right) along with additional selection criteria.}
\label{fig:llpdf}
\end{figure}

Here, we take three benchmark points (BPs) with a LLCP $\chonepm$, which satisfy all the collider constraints as discussed in section~\ref{sec:setup}. It should be noted that {\tt SModelS} database currently does not include disappearing track analyses results. These results may pose additional constraints on the parameter space which are not taken into account here. For illustrative purposes, we consider the $\chonepm$ decay lengths of $3$ mm, $3$ cm and $30$ cm corresponding to the benchmarks in equation \ref{eq:6.1}, \ref{eq:6.2} and \ref{eq:6.3} respectively. The LLCP is produced via the process, $pp\to H\to \chonemp\chtwopm$, $\chtwopm\to W^{\pm}\lspone$ (Fig.~\ref{FD:llcp}). We trigger these events by applying cuts on the transverse momentum, $p_T$ and pseudorapidity, $\eta$, of W boson decay products. The analysis is divided into three parts depending on the trigger. The first one is on the lepton from the decay of W boson which must satisfy $p_{T,\ell}>30~\text{GeV},~|\eta_{\ell}|<2.5$. For the second trigger, the events must contain at least one jet with $p_{T,j}>200~\text{GeV}$ and $|\eta_{j}|<2.5$. The events with at least two jets with $p_{T,j}>150~\text{GeV}$ and $|\eta_{j}|<2.5$ are selected in the third trigger. We also demand that the $\chonepm$ must be produced with $p_{T,\chonepm}>100~\text{GeV}$ within $|\eta|<2.5$ in all the trigger choices. These choices are summarised in Table~\ref{tab:llptriggercuts}. The whole set-up and the analysis is done in the \texttt{Pythia-6} framework. We expect use of \texttt{Pythia8} for event generation to produce largely identical results because the kinematic distribution of observables are identical in \texttt{Pythia-6} and \texttt{Pythia-8} which makes the final result unaltered. The event yield of this process after applying these trigger cuts along with the production cross section and the decay branching ratios for different benchmark points are listed in Table~\ref{tab:llpyield} at the HL-LHC with 3 ab$^{-1}$ of integrated luminosity. We also show the normalised distributions of the mean decay length, $\beta c\gamma\tau$, obtained after applying trigger cuts, in case of the three chosen BP's corresponding to different decay lengths in Fig.~\ref{fig:llpdf}.

\begin{equation} 
\label{eq:6.1}
\begin{gathered}
M_A=1.8~{\rm TeV},~tan\beta=16.3,~M_1 = 387~{\rm GeV},~M_2 = 124~{\rm GeV},~\mu = 303~{\rm GeV},\\
M_3 = 4.7~{\rm TeV},~M_{\tilde{Q}_{1_L},\tilde{Q}_{2_L}} = M_{\tilde{u}_R,\tilde{d}_R,\tilde{c}_R,\tilde{s}_R} = M_{\tilde{e}_L,\tilde{\mu}_L,\tilde{e}_R,\tilde{\mu}_R} = 3~{\rm TeV},~M_{\tilde{Q}_{3_L},} = 7.8~{\rm TeV},\\
A_t =-2.4~{\rm TeV},~A_b =0.5~{\rm TeV},~A_\tau =-1~{\rm TeV},~A_{e,\mu,u,d,c,s} = 0,\\
M_{\tilde{\tau}_L} = 1.8~{\rm TeV},~M_{\tilde{\tau}_R} = 2~{\rm TeV},~M_{\tilde{t}_R} = 3.7~{\rm TeV},~M_{\tilde{b}_R,} = 2.9~{\rm TeV}
\end{gathered}
\end{equation} 

\begin{equation} 
\label{eq:6.2}
\begin{gathered}
M_A=1.6~{\rm TeV},~tan\beta=43.7,~M_1 = 913~{\rm GeV},~M_2 = 154~{\rm GeV},~\mu = 347~{\rm GeV},\\
M_3 = 2.4~{\rm TeV},~M_{\tilde{Q}_{1_L},\tilde{Q}_{2_L}} = M_{\tilde{u}_R,\tilde{d}_R,\tilde{c}_R,\tilde{s}_R} = M_{\tilde{e}_L,\tilde{\mu}_L,\tilde{e}_R,\tilde{\mu}_R} = 3~{\rm TeV},~M_{\tilde{Q}_{3_L},} = 7.3~{\rm TeV},\\
A_t =-6.2~{\rm TeV},~A_b =-86~{\rm GeV},~A_\tau =-1.6~{\rm TeV},~A_{e,\mu,u,d,c,s} = 0,\\
M_{\tilde{\tau}_L} = 1~{\rm TeV},~M_{\tilde{\tau}_R} = 1.9~{\rm TeV},~M_{\tilde{t}_R} = 6.4~{\rm TeV},~M_{\tilde{b}_R,} = 8.2~{\rm TeV}
\end{gathered}
\end{equation} 

\begin{equation} 
\label{eq:6.3}
\begin{gathered}
M_A=1.5~{\rm TeV},~tan\beta=5.8,~M_1 = 618~{\rm GeV},~M_2 = 308~{\rm GeV},~\mu = 627~{\rm GeV},\\
M_3 = 4.9~{\rm TeV},~M_{\tilde{Q}_{1_L},\tilde{Q}_{2_L}} = M_{\tilde{u}_R,\tilde{d}_R,\tilde{c}_R,\tilde{s}_R} = M_{\tilde{e}_L,\tilde{\mu}_L,\tilde{e}_R,\tilde{\mu}_R} = 3~{\rm TeV},~M_{\tilde{Q}_{3_L},} = 6.2~{\rm TeV},\\
A_t =4.4~{\rm TeV},~A_b =-264~{\rm GeV},~A_\tau =1.9~{\rm TeV},~A_{e,\mu,u,d,c,s} = 0,\\
M_{\tilde{\tau}_L} = 0.8~{\rm TeV},~M_{\tilde{\tau}_R} = 1.1~{\rm TeV},~M_{\tilde{t}_R} = 4.2~{\rm TeV},~M_{\tilde{b}_R,} = 0.8~{\rm TeV}
\end{gathered}
\end{equation} 

\begin{table}[htb!]
\begin{bigcenter}\scalebox{0.85}{
\begin{tabular}{|c||c|}
\hline
Trigger & Cuts \\\hline

Trigger 1 & \makecell{$p_{T,\ell}>30~\text{GeV}$, $|\eta_{\ell}|<2.5$} \\\hline

Trigger 2 & \makecell{At least one jet with $p_{T}>200~\text{GeV}$ and $|\eta|<2.5$} \\\hline

Trigger 3 & \makecell{At least two jets with $p_{T}>150~\text{GeV}$ and $|\eta|<2.5$} \\

\hline
\end{tabular}}
\end{bigcenter}
\caption{\it Summarizing trigger cuts for LLCP scenario. Along with these trigger cuts, additional constraints are applied on $\chonepm$ which are $p_{T,\chonepm}>100~\text{GeV}$ and $|\eta_{\chonepm}|<2.5$.}
\label{tab:llptriggercuts}
\end{table}

\begin{table}[htb!]
\begin{bigcenter}\scalebox{0.8}{
\begin{tabular}{|c||c|c||c|c|c|}
\hline

$\beta c\gamma\tau$ of & Cross-section of & Branching ratio & Trigger cuts & \multicolumn{2}{c|}{Event yield at 3 ab$^{-1}$} \\\cline{5-6}

$\chonepm$ & $gg/b\bar{b}\to H/A$ (fb) & ($\%$) & & before trigger & after trigger \\\hline\hline

\multirow{3}{*}{$3$ mm} & \multirow{3}{*}{$1.51$} & \multirow{3}{*}{\makecell{$BR(H\to \chonepm\chtwomp)=40.23$\\$BR(A\to \chonepm\chtwomp)=38.33$\\$BR(\chtwopm \to W^{\pm}\lspone)=37.75$}} & Trigger 1 & $437.56$ & $314.52$ \\\cline{4-4}\cline{5-6}

                        & & & Trigger 2 & \multirow{2}{*}{$905.87$} & $403.73$ \\\cline{4-4}\cline{6-6}

                        & & & Trigger 3 &  & $125.30$ \\\hline\hline

\multirow{3}{*}{$3$ cm} & \multirow{3}{*}{$18.97$} & \multirow{3}{*}{\makecell{$BR(H\to \chonepm\chtwopm)=16.82$\\$BR(A\to \chonepm\chtwopm)=16.42$\\$BR(\chtwopm \to W^{\pm}\lspone)=37.13$}} & Trigger 1 & $2287.66$ & $1617.88$ \\\cline{4-4}\cline{5-6}

                        & & & Trigger 2 & \multirow{2}{*}{$4736.18$} & $1796.15$ \\\cline{4-4}\cline{6-6}

                        & & & Trigger 3 &  & $460.36$ \\\hline\hline

\multirow{3}{*}{$30$ cm} & \multirow{3}{*}{$0.93$} & \multirow{3}{*}{\makecell{$BR(H\to \chonepm\chtwopm)=48.77$\\$BR(A\to \chonepm\chtwopm)=45.07$\\$BR(\chtwopm \to W^{\pm}\lspone)=33.72$}} & Trigger 1 & $287.54$ & $221.41$ \\\cline{4-4}\cline{5-6}

                        & & & Trigger 2 & \multirow{2}{*}{$595.30$} & $241.04$ \\\cline{4-4}\cline{6-6}

                        & & & Trigger 3 &  & $49.49$ \\

\hline
\end{tabular}}
\end{bigcenter}
\caption{\it Production cross-section and branching ratios for all the three benchmark points along with the yield after putting trigger cuts at 3 ab$^{-1}$. For the leptonic case, the W boson from $\chtwopm$ is decayed leptonically ($\ell=e,\mu,\tau$), and the W boson decays to jets for the case of jet trigger.}
\label{tab:llpyield}
\end{table}

We compute the fractional number of events where chargino decays within various distances inside detector as the ratio of number of events within that range divided by the number of events passing trigger criterion. We quote these numbers in Table~\ref{tab:llpdecayfrac} for different $\chonepm$ decay lengths. From Table~\ref{tab:llpdecayfrac}, it is evident that due to the Lorentz factors, the $\chonepm$ decays mostly at larger distances with respect to its decay length. Such highly boosted chargino can improve the sensitivity of the disappearing track searches as they live for longer time in the detector. The existing disappearing track searches fail below a chargino decay length below approximately 3 mm, however the boosted chargino produced via heavy Higgs with such low decay length can lead to an additional handle for such scenarios.

\begin{table}[htb!]
\begin{bigcenter}\scalebox{0.85}{
\begin{tabular}{|c||c|c|c|c|c|c|c|}
\hline

Trigger cuts & $\beta c\gamma\tau$ of & \multicolumn{6}{c|}{Fraction of events after trigger in $\%$ within} \\\cline{3-8}

 &  $\chonepm$ & $0-3$ mm & $3-30$ mm & $30$ mm - $10$ cm & $10-30$ cm & $30-100$ cm & $>100$ cm\\\hline

\multirow{3}{*}{Trigger 1} & $3$ mm 
 & $26.61$ & $59.49$ & $13.58$ & $0.32$ & $0.0$ & $0.0$ \\ \cline{2-8}

 & $3$ cm & $9.83$ & $22.56$ & $32.53$ & $28.48$ & $6.57$ & $0.03$ \\ \cline{2-8}

 & $30$ cm & $6.16$ & $6.27$ & $14.07$ & $27.99$ & $35.27$ & $10.24$ \\ \hline\hline
 
\multirow{3}{*}{Trigger 2} & $3$ mm 
 & $24.07$ & $60.49$ & $15.10$ & $0.34$ & $0.0$ & $0.0$ \\ \cline{2-8}

 & $3$ cm & $9.49$ & $19.44$ & $31.62$ & $31.89$ & $7.51$ & $0.05$ \\ \cline{2-8}

 & $30$ cm & $5.86$ & $6.01$ & $13.64$ & $27.66$ & $36.39$ & $10.44$ \\ \hline\hline
 
\multirow{3}{*}{Trigger 3} & $3$ mm
 & $23.06$ & $60.75$ & $15.86$ & $0.33$ & $0.0$ & $0.0$ \\ \cline{2-8}

 & $3$ cm & $9.03$ & $18.01$ & $31.58$ & $32.79$ & $8.53$ & $0.06$ \\ \cline{2-8}

 & $30$ cm & $5.90$ & $5.63$ & $13.04$ & $27.07$ & $37.53$ & $10.83$ \\
 
\hline
\end{tabular}}
\end{bigcenter}
\caption{\it After triggering event with $p_T$ and $\eta$ cut, the fraction of charginos which decay in different regions of tracker (or outside) in detector with decay length of a few mm upto few centimeters.}
\label{tab:llpdecayfrac}
\end{table}

\subsection{Probing charged Higgs via LLCP signature}
\label{sec:Hc}
\begin{figure}[htb!]
\centering
\includegraphics[trim=0 600 0 70,clip,width=\textwidth]{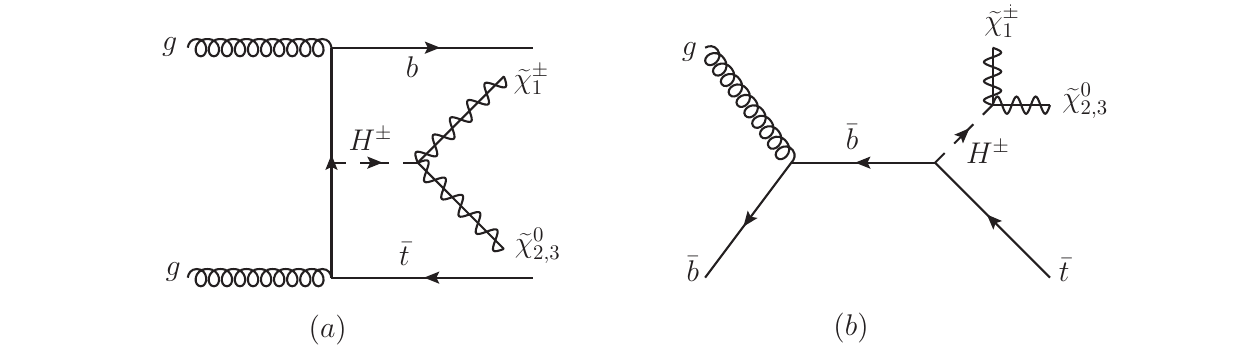}
\caption{The Feynman diagram of (a) $4F$ and (b) $5F$ charged Higgs production at LO, which decays to electroweakinos yielding LLCP signature at the collider.}
\label{FD:cH}
\end{figure}

Finally, we introduce another probe of new physics, namely supersymmetric decays of charged Higgs. The supersummetric final states arising from charged Higgs decays have so far not been analysed in the literature. A complete overview of these decays is beyond the scope of this work. However, we illustrate an example of charged Higgs decays to LLCP. At the LHC, the $H^{\pm}$ is already being searched for by its decay into various Standard Model (SM) particles, \textit{viz.} $H^{\pm}\to \tau^{\pm}\nu$~\cite{Aad:2014kga,Khachatryan:2015qxa,CMS-PAS-HIG-16-031,Aaboud:2016dig} and $H^{\pm}\to t\bar{b}$~\cite{Aad:2015typ,ATLAS-CONF-2016-089}. Below top quark mass, it is mainly produced from the top quark decay in $t\bar{t}$ production, $t\to H^{+}b$. For the case of $m_{H^{\pm}}>m_t$, the charged Higgs production happens via two processes, \textit{viz.} four-flavour ($4F$), $gg/q\bar{q}\to t\bar{b}H^{-}$ and five-flavour ($5F$), $gb\to tH^{-}$. Then, these two processes are matched to get the total inclusive cross-section, in the same way as we have discussed at the beginning of section~\ref{sec:collider}.

In the MSSM, the charged Higgs can decay into a pair of electroweakinos, \textit{viz.} $H^{\pm}\to \chonepm \lsptwo$ (Fig.~\ref{FD:cH}). For a winolike LSP scenario, the $\chonepm$ produced from the decay of $H^{\pm}$ can be long-lived. Therefore, the charged Higgs can be probed via missing charged track signature. A demonstration of possible analysis is our main goal in this section. For this, we choose the following three benchmark points from our scan, equation \ref{eq:Hc600}, \ref{eq:Hc800} and \ref{eq:Hc1000} which corresponds to charged Higgs mass around $600,~800$ and $1000$ GeV respectively. The $4F$ and $5F$ production cross-section for charged Higgs and various electroweakino branching ratios for these benchmark points are given in Table~\ref{tab:cHcsbr}.

\begin{equation} 
\label{eq:Hc600}
\begin{gathered}
M_A=615~{\rm GeV},~tan\beta=13.1,~M_1 = 485~{\rm GeV},~M_2 = 138~{\rm GeV},~\mu = 370~{\rm GeV},\\
M_3 = 4.1~{\rm TeV},~M_{\tilde{Q}_{1_L},\tilde{Q}_{2_L}} = M_{\tilde{u}_R,\tilde{d}_R,\tilde{c}_R,\tilde{s}_R} = M_{\tilde{e}_L,\tilde{\mu}_L,\tilde{e}_R,\tilde{\mu}_R} = 3~{\rm TeV},~M_{\tilde{Q}_{3_L},} = 7.5~{\rm TeV},\\
A_t =4.5~{\rm TeV},~A_b =-1.3~{\rm TeV},~A_\tau =-1.6~{\rm TeV},~A_{e,\mu,u,d,c,s} = 0,\\
M_{\tilde{\tau}_L} = 297~{\rm GeV},~M_{\tilde{\tau}_R} = 1.4~{\rm TeV},~M_{\tilde{t}_R} = 3.5~{\rm TeV},~M_{\tilde{b}_R,} = 4.4~{\rm TeV}
\end{gathered}
\end{equation} 

\begin{equation} 
\label{eq:Hc800}
\begin{gathered}
M_A=815~{\rm GeV},~tan\beta=17.5,~M_1 = 734~{\rm GeV},~M_2 = 105~{\rm GeV},~\mu = 335~{\rm GeV},\\
M_3 = 4~{\rm TeV},~M_{\tilde{Q}_{1_L},\tilde{Q}_{2_L}} = M_{\tilde{u}_R,\tilde{d}_R,\tilde{c}_R,\tilde{s}_R} = M_{\tilde{e}_L,\tilde{\mu}_L,\tilde{e}_R,\tilde{\mu}_R} = 3~{\rm TeV},~M_{\tilde{Q}_{3_L},} = 7.1~{\rm TeV},\\
A_t =695~{\rm GeV},~A_b =1~{\rm TeV},~A_\tau =0.2~{\rm TeV},~A_{e,\mu,u,d,c,s} = 0,\\
M_{\tilde{\tau}_L} = 1~{\rm TeV},~M_{\tilde{\tau}_R} = 1.8~{\rm TeV},~M_{\tilde{t}_R} = 4.8~{\rm TeV},~M_{\tilde{b}_R,} = 4.3~{\rm TeV}
\end{gathered}
\end{equation} 

\begin{equation} 
\label{eq:Hc1000}
\begin{gathered}
M_A=1.1~{\rm TeV},~tan\beta=21.4,~M_1 = 982~{\rm GeV},~M_2 = 119~{\rm GeV},~\mu = 311~{\rm GeV},\\
M_3 = 4.4~{\rm TeV},~M_{\tilde{Q}_{1_L},\tilde{Q}_{2_L}} = M_{\tilde{u}_R,\tilde{d}_R,\tilde{c}_R,\tilde{s}_R} = M_{\tilde{e}_L,\tilde{\mu}_L,\tilde{e}_R,\tilde{\mu}_R} = 3~{\rm TeV},~M_{\tilde{Q}_{3_L},} = 5.8~{\rm TeV},\\
A_t =5.3~{\rm TeV},~A_b =92~{\rm GeV},~A_\tau =-985~{\rm GeV},~A_{e,\mu,u,d,c,s} = 0,\\
M_{\tilde{\tau}_L} = 1.3~{\rm TeV},~M_{\tilde{\tau}_R} = 1~{\rm TeV},~M_{\tilde{t}_R} = 2.2~{\rm TeV},~M_{\tilde{b}_R,} = 1.8~{\rm TeV}
\end{gathered}
\end{equation} 

\begin{table}[htb!]
\begin{bigcenter}\scalebox{0.9}{
\begin{tabular}{|c|c|c|c|c|}
\hline
 $m_{H^\pm}$ & \multicolumn{3}{c|}{Cross-section at NLO (fb)} & Branching ratio \\\cline{2-4}

 (GeV) & $(4F)$ & $(5F)$ & Matched & ($\%$)  \\\hline\hline

 \makecell{$620.29$\\eq.(\ref{eq:Hc600})} & $11.92$ & $16.72$ & $15.48$ & \makecell{$BR(H^{\pm}\to \chonepm\lsptwo)=21.70$, $BR(H^{\pm}\to \chonepm\lspthree)=19.76$\\
                $BR(H^{\pm}\to\chtwopm \lspone)=19.93$, $BR(\lsptwo\to Z\lspone)=5.55$\\
                $BR(\lsptwo\to h\lspone)=23.16$, $BR(\lspthree\to Z\lspone)=27.88$\\
                $BR(\lspthree\to h\lspone)=3.80$, $BR(\chtwopm\to Z\chonepm)=34.89$\\
                $BR(\chtwopm\to h\chonepm)=27.79$}\\\hline\hline

 \makecell{$819.77$\\eq.(\ref{eq:Hc800})} & $6.48$ & $9.33$ & $8.64$ & \makecell{$BR(H^{\pm}\to \chonepm\lsptwo)=20.49$, $BR(H^{\pm}\to \chonepm\lspthree)=21.79$\\
                $BR(H^{\pm}\to\chtwopm \lspone)=22.52$, $BR(\lsptwo\to Z\lspone)=26.68$\\
                $BR(\lsptwo\to h\lspone)=4.54$, $BR(\lspthree\to Z\lspone)=7.00$\\
                $BR(\lspthree\to h\lspone)=21.81$, $BR(\chtwopm\to Z\chonepm)=35.74$\\
                $BR(\chtwopm\to h\chonepm)=27.74$} \\\hline\hline

 \makecell{$1077.18$\\eq.(\ref{eq:Hc1000})} & $2.46$ & $3.77$ & $3.47$ & \makecell{$BR(H^{\pm}\to \chonepm\lsptwo)=19.24$, $BR(H^{\pm}\to \chonepm\lspthree)=20.52$\\
                $BR(H^{\pm}\to\chtwopm \lspone)=22.23$, $BR(\lsptwo\to Z\lspone)=26.72$\\
                $BR(\lsptwo\to h\lspone)=3.55$, $BR(\lspthree\to Z\lspone)=6.59$\\
                $BR(\lspthree\to h\lspone)=20.62$, $BR(\chtwopm\to Z\chonepm)=36.17$\\
                $BR(\chtwopm\to h\chonepm)=26.04$} \\

\hline
\end{tabular}}
\end{bigcenter}
\caption{\it The production cross-section and branching ratios for the selected benchmark points in charged Higgs analysis.}
\label{tab:cHcsbr}
\end{table}

Again, we use \texttt{Pythia-6} to generate charged Higgs production in both the $4F$ and $5F$ scheme, whereas we compute the cross-sections at NLO using {\tt MadGraph-2.6.5} with the model file~\cite{cHModel} made by the authors of \cite{Degrande:2015vpa}. While generating the cross-sections, we set the factorisation and renormalisation scales at $\mu=(m_{H^{\pm}}~+~m_t)/2$. This cross-section depends strongly on the scale variation of the bottom mass which has not been included in the above model file. We take care of it by rescaling the cross-section according to the running of bottom quark mass given in \cite{Bednyakov:2016onn}. 

\subsubsection{$\ell\ell$ + LLCP}
\label{sec:llHc}

Here, we consider the following decay processes of the $H^{\pm}$, \textit{viz.} 
\begin{equation} 
\label{eq:llprocess}
\begin{gathered}
H^{\pm}\to \chonepm~{\wt\chi_{2,3}^0}~,~{\wt\chi_{2,3}^0}\to \lspone~+~(Z\to \ell\ell),\\H^{\pm}\to \chtwopm~\lspone~,~\chtwopm\to \chonepm~+~(Z\to \ell\ell).
\end{gathered}
\end{equation}  
We choose charged Higgs decays with high branching ratio to electroweakino pairs with the long-lived $\chonepm$. These events are triggered with the Z decay products arising from the decay of electroweakinos. There should be exactly two same flavour opposite sign leptons with $p_{T}>25~\text{GeV}$ and $|\eta|<2.5$. The di-lepton invariant mass should be within $15$ GeV window around $Z$ boson mass. We also demand at least one b-tagged jet with $p_{T}>30~\text{GeV}$ and $|\eta|<2.5$. In addition, the long-lived charged track must be within $|\eta|<2.5$ with $p_{T}>100~\text{GeV}$. Table~\ref{tab:cHllyield} summarises these trigger cuts along with the number of events at $3~ab^{-1}$, before and after the cuts. We also calculate the decay length and the decay fractions of the LLCP within different parts of the tracker for all the benchmark points, which are shown in Table~\ref{tab:cHlldf}.

\begin{table}[htb!]
\begin{bigcenter}\scalebox{0.9}{
\begin{tabular}{|c|c|c|c|}
\hline
 \multirow{3}{*}{Trigger Z} & $m_{H^\pm}$ & \multicolumn{2}{c|}{\multirow{2}{*}{\makecell{Total event yield from the processes in\\ equation \ref{eq:llprocess} at 3 ab$^{-1}$}}}\\
 
 & (GeV) &  \multicolumn{2}{c|}{}  \\\cline{3-4}

 &  & before Trigger Z & after Trigger Z \\\hline

\multirow{3}{*}{\makecell{$p_{T,\ell_{1,2}}>25~\text{GeV}$, ~$|\eta_{\ell_{1,2}}|<2.5$, \\ 
                      $76$ GeV $<m_{\ell\ell}<106$ GeV,\\
                      $p_{T,b}>30~\text{GeV},~|\eta_{b}|<2.5$}} &
                      $620.29$ & \makecell{$427.04$} & \makecell{$173.12$} \\\cline{2-4}

 & $819.77$ & \makecell{$262.40$} & \makecell{$119.54$} \\\cline{2-4}

 & $1077.18$ & \makecell{$101.91$} & \makecell{$47.98$} \\

\hline
\end{tabular}}
\end{bigcenter}
\caption{\it The event yield at 3 ab$^{-1}$ from all the processes before and after, applying trigger cuts and $p_{T,\chonepm}>100~\text{GeV}$, $|\eta_{\chonepm}|<2.5$.}
\label{tab:cHllyield}
\end{table}

\begin{table}[htb!]
\begin{bigcenter}\scalebox{0.85}{
\begin{tabular}{|c|c||c|c|c|c|c|c|}
\hline

 $m_{H^\pm}$ & $\beta c\gamma\tau$ of & \multicolumn{6}{c|}{Fraction of events after Trigger Z in $\%$ within} \\\cline{3-8}

 (GeV) &  $\chonepm$ (cm) & $0-3$ mm & $3-30$ mm & $30$ mm - $10$ cm & $10-30$ cm & $30-100$ cm & $>100$ cm\\\hline\hline

 $620.29$ & $27.66$ & $0.91$ & $6.90$ & $16.22$ & $31.31$ & $36.09$ & $8.57$ \\\hline
 
 $819.77$ & $18.10$ & $0.90$ & $7.94$ & $16.97$ & $32.75$ & $33.86$ & $7.58$ \\\hline

 $1077.18$ & $3.7$ & $4.07$ & $28.75$ & $37.55$ & $24.98$ & $4.60$ & $0.05$ \\
 
\hline
\end{tabular}}
\end{bigcenter}
\caption{\it The fractional number of events which decay at different parts inside tracker for the $4F$ production process with $H^{\pm}\to\chtwopm~\lspone$ in $\ell\ell$ + LLCP category.}
\label{tab:cHlldf}
\end{table}

\subsubsection{$b\bar{b}$ + LLCP}
\label{sec:bbHc}

In this case we consider the following charged Higgs decay cascades :
\begin{equation} 
\label{eq:bbprocess}
\begin{gathered}
H^{\pm}\to \chonepm~{\wt\chi_{2,3}^0}~,~{\wt\chi_{2,3}^0}\to \lspone~+~(h\to b\bar{b}),\\H^{\pm}\to \chtwopm~\lspone~,~\chtwopm\to \chonepm~+~(h\to b\bar{b}).
\end{gathered}
\end{equation}  
This channel has the advantage of having higher event yield because of large $h\to b\bar{b}$ branching ratio but may also suffer from huge QCD backgrounds. Since the b-jets will have smearing effect from the detector at collider, we simulate the detector effect with {\tt Delphes-3.4.1} with the same configuration as discussed in section~\ref{sec:collider}. The events should contain atleast two b-jets with $p_{T}>30~\text{GeV}$ and $|\eta|<2.5$. Since the b-jets are coming from the SM Higgs boson, the invariant mass of the two b-jets must be in the range, $[90,130]$ GeV with the separation in the $\eta-\phi$ plane as $\Delta R_{bb}=[0.4,2.0]$. As before, we allow only those LLCP track which are within $|\eta|<2.5$ with $p_{T}>100~\text{GeV}$. We show these trigger cuts along with the event yield at $3~ab^{-1}$ in Table~\ref{tab:cHbbyield}. In Table~\ref{tab:cHbbdf}, we list the fractions of charginos which decay at different ranges inside tracker.

\begin{table}[htb!]
\begin{bigcenter}\scalebox{0.9}{
\begin{tabular}{|c|c|c|c|}
\hline
 \multirow{3}{*}{Trigger hbb} & $m_{H^\pm}$ & \multicolumn{2}{c|}{\multirow{2}{*}{\makecell{Total event yield from the processes in\\ equation \ref{eq:bbprocess} at 3 ab$^{-1}$}}}\\
 
 & (GeV) &  \multicolumn{2}{c|}{}  \\\cline{3-4}

 &  & before Trigger hbb & after Trigger hbb \\\hline

\multirow{3}{*}{\makecell{$p_{T,b_{1,2}}>30~\text{GeV}$, ~$|\eta_{b_{1,2}}|<2.5$, \\ 
                      $90$ GeV $<m_{bb}<130$ GeV,\\
                      $0.4<\Delta R_{bb}<2.0$}} &
                      $620.29$ & \makecell{$3060.04$} & \makecell{$238.06$} \\\cline{2-4}

 & $819.77$ & \makecell{$1801.37$} & \makecell{$169.75$} \\\cline{2-4}

 & $1077.18$ & \makecell{$649.55$} & \makecell{$68.58$} \\

\hline
\end{tabular}}
\end{bigcenter}
\caption{\it Summarising the trigger cuts (additional cut: $p_{T,\chonepm}>100~\text{GeV}$ and $|\eta_{\chonepm}|<2.5$) and the event yield at 3 ab$^{-1}$ in the $b\bar{b}$ + LLCP category.}
\label{tab:cHbbyield}
\end{table}

\begin{table}[htb!]
\begin{bigcenter}\scalebox{0.85}{
\begin{tabular}{|c|c||c|c|c|c|c|c|}
\hline

 $m_{H^\pm}$ & $\beta c\gamma\tau$ of & \multicolumn{6}{c|}{Fraction of events after Trigger hbb in $\%$ within} \\\cline{3-8}

 (GeV) &  $\chonepm$ (cm) & $0-3$ mm & $3-30$ mm & $30$ mm - $10$ cm & $10-30$ cm & $30-100$ cm & $>100$ cm\\\hline\hline

 $620.29$ & $27.66$ & $6.43$ & $7.86$ & $16.55$ & $31.49$ & $32.08$ & $5.59$ \\\hline
 
 $819.77$ & $18.10$ & $6.09$ & $8.96$ & $17.59$ & $31.27$ & $30.00$ & $6.09$ \\\hline

 $1077.18$ & $3.7$ & $10.55$ & $31.13$ & $34.07$ & $20.67$ & $3.54$ & $0.04$ \\
 
\hline
\end{tabular}}
\end{bigcenter}
\caption{\it The fractional number of events which decay at different parts inside tracker for the $4F$ production process with $H^{\pm}\to\chtwopm~\lspone$ in $b\bar{b}$ + LLCP category.}
\label{tab:cHbbdf}
\end{table}

\subsubsection{$\gamma\gamma$ + LLCP}
\label{sec:aaHc}
Finally we consider the following decay chain of the charged Higgs :
\begin{equation} 
\label{eq:aaprocess}
\begin{gathered}
H^{\pm}\to \chonepm~{\wt\chi_{2,3}^0}~,~{\wt\chi_{2,3}^0}\to \lspone~+~(h\to \gamma\gamma),\\H^{\pm}\to \chtwopm~\lspone~,~\chtwopm\to \chonepm~+~(h\to \gamma\gamma).
\end{gathered}
\end{equation}  
This is the cleanest channel because of the photons in the final state at the cost of event yield. In this channel, we demand exactly two photons and at least one b-tagged jet with $p_{T}>30~\text{GeV}$ and $|\eta|<2.5$. The di-photon invariant mass must fall in the range, $[122,128]$ GeV with the $\Delta R$ separation between the photons, $\Delta R_{\gamma\gamma}=[0.4,2.0]$. Here, the event yield (Table~\ref{tab:cHaayield}) is negligible even at $14$ TeV with 3 ab$^{-1}$ of integrated luminosity. We would like to mention here that the matched cross-sections in Table~\ref{tab:cHcsbr} becomes $113.84$ fb, $75.50$ fb and $38.44$ fb for $m_{H^\pm}=620.29,~819.77,~\text{and}~1077.18$ GeV respectively at the proposed HE-LHC (High Energy LHC) with $\sqrt{s}=27$ TeV. This will increase the event yield an order of magnitude higher to have a better prospect of observing charged Higgs in this channel. We show the decay fractions of charginos at different track ranges in Table~\ref{tab:cHaadf}.

\begin{table}[htb!]
\begin{bigcenter}\scalebox{0.9	}{
\begin{tabular}{|c|c|c|c|}
\hline
 \multirow{3}{*}{Trigger $h\gamma\gamma$} & $m_{H^\pm}$ & \multicolumn{2}{c|}{\multirow{2}{*}{\makecell{Total event yield from the processes in\\ equation \ref{eq:aaprocess} at 3 ab$^{-1}$}}}\\
 
 & (GeV) &  \multicolumn{2}{c|}{}  \\\cline{3-4}

 &  & before Trigger $h\gamma\gamma$ & after Trigger $h\gamma\gamma$ \\\hline

\multirow{3}{*}{\makecell{$p_{T,\gamma_{1,2},b}>30~\text{GeV}$, ~$|\eta_{\gamma_{1,2},b}|<2.5$, \\
                      $122$ GeV $<m_{\gamma\gamma}<128$ GeV,\\
                      $0.4<\Delta R_{\gamma\gamma}<2.0$}} &
                      $620.29$ & \makecell{$11.93$} & \makecell{$3.65$} \\\cline{2-4}

 & $819.77$ & \makecell{$7.03$} & \makecell{$2.38$} \\\cline{2-4}

 & $1077.18$ & \makecell{$2.53$} & \makecell{$0.87$} \\

\hline
\end{tabular}}
\end{bigcenter}
\caption{\it Summarising the trigger cuts (additional cuts applied on $\chonepm$: $p_{T,\chonepm}>100~\text{GeV}$ and $|\eta_{\chonepm}|<2.5$) and the event yield at 3 ab$^{-1}$ in the $\gamma\gamma$ + LLCP category.}
\label{tab:cHaayield}
\end{table}

\begin{table}[htb!]
\begin{bigcenter}\scalebox{0.85}{
\begin{tabular}{|c|c||c|c|c|c|c|c|}
\hline

 $m_{H^\pm}$ & $\beta c\gamma\tau$ of & \multicolumn{6}{c|}{Fraction of events after Trigger $h\gamma\gamma$ in $\%$ within} \\\cline{3-8}

 (GeV) &  $\chonepm$ (cm) & $0-3$ mm & $3-30$ mm & $30$ mm - $10$ cm & $10-30$ cm & $30-100$ cm & $>100$ cm\\\hline\hline

 $620.29$ & $27.66$ & $0.89$ & $7.71$ & $16.30$ & $31.60$ & $35.56$ & $7.94$ \\\hline
 
 $819.77$ & $18.10$ & $0.90$ & $8.22$ & $17.46$ & $31.95$ & $34.28$ & $7.19$ \\\hline

 $1077.18$ & $3.7$ & $3.93$ & $29.98$ & $37.70$ & $24.41$ & $3.94$ & $0.04$ \\
 
\hline
\end{tabular}}
\end{bigcenter}
\caption{\it The fractional number of events which decay at different parts inside tracker for the $4F$ production process with $H^{\pm}\to\chtwopm~\lspone$ in $\gamma\gamma$ + LLCP category.}
\label{tab:cHaadf}
\end{table}
\section*{Summary LLCP final states}
\label{sec:summllcp}

To summarise this section, we find possible interesting signatures for discovering heavy neutral and charged Higgs boson at the collider. Because of the boost received from the heavy Higgs decay, the charginos can travel larger distance inside the detector compared to their decay length. This can improve the existing sensitivity on disappearing track searches and look for possible signature of heavy Higgs production. If the decay length of long-lived particle is less than 1 cm or a few mm, a short track is formed, called tracklet. Their  search has been proposed in the literature~\cite{Aaboud:2017mpt, Fukuda:2020uva, Fukuda:2017jmk, PhysRevD.87.015028}~\footnote{In case of disappearing track, one has to measure the lifetime of long-lived particle. This can be looked up for example in Ref.~\cite{Banerjee:2019ktv} and the references therein.}. In case of charged Higgs production, one can look for $\ell\ell+\met$, $b\bar{b}+\met$ and $\gamma\gamma+\met$ final states along with a disappearing charged track with large transverse momentum. Background contamination to these final states mainly comes in the form of fake track signature due to incorrect reconstruction of hits inside tracker, along with SM particles giving rise to similar final states. Data-driven techniques could play an important role in estimating these backgrounds which is beyond the scope of our work. Instead we compute the event yield for such signal processes at the HL-LHC and calculate the fraction of events which decay at different parts of the tracker. In short, the features of these final states opens a new avenue to search for heavy Higgses at the collider.


\section{Conclusion}
\label{sec:conclusion}
In this work, we visited the decays to heavy Higgs to supersymmetric particles, particularly concentrating on the electroweakino sector. After performing a survey of available MSSM parameter space, we chose a few benchmark scenarios and analysed in detail the reach of HL-LHC for these benchmark in the mono-X category, particularly concentrating on mono-Z (dilepton) and mono-h ($b\bar{b}, \gamma\gamma$) final states. These were further split into analyses depending on the Higgs production mechanisms which corresponded to b-tag or b-veto category analysis. This resulted in a total of six analyses categories. 

In order to perform signal optimisation, we considered events originating from SM backgrounds as well as direct susy production at the LHC. We demonstrated by constructing specific kinematic variables, that it is possible to discriminate between direct electroweakino production and electroweakino production via Higgs decays. Such discrimination relies on the fact that electroweakino production via heavy Higgs decays carries an imprint of the resonance, while the direct electroweakino production takes place via off-shell SM mediators. 

In terms of optimised analysis, we get the largest significances for the dilepton + MET final state due to the cleanliness of the signal at the LHC. Within this category, we demonstrate that exploiting additional b-tagged jet helps improve the significance of the signal. The second most important channel is the $b\bar{b} + \met$ final state, which results in the mono-h events. In this case, tagging the additional jet actually reduces the significance due to increased backgrounds. The least promising final state is the $\gamma\gamma + \met$ final state. This is understandable as the SM Higgs to photon branching ratio is very small, therefore even if the channel is clean, it is not helpful at the LHC. 

Complementing our analysis in the missing energy final state, we also explored the possibility of heavy Higgs decays to long lived chargino. These chargino travel a finite distance in the detector before decaying, therefore producing the disappearing track signature. Heavy Higgs decays to long lived chargino can involve visible states along with disappearing track in the detector e.g. jets or leptons. We demonstrate that boost gained by the chargino due to on-shell Higgs mediator might be exploited for such searches potentially increasing the reach of LHC searches for disappearing track analyses in addition opening another channel for heavy Higgs searches. 

While the existence of susy at the LHC is increasingly being doubted, all attempts should be made to search for a possible signature before abandoning the idea of SUSY. With this in mind, the heavy Higgs decays to susy present an interesting opportunity to search for new physics scenarios at the HL-LHC.

\appendix

\section{Random scan results}
\label{sec:appendixA}

\begin{figure}[htb!]
\centering
\includegraphics[scale=0.18]{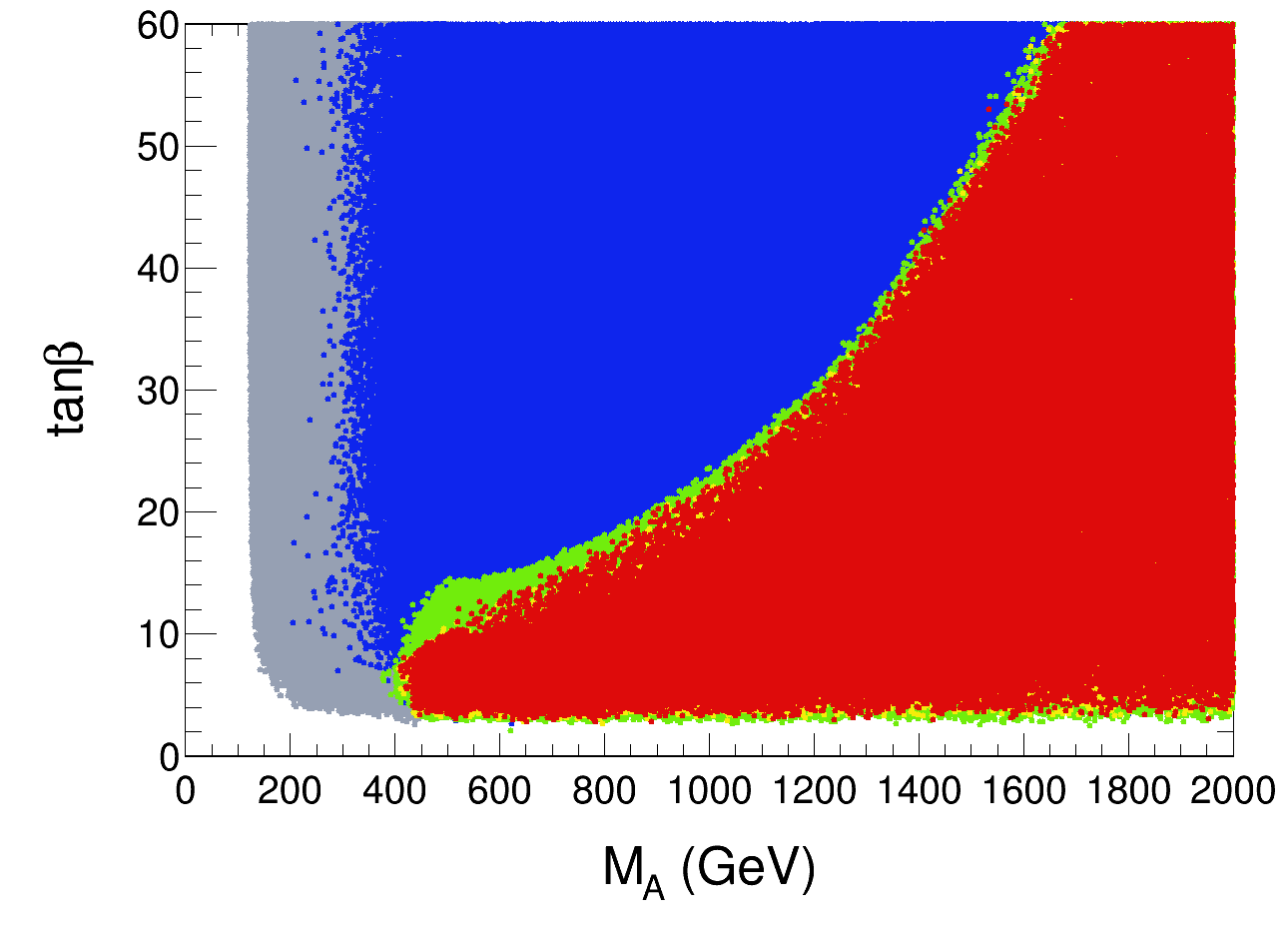}\includegraphics[scale=0.18]{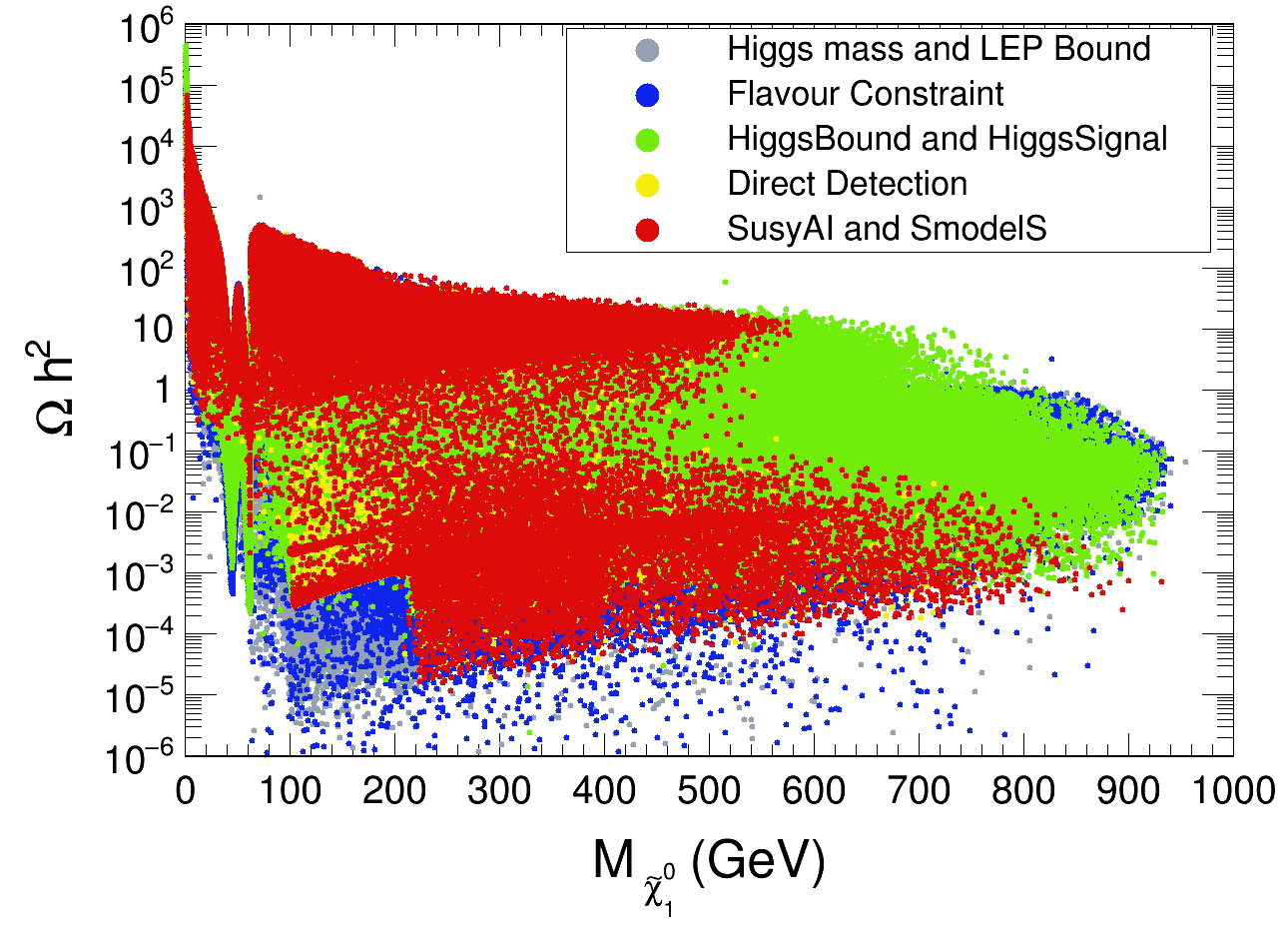}\\
\includegraphics[scale=0.18]{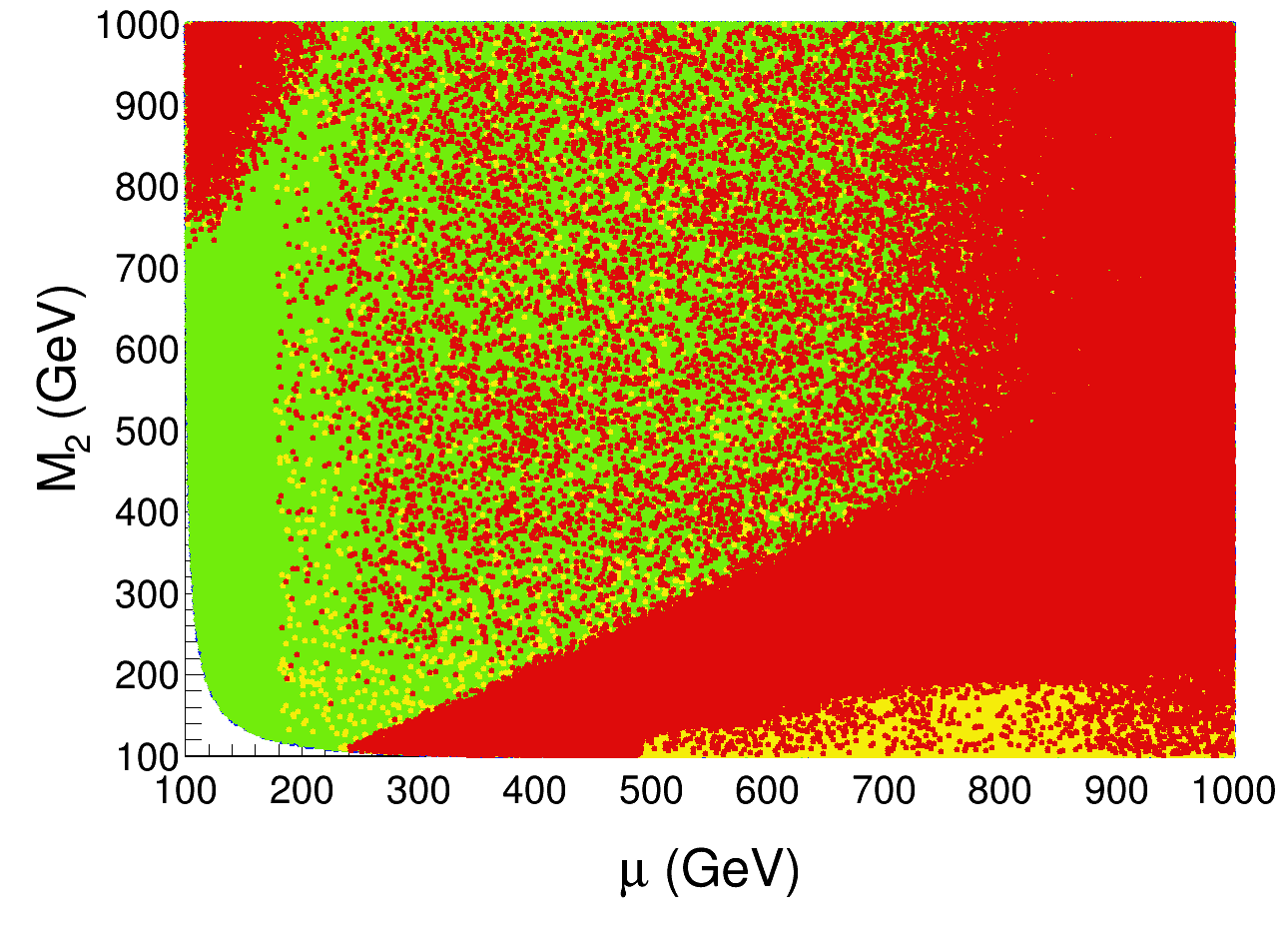}\includegraphics[scale=0.18]{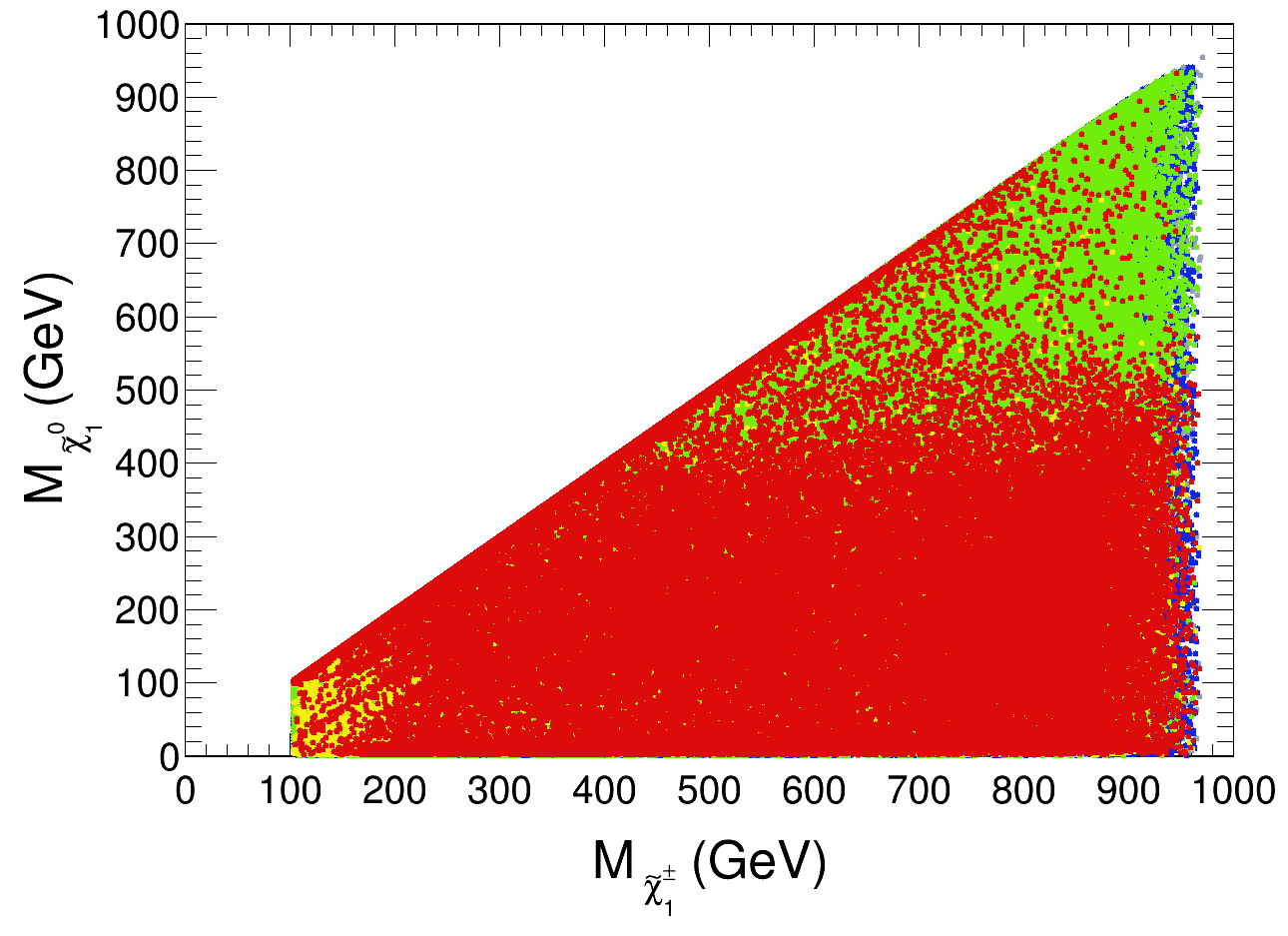}\\
\includegraphics[scale=0.18]{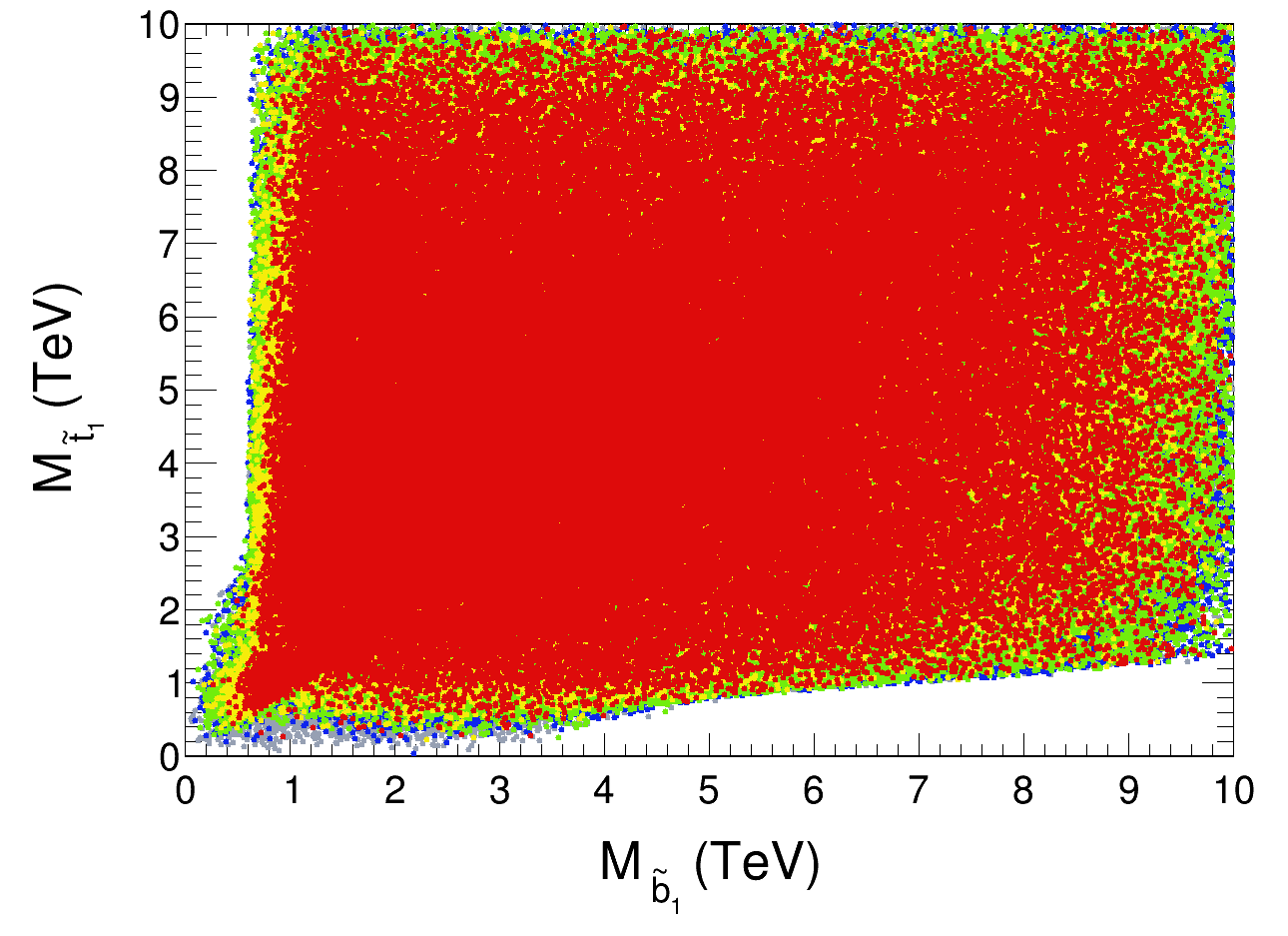}\includegraphics[scale=0.18]{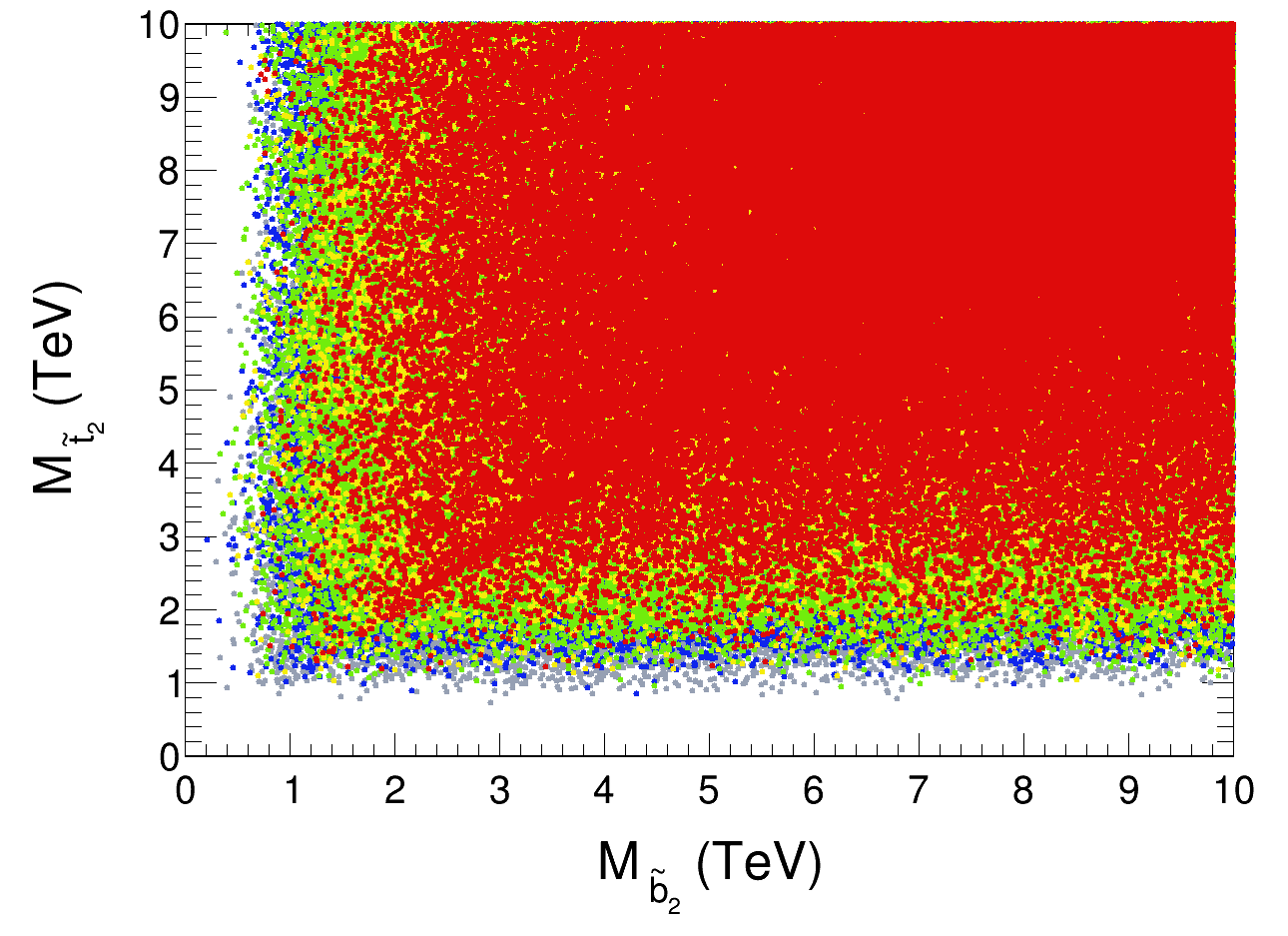}
\caption{\it Scatter plots in the plane of various MSSM parameters and masses.}
\label{fig:scanvar}
\end{figure}
In this appendix, we discuss the results of our flat random scan in the plane of various MSSM parameters. In Fig.~\ref{fig:scanvar}, we show different mass planes to this effect. They are $m_A-tan\beta$ (top left), $\Omega h^2-M_{\lspone}$ where $\Omega h^2$ is the relic abundance of dark matter (top right), $\mu-M_2$ (middle left), $M_{\chonepm}-M_{\lspone}$ (middle right), $M_{\tilde{b_1}}-M_{\tilde{t_1}}$ (bottom left) and $M_{\tilde{b_2}}-M_{\tilde{t_2}}$ (bottom right) plane. We plot the allowed points after each of the experimental constraints i.e. SM Higgs boson mass range $(122,128)$ GeV, LEP constraints (grey), the flavour physics constraint (blue), Higgs signal strengths and heavy Higgs searches (green), dark matter direct detection constraint (yellow) and finally LHC constraints (red). The exact codes tools and experimental constraints used here, has been mentioned in section~\ref{sec:setup}, In general there are a few take home messages here. First, we see that in general points with light pseudo-scalar Higgs are ruled out primarily by the heavy Higgs searches in combination with the SM Higgs signal strength measurements. Second, in general light electroweakinos are in general allowed at the LHC, such points are either strongly wino-like or strongly higgsino-like if at least one of the electroweakino is to be light. For heavier electroweakinos in general an arbitrary combination can be obtained. Such an observation has important consequences at the LHC, as a wino-like LSP is often accompanied with a long lived chargino. This is reflected in our benchmark points. Finally, we also see that generally it is difficult to obtain light stop and sbottoms at the LHC. 

\section{Parton level kinematics}
\label{sec:appendixB}

\begin{figure}[htb!]
\centering
\includegraphics[scale=0.37]{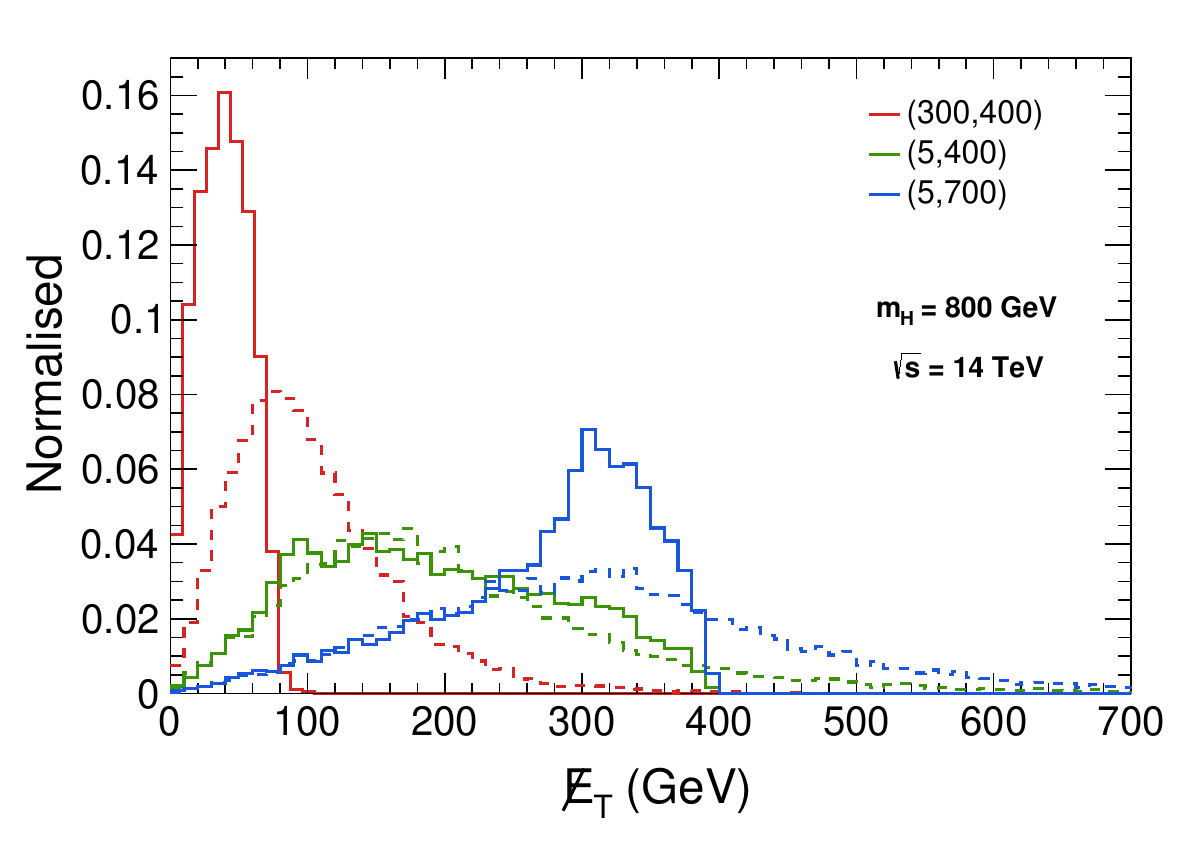}
\caption{\it Normalised distribution of $\met$ at the parton level. The other details are same as the left plot of Fig.~\ref{fig:kin_met_deltaR}.}
\label{fig:metparton}
\end{figure}

At the parton level, the heavy Higgs is produced almost at rest in case of resonant production and the leptons from Z boson gets boost only from the mass difference between neutralinos ($m_{\lsptwo} - m_{\lspone}$). In case of direct production, the neutralinos are produced with large transverse momentum ($p_T$) which goes into the final state leptons along with the contribution coming from the mass gap between neutralinos. Therefore, the direct susy production creates more missing transverse momentum ($\met$) compared to the resonant production. Fig.~\ref{fig:metparton} describes this feature which has an endpoint for the resonant susy production. Now, at the detector level the whole system of neutralinos gets recoiled against jets. The heavy Higgs in the resonant production is now produced with some boost as compared with parton level case and give similar $\met$ distribution from direct susy production for ($m_{\lsptwo}, m_{\lspone}$) = (400, 300) and (400,5) GeV cases which is depicted in Fig.~\ref{fig:kin_met_deltaR}.

\newpage

\section{Summarising the cross sections and generator level cuts for the SM backgrounds}
\label{sec:appendixC}

Here, we give a summary of the production cross section of the SM backgrounds and the cuts used while generating these backgrounds in {\tt MadGraph-2.6.5}.

\begin{table}[h!]
\centering
\begin{bigcenter}
\scalebox{0.75}{%
\begin{tabular}{|c|c|c|c|c|}
\hline
Process & Backgrounds & \makecell{Generation-level cuts ($\ell=e^\pm,\mu^\pm,\tau^\pm$)\\ (NA : Not Applied)} & Cross section (fb)  \\ \hline\hline

\multicolumn{4}{|c|}{$pp\to H\to \lspone~+~(\widetilde{\chi}_{ 2,3}^{0}) \to \lspone~+~(\lspone~+~Z,~Z\to \ell\ell)$ final state} \\ \hline\hline
\multirow{9}{*}{$\ell\ell+\met$}
                                           & {$ZZ$ + jets} & {\makecell{$p_{T,j/b}>20~\text{GeV}$, $|\eta_j|<5.0$, $|\eta_{b}|<3.0$, \\ $\Delta R_{b,j}$\footnote{$\Delta R_{b,j}$ means $\Delta R$ between all possible combination of $b$ and light jet.}$>0.2$}} & {$11427.77$}   \\\cline{2-4}                                                                                      

                                           & {$WZ$ + jets} & {$same$ as $ZZ$ + jets} & {$39684.46$}\\\cline{2-4}  

                                           & {$VVV$} & {NA} & {$266.61$}\\\cline{2-4}                                                                                                                             
                                           
                                           & {$t\bar{t}Z$} & {NA} & {$851.86$}\\\cline{2-4}                                                                                                                                                                      

                                           & {$t\bar{t}$ leptonic} & {\makecell{$p_{T,j/b}>20~\text{GeV}$, $p_{T,\ell}>15~\text{GeV}$, $|\eta_j|<5.0$, \\ $|\eta_{b/\ell}|<3.0$, $\Delta R_{b,j,\ell}>0.2$, $\met >100~\text{GeV}$}}  & {$12013.93$}  \\\cline{2-4}                                                                                  
   
                                           & {$\ell\ell$ + jets} & {$same$ as $t\bar{t}$ leptonic} & $3154714.02$   \\\cline{2-4}                                                                                  
                                                                                                                             
                                           & {$t\bar{t}h$} & {NA} & {$611.30$}\\\cline{2-4}                                                                                                                                                                      
                                           
                                           & {$t\bar{t}W$} & {NA} & {$520.03$}\\\cline{2-4}                                                                                                                                                                      
                                           
                                           & {$WW$ + jets} & {\makecell{$p_{T,j/b}>20~\text{GeV}$, $|\eta_j|<5.0$, $|\eta_{b}|<3.0$,\\ $\Delta R_{b,j}>0.2$}} & {$478670.32$}\\
                                           
                                           \hline

\end{tabular}}
\end{bigcenter}
\caption{Generation level cuts and cross-sections for the various Standard Model backgrounds used in the analyses.}
\label{app1:1}
\end{table}

\begin{table}[h!]
\centering
\begin{bigcenter}
\scalebox{0.75}{%
\begin{tabular}{|c|c|c|c|c|}
\hline
Process & Backgrounds & \makecell{Generation-level cuts ($\ell=e^\pm,\mu^\pm,\tau^\pm$)\\ (NA : Not Applied)} & Cross section (fb)  \\ \hline\hline


\multicolumn{4}{|c|}{$pp\to H\to \lspone~+~\widetilde{\chi}_{ 2,3}^{0} \to \lspone~+~(\lspone~+~h,~h\to b\bar{b}/\gamma\gamma)$ final states} \\ \hline\hline
\multirow{9}{*}{$b\bar{b}+\met$}          
                                           & {$t\bar{t}$ hadronic} & {\makecell{$p_{T,j/b}>20~\text{GeV}$, $|\eta_j|<5.0$, $|\eta_{b}|<3.0$, \\ $\Delta R_{b,j,\ell}>0.2$, $m_{bb}>30$ GeV, $\met >100~\text{GeV}$}}  & {$165994.96$}  \\\cline{2-4}
                                           
                                           & {$t\bar{t}$ semi-leptonic} & {$same$ as $t\bar{t}$ hadronic}  & {$32282.76$}  \\\cline{2-4}
                                                                                      
                                           & {$t\bar{t}$ leptonic} & {$same$ as $t\bar{t}$ hadronic}  & {$16340.61$}  \\\cline{2-4}
                                           
                                           & {$Zb\bar{b}$, $Z\to \nu\nu$} & {\makecell{$p_{T,b}>20~\text{GeV}$, $|\eta_{b}|<3.0$, $\Delta R_{bb}>0.2$,\\ $m_{bb}>30$ GeV, $\met >100~\text{GeV}$}}  & {$2158.69$}   \\\cline{2-4}       
                                                                                 
                                           & {$Zh$ + jets} & {\makecell{$p_{T,j/b}>20~\text{GeV}$, $|\eta_j|<5.0$, $|\eta_{b}|<3.0$,\\ $\Delta R_{j,b}>0.2$}}  & {$969.00$}  \\\cline{2-4}
                                                                                      
                                           & {$Wh$ + jet, $W\to\ell\nu$, $h\to b\bar{b}$} & {$same$ as $t\bar{t}$ hadronic}  & {$0.55$}   \\\cline{2-4}
                                         
                                           & {$t\bar{t}h$} & {NA}  & {$611.30$}   \\\cline{2-4}
                                           
                                           & {$t\bar{t}Z$} & {NA}  & {$851.86$}  \\\cline{2-4}                                             
                                           
                                           & {$t\bar{t}W$} & {NA}  & {$520.03$}   \\\hline

\multirow{4}{*}{$\gamma\gamma+\met$ }      & {$Zh$ + jet, $h\to\gamma\gamma$} & {\makecell{$p_{T,j/b/\gamma}>20~\text{GeV}$, $|\eta_j|<5.0$, $|\eta_{b/\gamma}|<3.0$,\\ $\Delta R_{j/b/\gamma}>0.2$, 110 GeV  $<m_{\gamma\gamma}<$ 140 GeV}}  & {$1.65$}  \\\cline{2-4}
                                               
                                           & {$Wh$ + jet, $h\to\gamma\gamma$} & {$same$ as $Zh$ + jet}  & {$2.54$}   \\\cline{2-4}    
                                               
                                           & {$Z\gamma\gamma$ + jet, $Z\to \nu\nu$} & {$same$ as $Zh$ + jet}  & {$1.87$}   \\\cline{2-4}
                                                                                                                                                                             
                                           & {$t\bar{t}h$, $h\to \gamma\gamma$} & {$same$ as $Zh$ + jet}  & {$1.19$}   \\                 
\hline 
\end{tabular}}
\end{bigcenter}
\caption{Generation level cuts and cross-sections for the various Standard Model backgrounds used in the analyses.}
\label{app1:1}
\end{table}

\newpage

\acknowledgments
We thank Rahool kumar Barman, Aravind Vijay, Alberto Escalante del Valle, Nishita Desai, Jean-Loic Kneur and Dirk Zerwas for helpful discussions at various stages of the work. The work of BB is supported by the Department of Science and Technology, Government of India, under the Grant Agreement number IFA13-PH-75 (INSPIRE Faculty Award). The work of RMG is supported by the Department of Science and Technology, India under Grant No. SR/S2/JCB-64/2007. The work of Najimuddin Khan is supported by the Department of Science and Technology, Government of INDIA under the SERB-Grant PDF/2017/00372. SK is supported by Elise-Richter grant project number V592-N27. We acknowledge Austrian - India WTZ-DAE exchange project number IN 15/2018. S.K. thanks IISc CHEP department for hospitality and WHEPP workshop, IIT Guwahati for hospitality where part of this work was completed.

\providecommand{\href}[2]{#2}
\addcontentsline{toc}{section}{References}
\bibliographystyle{unsrt}
\bibliographystyle{JHEP}
\bibliography{refs}

\begin{thebibliography}{100}

\bibitem{Dreesbook}
R.~Godbole M.~Drees and P.~Roy.
\newblock {\em Theory and Phenomenology of Sparticles}, (world scientific,
  2004).

\bibitem{Baerbook}
H.~Baer and X.~Tata.
\newblock {\em Weak Scale Supersymmetry}, (cambridge university press, 2006).

\bibitem{Martin:1997ns}
Stephen~P. Martin.
\newblock {A Supersymmetry primer}.
\newblock pages 1--98, 1997.
\newblock [Adv. Ser. Direct. High Energy Phys.18,1(1998)].

\bibitem{Djouadi:2005gj}
Abdelhak Djouadi.
\newblock {The Anatomy of electro-weak symmetry breaking. II. The Higgs bosons
  in the minimal supersymmetric model}.
\newblock {\em Phys. Rept.}, 459:1--241, 2008.

\bibitem{Aaboud:2017sjh}
Morad Aaboud et~al.
\newblock {Search for additional heavy neutral Higgs and gauge bosons in the
  ditau final state produced in 36 fb$^{−1}$ of pp collisions at $
  \sqrt{s}=13 $ TeV with the ATLAS detector}.
\newblock {\em JHEP}, 01:055, 2018.

\bibitem{CMS-PAS-HIG-17-008}
{Search for Higgs boson pair production in the final state containing two
  photons and two bottom quarks in proton-proton collisions at
  $\sqrt{s}=13~\mathrm{TeV}$}.
\newblock Technical Report CMS-PAS-HIG-17-008, CERN, Geneva, 2017.

\bibitem{CMS-PAS-HIG-17-009}
{Search for resonant pair production of Higgs bosons decaying to bottom
  quark-antiquark pairs in proton-proton collisions at 13 TeV}.
\newblock Technical Report CMS-PAS-HIG-17-009, CERN, Geneva, 2017.

\bibitem{Sirunyan:2017djm}
Albert~M Sirunyan et~al.
\newblock {Search for Higgs boson pair production in events with two bottom
  quarks and two tau leptons in proton–proton collisions at $\sqrt s$
  =13TeV}.
\newblock {\em Phys. Lett.}, B778:101--127, 2018.

\bibitem{Aaboud:2017hnm}
Morad Aaboud et~al.
\newblock {Search for Heavy Higgs Bosons $A/H$ Decaying to a Top Quark Pair in
  $pp$ Collisions at $\sqrt{s}=8\text{ }\text{ }\mathrm{TeV}$ with the ATLAS
  Detector}.
\newblock {\em Phys. Rev. Lett.}, 119(19):191803, 2017.

\bibitem{Aaboud:2018mjh}
Morad Aaboud et~al.
\newblock {Search for heavy particles decaying into top-quark pairs using
  lepton-plus-jets events in proton–proton collisions at $\sqrt{s} = 13$
  $\text {TeV}$ with the ATLAS detector}.
\newblock {\em Eur. Phys. J.}, C78(7):565, 2018.

\bibitem{Adhikary:2018ise}
Amit Adhikary, Shankha Banerjee, Rahool Kumar~Barman, and Biplob Bhattacherjee.
\newblock {Resonant heavy Higgs searches at the HL-LHC}.
\newblock {\em JHEP}, 09:068, 2019.

\bibitem{Bisset:2007mk}
Mike Bisset, Jun Li, and Nick Kersting.
\newblock {How to Detect `Decoupled' Heavy Supersymmetric Higgs Bosons}.
\newblock 2007.

\bibitem{Bisset:2007mi}
Mike Bisset, Jun Li, Nick Kersting, Ran Lu, Filip Moortgat, and Stefano
  Moretti.
\newblock {Four-lepton LHC events from MSSM Higgs boson decays into neutralino
  and chargino pairs}.
\newblock {\em JHEP}, 08:037, 2009.

\bibitem{Arhrib:2011rp}
Abdesslam Arhrib, Rachid Benbrik, Mohamed Chabab, and Chuan-Hung Chen.
\newblock {Pair production of neutralinos and charginos at the LHC: the role of
  Higgs bosons exchange}.
\newblock {\em Phys. Rev.}, D84:115012, 2011.

\bibitem{Gunion:1987ki}
M.Drees~D.Karatas X.Tata~R.Godbole J.F.Gunion, H.E.Haber and N.Tracas.
\newblock {Workshop: From Colliders to SuperColliders Madison, Wisconsin, May
  11-22, 1987}.
\newblock {\em Int. J. Mod. Phys.}, A2:1035, 1987.

\bibitem{GUNION1988445}
John~F. Gunion and Howard~E. Haber.
\newblock Higgs bosons in supersymmetric models (iii). decays into neutralinos
  and charginos.
\newblock {\em Nuclear Physics B}, 307(3):445 -- 475, 1988.

\bibitem{Djouadi:1996mj}
A.~Djouadi, P.~Janot, J.~Kalinowski, and P.~M. Zerwas.
\newblock {SUSY decays of Higgs particles}.
\newblock {\em Phys. Lett.}, B376:220--226, 1996.

\bibitem{Belanger:2000tg}
G.~Belanger, F.~Boudjema, F.~Donato, R.~Godbole, and S.~Rosier-Lees.
\newblock {SUSY Higgs at the LHC: Effects of light charginos and neutralinos}.
\newblock {\em Nucl. Phys.}, B581:3--33, 2000.

\bibitem{Bisset:2000ud}
Mike Bisset, Monoranjan Guchait, and Stefano Moretti.
\newblock {Signatures of MSSM charged Higgs bosons via chargino neutralino
  decay channels at the LHC}.
\newblock {\em Eur. Phys. J.}, C19:143--154, 2001.

\bibitem{Choi:2002zp}
S.~Y. Choi, Manuel Drees, Jae~Sik Lee, and J.~Song.
\newblock {Supersymmetric Higgs boson decays in the MSSM with explicit CP
  violation}.
\newblock {\em Eur. Phys. J.}, C25:307--313, 2002.

\bibitem{Charlot_2006}
C~Charlot, R~Salerno, and Y~Sirois.
\newblock Observability of the heavy neutral {SUSY} higgs bosons decaying into
  neutralinos at the {LHC}.
\newblock {\em Journal of Physics G: Nuclear and Particle Physics},
  34(1):N1--N12, nov 2006.

\bibitem{Li:2013nma}
Tong Li.
\newblock {Decoupling MSSM Higgs Sector and Heavy Higgs Decay}.
\newblock {\em Phys. Lett.}, B728:77--84, 2014.

\bibitem{Belanger:2015vwa}
Genevieve Belanger, Diptimoy Ghosh, Rohini Godbole, and Suchita Kulkarni.
\newblock {Light stop in the MSSM after LHC Run 1}.
\newblock {\em JHEP}, 09:214, 2015.

\bibitem{Ananthanarayan:2015fwa}
B.~Ananthanarayan, Jayita Lahiri, and P.~N. Pandita.
\newblock {Invisible decays of the heavier Higgs boson in the minimal
  supersymmetric standard model}.
\newblock {\em Phys. Rev.}, D91:115025, 2015.

\bibitem{Djouadi:2015jea}
A.~Djouadi, L.~Maiani, A.~Polosa, J.~Quevillon, and V.~Riquer.
\newblock {Fully covering the MSSM Higgs sector at the LHC}.
\newblock {\em JHEP}, 06:168, 2015.

\bibitem{Barman:2016kgt}
Rahool~K. Barman, Biplob Bhattacherjee, Amit Chakraborty, and Arghya Choudhury.
\newblock {Study of MSSM heavy Higgs bosons decaying into charginos and
  neutralinos}.
\newblock {\em Phys. Rev.}, D94(7):075013, 2016.

\bibitem{Gori:2018pmk}
Stefania Gori, Zhen Liu, and Bibhushan Shakya.
\newblock {Heavy Higgs as a Portal to the Supersymmetric Electroweak Sector}.
\newblock {\em JHEP}, 04:049, 2019.

\bibitem{Baum:2019uzg}
Sebastian Baum, Nausheen~R. Shah, and Katherine Freese.
\newblock {The NMSSM is within Reach of the LHC: Mass Correlations \& Decay
  Signatures}.
\newblock {\em JHEP}, 04:011, 2019.

\bibitem{Profumo:2017ntc}
Stefano Profumo, Tim Stefaniak, and Laurel Stephenson~Haskins.
\newblock {The Not-So-Well Tempered Neutralino}.
\newblock {\em Phys. Rev.}, D96(5):055018, 2017.

\bibitem{Bahl:2018zmf}
Emanuele Bagnaschi et~al.
\newblock {MSSM Higgs Boson Searches at the LHC: Benchmark Scenarios for Run 2
  and Beyond}.
\newblock {\em Eur. Phys. J.}, C79(7):617, 2019.

\bibitem{Bahl:2019ago}
Henning Bahl, Stefan Liebler, and Tim Stefaniak.
\newblock {MSSM Higgs benchmark scenarios for Run 2 and beyond: the low $\tan
  \beta $ region}.
\newblock {\em Eur. Phys. J.}, C79(3):279, 2019.

\bibitem{Sirunyan:2017qfc}
A.~M. Sirunyan et~al.
\newblock {Search for new physics in events with a leptonically decaying Z
  boson and a large transverse momentum imbalance in proton-proton collisions
  at $\sqrt{s} $ = 13 $\,\text {TeV}$}.
\newblock {\em Eur. Phys. J.}, C78(4):291, 2018.

\bibitem{Giudice:1998xp}
Gian~F. Giudice, Markus~A. Luty, Hitoshi Murayama, and Riccardo Rattazzi.
\newblock {Gaugino mass without singlets}.
\newblock {\em JHEP}, 12:027, 1998.

\bibitem{Randall:1998uk}
Lisa Randall and Raman Sundrum.
\newblock {Out of this world supersymmetry breaking}.
\newblock {\em Nucl. Phys.}, B557:79--118, 1999.

\bibitem{Allanach:2004rh}
B.~C. Allanach, A.~Djouadi, J.~L. Kneur, W.~Porod, and P.~Slavich.
\newblock {Precise determination of the neutral Higgs boson masses in the
  MSSM}.
\newblock {\em JHEP}, 09:044, 2004.

\bibitem{Djouadi:2002ze}
Abdelhak Djouadi, Jean-Loic Kneur, and Gilbert Moultaka.
\newblock {SuSpect: A Fortran code for the supersymmetric and Higgs particle
  spectrum in the MSSM}.
\newblock {\em Comput. Phys. Commun.}, 176:426--455, 2007.

\bibitem{Chowdhury:2013dka}
Debtosh Chowdhury, Rohini~M. Godbole, Kirtimaan~A. Mohan, and Sudhir~K.
  Vempati.
\newblock {Charge and Color Breaking Constraints in MSSM after the Higgs
  Discovery at LHC}.
\newblock {\em JHEP}, 02:110, 2014.
\newblock [Erratum: JHEP 03, 149 (2018)].

\bibitem{GHERGHETTA199927}
Tony Gherghetta, Gian~F. Giudice, and James~D. Wells.
\newblock Phenomenological consequences of supersymmetry with anomaly induced
  masses.
\newblock {\em Nuclear Physics B}, 559(1):27 -- 47, 1999.

\bibitem{Ibe:2012sx}
Masahiro Ibe, Shigeki Matsumoto, and Ryosuke Sato.
\newblock {Mass Splitting between Charged and Neutral Winos at Two-Loop Level}.
\newblock {\em Phys. Lett.}, B721:252--260, 2013.

\bibitem{Gladyshev:2008ag}
A.~V. Gladyshev, D.~I. Kazakov, and M.~G. Paucar.
\newblock {Long-lived Charginos in the Focus-point Region of the MSSM Parameter
  Space}.
\newblock {\em J. Phys.}, G36:125009, 2009.

\bibitem{Kulkarni:2017xtf}
Suchita Kulkarni and Lukas Lechner.
\newblock {Characterizing simplified models for heavy Higgs decays to
  supersymmetric particles}.
\newblock 2017.

\bibitem{Arbey:2013jla}
Alexandre Arbey, Marco Battaglia, and Farvah Mahmoudi.
\newblock {Supersymmetric Heavy Higgs Bosons at the LHC}.
\newblock {\em Phys. Rev.}, D88(1):015007, 2013.

\bibitem{Aprile:2018dbl}
E.~Aprile et~al.
\newblock {Dark Matter Search Results from a One Ton-Year Exposure of XENON1T}.
\newblock {\em Phys. Rev. Lett.}, 121(11):111302, 2018.

\bibitem{Amhis:2016xyh}
Y.~Amhis et~al.
\newblock {Averages of $b$-hadron, $c$-hadron, and $\tau$-lepton properties as
  of summer 2016}.
\newblock {\em Eur. Phys. J.}, C77(12):895, 2017.

\bibitem{Aaij:2017vad}
Roel Aaij et~al.
\newblock {Measurement of the $B^0_s\to\mu^+\mu^-$ branching fraction and
  effective lifetime and search for $B^0\to\mu^+\mu^-$ decays}.
\newblock {\em Phys. Rev. Lett.}, 118(19):191801, 2017.

\bibitem{Abbiendi:2003sc}
G.~Abbiendi et~al.
\newblock {Search for chargino and neutralino production at s**(1/2) = 192-GeV
  to 209 GeV at LEP}.
\newblock {\em Eur. Phys. J.}, C35:1--20, 2004.

\bibitem{Belanger:2018mqt}
Genevieve Belanger, Fawzi Boudjema, Andreas Goudelis, Alexander Pukhov, and
  Bryan Zaldivar.
\newblock {micrOMEGAs5.0 : Freeze-in}.
\newblock {\em Comput. Phys. Commun.}, 231:173--186, 2018.

\bibitem{Bechtle:2013wla}
Philip Bechtle, Oliver Brein, Sven Heinemeyer, Oscar Stal, Tim Stefaniak, Georg
  Weiglein, and Karina~E. Williams.
\newblock {$\mathsf{HiggsBounds}-4$: Improved Tests of Extended Higgs Sectors
  against Exclusion Bounds from LEP, the Tevatron and the LHC}.
\newblock {\em Eur. Phys. J.}, C74(3):2693, 2014.

\bibitem{Bechtle:2013gu}
Philip Bechtle, Oliver Brein, Sven Heinemeyer, Oscar Stal, Tim Stefaniak, Georg
  Weiglein, and Karina Williams.
\newblock {Recent Developments in HiggsBounds and a Preview of HiggsSignals}.
\newblock {\em PoS}, CHARGED2012:024, 2012.

\bibitem{Bechtle:2011sb}
Philip Bechtle, Oliver Brein, Sven Heinemeyer, Georg Weiglein, and Karina~E.
  Williams.
\newblock {HiggsBounds 2.0.0: Confronting Neutral and Charged Higgs Sector
  Predictions with Exclusion Bounds from LEP and the Tevatron}.
\newblock {\em Comput. Phys. Commun.}, 182:2605--2631, 2011.

\bibitem{Bechtle:2008jh}
Philip Bechtle, Oliver Brein, Sven Heinemeyer, Georg Weiglein, and Karina~E.
  Williams.
\newblock {HiggsBounds: Confronting Arbitrary Higgs Sectors with Exclusion
  Bounds from LEP and the Tevatron}.
\newblock {\em Comput. Phys. Commun.}, 181:138--167, 2010.

\bibitem{Bechtle:2013xfa}
Philip Bechtle, Sven Heinemeyer, Oscar Stal, Tim Stefaniak, and Georg Weiglein.
\newblock {$HiggsSignals$: Confronting arbitrary Higgs sectors with
  measurements at the Tevatron and the LHC}.
\newblock {\em Eur. Phys. J.}, C74(2):2711, 2014.

\bibitem{Caron:2016hib}
Sascha Caron, Jong~Soo Kim, Krzysztof Rolbiecki, Roberto Ruiz~de Austri, and
  Bob Stienen.
\newblock {The BSM-AI project: SUSY-AI-generalizing LHC limits on supersymmetry
  with machine learning}.
\newblock {\em Eur. Phys. J.}, C77(4):257, 2017.

\bibitem{Kraml:2013mwa}
Sabine Kraml, Suchita Kulkarni, Ursula Laa, Andre Lessa, Wolfgang Magerl, Doris
  Proschofsky, and Wolfgang Waltenberger.
\newblock {SModelS: a tool for interpreting simplified-model results from the
  LHC and its application to supersymmetry}.
\newblock {\em Eur.Phys.J.}, C74:2868, 2014.

\bibitem{Ambrogi:2017neo}
Federico Ambrogi, Sabine Kraml, Suchita Kulkarni, Ursula Laa, Andre Lessa,
  Veronika Magerl, Jory Sonneveld, Michael Traub, and Wolfgang Waltenberger.
\newblock {SModelS v1.1 user manual}.
\newblock 2017.

\bibitem{Ambrogi:2018ujg}
Federico Ambrogi et~al.
\newblock {SModelS v1.2: long-lived particles, combination of signal regions,
  and other novelties}.
\newblock 2018.

\bibitem{Heisig:2018kfq}
Jan Heisig, Sabine Kraml, and Andre Lessa.
\newblock {Constraining new physics with searches for long-lived particles:
  Implementation into SModelS}.
\newblock 2018.

\bibitem{Dutta:2018ioj}
Juhi Dutta, Sabine Kraml, Andre Lessa, and Wolfgang Waltenberger.
\newblock {SModelS extension with the CMS supersymmetry search results from Run
  2}.
\newblock {\em LHEP}, 1(1):5--12, 2018.

\bibitem{Aaboud:2016nwl}
Morad Aaboud et~al.
\newblock {Search for bottom squark pair production in proton--proton
  collisions at $\sqrt{s}=13$ TeV with the ATLAS detector}.
\newblock {\em Eur. Phys. J.}, C76(10):547, 2016.

\bibitem{Aaboud:2016lwz}
Morad Aaboud et~al.
\newblock {Search for top squarks in final states with one isolated lepton,
  jets, and missing transverse momentum in $\sqrt{s}=13$ TeV $pp$ collisions
  with the ATLAS detector}.
\newblock {\em Phys. Rev.}, D94(5):052009, 2016.

\bibitem{Aaboud:2016zdn}
Morad Aaboud et~al.
\newblock {Search for squarks and gluinos in final states with jets and missing
  transverse momentum at $\sqrt{s} =$ 13 TeV with the ATLAS detector}.
\newblock {\em Eur. Phys. J.}, C76(7):392, 2016.

\bibitem{Aad:2016tuk}
Georges Aad et~al.
\newblock {Search for supersymmetry at $\sqrt{s}=13$ TeV in final states with
  jets and two same-sign leptons or three leptons with the ATLAS detector}.
\newblock {\em Eur. Phys. J.}, C76(5):259, 2016.

\bibitem{Aaboud:2018kya}
Morad Aaboud et~al.
\newblock {Search for top squarks decaying to tau sleptons in $pp$ collisions
  at $\sqrt{s}= 13$ TeV with the ATLAS detector}.
\newblock {\em Phys. Rev.}, D98(3):032008, 2018.

\bibitem{Aaboud:2018zjf}
Morad Aaboud et~al.
\newblock {Search for supersymmetry in final states with charm jets and missing
  transverse momentum in 13 TeV $pp$ collisions with the ATLAS detector}.
\newblock {\em JHEP}, 09:050, 2018.

\bibitem{Aaboud:2018ujj}
Morad Aaboud et~al.
\newblock {Search for new phenomena using the invariant mass distribution of
  same-flavour opposite-sign dilepton pairs in events with missing transverse
  momentum in $\sqrt{s}=13$ $\text {Te}\text {V}$ pp collisions with the ATLAS
  detector}.
\newblock {\em Eur. Phys. J.}, C78(8):625, 2018.

\bibitem{Aaboud:2018sua}
Morad Aaboud et~al.
\newblock {Search for chargino-neutralino production using recursive jigsaw
  reconstruction in final states with two or three charged leptons in
  proton-proton collisions at $\sqrt{s}=13$ TeV with the ATLAS detector}.
\newblock 2018.

\bibitem{ATLAS-CONF-2013-007}
{Search for strongly produced superpartners in final states with two same sign
  leptons with the ATLAS detector using 21 fb$^{-1}$ of proton-proton
  collisions at sqrt(s)=8 TeV.}
\newblock Technical Report ATLAS-CONF-2013-007, CERN, Geneva, Mar 2013.

\bibitem{ATLAS-CONF-2013-061}
{Search for strong production of supersymmetric particles in final states with
  missing transverse momentum and at least three b-jets using 20.1 fb$^{-1}$ of
  pp collisions at sqrt(s) = 8 TeV with the ATLAS Detector.}
\newblock Technical Report ATLAS-CONF-2013-061, CERN, Geneva, Jun 2013.

\bibitem{ATLAS-CONF-2013-089}
{Search for strongly produced supersymmetric particles in decays with two
  leptons at $\sqrt{s}$ = 8 TeV}.
\newblock Technical Report ATLAS-CONF-2013-089, CERN, Geneva, Aug 2013.

\bibitem{Aad:2014wea}
Georges Aad et~al.
\newblock {Search for squarks and gluinos with the ATLAS detector in final
  states with jets and missing transverse momentum using $\sqrt{s}=8$ TeV
  proton--proton collision data}.
\newblock {\em JHEP}, 09:176, 2014.

\bibitem{Aad:2013wta}
Georges Aad et~al.
\newblock {Search for new phenomena in final states with large jet
  multiplicities and missing transverse momentum at $\sqrt{s}$=8 TeV
  proton-proton collisions using the ATLAS experiment}.
\newblock {\em JHEP}, 10:130, 2013.
\newblock [Erratum: JHEP01,109(2014)].

\bibitem{Aad:2013ija}
Georges Aad et~al.
\newblock {Search for direct third-generation squark pair production in final
  states with missing transverse momentum and two $b$-jets in $\sqrt{s} =$ 8
  TeV $pp$ collisions with the ATLAS detector}.
\newblock {\em JHEP}, 10:189, 2013.

\bibitem{Aad:2014mha}
Georges Aad et~al.
\newblock {Search for direct top squark pair production in events with a Z
  boson, b-jets and missing transverse momentum in sqrt(s)=8 TeV pp collisions
  with the ATLAS detector}.
\newblock {\em Eur. Phys. J.}, C74(6):2883, 2014.

\bibitem{Aad:2014pda}
Georges Aad et~al.
\newblock {Search for supersymmetry at $\sqrt{s}$=8 TeV in final states with
  jets and two same-sign leptons or three leptons with the ATLAS detector}.
\newblock {\em JHEP}, 06:035, 2014.

\bibitem{Aad:2014vma}
Georges Aad et~al.
\newblock {Search for direct production of charginos, neutralinos and sleptons
  in final states with two leptons and missing transverse momentum in $pp$
  collisions at $\sqrt{s} =$ 8 TeV with the ATLAS detector}.
\newblock {\em JHEP}, 05:071, 2014.

\bibitem{Aad:2014nua}
Georges Aad et~al.
\newblock {Search for direct production of charginos and neutralinos in events
  with three leptons and missing transverse momentum in $\sqrt{s} =$ 8 TeV $pp$
  collisions with the ATLAS detector}.
\newblock {\em JHEP}, 04:169, 2014.

\bibitem{Aad:2014kra}
Georges Aad et~al.
\newblock {Search for top squark pair production in final states with one
  isolated lepton, jets, and missing transverse momentum in $\sqrt s =$8 TeV
  $pp$ collisions with the ATLAS detector}.
\newblock {\em JHEP}, 11:118, 2014.

\bibitem{Aad:2014bva}
Georges Aad et~al.
\newblock {Search for direct pair production of the top squark in all-hadronic
  final states in proton-proton collisions at $\sqrt{s}=8$ TeV with the ATLAS
  detector}.
\newblock {\em JHEP}, 09:015, 2014.

\bibitem{Aad:2014lra}
Georges Aad et~al.
\newblock {Search for strong production of supersymmetric particles in final
  states with missing transverse momentum and at least three $b$-jets at
  $\sqrt{s}$= 8 TeV proton-proton collisions with the ATLAS detector}.
\newblock {\em JHEP}, 10:024, 2014.

\bibitem{Aad:2014qaa}
Georges Aad et~al.
\newblock {Search for direct top-squark pair production in final states with
  two leptons in pp collisions at $\sqrt{s} =$ 8 TeV with the ATLAS detector}.
\newblock {\em JHEP}, 06:124, 2014.

\bibitem{Aad:2014nra}
Georges Aad et~al.
\newblock {Search for pair-produced third-generation squarks decaying via charm
  quarks or in compressed supersymmetric scenarios in $pp$ collisions at
  $\sqrt{s}=8~$TeV with the ATLAS detector}.
\newblock {\em Phys. Rev.}, D90(5):052008, 2014.

\bibitem{Aad:2015jqa}
Georges Aad et~al.
\newblock {Search for direct pair production of a chargino and a neutralino
  decaying to the 125 GeV Higgs boson in $\sqrt{s} = 8$ TeV ${pp}$ collisions
  with the ATLAS detector}.
\newblock {\em Eur. Phys. J.}, C75(5):208, 2015.

\bibitem{Aad:2015gna}
Georges Aad et~al.
\newblock {Search for Scalar Charm Quark Pair Production in $pp$ Collisions at
  $\sqrt{s}=$ 8 TeV with the ATLAS Detector}.
\newblock {\em Phys. Rev. Lett.}, 114(16):161801, 2015.

\bibitem{CMS-PAS-EXO-16-036}
{Search for heavy stable charged particles with $12.9~\mathrm{fb}^{-1}$ of 2016
  data}.
\newblock Technical Report CMS-PAS-EXO-16-036, CERN, Geneva, 2016.

\bibitem{CMS-PAS-SUS-16-022}
{Search for SUSY with multileptons in 13 TeV data}.
\newblock Technical Report CMS-PAS-SUS-16-022, CERN, Geneva, 2016.

\bibitem{CMS-PAS-SUS-16-052}
{Search for supersymmetry in events with at least one soft lepton, low jet
  multiplicity, and missing transverse momentum in proton-proton collisions at
  $\sqrt{s}=13~\mathrm{TeV}$}.
\newblock Technical Report CMS-PAS-SUS-16-052, CERN, Geneva, 2017.

\bibitem{CMS-PAS-SUS-17-004}
{Combined search for electroweak production of charginos and neutralinos in pp
  collisions at $\sqrt{s} = 13~\mathrm{TeV}$}.
\newblock Technical Report CMS-PAS-SUS-17-004, CERN, Geneva, 2017.

\bibitem{Sirunyan:2275490}
{CMS collaboration}.
\newblock {Search for the pair production of third-generation squarks with
  two-body decays to a bottom or charm quark and a neutralino in proton-proton
  collisions at $ \sqrt{s} = $ 13 TeV.}
\newblock {\em Phys. Lett. B}, 778(CMS-SUS-16-032.
  CMS-SUS-16-032-003):263--291. 29 p, Jul 2017.

\bibitem{Sirunyan:2261105}
{CMS collaboration}.
\newblock {Search for supersymmetry in multijet events with missing transverse
  momentum in proton-proton collisions at 13 TeV.}
\newblock {\em Phys. Rev. D}, 96(CMS-SUS-16-033. CMS-SUS-16-033-003):032003. 38
  p, Apr 2017.

\bibitem{Sirunyan:2285878}
{CMS collaboration}.
\newblock {Search for new phenomena in final states with two opposite-charge,
  same-flavor leptons, jets, and missing transverse momentum in pp collisions
  at $\sqrt{s} = $ 13 TeV}.
\newblock {\em JHEP}, 03(arXiv:1709.08908. CMS-SUS-16-034-003):076. 52 p, Sep
  2017.

\bibitem{Sirunyan:2260986}
{CMS collaboration}.
\newblock {Search for physics beyond the standard model in events with two
  leptons of same sign, missing transverse momentum, and jets in proton--proton
  collisions at $\sqrt{s} = 13\,\text {TeV} $.}
\newblock {\em Eur. Phys. J. C}, 77(CMS-SUS-16-035. CMS-SUS-16-035-003. 9):578.
  42 p, Apr 2017.

\bibitem{Sirunyan:2264381}
{CMS collaboration}.
\newblock {Search for new phenomena with the $ \mathrm{ M_{\rm T2} } $ variable
  in the all-hadronic final state produced in proton-proton collisions at
  $\sqrt{s} = $ 13 TeV.}
\newblock {\em Eur. Phys. J. C}, 77(CMS-SUS-16-036. CMS-SUS-16-036-003.
  10):710. 46 p, May 2017.

\bibitem{Sirunyan:2264382}
{CMS collaboration}.
\newblock {Search for supersymmetry in pp collisions at sqrt(s) = 13 TeV in the
  single-lepton final state using the sum of masses of large-radius jets.}
\newblock {\em Phys. Rev. Lett.}, 119(CMS-SUS-16-037. CMS-SUS-16-037-004.
  15):151802. 18 p, May 2017.

\bibitem{Sirunyan:2284431}
{CMS collaboration}.
\newblock {Search for electroweak production of charginos and neutralinos in
  multilepton final states in proton-proton collisions at $\sqrt{s} = $ 13
  TeV.}
\newblock {\em JHEP}, 03(CMS-SUS-16-039. CMS-SUS-16-039-003):166. 58 p, Sep
  2017.

\bibitem{Sirunyan:2290511}
{CMS collaboration}.
\newblock {Search for supersymmetry in events with at least three electrons or
  muons, jets, and missing transverse momentum in proton-proton collisions at $
  \sqrt{s}=13 $ TeV}.
\newblock {\em JHEP}, 02(CMS-SUS-16-041. CMS-SUS-16-041-003):067. 44 p, Oct
  2017.

\bibitem{Sirunyan:2286124}
{CMS collaboration}.
\newblock {Search for supersymmetry in events with one lepton and multiple jets
  exploiting the angular correlation between the lepton and the missing
  transverse momentum in proton-proton collisions at $\sqrt{s} = $ 13 TeV.}
\newblock {\em Phys. Lett. B}, 780(CMS-SUS-16-042. CMS-SUS-16-042-003):384. 26
  p, Sep 2017.

\bibitem{Sirunyan:2272346}
{CMS collaboration}.
\newblock {Search for electroweak production of charginos and neutralinos in WH
  events in proton-proton collisions at $ \sqrt{s}=13 $ TeV.}
\newblock {\em JHEP}, 11(CMS-SUS-16-043. CMS-SUS-16-043-003):029. 38 p, Jun
  2017.

\bibitem{Sirunyan:2282000}
{CMS collaboration}.
\newblock {Search for supersymmetry with Higgs boson to diphoton decays using
  the razor variables at $\sqrt{s} = $ 13 TeV.}
\newblock {\em Phys. Lett. B}, 779(CMS-SUS-16-045.
  CMS-SUS-16-045-003):166--190. 25 p, Sep 2017.

\bibitem{Sirunyan:2293644}
{CMS collaboration}.
\newblock {Search for gauge-mediated supersymmetry in events with at least one
  photon and missing transverse momentum in pp collisions at $\sqrt{s} = $ 13
  TeV}.
\newblock {\em Phys. Lett. B}, 780(CMS-SUS-16-046.
  CMS-SUS-16-046-003):118--143. 26 p, Nov 2017.

\bibitem{Sirunyan:2275103}
{CMS collaboration}.
\newblock {Search for supersymmetry in events with at least one photon, missing
  transverse momentum, and large transverse event activity in proton-proton
  collisions at $ \sqrt{s}=13 $ TeV.}
\newblock {\em JHEP}, 12(CMS-SUS-16-047. CMS-SUS-16-047-003):142. 32 p, Jul
  2017.

\bibitem{Sirunyan:2274031}
{CMS collaboration}.
\newblock {Search for direct production of supersymmetric partners of the top
  quark in the all-jets final state in proton-proton collisions at $
  \sqrt{s}=13 $ TeV}.
\newblock {\em JHEP}, 10(CMS-SUS-16-049. CMS-SUS-16-049-003):005. 58 p, Jul
  2017.

\bibitem{Sirunyan:2291344}
{CMS collaboration}.
\newblock {Search for supersymmetry in proton-proton collisions at 13 TeV using
  identified top quarks.}
\newblock {\em Phys. Rev. D}, 97(CMS-SUS-16-050):012007. 29 p, Oct 2017.

\bibitem{Sirunyan:2269047}
{CMS collaboration}.
\newblock {Search for top squark pair production in pp collisions at $ \sqrt{s}
  = $ 13 TeV using single lepton events.}
\newblock {\em JHEP}, 10(CMS-SUS-16-051. CMS-SUS-16-051-004):019. 40 p, Jun
  2017.

\bibitem{Sirunyan:2291416}
{CMS collaboration}.
\newblock {Search for top squarks and dark matter particles in opposite-charge
  dilepton final states at $\sqrt{s}=$ 13 TeV}.
\newblock {\em Phys. Rev. D}, 97(CMS-SUS-17-001. CMS-SUS-17-001-003. 3):032009.
  29 p, Nov 2017.

\bibitem{Chatrchyan:1545325}
{CMS collaboration}.
\newblock {Searches for long-lived charged particles in pp collisions at
  $\sqrt{s}$=7 and 8 TeV}.
\newblock {\em JHEP}, 07(CMS-EXO-12-026. CMS-EXO-12-026.
  CERN-PH-EP-2013-073):122. 48 p, May 2013.

\bibitem{Khachatryan:1987723}
{CMS collaboration}.
\newblock {Constraints on the pMSSM, AMSB model and on other models from the
  search for long-lived charged particles in proton-proton collisions at
  $\sqrt{s}$ = 8 TeV}.
\newblock {\em Eur. Phys. J. C}, 75(CMS-EXO-13-006. CMS-EXO-13-006.
  CERN-PH-EP-2015-014):325. 39 p, Feb 2015.

\bibitem{CMS-PAS-SUS-13-015}
{Search for top squarks in multijet events with large missing momentum in
  proton-proton collisions at 8 TeV}.
\newblock Technical Report CMS-PAS-SUS-13-015, CERN, Geneva, 2013.

\bibitem{CMS-PAS-SUS-13-016}
{Search for supersymmetry in pp collisions at sqrt(s) = 8 TeV in events with
  two opposite sign leptons, large number of jets, b-tagged jets, and large
  missing transverse energy.}
\newblock Technical Report CMS-PAS-SUS-13-016, CERN, Geneva, 2013.

\bibitem{CMS-PAS-SUS-13-018}
{Search for direct production of bottom squark pairs}.
\newblock Technical Report CMS-PAS-SUS-13-018, CERN, Geneva, 2014.

\bibitem{CMS-PAS-SUS-13-023}
{A Search for Scalar Top Quark Production and Decay to All Hadronic Final
  States in pp Collisions at sqrt(s) = 8 TeV}.
\newblock Technical Report CMS-PAS-SUS-13-023, CERN, Geneva, 2015.

\bibitem{Chatrchyan:1546693}
{CMS collaboration}.
\newblock {Search for gluino mediated bottom- and top-squark production in
  multijet final states in pp collisions at 8 TeV}.
\newblock {\em Phys. Lett. B}, 725(CMS-SUS-12-024. CMS-SUS-12-024.
  CERN-PH-EP-2013-076):243--270. 28 p, May 2013.

\bibitem{Chatrchyan:1527115}
{CMS collaboration}.
\newblock {Search for supersymmetry in hadronic final states with missing
  transverse energy using the variables $\alpha_T$ and b-quark multiplicity in
  pp collisions at $\sqrt{s}$ = 8 TeV}.
\newblock {\em Eur. Phys. J. C}, 73(CMS-SUS-12-028. CMS-SUS-12-028.
  CERN-PH-EP-2013-037):2568. 45 p, Mar 2013.

\bibitem{Chatrchyan:1696925}
{CMS collaboration}.
\newblock {Search for anomalous production of events with three or more leptons
  in pp collisions at $\sqrt{s}$=8 TeV}.
\newblock {\em Phys. Rev. D}, 90(CMS-SUS-13-002. CMS-SUS-13-002-003.
  CERN-PH-EP-2014-039):032006. 27 p, Apr 2014.

\bibitem{Khachatryan:1984165}
{CMS collaboration}.
\newblock {Search for supersymmetry using razor variables in events with
  b-tagged jets in pp collisions at $\sqrt{s}$ = 8 TeV}.
\newblock {\em Phys. Rev. D}, 91(CMS-SUS-13-004. CERN-PH-EP-2015-005.
  CMS-SUS-13-004):052018. 45 p, Feb 2015.

\bibitem{Khachatryan:1704963}
{CMS collaboration}.
\newblock {Searches for electroweak production of charginos, neutralinos, and
  sleptons decaying to leptons and W, Z, and Higgs bosons in pp collisions at 8
  TeV}.
\newblock {\em Eur. Phys. J. C}, 74(CMS-SUS-13-006. CMS-SUS-13-006.
  CERN-PH-EP-2014-098):3036. 42 p, May 2014.

\bibitem{CMS-PAS-SUS-13-007}
{Search for Supersymmetry in pp collisions at 8 TeV in events with a single
  lepton, multiple jets and b-tags}.
\newblock Technical Report CMS-PAS-SUS-13-007, CERN, Geneva, Mar 2013.

\bibitem{Chatrchyan:1567175}
{CMS collaboration}.
\newblock {Search for top-squark pair production in the single-lepton final
  state in pp collisions at $\sqrt{s}$ = 8 TeV}.
\newblock {\em Eur. Phys. J. C}, 73(CMS-SUS-13-011. CMS-SUS-13-011.
  CERN-PH-EP-2013-148):2677. 46 p, Aug 2013.

\bibitem{Chatrchyan:1662652}
{CMS collaboration}.
\newblock {Search for new physics in the multijet and missing transverse
  momentum final state in proton-proton collisions at $\sqrt{s}$ = 8 TeV}.
\newblock {\em JHEP}, 06(CMS-SUS-13-012. CMS-SUS-13-012.
  CERN-PH-EP-2014-015):055. 38 p, Feb 2014.

\bibitem{Chatrchyan:1631468}
{CMS collaboration}.
\newblock {Search for new physics in events with same-sign dileptons and jets
  in pp collisions at $\sqrt{s}$=8 TeV}.
\newblock {\em JHEP}, 01(arXiv:1311.6736. CMS-SUS-13-013.
  CERN-PH-EP-2013-213):163. 45 p, Nov 2013.

\bibitem{Khachatryan:1989788}
{CMS collaboration}.
\newblock {Searches for supersymmetry using the $M_\mathrm{T2}$ variable in
  hadronic events produced in pp collisions at 8 TeV}.
\newblock {\em JHEP}, 05(CMS-SUS-13-019. CMS-SUS-13-019.
  CERN-PH-EP-2015-017):078. 51 p, Feb 2015.

\bibitem{Khachatryan:1976453}
{CMS collaboration}.
\newblock {Searches for supersymmetry based on events with b jets and four W
  bosons in pp collisions at 8 TeV}.
\newblock {\em Phys. Lett. B}, 745(CMS-SUS-14-010. CERN-PH-EP-2014-286.
  CMS-SUS-14-010):5. 24 p, Dec 2014.

\bibitem{Khachatryan:2117955}
{CMS collaboration}.
\newblock {Search for supersymmetry in events with soft leptons, low jet
  multiplicity, and missing transverse energy in proton-proton collisions at
  $\sqrt{s}= $ 8 TeV}.
\newblock {\em Phys. Lett. B}, 759(CMS-SUS-14-021. CERN-PH-EP-2015-307.
  CMS-SUS-14-021):9--35. 27 p, Dec 2015.

\bibitem{Sjostrand:2006za}
Torbjorn Sjostrand, Stephen Mrenna, and Peter~Z. Skands.
\newblock {PYTHIA 6.4 Physics and Manual}.
\newblock {\em JHEP}, 0605:026, 2006.

\bibitem{Sjostrand:2014zea}
Torbj{\"o}rn Sj{\"o}strand, Stefan Ask, Jesper~R. Christiansen, Richard Corke,
  Nishita Desai, Philip Ilten, Stephen Mrenna, Stefan Prestel, Christine~O.
  Rasmussen, and Peter~Z. Skands.
\newblock {An Introduction to PYTHIA 8.2}.
\newblock {\em Comput. Phys. Commun.}, 191:159--177, 2015.

\bibitem{Djouadi:2006bz}
A.~Djouadi, M.~M. Muhlleitner, and M.~Spira.
\newblock {Decays of supersymmetric particles: The Program SUSY-HIT
  (SUspect-SdecaY-Hdecay-InTerface)}.
\newblock {\em Acta Phys. Polon.}, B38:635--644, 2007.

\bibitem{Beenakker:1996ch}
W.~Beenakker, R.~Hopker, M.~Spira, and P.M. Zerwas.
\newblock {Squark and gluino production at hadron colliders}.
\newblock {\em Nucl.Phys.}, B492:51--103, 1997.

\bibitem{Beenakker:1997ut}
W.~Beenakker, M.~Kramer, T.~Plehn, M.~Spira, and P.M. Zerwas.
\newblock {Stop production at hadron colliders}.
\newblock {\em Nucl.Phys.}, B515:3--14, 1998.

\bibitem{Kulesza:2008jb}
A.~Kulesza and L.~Motyka.
\newblock {Threshold resummation for squark-antisquark and gluino-pair
  production at the LHC}.
\newblock {\em Phys.Rev.Lett.}, 102:111802, 2009.

\bibitem{Kulesza:2009kq}
A.~Kulesza and L.~Motyka.
\newblock {Soft gluon resummation for the production of gluino-gluino and
  squark-antisquark pairs at the LHC}.
\newblock {\em Phys.Rev.}, D80:095004, 2009.

\bibitem{Beenakker:2009ha}
Wim Beenakker, Silja Brensing, Michael Kramer, Anna Kulesza, Eric Laenen,
  et~al.
\newblock {Soft-gluon resummation for squark and gluino hadroproduction}.
\newblock {\em JHEP}, 0912:041, 2009.

\bibitem{Beenakker:2010nq}
Wim Beenakker, Silja Brensing, Michael Kramer, Anna Kulesza, Eric Laenen,
  et~al.
\newblock {Supersymmetric top and bottom squark production at hadron
  colliders}.
\newblock {\em JHEP}, 1008:098, 2010.

\bibitem{Beenakker:2011fu}
W.~Beenakker, S.~Brensing, M.n Kramer, A.~Kulesza, E.~Laenen, et~al.
\newblock {Squark and Gluino Hadroproduction}.
\newblock {\em Int.J.Mod.Phys.}, A26:2637--2664, 2011.

\bibitem{Aad:2015baa}
Georges Aad et~al.
\newblock {Summary of the ATLAS experiment's sensitivity to supersymmetry after
  LHC Run 1-interpreted in the phenomenological MSSM}.
\newblock {\em JHEP}, 10:134, 2015.

\bibitem{Aghanim:2018eyx}
N.~Aghanim et~al.
\newblock {Planck 2018 results. VI. Cosmological parameters}.
\newblock 2018.

\bibitem{Harlander:2016hcx}
Robert~V. Harlander, Stefan Liebler, and Hendrik Mantler.
\newblock {SusHi Bento: Beyond NNLO and the heavy-top limit}.
\newblock {\em Comput. Phys. Commun.}, 212:239--257, 2017.

\bibitem{suspect3}
{\url{http://suspect.in2p3.fr/}}.

\bibitem{deFavereau:2013fsa}
J.~de~Favereau, C.~Delaere, P.~Demin, A.~Giammanco, V.~Lemaître, A.~Mertens,
  and M.~Selvaggi.
\newblock {DELPHES 3, A modular framework for fast simulation of a generic
  collider experiment}.
\newblock {\em JHEP}, 02:057, 2014.

\bibitem{Binosi:2008ig}
D.~Binosi, J.~Collins, C.~Kaufhold, and L.~Theussl.
\newblock {JaxoDraw: A Graphical user interface for drawing Feynman diagrams.
  Version 2.0 release notes}.
\newblock {\em Comput. Phys. Commun.}, 180:1709--1715, 2009.

\bibitem{Angloher:2015ewa}
G.~Angloher et~al.
\newblock {Results on light dark matter particles with a low-threshold
  CRESST-II detector}.
\newblock {\em Eur. Phys. J.}, C76(1):25, 2016.

\bibitem{Dittmaier:2003ej}
Stefan Dittmaier, Michael Krämer, and Michael Spira.
\newblock {Higgs radiation off bottom quarks at the Tevatron and the CERN LHC}.
\newblock {\em Phys. Rev.}, D70:074010, 2004.

\bibitem{Dawson:2005vi}
S.~Dawson, C.~B. Jackson, L.~Reina, and D.~Wackeroth.
\newblock {Higgs production in association with bottom quarks at hadron
  colliders}.
\newblock {\em Mod. Phys. Lett.}, A21:89--110, 2006.

\bibitem{Dawson:2004wq}
S.~Dawson, C.~B. Jackson, L.~Reina, and D.~Wackeroth.
\newblock {Higgs boson production with bottom quarks at hadron colliders}.
\newblock {\em Int. J. Mod. Phys.}, A20:3353--3355, 2005.

\bibitem{Dawson:2003kb}
S.~Dawson, C.~B. Jackson, L.~Reina, and D.~Wackeroth.
\newblock {Exclusive Higgs boson production with bottom quarks at hadron
  colliders}.
\newblock {\em Phys. Rev.}, D69:074027, 2004.

\bibitem{Harlander:2003ai}
Robert~V. Harlander and William~B. Kilgore.
\newblock {Higgs boson production in bottom quark fusion at next-to-next-to
  leading order}.
\newblock {\em Phys. Rev.}, D68:013001, 2003.

\bibitem{Dawson:2004sh}
S.~Dawson, C.~B. Jackson, L.~Reina, and D.~Wackeroth.
\newblock {Higgs boson production with one bottom quark jet at hadron
  colliders}.
\newblock {\em Phys. Rev. Lett.}, 94:031802, 2005.

\bibitem{Dawson:2010yz}
S.~Dawson and P.~Jaiswal.
\newblock {Weak Corrections to Associated Higgs-Bottom Quark Production}.
\newblock {\em Phys. Rev.}, D81:073008, 2010.

\bibitem{Maltoni:2003pn}
F.~Maltoni, Z.~Sullivan, and S.~Willenbrock.
\newblock {Higgs-Boson Production via Bottom-Quark Fusion}.
\newblock {\em Phys. Rev.}, D67:093005, 2003.

\bibitem{Boos:2003yi}
Eduard Boos and Tilman Plehn.
\newblock {Higgs boson production induced by bottom quarks}.
\newblock {\em Phys. Rev.}, D69:094005, 2004.

\bibitem{Plehn:2002vy}
Tilman Plehn.
\newblock {Charged Higgs boson production in bottom gluon fusion}.
\newblock {\em Phys. Rev.}, D67:014018, 2003.

\bibitem{Harlander:2011}
M.~Krämer R.~Harlander and M.~Schumacher.
\newblock {Bottom-quark associated Higgs-boson production: reconciling the
  four- and five-flavour scheme approach}.
\newblock page~8.

\bibitem{Alwall:2014hca}
J.~Alwall, R.~Frederix, S.~Frixione, V.~Hirschi, F.~Maltoni, O.~Mattelaer,
  H.~S. Shao, T.~Stelzer, P.~Torrielli, and M.~Zaro.
\newblock {The automated computation of tree-level and next-to-leading order
  differential cross sections, and their matching to parton shower
  simulations}.
\newblock {\em JHEP}, 07:079, 2014.

\bibitem{Sirunyan:2017ezt}
A.~M. Sirunyan et~al.
\newblock {Identification of heavy-flavour jets with the CMS detector in pp
  collisions at 13 TeV}.
\newblock {\em JINST}, 13(05):P05011, 2018.

\bibitem{Cacciari:2011ma}
Matteo Cacciari, Gavin~P. Salam, and Gregory Soyez.
\newblock {FastJet User Manual}.
\newblock {\em Eur. Phys. J.}, C72:1896, 2012.

\bibitem{Beenakker:1996ed}
W.~Beenakker, R.~Hopker, and M.~Spira.
\newblock {PROSPINO: A Program for the production of supersymmetric particles
  in next-to-leading order QCD}.
\newblock 1996.

\bibitem{Wiesemann:2014ioa}
M.~Wiesemann, R.~Frederix, S.~Frixione, V.~Hirschi, F.~Maltoni, and
  P.~Torrielli.
\newblock {Higgs production in association with bottom quarks}.
\newblock {\em JHEP}, 02:132, 2015.

\bibitem{Mangano:2006rw}
Michelangelo~L. Mangano, Mauro Moretti, Fulvio Piccinini, and Michele Treccani.
\newblock {Matching matrix elements and shower evolution for top-quark
  production in hadronic collisions}.
\newblock {\em JHEP}, 01:013, 2007.

\bibitem{Lazopoulos:2008de}
Achilleas Lazopoulos, Thomas McElmurry, Kirill Melnikov, and Frank Petriello.
\newblock {Next-to-leading order QCD corrections to $t \bar{t} Z$ production at
  the LHC}.
\newblock {\em Phys. Lett.}, B666:62--65, 2008.

\bibitem{ttbarNNLO}
{\url{https://twiki.cern.ch/twiki/bin/view/LHCPhysics/TtbarNNLO}}.

\bibitem{Bonvini:2016fgf}
Marco Bonvini, Andrew~S. Papanastasiou, and Frank~J. Tackmann.
\newblock {Matched predictions for the $ b\overline{b}H $ cross section at the
  13 TeV LHC}.
\newblock {\em JHEP}, 10:053, 2016.

\bibitem{Adhikary:2017jtu}
Amit Adhikary, Shankha Banerjee, Rahool~Kumar Barman, Biplob Bhattacherjee, and
  Saurabh Niyogi.
\newblock {Revisiting the non-resonant Higgs pair production at the HL-LHC}.
\newblock {\em JHEP}, 07:116, 2018.

\bibitem{bkg_twiki_cs}
{\url{https://twiki.cern.ch/twiki/bin/view/LHCPhysics/CERNYellowReportPageAt1314TeV2014}}.

\bibitem{Campbell:2012dh}
John~M. Campbell and R.~Keith Ellis.
\newblock {$t \bar{t} W^{+-}$ production and decay at NLO}.
\newblock {\em JHEP}, 07:052, 2012.

\bibitem{Aaboud:2017mpt}
Morad Aaboud et~al.
\newblock {Search for long-lived charginos based on a disappearing-track
  signature in pp collisions at $ \sqrt{s}=13 $ TeV with the ATLAS detector}.
\newblock {\em JHEP}, 06:022, 2018.

\bibitem{Sirunyan:2018ldc}
Albert~M Sirunyan et~al.
\newblock {Search for disappearing tracks as a signature of new long-lived
  particles in proton-proton collisions at $\sqrt{s} =$ 13 TeV}.
\newblock {\em JHEP}, 08:016, 2018.

\bibitem{Aad:2014kga}
Georges Aad et~al.
\newblock {Search for charged Higgs bosons decaying via $H^{\pm} \rightarrow
  \tau^{\pm}\nu$ in fully hadronic final states using $pp$ collision data at
  $\sqrt{s} = 8$ TeV with the ATLAS detector}.
\newblock {\em JHEP}, 03:088, 2015.

\bibitem{Khachatryan:2015qxa}
Vardan Khachatryan et~al.
\newblock {Search for a charged Higgs boson in pp collisions at $ \sqrt{s}=8 $
  TeV}.
\newblock {\em JHEP}, 11:018, 2015.

\bibitem{CMS-PAS-HIG-16-031}
{Search for charged Higgs bosons with the $\mathrm{H}^{\scriptscriptstyle
  \pm}\rightarrow \tau^{\scriptscriptstyle \pm}\nu_{\tau}$ decay channel in the
  fully hadronic final state at $\sqrt{s} = 13~\mathrm{TeV}$}.
\newblock Technical Report CMS-PAS-HIG-16-031, CERN, Geneva, 2016.

\bibitem{Aaboud:2016dig}
Morad Aaboud et~al.
\newblock {Search for charged Higgs bosons produced in association with a top
  quark and decaying via $H^{\pm} \rightarrow \tau\nu$ using $pp$ collision
  data recorded at $\sqrt{s} = 13$ TeV by the ATLAS detector}.
\newblock {\em Phys. Lett.}, B759:555--574, 2016.

\bibitem{Aad:2015typ}
Georges Aad et~al.
\newblock {Search for charged Higgs bosons in the $H^{\pm} \rightarrow tb$
  decay channel in $pp$ collisions at $\sqrt{s}=8 $ TeV using the ATLAS
  detector}.
\newblock {\em JHEP}, 03:127, 2016.

\bibitem{ATLAS-CONF-2016-089}
{Search for charged Higgs bosons in the $H^{\pm}\to tb$ decay channel in $pp$
  collisions at $\sqrt{s}=13$ TeV using the ATLAS detector}.
\newblock Technical Report ATLAS-CONF-2016-089, CERN, Geneva, Aug 2016.

\bibitem{cHModel}
{\url{https://cp3.irmp.ucl.ac.be/projects/madgraph/wiki/chargedHiggs#no1}}.

\bibitem{Degrande:2015vpa}
Celine Degrande, Maria Ubiali, Marius Wiesemann, and Marco Zaro.
\newblock {Heavy charged Higgs boson production at the LHC}.
\newblock {\em JHEP}, 10:145, 2015.

\bibitem{Bednyakov:2016onn}
A.~V. Bednyakov, B.~A. Kniehl, A.~F. Pikelner, and O.~L. Veretin.
\newblock {On the $b$-quark running mass in QCD and the SM}.
\newblock {\em Nucl. Phys.}, B916:463--483, 2017.

\bibitem{Fukuda:2020uva}
Hajime Fukuda, Natsumi Nagata, Hideyuki Oide, Hidetoshi Otono, and Satoshi
  Shirai.
\newblock {Cornering Higgsinos Using Soft Displaced Tracks}.
\newblock {\em Phys. Rev. Lett.}, 124(10):101801, 2020.

\bibitem{Fukuda:2017jmk}
Hajime Fukuda, Natsumi Nagata, Hidetoshi Otono, and Satoshi Shirai.
\newblock {Higgsino Dark Matter or Not: Role of Disappearing Track Searches at
  the LHC and Future Colliders}.
\newblock {\em Phys. Lett.}, B781:306--311, 2018.

\bibitem{PhysRevD.87.015028}
Biplob Bhattacherjee, Brian Feldstein, Masahiro Ibe, Shigeki Matsumoto, and
  Tsutomu~T. Yanagida.
\newblock Pure gravity mediation of supersymmetry breaking at the large hadron
  collider.
\newblock {\em Phys. Rev. D}, 87:015028, Jan 2013.

\bibitem{Banerjee:2019ktv}
Shankha Banerjee, Biplob Bhattacherjee, Andreas Goudelis, Björn Herrmann,
  Dipan Sengupta, and Rhitaja Sengupta.
\newblock {Determining the lifetime of long-lived particles at the LHC}.
\newblock 2019.

\end{thebibliography}

\end{document}